\documentclass[rmp,aps,nofootinbib,twocolumn,amsmath,10pt]{revtex4-1}
\usepackage{graphicx}
\usepackage{color}
\usepackage{siunitx}
\usepackage[usenames,dvipsnames]{xcolor}

\usepackage{longtable}
\usepackage{calc}
\newlength{\tenchars}
\newlength{\restofcolumn}

\newcommand{\be}{\begin{equation}}
\newcommand{\ee}{\end{equation}}
\newcommand \bs{\begin{subequations}}
\newcommand \es{\end{subequations}}
\newcommand \bea{\begin{eqnarray}}
\newcommand \eea{\end{eqnarray}}
\newcommand \nn{\nonumber}

\usepackage[colorlinks=true,urlcolor=blue,linkcolor=blue,citecolor=blue]{hyperref}

\begin{document}

\title{Interfacing single photons and single quantum dots with photonic nanostructures}
\author{Peter~Lodahl}\email{lodahl@nbi.ku.dk} \homepage{www.quantum-photonics.dk}
\author{Sahand~Mahmoodian}
\author{S{\o}ren~Stobbe}
\affiliation{Niels Bohr Institute, University of Copenhagen, Blegdamsvej 17, DK-2100 Copenhagen, Denmark}

\date{\today}



\begin{abstract}
Photonic nanostructures provide means of tailoring the interaction between light and matter and the past decade has witnessed a tremendous experimental and theoretical progress in this subject. In particular, the combination with semiconductor quantum dots has proven successful. This manuscript reviews quantum optics with excitons in single quantum dots embedded in photonic nanostructures. The ability to engineer the light-matter interaction strength in integrated photonic nanostructures enables a range of fundamental quantum-electrodynamics experiments on, e.g., spontaneous-emission control, modified Lamb shifts, and enhanced dipole-dipole interaction. Furthermore, highly efficient single-photon sources and giant photon nonlinearities may be implemented with immediate applications for photonic quantum-information processing. The review summarizes the general theoretical framework of photon emission including the role of dephasing processes, and applies it to photonic nanostructures of current interest, such as photonic-crystal cavities and waveguides, dielectric nanowires, and plasmonic waveguides. The introduced concepts are generally applicable in quantum nanophotonics and apply to a large extent also to other quantum emitters, such as molecules, nitrogen vacancy ceters, or atoms. Finally, the progress and future prospects of applications in quantum-information processing are considered.

\end{abstract}
\maketitle
\tableofcontents
\section{Introduction}
Quantum electrodynamics (QED) studies the interaction between light and matter at the most fundamental level where single quanta of light (photons) and single entities of matter (quantum emitters) are controllably coupled. Founded by Paul Dirac in the 1920s, the theory of QED encompasses intriguing quantum phenomena such as quantum superposition states and entanglement, and has proven to be remarkably precise. The progress on QED experiments with atoms and photons have spun out of research on atomic spectroscopy. A remarkable experimental frontrunner was the demonstration of the Lamb shift \cite{Lamb1947PR} proving that the anticipated degeneracy of the $2S_{1/2}$ and $2P_{1/2}$ states of atomic hydrogen was lifted. The Lamb shift can be interpreted as being due to the interaction of the emitter with the quantum vacuum and its experimental demonstration stimulated further developments of QED, notably leading to renormalization theory. The Casimir effect constitutes another landmark in the history of QED; it describes how two mirrors experience an attractive force due to the radiation pressure of vacuum \cite{Casimir1948}. The development of tools for experimenting with single photons and single atoms started in the 1970s following the invention of the laser. The first experimental demonstration that an excited atom emits a single photon at a time was reported by \citet{Kimble1977PRL}, which marked the birth of experimental quantum optics. Since then a range of exciting experiments on fundamental aspects of light-matter interaction at the single-photon level have appeared using atoms where, e.g., cavities can be exploited for increasing the interaction strength. For a recent review of atom-based cavity QED including a historical account of the field, see \citet{Haroche2013RMP}.

In parallel with the development of atomic QED, major research efforts have been focused on solid-state alternatives. Solid-state systems have the obvious experimental asset that the elaborate experimental techniques needed for trapping and cooling single atoms are not required. Both the emitter and the optical environment can be engineered to enhance the photon-emitter coupling, and consequently two different research disciplines have merged into solid-state QED. The first is material-science research, which has developed methods to synthesize solid-state single-photon emitters with excellent optical properties. In particular, the discovery of photoluminescence from single self-assembled quantum dots \cite{Marzin1994PRL} embedded in GaAs where atomically smooth heterostructures can be grown with molecular beam epitaxy was the first in series of major breakthroughs. Since then growth methods have developed tremendously and today quantum dots can be tailored to have excellent optical properties. The second is nanophotonics research where the optical environment of the emitter is engineered by nanofabrication methods, which builds on the original insight of Purcell that radiative processes are not immutable properties of the emitter, but can be controlled by the environment \cite{Purcell1946PR}. Pioneering solid-state QED experiments were carried out by \citet{Drexhage1970JL}, who showed that the radiative lifetime of europium ions is influenced by the presence of a nearby dielectric interface. This experiment constitutes the first example that the light-matter coupling efficiency can be tailored by structuring the environment of the emitter, which is the essence of modern research in solid-state QED. Today experimental techniques have matured very significantly, and experiments are routinely performed on single solid-state quantum emitters in highly complex photonic nanostructures; the present manuscript reviews this progress. We are here concerned with quantum emitters at optical frequencies and the majority of the considered examples pertain to quantum dots. Nonetheless most of the concepts are of much broader scope and apply equally well to other solid-state emitters and atoms, or to two-level emitters implemented in superconducting circuits.

The research field of solid-state QED has widespread and far-ranging implications and applications. For instance, QED systems are widely proposed for quantum-information processing and optimized photon-matter interfaces are at the heart of photonic devices such as lasers and solar cells. An outstanding challenge in quantum physics today is to construct scalable quantum networks that exploit quantum parallelism for encoding and processing information. To this end, a variety of different physical systems exploiting either photon or matter degrees of freedom have been considered \cite{ChuangNielsen}, and each system has pros and cons: photons are robust carriers of quantum information over long distances but tend to interact weakly, which makes quantum-computing protocols experimentally demanding \cite{Kok2007RMP}. In contrast, e.g., electrons confined in quantum dots may interact very strongly but this also makes the system vulnerable to decoherence processes from the environment and the interaction has limited spatial range \cite{Hanson2007RMP}. An efficient quantum interface between light and matter implemented in a scalable quantum architecture is therefore expected to have wide applications in quantum-information processing \cite{Kimble2008QuantumInternet} since it allows encoding quantum information in both light and matter variables, thus potentially benefiting from the advantages of each system.

The main challenge in order to realize efficient light-matter interfaces is that the interaction between a single photon and a single emitter tends to be very weak. A natural photon-emitter process is that of spontaneous emission: a single two-level emitter with an electron prepared in the excited state undergoes a transition to the ground state by emitting a single photon. Spontaneous emission constitutes an example of a general class of quantum-mechanics problems where a simple quantum system is coupled to a large reservoir since the emitter comprises only two states while the photon can be emitted to any of a continuum of optical states each characterized by different wave vectors. Consequently, the challenge for efficient photon-emitter interfacing is to strongly enhance the coupling to one preferred mode and/or suppress the coupling to all unwanted modes. High-finesse optical cavities have been a popular approach for accomplishing this and very significant experimental progress has been obtained within the last few decades with single atoms \cite{Haroche2013RMP}. In recent years, solid-state alternatives have emerged with the benefit that the systems can be engineered by modern nanofabrication techniques. This leads to a whole range of new design strategies based on dielectric and metallic nanostructures including waveguides and nanocavities. Two different classes of solid-state single-photon emitters have been considered, using either superconducting circuits emitting in the microwave regime \cite{Wallraff2004Nature} or two-level quantum emitters with optical transitions, e.g., semiconductor quantum dots \cite{Michler2000Science}, nitrogen vacancy centers in diamond \cite{Kurtsiefer2000PRL}, or single molecules \cite{Lounis2000Nature}.  A benefit of optical methods is that many functionalities can be highly integrated on a photonic chip \cite{OBrien2009NPHOT} since the wavelength of the electromagnetic radiation determines the typical length scale of the building blocks. As a consequence, highly integrated and engineered photonic circuits could potentially be constructed, which is a quest in the burgeoning research field of quantum nanophotonics.

A variety of different optical emitters have been studied in nanophotonic experiments including laser-cooled and trapped atoms, rare-earth ions, single molecules, impurity centers (e.g., nitrogen vacancy centers), colloidal quantum dots, and self-assembled quantum dots. In the present Review, most of the discussed examples concern epitaxially grown III-V semiconductor quantum dots (primarily InGaAs or GaAs) that typically emit in the near infrared. These emitters constitute many-particle mesoscopic systems that contain rich and exciting quantum physics. The present knowledge of and control over InGaAs quantum dots, regarding level structure, emission dynamics, stability, and coherence, is now sufficiently mature that a serious exploration of their potential in quantum optics and quantum-information processing seems viable. It should be stressed that such experiments are carried out at cryogenic conditions, typically at a temperature of \SI{4}{\kelvin}, which is necessary to prevent thermal depopulation and achieve sufficiently good coherence properties as required in most quantum-optics experiments. In contrast, room-temperature experiments with solid-state emitters typically suffer from severe decoherence due to Coulomb and phonon-induced scattering that makes coherent quantum-optics experiments infeasible. The ability to tailor light-matter interfaces at room temperature may nonetheless find a range of other important applications, e.g., in the context of light harvesting or efficient light sources. We emphasize that while the present review in focused on the quantum aspects of light-matter interaction, many of the discussed concepts are generally applicable and could equally well be exploited in room-temperature applications.
\section{Semiconductor quantum dots\label{sec:sst:QDs}}
\begin{figure}
\begin{center}
\includegraphics[width=\columnwidth]{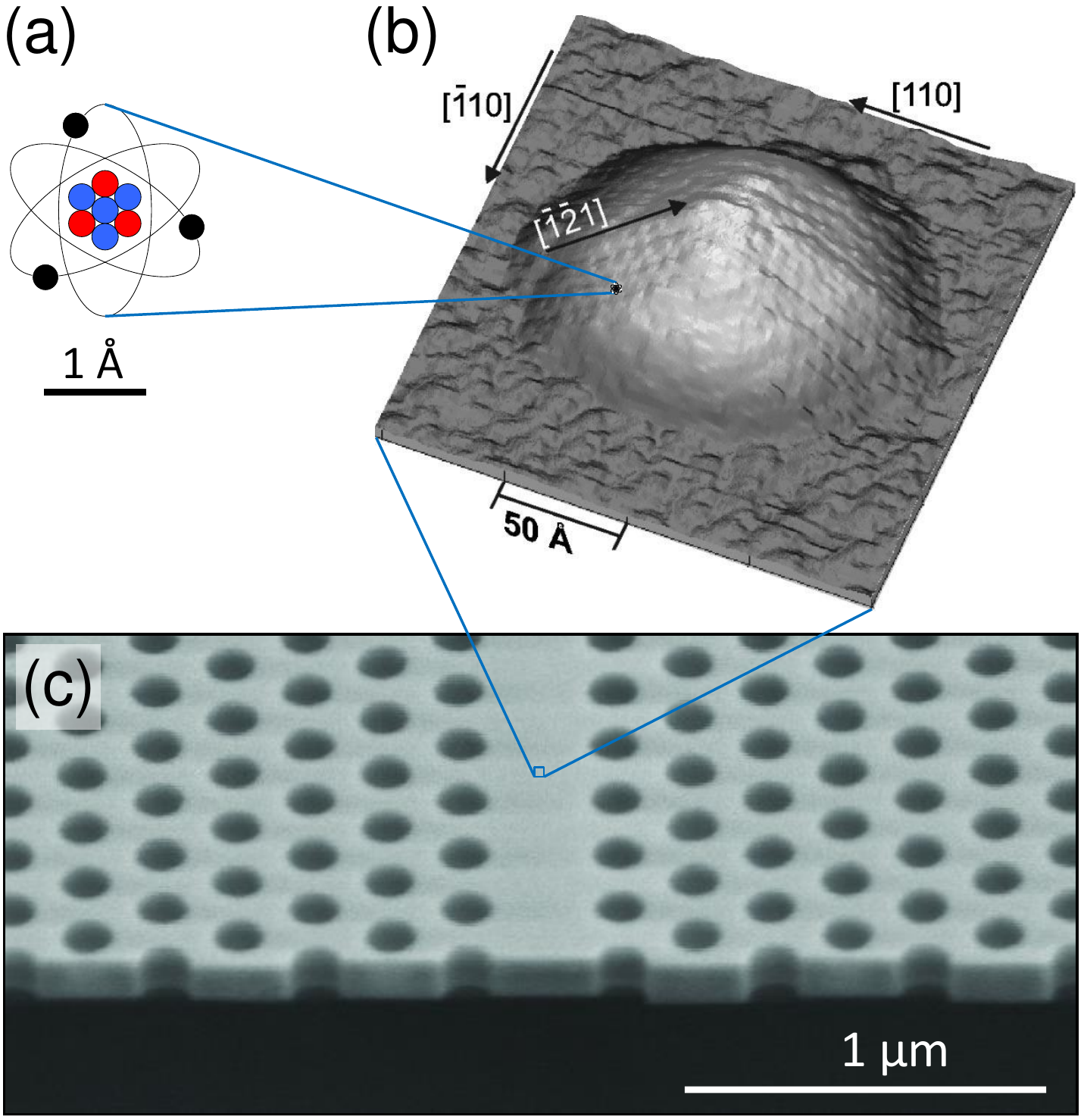}
\caption{Characteristic size of a quantum dot relative to a single atom and a photonic crystal. A single atom (a) measures a few {\AA}ngstr\"{o}m while self-assembled InGaAs quantum dots (b) typically have dimensions of tens of nanometer and consist of approximately $10^5$ atoms. The micrograph in (b) shows an uncapped quantum dot obtained by scanning tunneling microcopy. Single quantum dots can be embedded in photonic nanostructures for quantum-optics experiments, an example of which is (c) that shows a scanning electron micrograph image of a photonic-crystal waveguide, where the photonic lattice constant typically is around \SI{250}{\nano\meter}. (b) Reprinted with permission from \citet{Marquez2001APL}.
\label{fig:sst:Fig1}
}
\end{center}
\end{figure}

The existence of a discrete and anharmonic electronic spectrum is the prerequisite for many quantum-optics experiments since it enables generating single photons when an electron undergoes a transition between two levels. An obvious choice is a single atom, which represents a clean quantum system with discrete electronic states. The ability to create discrete electronic states in a solid-state system enables a range of new opportunities for integrated quantum-optics experiments, and this can be achieved in a quantum dot. A quantum dot is a semiconductor ``artifical atom'' that although consisting of tens of thousands of atoms has optical properties similar to single atoms due to the quantum confinement of electrons to a nanometer length scale. Since quantum dots are solid-state emitters they can readily be implemented in photonic nanostructures such as nanowires, plasmonic nanoantennas, and photonic crystals, as discussed in the following sections. Figure \ref{fig:sst:Fig1} indicates the size of a quantum dot relative to a single atom and a typical photonic nanostructure. In the present section we briefly review the structural and optical properties of quantum dots. Unless explicitly indicated, we are solely considering quantum dots in the family of InAs/GaAs/AlAs III-V semiconductors, which are grown by epitaxial methods.

\subsection{Growth and structural properties\label{sec:sst:growth}}

\begin{figure*}
\begin{center}
\includegraphics[width=\textwidth]{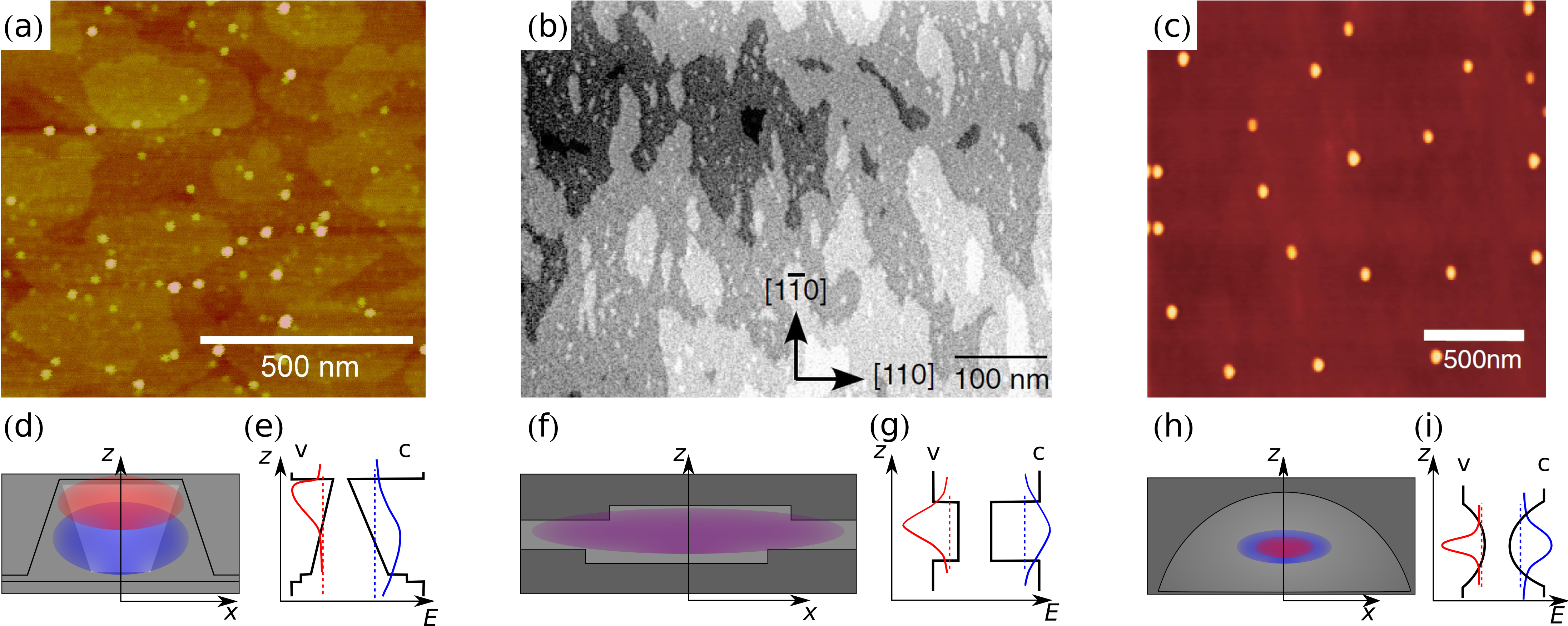}
\caption{Structural properties of quantum dots. (a)-(c) Atomic-force micrographs (AFMs) of uncapped quantum dots for (a) Stranski-Krastanov quantum dots of InAs, (b) interface fluctuations of GaAs, and (c) droplet epitaxy quantum dots of GaAs displaying the surface topography with bright (dark) colors indicating high (low) features. (d), (f), and (h) illustrate the confinement potentials, where dark, neutral, and bright gray indicate AlGaAs, GaAs, and InAs, respectively for the three types of quantum dots shown  above. The electron (hole) wave function is shown in blue (red). For interface-fluctuation quantum dots (f) the in-plane motion of electrons and holes may become correlated (purple exciton wave function) and could lead to a giant oscillator strength. (e), (g), and (i) illustrate the wave functions along the growth axis $z$ for the respective type of quantum dots. For Stranski-Krastanov quantum dots, an asymmetric confinement potential leads to a significant offset between the electron and hole. (b) Reprinted with permission from \citet{Peter2005PRL}. (c) Reprinted with permission from \citet{Mano2009Nanotechnology}.
\label{fig:sst:structural_properties_of_QDs}}
\end{center}
\end{figure*}

One of the important experimental advantages of qu\-an\-tum dots is that they are made from semiconductor materials for which a wealth of growth and processing technology has been developed over the past decades. Sophisticated crystal-growth procedures combined with semiconductor processing methods such as electron-beam lithography, etching, and deposition constitute the generic nanofabrication platform on which the significant experimental progress within quantum nanophotonics during the past decades is built. In this section we discuss the most common methods for growing quantum dots and the impact of the growth method on their optical properties.

Since semiconductors are very sensitive to impurities and defects, quantum dots are fabricated by epitaxial methods such as molecular-beam epitaxy where heterostructures are grown with monolayer precision under ultra-high-vacuum conditions \cite{Shchukin1999RMP,Stangl2004RMP,Biasiol2011PhysRep}. The most common approach for InGaAs quantum dots is the Stranski-Krastanov method that relies on the self-assembly of InAs or InGaAs quantum dots on a GaAs surface due to the 7\% larger lattice constant of InAs compared to that of GaAs. As a consequence, only a thin wetting layer of InAs can be deposited on GaAs before the strain is relaxed by the nucleation of quantum dots in the form of randomly positioned islands as shown in Fig.\ \ref{fig:sst:structural_properties_of_QDs}(a). In order to protect the quantum dots from oxidation and to prevent interaction with surface states, a GaAs capping layer is grown atop the quantum dots. While Stranski-Krastanov quantum dots have a pyramidal shape before capping, cf.\ Fig.\ \ref{fig:sst:Fig1}(b), they develop the shape of a truncated pyramid after capping \cite{Eisele2008JAP} due to a significant material intermixing as shown in Fig.\ \ref{fig:sst:structural_properties_of_QDs}(d). This in turn leads to an inhomogeneous indium distribution and a strain that varies throughout the quantum dot. Typically, quantum dots are grown with heights in the range of 1-\SI{10}{\nano\meter} and in-plane sizes in the range of 10-\SI{70}{\nano\meter}. Controlling the size and therefore the quantum confinement as well as the material composition enable tailoring the emission wavelength. Size variations between different quantum dots within a single growth run are inevitable, i.e., a quantum¨-dot ensemble will be inhomogeneously broadened, implying that individual tuning of single quantum dots would generally be required in order to couple them mutually. It has been found that the indium is concentrated in an inverted pyramid inside the quantum dot \cite{Liu2000PRL}, which leads to variations in the confinement potential along the growth axis as indicated in Fig.\ \ref{fig:sst:structural_properties_of_QDs}(e) where also schematic electron and hole envelope wave functions are shown. Notably, the hole resides above the electron and this leads to a significant static electric dipole along the growth axis \cite{Fry2000PRL}.

Quantum dots may also form by growing a thin quantum well but terminating the growth while the lower and upper monolayers of the quantum well are formed \cite{Gammon1996Science} leading to monolayer fluctuations as shown in Fig.\ \ref{fig:sst:structural_properties_of_QDs}(b) and (f) with a confinement potential as indicated in Fig.\ \ref{fig:sst:structural_properties_of_QDs}(g). This allows growing practically unstrained GaAs quantum dots embedded in AlGaAs and these interface-fluctuation quantum dots can be larger in the plane perpendicular to the growth direction than typical Stranski-Krastanov quantum dots. Interface-fluctuation quantum dots benefit from a relatively narrow inhomogeneous broadening and predictable wave functions along the growth direction as compared to Stranski-Krastanov quantum dots since intermixing is absent. However, the integration in optical nanostructures is more challenging since they are embedded in AlGaAs, whose surface is prone to oxidation. They are particularly promising for obtaining quantum dots with an enhanced oscillator strength as discussed in Sec.\ \ref{sec:sst:OS}.

Droplet epitaxy is an emerging growth technique, where droplets of gallium are saturated with arsenic, resulting in relatively large and low-density GaAs quantum dots in AlGaAs as shown in Fig.\ \ref{fig:sst:structural_properties_of_QDs}(c). Since droplet epitaxy quantum dots are embedded in AlGaAs surroundings, the above mentioned challenges pertain to their integration in photonic nanostructures as well. Since the AlGaAs capping layer is often grown at low temperatures, a high-temperature post-growth annealing is required to make them optically active and the demonstrated quantum efficiency cannot yet compete with Stranski-Krastanov quantum dots \cite{Tighineanu2013PRB} although narrow linewidths have been reported \cite{Mano2009Nanotechnology,Sallen2011PRL}. The post-growth annealing results in intermixing and the resulting confinement is therefore smaller than the apparent size as shown in Fig.\ \ref{fig:sst:structural_properties_of_QDs}(h) and (i).

In quantum-optics experiments it is often required to isolate a single quantum dot, which may be achieved by various methods, such as etching away the material surrounding a single quantum dot or evaporating an opaque metallic aperture masking other quantum dots. Such methods may modify the optical properties in adverse ways and are therefore not generally compatible with nanophotonic devices, such as waveguides or cavities. The most versatile experimental approach at present is to use samples with a low quantum-dot density combined with confocal microscopy. A main drawback of the standard growth methods is that the lateral positions of the quantum dots are not controlled and it has proven a successful strategy to deterministically align nanophotonic structures to a single quantum dot that is first located by microscopy techniques \cite{Badolato2005Science,Hennessy2007Nature,Dousse2008PRL,Thon2009APL,Kojima2013APL}.

It would generally be ideal to grow periodic arrays of quantum dots for deterministic and scalable integration with photonic nanostructures and many approaches are presently studied. This is currently an active field of crystal-growth research \cite{Kiravittaya2009RepProgPhys} and relies on growing quantum dots on patterned substrates obtained, e.g., by electron-beam lithography, and subsequently etching small recesses creating nucleation sites for quantum dots. This processing leads to challenges because impurities and defects are introduced and growing high-quality material immediately atop the regrowth interface is difficult. Therefore a common approach involves growing a seed layer of quantum dots that is optically inactive due to defects at the regrowth interface but introduces a suitable amount of strain so that another quantum-dot layer can be grown atop. The optical quality of positioned quantum dots tends to be lower than that of Stranski-Krastanov quantum dots giving broader linewidths and lower quantum efficiency \cite{Albert2010APL} and patterned regrowth of quantum dots is not yet widely applied in quantum-nanophotonics experiments. Recent promising progress includes the demonstration of emission of indistinguishable \cite{Joens2013NL} and polarization-entangled \cite{Juska2013NPHOT} photons although the yield, i.e., the fraction of sites with an optically active high-quality quantum dot is so far below unity. It is nonetheless anticipated that future progress eventually will render positioned quantum dots indispensable for more controlled experiments \cite{Gallo2008APL} and for scaling up quantum architectures. It should also be mentioned that even if the inherent optical properties of positioned quantum dots do not reach the quality of Stranski-Krastanov material, the Purcell effect, see Sec.\ \ref{Section-spontaneous-emission-control}, or resonant excitation, see Sec.\ \ref{Section-coherent-single-photons}, can be employed to enhance the effective quantum efficiency or reduce the effects of linewidth broadening, respectively.

The material composition and size determine the energy range of the emission spectrum of quantum dots. In particular, the lowest-energy transition of InAs quantum dots typically falls in the range of 850 to \SI{1000}{\nano\meter} while GaAs quantum dots typically emit at around 670 to \SI{760}{\nano\meter}. The choice of material affects also photonic properties, in particular through the real part of the index of refraction. Also bulk and surface absorption are strongly material dependent and play an important role for the performance of nanophotonic devices \cite{Michael2007APL}. In general, longer wavelengths imply that fabrication of nanostructures is easier and that the relative importance of fabrication imperfections is reduced. Furthermore, the material absorption is smaller at longer wavelengths, which therefore allow for, e.g., higher $Q$-factors of optical nanocavities. On the other hand, silicon-based photodetectors such as avalanche photodiodes are much more efficient at shorter wavelengths and therefore lead to higher photon-count rates, although this obstacle may be overcome by the use of superconducting detectors. A tradeoff between these factors determine the most ideal wavelength for a given experimental setting.

\subsection{Excitons in quantum dots\label{sec:sst:excitons}}
The fundamental optical excitation in a quantum dot consists of an electron in the conduction band and a hole in the valence band. In bulk semiconductors and quantum wells there is an important distinction between an uncorrelated electron-hole pair and an exciton, which is an electron-hole pair bound by direct and exchange Coulomb interaction. In quantum dots this distinction is often not required. In small quantum dots, the motion of electrons and holes is dominated by quantum confinement, which implies that they are mutually independent, but exciton effects are required to explain the fine structure. The optically active states in quantum dots are therefore always excitonic. A further discussion of the effects of Coulomb interaction and confinement is given in Sec.\ \ref{sec:sst:OS}.

The unfilled orbital shells of atomic Al, Ga, In, and As are $3\text{s}^2 3\text{p}$, $4\text{s}^2 4\text{p}$, $5\text{s}^2 5\text{p}$, and $4\text{s}^2 4\text{p}^3$, respectively, so predominantly covalent bonds are formed in GaAs and in the other relevant binary and ternary alloys. This leads to tetrahedral bonds and a zincblende crystal structure and in the absence of spin-orbit effects there would be three degenerate valence bands. Including spin orbit, the split-off valence band is shifted to lower energy but the light- and heavy-hole bands remain degenerate \cite{YuCardonaBook}. The heterojunctions of InGaAs in GaAs and GaAs in AlGaAs have type-I energy-band alignment, which ensures carrier confinement at both the conduction- and the valence-band edge. This is essential for the efficient interaction with light and allows for quantized states for both electrons and holes. The aspect ratio of quantum dots is larger than unity and the dominant quantization axis is the growth direction. This lifts the heavy-/light-hole degeneracy so that the transitions from the conduction band to the heavy-hole band have the lowest energy. Strain plays a major role in InGaAs quantum dots and lifts the degeneracy further. While neglecting the light-hole band is often a good approximation, a substantial band mixing can occur in quantum dots with a pronounced structural asymmetry \cite{Belhadj2010APL}.  Sophisticated numerical models have been developed in order to encompass these effects, such as the $\mathbf{k}\cdot\mathbf{p}$ method \cite{Stier1999PRB}, which is a continuum theory and the empirical pseudo-potential theory \cite{Bester2009JPhysC}, which is an atomistic approach. Despite these significant advances, the comparison to experiments is limited by the lack of knowledge about the exact atomic configuration of quantum dots. Such information could be extracted using, e.g., high-resolution transmission electron microscopy or scanning-tunneling microscopy but these techniques are time consuming and often destructive and therefore challenging to combine with optical spectroscopy. Fortunately, under the conditions relevant for quantum-optics experiments, i.e., quantum dots with a large aspect ratio at low temperatures and small carrier populations, many features of quantum dots can be described remarkably well by using a simple two-band effective-mass model where only the heavy-hole valence band and the conduction band are included. We restrict the discussion in this review to such a model. In the effective-mass approximation, the electronic energy bands are taken into account by the band-edge effective mass of the carriers. We are mainly concerned with quantum dots of sizes where the energy-level spacing is large compared to the Coulomb energy. In this case, Coulomb effects may be included perturbatively and the motion of the carriers in the conduction and valence band may be considered independent. This is known as the strong-confinement regime and throughout the review we restrict the discussion to this regime unless explicitly noted. It should be noted that in quantum dots the energy-level spacing may actually be comparable to the Coulomb energy leading to a complex interplay between confinement and Coulomb interaction, which is denoted the intermediate-confinement regime. However, the strong-confinement model has so far been able to describe experiments successfully. As already mentioned and discussed further below, even in the strong-confinement regime, Coulomb exchange determines the fine structure.

\subsection{The transition matrix element\label{sec:sst:TransitionMatrixElement}}

The key property of a quantum emitter determining the strength of its interaction with light is the transition matrix element between the ground and excited states. This may also be expressed as a transition dipole moment or an oscillator strength. For atoms this is an immutable intrinsic property but for quantum dots it may be controlled by modifying the exciton wave function and changing the selection rules of optical transitions. Here we discuss the physics underlying the transition matrix element of quantum dots.

The quantum state of an electron in the conduction band, $\text{c}$, or heavy-hole valence band, $\text{v}$, consists of three parts,
\begin{align}
|\Psi_\text{c/v}\rangle &= |F_\text{c/v}\rangle |u_\text{c/v}\rangle |\alpha_\text{c/v}\rangle,\label{eq:sst:SingleParticleWF}
\end{align}
where $|F_\text{c/v}\rangle$, $|u_\text{c/v}\rangle$, and $|\alpha_\text{c/v}\rangle$ is the envelope wave function, the electronic Bloch function evaluated at the $\Gamma$-point of the bandstructure, and the spin state, respectively. The envelope wave function is obtained from the effective-mass Schr\"odinger equation $-\frac{\hbar^2}{2m_0} \nabla \cdot \left( \frac{1}{m^\ast(\mathbf{r})} \nabla F_\text{c/v}(\mathbf{r})\right) + V_\text{c/v}(\mathbf{r})F_\text{c/v}(\mathbf{r}) = (E-E_\text{c/v})F_\text{c/v}(\mathbf{r})$, where $m_0$ is the electron rest mass and $m^\ast(\mathbf{r})$ is the anisotropic effective mass, which in general is a tensor but here taken to be scalar for simplicity. $V_\text{c/v}(\mathbf{r})$ is the confinement potential, $E$ is the electron eigenenergy, and $E_\text{c/v}$ is the band-edge energy.

Optical transitions are induced by the minimal-coup\-ling Hamiltonian, which in the generalized Coulomb gauge and the dipole approximation can be written as $H(\mathbf{r}_0,t)=-\frac{q}{m_0}\mathbf{p}\cdot\mathbf{A}(\mathbf{r}_0,t)$, where $q$ is the elementary charge, $\mathbf{p}=-i\hbar\nabla$ is the momentum operator, and $\mathbf{A}(\mathbf{r}_0,t)$ is the vector potential of the electromagnetic field that is evaluated at the position of the quantum dot, $\mathbf{r}_0$. The relevant quantity for spontaneous emission is therefore the momentum matrix element $\mathbf{P} = \langle\Psi_\text{v}|\mathbf{p}|\Psi_\text{c}\rangle$ that follows from Eq.\ (\ref{eq:sst:SingleParticleWF}),
\begin{align}
\mathbf{P} = \langle F_\text{v}|F_\text{c}\rangle \langle u_\text{v}|\mathbf{p}|u_\text{c}\rangle_\text{uc} \langle\alpha_\text{v}|\alpha_\text{c}\rangle,\label{eq:sst:DipoleMatrixElement}
\end{align}
leading to three selection rules for optical transitions: i) the envelope wave functions must have the same parity, ii) the Bloch functions must have opposite parity, and iii) the electron spin must remain unchanged. The Bloch matrix element is $\langle u_\text{v}|\mathbf{p}|u_\text{c}\rangle_\text{uc} = \frac{1}{V_\text{uc}} \int_\text{uc} \text{d}^3\mathbf{r} u_\mathrm{v}^\ast(\mathbf{r}) \mathbf{p} u_\text{c}(\mathbf{r})$, where $V_\text{uc}$ is the volume of a unit cell, and depends only on bulk material properties as is quantified by the Kane energy to be introduced below. Formally the selection rules can be relaxed by band mixing, but experimentally transitions obeying them have been observed to dominate \cite{Johansen2008PRB}. The first selection rule is only approximately valid since parity often is not a good quantum number for quantum-dot envelope functions.

The electronic states are commonly described in the equivalent electron-hole picture in which the valence-band states are transformed to the hole picture according to
$|F_\text{h}\rangle = |F_\text{v}\rangle^\ast$, $|u_\text{h}\rangle = |u_\text{v}\rangle^\ast$, and $|\alpha_\text{h}\rangle = |\alpha_\text{v}\rangle^\ast$.
In this picture the decay of an electron from the conduction band to the valence band is viewed as the recombination of an electron and a hole and we define the electron and hole pseudo-spin states, which describe the total angular momentum of the Bloch functions and the spin, i.e., $\left|\uparrow\right> = |u_\text{c}\rangle \left|\uparrow_\text{e} \right> $, $\left|\downarrow\right> = |u_\text{c}\rangle\left|\downarrow_\text{e}\right>$, $\left|\Uparrow\right> = |u_\text{h}\rangle \left| \uparrow_\text{h} \right> $, and $\left|\Downarrow\right> = |u_\text{h}\rangle\left|\downarrow_\text{h}\right>$. Here the arrows with subscripts denote the projected spin so that, e.g., $|\uparrow_\text{e}\rangle$ is an electron with $S_z = + 1/2$, where $z$ is along the growth direction. A quadruplet of exciton states are formed enabling dipole-allowed transitions for the states $\left|\Uparrow \downarrow\right>$ and $\left|\Downarrow \uparrow\right>$, and dipole-forbidden transitions for $\left|\Uparrow \uparrow\right>$ and $\left|\Downarrow \downarrow\right>$. It is common practice to suppress the envelope wave function in the notation because they only contribute to the matrix element with a pre-factor, i.e., the wave-function overlap. The evaluation of the corresponding dipole matrix elements is obtained by converting back to the electron picture and using Eq.~(\ref{eq:sst:DipoleMatrixElement}). To summarize the implications of the notation with an example, the matrix element $\mathbf{P} = \langle g | \mathbf{p} \left|\Uparrow \downarrow\right>$ for transitions between the state $\left|\Uparrow \downarrow\right>$ and the state void of excitations, i.e., the quantum-dot ground state, $|g\rangle$, corresponds to $\mathbf{P} = \langle F_\text{v}|F_\text{c}\rangle \langle u_\text{v}|\mathbf{p}|u_\text{c}\rangle_\text{uc}$.

The Bloch functions inherit the symmetry of the atomic orbitals: the conduction-band Bloch functions have $\text{s}$-symmetry and the valence band has $\text{p}$-symmetry. Hence we define
$\left|\uparrow\right> = |u_\text{s}\rangle \left|\uparrow_\text{e}\right>$, $\left|\downarrow\right> = |u_\text{s}\rangle\left|\downarrow_\text{e}\right>$, $\left|\Uparrow\right> = - \frac{1}{\sqrt{2}}(|u_x\rangle + i|u_y\rangle ) \left|\uparrow_\text{h}\right>$, and $\left|\Downarrow\right> = \frac{1}{\sqrt{2}}(|u_x\rangle - i|u_y\rangle ) \left|\downarrow_\text{h}\right>$, where $u_\text{s}$, $u_x$, and $u_y$ denote functions with even parity, odd parity along $x$, and odd parity along $y$, respectively. With these definitions we obtain, e.g., for $\left|\Uparrow \downarrow\right>$ that $\mathbf{P} = - \langle F_\text{v}|F_\text{c}\rangle \langle u_x|p_x|u_\text{s}\rangle_\text{uc} \frac{1}{\sqrt{2}}\left(\hat{\mathbf{e}}_x + i\hat{\mathbf{e}}_y\right)$, where $\hat{\mathbf{e}}_x$ ($\hat{\mathbf{e}}_y$) is a unit vector along $x$ ($y$). Evidently, the polarization of the dipole moment is given by the term $\frac{1}{\sqrt{2}}\left(\hat{\mathbf{e}}_x + i\hat{\mathbf{e}}_y\right)$, which describes circular polarization. The magnitude of the dipole moment depends on the wave-function overlap $\langle F_\text{v}|F_\text{c}\rangle$, which can be controlled by, e.g., external static electric fields, and the Bloch matrix element $\langle u_x|p_x|u_\text{s}\rangle_\text{uc}$. The latter is an entirely bulk semiconductor quantity given by the Kane energy $E_\text{P} = 2 |\langle u_x|\mathbf{p}|u_\text{s}\rangle_\text{uc}|^2/m_0 = 2 |\langle u_y|\mathbf{p}|u_\text{s}\rangle_\text{uc}|^2/m_0$ \cite{Vurgaftman2001JAP}.

Quantum dots are most often grown on (001) substrates where the symmetry leads to in-plane anisotropic confinement potentials even for rotationally symmetric quantum dots \cite{Bester2003PRB} and typically an in-plane structural asymmetry is also present. The preferential elongation is along $[1\bar{1}0]$, which is also apparent from Fig.\ \ref{fig:sst:structural_properties_of_QDs}(b). The exchange interaction splits the four excitonic states into two doublets, which are separated by the dark-bright energy splitting \cite{Bayer2002PRBb}: the dark-exciton doublet, $|X_\text{d}\rangle = \frac{1}{\sqrt{2}}(\left|\Uparrow\uparrow\right> - \left|\Downarrow\downarrow\right>)$ and $|Y_\text{d}\rangle = \frac{1}{\sqrt{2}}(\left|\Uparrow\uparrow\right> + \left|\Downarrow\downarrow\right>)$, and the bright-exciton doublet,
$|X_\text{b}\rangle=\frac{1}{\sqrt{2}}(\left|\Uparrow\downarrow\right> - \left|\Downarrow\uparrow\right>)$ and $|Y_\text{b}\rangle=\frac{1}{\sqrt{2}}(\left|\Uparrow\downarrow\right> + \left|\Downarrow\uparrow\right>)$. Furthermore, the exchange interaction splits the bright doublet due to the reduced in-plane symmetry with a fine-structure splitting $\Delta E_\text{fss}$ typically on the order of $10-\SI{100}{\micro\eV}$ \cite{Seguin2005PRL}. The splitting between the two dark excitons is of the order of $\SI{1}{\micro\eV}$ \cite{Poem2010NPHYS}. The resulting excitonic states are shown in Fig.\ \ref{fig:sst:QD_decay_cascade} along with the radiative transitions as well as the spin-flip and non-radiative transitions that are discussed further below. Using the procedure outlined above, it is straightforward to show that $|X_\text{b}\rangle$ and $|Y_\text{b}\rangle$ have linearly polarized dipole transitions to the ground state with orientations along $x$ and $y$, respectively.

\subsection{Multiexcitonic states\label{sec:sst:multiexcitons}}
\begin{figure*}
\begin{center}
\includegraphics[width=0.9\textwidth]{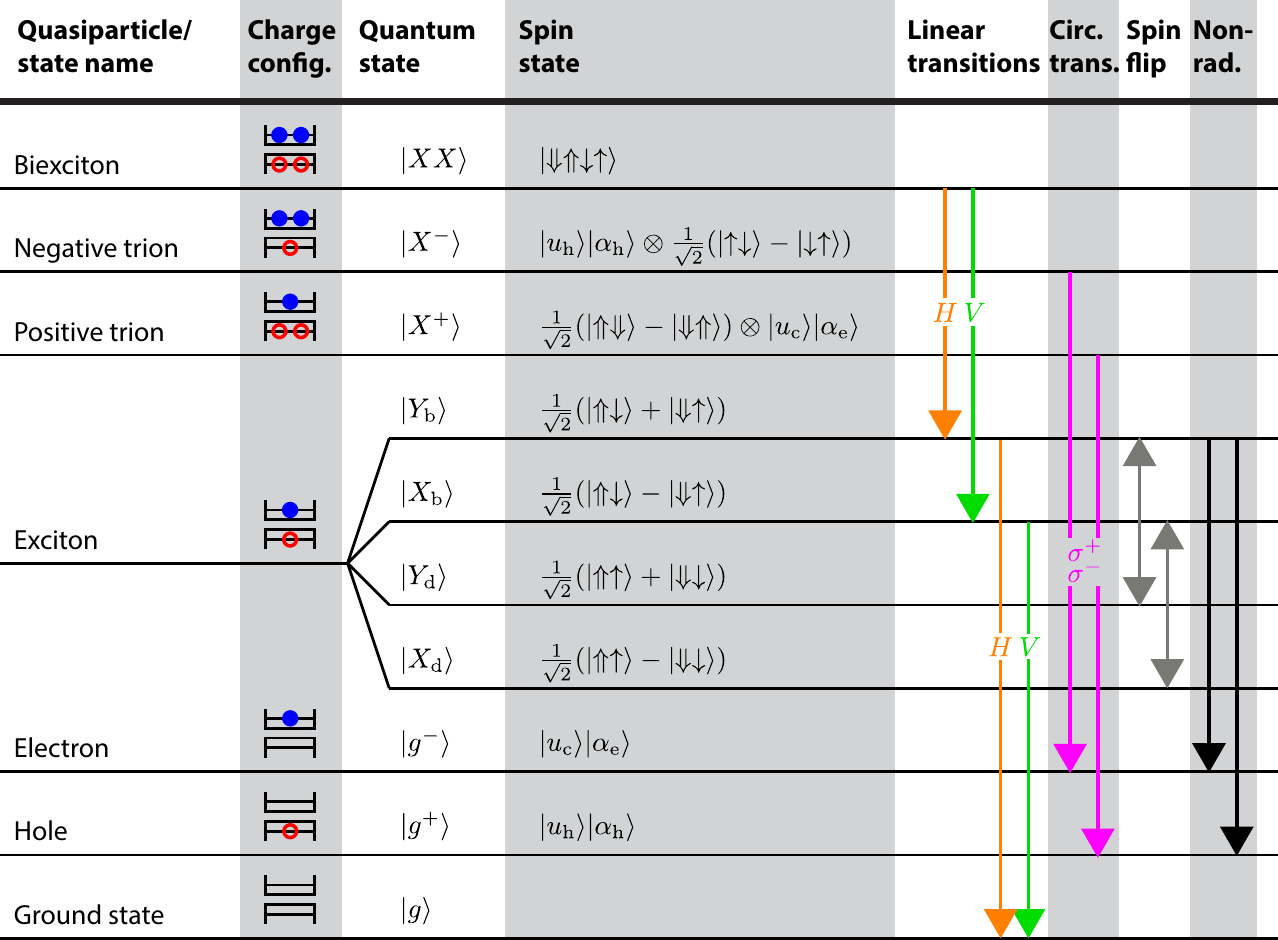}
\caption{The lowest-energy confined states in quantum dots and the transitions between them. The full blue (empty red) circles indicate the electron (hole) configuration in the conduction (valence) band $s$-shells of the quantum dot. The pseudo-spin states are discussed in the text. The biexciton may decay to one of the two bright exciton states by emission of a horizontally (H) or vertically (V) polarized photon, respectively. The negative (positive) trion decays to a single electron (hole) by emission of circularly polarized light with the helicity depending on the additional carrier. Furthermore, spin-flip processes (gray arrows) couple bright and dark excitons. Non-radiative processes (black arrows) are generally present and can for some transitions be dominant; here only the non-radiative decay of the bright excitons are indicated explicitly. Note that the ordering of the states in the figure follows the occupancy, while the emission energies of the radiative excitonic complexes depend not only on the occupancy but also on confinement and correlation effects.}
\label{fig:sst:QD_decay_cascade}
\end{center}
\end{figure*}

Quantum dots can contain multiple electrons and/or holes leading to additional transitions with different optical properties than single neutral excitons. The simplest examples of multiexcitonic states are trions and biexcitons that are useful in quantum-optics experiments requiring multi-level schemes.

Multiple bound states may exist as solutions to the effective-mass equation for both the conduction and valence bands and these eigenstates are commonly denoted $s$-, $p$-, and $d$-shells, etc., in analogy with the convention in atomic physics and not to be confused with the symmetries of the Bloch functions discussed above. We note that the approximate selection rule for the envelope functions imply that only transitions between the same shell in the conduction and the valence band need to be considered. Excitons in higher-lying states are often not relevant for optical emission because they decay to the $s$-shells on a few-picosecond time scale by emission of phonons but depending on the type of excitation in a given experiment they may affect the decay cascade towards the $s$-shells. In the following we consider only the $s$-shells. Here only four types of excitonic quasiparticles exist due to the Pauli exclusion principle: the neutral exciton, the negative trion (two electrons and one hole), the positive trion (one electron and two holes), and the biexciton (two electrons and two holes). The charge configuration, pseudo-spin state, and dipole-allowed transitions of these states are shown in Fig.\ \ref{fig:sst:QD_decay_cascade} and are discussed further below. We note that non-radiative processes have not been studied systematically for trions and biexcitons.

The exchange interaction vanishes for the trions due to Kramer's theorem according to which a system with an odd number of fermions is at least two-fold degenerate if it is governed by a Hamiltonian that is symmetric under time reversal \cite{Messiah}, which is the case for a trion in the absence of external magnetic fields. The state of the two identical carriers must be antisymmetrized because trions are fermions. The negative trion is given by  $|X^-\rangle=\frac{1}{\sqrt{2}}(\left|\uparrow\downarrow\right> - \left|\downarrow\uparrow\right>) |u_\text{h}\rangle|\alpha_\text{h}\rangle$, where $|\alpha_\text{h}\rangle$ is the hole-spin state. For the positive trion, $|X^+\rangle=\frac{1}{\sqrt{2}}(\left|\Uparrow\Downarrow\right> - \left|\Downarrow\Uparrow\right>) |u_\text{c}\rangle |\alpha_\text{e}\rangle$, where $|\alpha_\text{e}\rangle$ is the electron-spin state. The transition matrix elements for the decay of a trion to a single electron or hole are circularly polarized with the helicity depending on the spin of the additional carrier. As opposed to excitons they have no corresponding dark states. The additional carrier in trions may be prepared in a spin eigenstate and trions are therefore particularly relevant for spin physics. Quantum-dot spin physics was recently reviewed by \citet{Warburton2013NMAT} and \citet{Urbaszek2013RMP}.

Biexcitons have the pseudo-spin configuration $|XX\rangle=\left|\Uparrow\Downarrow\uparrow\downarrow\right>$ and they can decay radiatively to either of the bright excitons. Depending on which exciton it decays to, a cascade of either two horizontally polarized or two vertically polarized photons is triggered, i.e., the emitted photonic state is $|H_\text{XX}\rangle|H_\text{X}\rangle$ or $|V_\text{XX}\rangle|V_\text{X}\rangle$, cf.\ Fig.\ \ref{fig:sst:Bright-dark_level_scheme}.(c). Since biexcitons contain two excitons they are observed at higher excitation densities and decay approximately twice as fast as excitons since they have twice the number of radiative decay channels. If the fine-structure splitting is much smaller than the natural linewidth, biexcitons are a source of polarization-entangled photons because the emission cascade leads to the entangled photonic state $\frac{1}{\sqrt{2}}\left(|\sigma^+_\text{XX}\rangle|\sigma^-_\text{X}\rangle+|\sigma^-_\text{XX}\rangle|\sigma^+_\text{XX}\rangle\right)$, where $\sigma^+$ and $\sigma^-$ denote left- and right-hand circular polarization, respectively \cite{Benson2000PRL}. The fine-structure splitting can be reduced by growing quantum dots on the higher-symmetry (111) substrates \cite{Juska2013NPHOT,Kuroda2013PRB} or by applying various tuning schemes, such as electric fields \cite{Bennett2010NaturePhys}.

\subsection{Optical properties of quantum dots\label{sec:sst:decaydynamics}}

\begin{figure*}
\begin{center}
\includegraphics[width=0.8\textwidth]{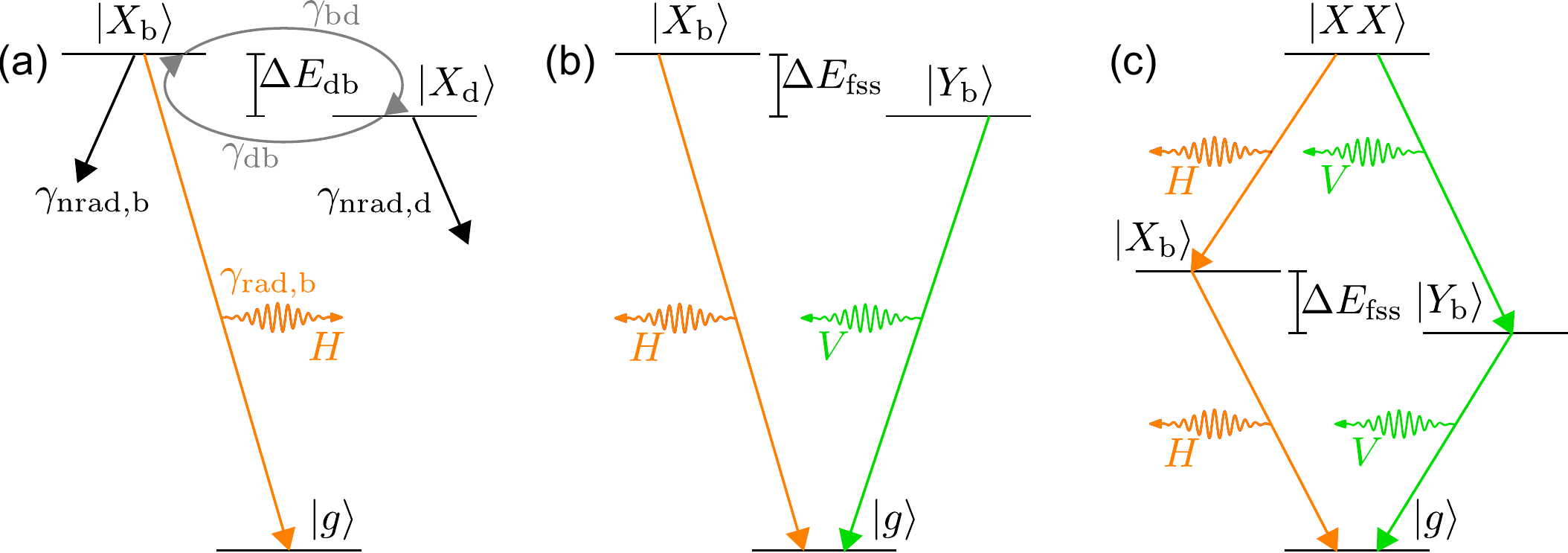}
\caption{ Examples of excitonic level schemes in quantum dots of relevance for quantum-optics experiments. (a) The most basic three-level optical transition scheme of a single bright exciton, $|X_\text{b}\rangle$ that can emit a photon by decaying radiatively  to the ground state, $|g\rangle$. Also non-radiative decay processes and coupling to the dark exciton state, $|X_\text{d}\rangle$, through spin-flip processes may occur. The dark state can also recombine non-radiatively. This level scheme leads to a bi-exponential decay of the emitted intensity and holds also for the other bright exciton, $|Y_\text{b}\rangle$ (not shown). (b) Optical $V$-scheme formed by the two bright exciton states that decay to the ground state. The two orthogonally polarized bright excitons are split by the fine-structure splitting, $\Delta E_\text{fss}$. The non-radiative processes and dark states indicated in (a) are excluded for simplicity. (c) Four-level scheme formed by the biexciton level, the two bright exciton levels, and the ground state. The biexciton decays through a cascaded process of either emitting two horizontally or two vertically polarized photons, respectively.
\label{fig:sst:Bright-dark_level_scheme}}
\end{center}
\end{figure*}

Quantum dots are reliable sources of photons for quantum optics and various level schemes can be realized by different excitation and detection strategies. Figure \ref{fig:sst:Bright-dark_level_scheme} summarizes the most relevant levels and decay processes. In many applications the three-level scheme of Fig. \ref{fig:sst:Bright-dark_level_scheme}(a) is applied, which suffices for, e.g., describing single-photon emission. We consider this level scheme throughout the review unless otherwise noted. In this case we define the excited state $|e\rangle = |X_\text{b}\rangle$. In the following we discuss the key figures of merit of the three-level system and elaborate on the single-photon-emission properties and coherence in Sec.\ \ref{sec:sst:singlephotonsources}.

\subsubsection{Quantum-dot decay dynamics}
A detailed understanding of the dynamics of quantum dots is essential in order to exploit them as reliable photon sources in quantum-photonics applications. The various exciton states can be coupled by spin-flip processes where, e.g., an exchange-mediated process between electron and hole \cite{Roszak2007PRB} or spin-orbit coupling \cite{Liao2011PRB}  flips the spin of the exciton while longitudinal acoustic (LA) phonons provide or remove the energy difference between the two states. Spin-flip processes are generally much slower than the radiative decay processes so it is a good approximation to include only spin flips between $|X_\text{b}\rangle$ and $|X_\text{d}\rangle$ as well as between $|Y_\text{b}\rangle$ and $|Y_\text{d}\rangle$. Transitions between $|X_\text{b/d}\rangle$ and $|Y_\text{b/d}\rangle$ require changing the spin of both the electron and the hole and are therefore negligible. This implies that $|X_\text{b/d}\rangle$ and $|Y_\text{b/d}\rangle$ are decoupled and leads to two identical three-level schemes of the form shown in Fig.\ \ref{fig:sst:Bright-dark_level_scheme}(a). This constitutes the basic level scheme relevant for many quantum-photonics experiments employing a single (neutral) exciton in a quantum dot. For the case of non-resonant excitation where both bright and dark exciton states are populated, the population $\rho_\text{b}$ of the bright exciton follows a bi-exponential decay \cite{Smith2005PRL,Wang2011PRL},
\begin{align}
\rho_\text{b}(t)=A_\text{f} e^{-\gamma_\text{f} t} + A_\text{s} e^{-\gamma_\text{s} t},
\label{eq:bi-ex-model}
\end{align}
with the fast and slow decay rates $\gamma_\text{f}=\gamma_\text{rad,b}/2+\gamma_\text{nrad,b}+\gamma_\text{db}+\sqrt{\gamma_\text{rad,b}^2/4+\gamma_\text{db}^2}$ and $\gamma_\text{s}=\gamma_\text{rad,b}/2+\gamma_\text{nrad,b}+\gamma_\text{db}-\sqrt{\gamma_\text{rad,b}^2/4+\gamma_\text{db}^2}$, and corresponding amplitudes, $A_\text{f}=\rho_\text{b}(0)\left(1+\frac{\gamma_\text{rad,b}}{\gamma_\text{f}-\gamma_\text{s}} \right)/2-\rho_\text{d}(0) \frac{\gamma_\text{db}}{\gamma_\text{f}-\gamma_\text{s}}$ and
$A_\text{s}=\rho_\text{b}(0)\left(1-\frac{\gamma_\text{rad,b}}{\gamma_\text{f}-\gamma_\text{s}} \right)/2+\rho_\text{d}(0) \frac{\gamma_\text{db}}{\gamma_\text{f}-\gamma_\text{s}}$. Here $\gamma_\text{rad,b}$ denotes the radiative decay rate for the bright exciton, $\gamma_\text{nrad,b}$ ($\gamma_\text{nrad,d}$) denotes the non-radiative decay rate of the bright (dark) exciton, and $\gamma_\text{db}$ is the bright-dark spin-flip rate. The initial population of bright and dark excitons can be assumed identical, i.e., $\rho_\text{b}(0)=\rho_\text{d}(0)=0.5$ for weak non-resonant excitation although corrections to this assumption may be imposed by the decay cascade involving higher-excited states in the quantum dot \cite{Poem2010NPHYS}. It has been found experimentally that the recombination of dark excitons is dominated by non-radiative processes and that $\gamma_\text{nrad,b} \approx \gamma_\text{nrad,d}$ \cite{Johansen2010PRB}, i.e., the intrinsic non-radiative decay rates of bright and dark exciton states are approximately equal, which reflects that the binding energy of bright and dark excitons are very similar. This approximation has been used to obtain Eq.\ (\ref{eq:bi-ex-model}). We note that with externally appplied strain, light- and heavy-hole mixing \cite{Huo2014NPHYS} may lead to a radiative contribution to the recombination of dark excitons. Since most experiments are carried out at temperatures where the thermal energy exceeds the energy splitting between dark and bright states $(k_\text{B} T > \Delta E_\text{db})$, the probability of emission and absorption of phonons is approximately equal and it is a good approximation to assume $\gamma_\text{bd} \approx \gamma_\text{db}$, which has also been assumed to reach the simplified dynamics of Eq.\ (\ref{eq:bi-ex-model}).

By fitting experimental decay curves with the bi-ex\-po\-nen\-ti\-al model of Eq.\ (\ref{eq:bi-ex-model}), the radiative, non-radiative, and spin-flip rates can be extracted for a single quantum dot. With this method, single quantum dots can be employed for mapping the local light-matter interaction strength, as discussed in Sec.\ \ref{Section-spontaneous-emission-control}. Considerable variations in the spin-flip and non-radiative rates are generally found across a quantum-dot ensemble and between different growth runs, and the described quantitative method of determining these processes is thus required. We note that trions and biexcitons do not have a fine structure implying that their population dynamics follows a single-exponential decay with no direct access to the non-radiative rates.

\subsubsection{The oscillator strength \label{sec:sst:OS}}
The intrinsic capability of a dipole emitter to interact with light is characterized by the magnitude of the transition momentum matrix element, $\mathbf{P}$, as introduced above. In the dipole and rotating-wave approximations this is equivalent to retaining only the dipole term in the multipolar gauge \cite{Thirunamachandran}. In this approximation, the interaction Hamiltonian can be expressed in terms of the transition dipole moment operator, $\mathbf{d}=q\mathbf{r}$, and the electric field amplitude, $\mathbf{E}(\mathbf{r}_0,t)$, evaluated at the position of the emitter, $\mathbf{r}_0$, as $H=-\mathbf{d}\cdot\mathbf{E}(\mathbf{r}_0,t)$. This description is common in atomic physics where the dipole approximation is an excellent approximation but verifying its validity for quantum dots is non-trivial as discussed in Sec.\ \ref{Section-beyond-dipole}. The dipole matrix element is related to the momentum matrix element via $\mathbf{P}=i m_0\omega_0\mathbf{r}$, where $\hbar\omega_0$ is the optical transition energy. The optical transition strength is commonly quantified by the oscillator strength, which is a dimensionless parameter defined as the decay rate in a homogeneous medium divided by the decay rate of a classical oscillator. It can be expressed as
\begin{align}
f = \frac{2}{\hbar\omega_0 m_0}|\mathbf{P}|^2 = \frac{E_\text{P}}{\hbar \omega_0}|\langle F_\text{v}|F_\text{c}\rangle|^2.\label{eq:sst:OS_DA_SC}
\end{align}
A large oscillator strength is desirable since it increases the light-matter interaction strength and therefore increases the generation rate of single photons. A large oscillator strength also reduces the relative impact of other interactions that may give rise to undesired effects, such as phonon-dephasing or non-radiative decay. Evidently the squared wave-function overlap,  $O_{\rm eh} = |\langle F_\text{v}|F_\text{c}\rangle|^2$, must be maximized to enhance the oscillator strength cf.\ the discussion of Fig.\ \ref{fig:sst:interface} below.

While the wave-function overlap cannot exceed unity and thus sets the upper limit for the oscillator strength, the strong-confinement model underlying Eq.~(\ref{eq:sst:OS_DA_SC}) is only valid when the Coulomb interaction can be neglected or included perturbatively. For large quantum dots with a radius larger than the exciton Bohr radius, this assumption breaks down. The exciton Bohr radius is $a_0 = \frac{4\pi \epsilon_0 \epsilon_\mathrm{r} \hbar^2}{q^2 m_0 m}$, where $\epsilon_0$ denotes the vacuum permittivity, $\epsilon_\mathrm{r}$ is the relative static permittivity of the material, and the reduced mass is defined as $m = \frac{m_\text{e} m_\text{hh}}{m_\text{e} + m_\text{hh}}$ with the electron (hole) effective mass, $m_\text{e}$ ($m_\text{hh}$). The exciton Bohr radius in nanostructures is modified by confinement and in quantum dots where the quantization axis is dominant the Bohr radius is reduced by a factor of $\simeq 2$ \cite{Bastard1982PRB}. When the quantum-dot radius is much larger than the exciton Bohr radius the exciton is weakly confined leading to a hydrogen-like state. In the weak-confinement regime, the oscillator strength turns out to be proportional to the exciton volume \cite{Hanamura1988PRB,Andreani1999PRB,Stobbe2012PRB} or, for a disc-shaped quantum dot, to the area, i.e.,
\begin{align}
f = 8\frac{E_\text{P}}{\hbar \omega}\left( \frac{L}{a_0} \right)^2.\label{eq:sst:OS_disc_DA}
\end{align}
This remarkable phenomenon is known as the giant-oscillator-strength effect and may be considered an analogue of superradiance. It implies that the oscillator strength can be significantly enhanced by increasing the lateral extent of the quantum dot. This prediction has lead to significant experimental efforts to observe the giant-oscillator-strength effect, which is challenging since non-radiative processes could mask the effect. The indirect extraction of the oscillator strength in large InGaAs quantum dots obtained from the founding work on strong coupling in cavity QED \cite{Reithmaier2004Nature} has turned out to be inconsistent with direct measurements of the oscillator strength from time-resolved spectroscopy, since non-radiative recombination was found to be relevant \cite{Stobbe2010PRB}. This discrepancy is likely a result of the influence of additional excitons feeding the cavity, as discussed in further details in Sec.\ \ref{Section-cavity-QED}. Indications of a large oscillator strength have been reported for GaAs interface-fluctuation quantum dots from time-resolved experiments \cite{Peter2005PRL,Hours2005PRB}.

\subsubsection{The quantum efficiency\label{sec:sst:QE}}
\begin{figure}
\begin{center}
\includegraphics[width=\columnwidth]{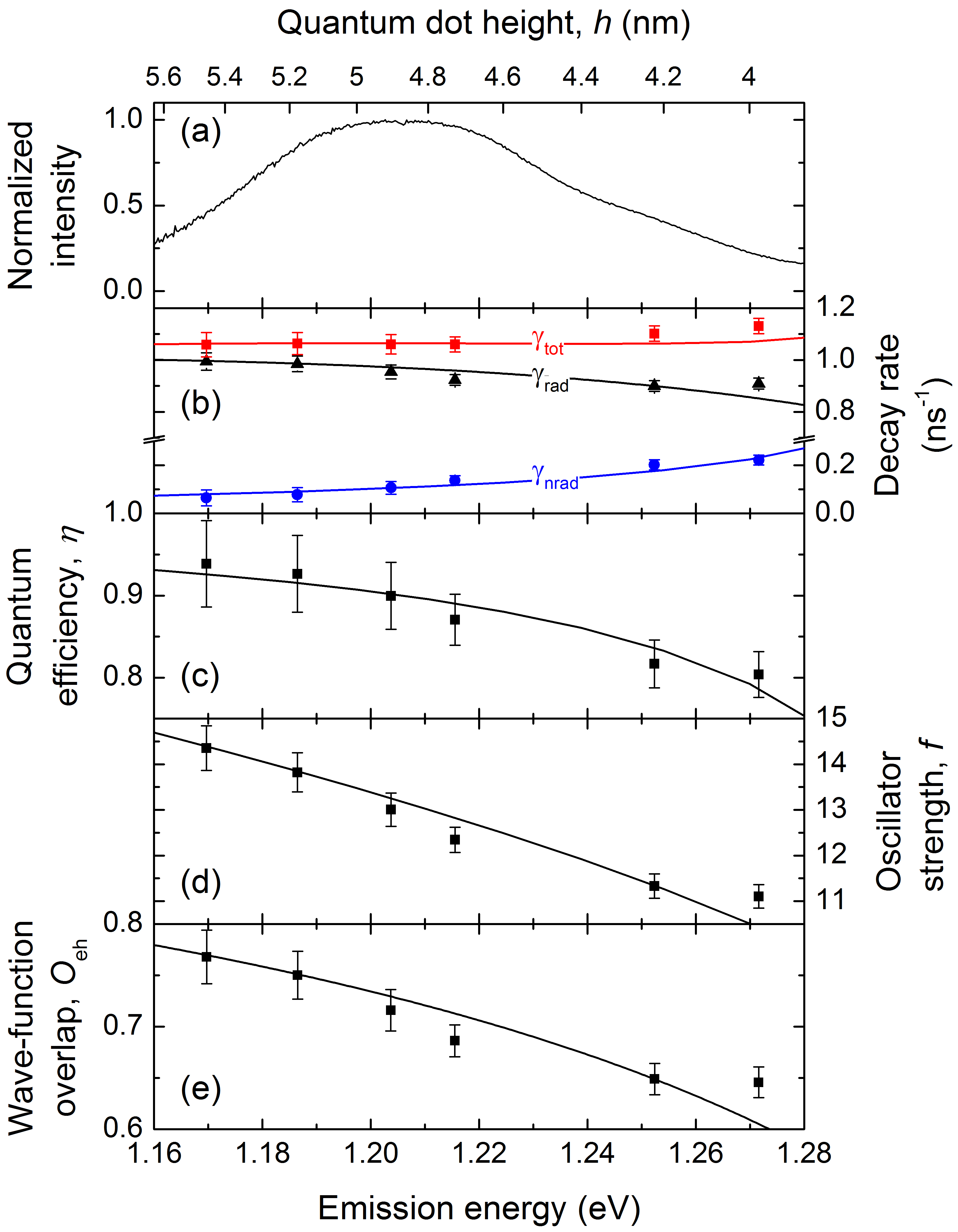}
\caption{Emission-energy dependence of fundamental quan\-tum-optical properties of quantum dots. (a) Emission spectrum of an InGaAs quantum-dot ensemble. (b) Total decay rate (red), which is the sum of the radiative (black) and non-radiative (blue) decay rates. The solid blue line is the predictions from a scaling model for non-radiative recombination at the quantum-dot surface. The solid black line is a fit using a numerical solution to the effective-mass equation with the quantum-dot heights shown in the upper scale. (c) Quantum efficiency, (d) oscillator strength, and (e) wave-function overlap where the data points (solid lines) are directly extracted from the data (theory) in (b). The wave-function overlap governs the radiative properties of excitons. Data reproduced from \citet{Johansen2008PRB} and \citet{Stobbe2009PRB}.\label{fig:sst:interface}}
\end{center}
\end{figure}

Exciton-recombination processes in quantum dots may suffer from non-radiative contributions, which add undesirable losses to the radiative decay. The relative strength of the radiative and the non-radiative decays is quantified by the fraction of recombination events leading to photon emission, i.e., the quantum efficiency,
\begin{align}
\eta=\frac{\gamma_\text{rad}^\text{hom}}{\gamma_\text{rad}^\text{hom}+\gamma_\text{nrad}},
\end{align}
where $\gamma_\text{rad}^\text{hom}$ is the radiative decay rate in a homogeneous reference medium, e.g., GaAs in the case of Stranski-Krastanov quantum dots. In optics applications it is desirable to have $\eta \simeq 1.$

Measuring the quantum efficiency can be challenging since it is non-trivial to separate radiative and non-radiative processes. For example, if $\gamma_\text{db}\ll\gamma_\text{rad}$, which is typically the case, the fast decay rate probed in a time-resolved photoluminescence experiment is $\gamma_\text{f} = \gamma_\text{rad} + \gamma_\text{nrad}$. Figure \ref{fig:sst:interface} shows experimental data on the key optical properties of quantum dots where a systematic variation in the projected local density of optical states (LDOS; discussed at length in Sec.\ \ref{Section-spontaneous-emission}) was used to modify the radiative rate while leaving the non-radiative decay unaffected \cite{Johansen2008PRB} and thus enabling their separation. The energy dependence of the total decay rate across the emission spectrum of the quantum-dot ensemble (Fig.\ \ref{fig:sst:interface}(a)) is shown in Fig.\ \ref{fig:sst:interface}(b) and turns out to be influenced by the variation in $\gamma_\text{nrad}$ as reflected in the drop in quantum efficiency for high energies shown in Fig.\ \ref{fig:sst:interface}(c). The radiative rate $\gamma_\text{rad}$ and thus also the oscillator strength, cf.\ Fig.\ \ref{fig:sst:interface}(d), were successfully modeled by Eq.\ (\ref{eq:sst:OS_DA_SC}) that predicts that they are proportional to the overlap of the envelope wavefunctions as shown in Fig.\ \ref{fig:sst:interface}(e). The observed energy dependence is due to the different effective masses of electrons and holes such that the wavefunction confinement inside the quantum dot differs with energy. While this detailed information was extracted for an ensemble of quantum dots by applying a controlled modification of the LDOS, the radiative and non-radiative rates can also be extracted at the single quantum-dot level by recording the bi-exponential decay dynamics.

Little is known about the physical mechanism of non-radiative processes in quantum dots but the few existing experiments \cite{Stobbe2009PRB} indicate that the non-radiative decay rate scales with the surface-to-volume ratio of the quantum dot. This shows that charge trapping at lattice defects at the interface between the quantum dot and the surroundings is likely a contribution to non-radiative recombination in Stranski-Krastanov quantum dots; this finding is consistent with the large variations in the non-radiative decay rate between different quantum dots that is typically found. Recently also mesoscopic light-matter interaction effects were found to contribute to this additional decay process \cite{Tighineanu2014Arxiv}; for a further discussion of these effects see Sec.\ \ref{Section-beyond-dipole}. Since the LDOS can be strongly altered in photonic nanostructures it is sometimes convenient to define an effective quantum efficiency, $\eta_\text{eff}=\frac{\gamma_\text{rad}}{\gamma_\text{rad}+\gamma_\text{nrad}}$, which is the quantum efficiency of the emitter at a particular position, wavelength, and orientation in a photonic structure. Since $\eta_\text{eff}$ may be significantly enhanced by enhancing the LDOS, the influence of non-radiative effects may be partially suppressed.

\subsubsection{Excitation schemes and tuning}

One of the simplest yet very powerful experimental techniques for assessing the optical properties of quantum dots is photoluminescence spectroscopy, where the quantum dot is optically excited  and the spontaneous emission recorded with a spectrometer equipped with a charged-coupled-device camera. The excitation energy can be either 1) above-band, i.e., above the band gap of the barrier material, 2) quasi-resonant, i.e., matching an excited excitonic state or continuum of states in the wetting layer, or 3) resonant, i.e., exactly matching the energy of the exciton transition under investigation. Detailed spectroscopic insight is obtained by monitoring the emission from the exciton while scanning the laser wavelength, which is known as photoluminescence excitation spectroscopy. Resonant excitation is particularly challenging and has been demonstrated only rather recently because the  strong excitation laser must be efficiently filtered. Resonant excitation is discussed in detail in Sec.\ \ref{Section-Resonance-Fluorescence}. An alternative approach is to embed the quantum dots in a diode and mask the wafer surface with opaque metal except a small aperture atop the quantum dot, which allows probing resonant-excitation properties in the photocurrent. The diode may also be used in forward bias to create a single-photon light-emitting diode \cite{Yuan2002Science,Salter2010Nature}, which is appealing for certain applications but the drawback of this electroluminescence is that the energy and polarization of the injected carriers is more difficult to control as compared to controlling an excitation laser although there is currently significant progress \cite{Conterio2013APL}. Nonetheless the ability to electrically excite quantum dot distinguish them from other optical emitters and may become of major practical relevance.

For many applications in quantum nanophotonics, precise spectral tuning of quantum dots is required. For instance, in the case of a cavity with a $Q$-factor of $10^4$, the cavity linewidth is approximately \SI{0.1}{\nano\meter} and the observation of cavity-QED effects requires tuning the quantum dot into resonance with the mode to this precision. A wide tuning range is generally required since a quantum-dot ensemble is usually strongly inhomogenously broadened (can be up to 50-\SI{100}{\nano\meter}) and fabrication imperfections also lead to a significant spread in the actual cavity resonance frequencies.  One possible mechanism exploits temperature tuning where both the cavity-resonance and exciton-transition energies shift in the same direction (red-shift with increasing temperature) but at a different rate so that they may be tuned into mutual resonance \cite{Englund2007Nature}. This has proven to be a very useful method that is easy to implement but may compromise coherence properties since increasing the temperature also increases the phonon population leading to dephasing. Another option is tuning with magnetic fields \cite{Stevenson2006Nature}, which depending on the orientation of the field relative to the quantum dot also changes the fine structure and the optical selection rules \cite{Bayer2002PRBb}. However, this method has rather limited tuning range and it appears unfeasible to apply large local magnetic fields to individual quantum dots. Another option is adsorption of inert gasses, which changes the dielectric environment that in turn spectrally shift the optical response of the nanostructures while leaving the quantum dots unaffected. This works well for photonic crystals due to the relatively large field amplitude in the holes where the gas is adsorbed.  However, for more advanced applications involving multiple quantum dots to be independently tuned, this method is not useful. Applying external strain enables not only tuning of the energy levels but also allows modifying the underlying electronic band structure to create light-hole excitons \cite{Huo2014NPHYS}. However, it appears challenging to incorporate integrate such strain tuning with high-quality photonic nanostructures. The most appealing tuning mechanism is to apply a static electric field to tune the quantum dots via the quantum-confined Stark effect. Electric fields may be readily applied between doped layers below and above the quantum dots forming a vertical p-i-n diode, which under reverse bias can be used to apply large fields. With a proper device design involving the tunneling barriers, a tuning range of \SI{25}{\milli\electronvolt} corresponding to \SI{18}{\nano\meter} \cite{Bennett2010NaturePhys} has been demonstrated. The doping levels can be so small that absorptive losses do not limit the $Q$-factor of cavities or the transmission of waveguides. Independent tuning of multiple quantum dots could be achieved by etching isolation trenches \cite{Winger2011OE} or by local dopant implantation \cite{Ellis2011NPHOT}.

\subsection{Coherent single-photon emission from quantum dots\label{sec:sst:singlephotonsources}}

Quantum dots are mesoscopic emitters embedded in a solid-state environment and a number of processes not encountered in atomic physics play an important role for the single-photon emission from quantum dots. In particular, phonons, charge fluctuations associated with lattice defects or impurities, and spin fluctuations in the ensemble of nuclei give rise to dephasing, spin-flip, and non-radiative decay processes. Such processes can influence the quality of the single-photon emission from quantum dots and in particular determine the coherence properties. Highly coherent single photons are required for quantum-information processing applications and in the present section we briefly outline the main decoherence processes and governing parameters for quantum dots.

In order to assess the quality of a single-photon source it is essential to develop precise experimental methods. To this end, the photon statistics of a quantum state of light is determined by recording the second-order correlation function defined as
\begin{align}
g^{(2)}(\tau)=\frac{\langle\hat{a}^\dagger(t) \hat{a}^\dagger(t+\tau) \hat{a}(t+\tau) \hat{a}(t) \rangle}{\langle\hat{a}^\dagger(t)\hat{a}(t)\rangle^2},
\end{align}
where $\hat{a}$ and $\hat{a}^\dagger$ are annihilation and creation operators for the optical mode probed in the experiment and  $\tau$ is a time delay introduced in between two subsequent measurements of the number of photons at time $t$ and $t+\tau$, respectively. Experimentally, $g^{(2)}(\tau)$ can be recorded by dividing the light beam on a beam splitter and recording with two different photodetectors the number of photons with an electronically controlled delay in each path. The second-order correlation function is independent of $t$ in a stationary experiment. This is referred to as the Hanbury-Brown-Twiss setup. The value at $\tau=0$ is particularly important because it directly determines the quality of the single-photon source. For Fock states, $g^{(2)}(0)=1-\frac{1}{n}$, where $n$ is the number of photons and thus $g^{(2)}(0)=0$ for a true single-photon source. For a coherent (thermal) state, $g^{(2)}(0)=1$ ($g^{(2)}(0)=2$). Consequently, $g^{(2)}(0)$ can be employed as a measure of the single-photon purity of a light source. In the case of quantum dots, the recombination of a single exciton generates a single photon, but the recorded signal may be polluted by photons emitted through other recombination processes on different transitions. In experiments, much effort is devoted to the selective excitation of a single exciton in a quantum dot and the subsequent spectral filtering of the emitted light in order to suppress any multi-photon contributions from other emission processes.

Information about the coherence of the emitter can be obtained from the spontaneous-emission spectrum. A number of different dephasing mechanisms will in general influence a quantum dot that can be included at various levels of sophistication when modelling experimental data. In the simplest approximation, the dephasing reservoir is assumed to be Markovian and described by a single rate $\gamma_{\rm dp}$. In this case the emission spectrum exhibits the well-known Lorentzian form with a linewidth determined by the spontaneous-emission lifetime and dephasing times according to
\begin{align}
\frac{1}{T_2}=\frac{1}{2T_1} + \frac{1}{T_2^\ast},\label{eq:sst:T2}
\end{align}
where $T_2$ is the total coherence time (i.e., the inverse linewidth), $T_1=1/\gamma_\text{tot}$ is the inverse total decay rate of the emitter, and $T_2^\ast = 1/\gamma_{\rm dp}$ is the pure-dephasing time. We note that due to dephasing, the coherence of the emitter usually decays faster than the population. In the absence of dephasing, $1/T_2^\ast = 0$, the two-fold slowdown of coherence relative to population reflects that the former depends linearly on the complex expansion coefficient of the excited state of the emitter while the latter depends quadratically. The theory of spontaneous-emission dynamics and coherence is treated in detail in Secs.\ \ref{LDOS-EOMs} and \ref{sec:Lamb}.  The pure-dephasing model captures the basic feature of a broadening of the emission spectrum and may be used to describe, e.g., the broadening of the zero-phonon line of the emission spectrum \cite{Muljarov2004PRL}. Other dephasing processes in quantum dots include charge-fluctuation noise and spin noise from the coupling of the exciton to the spins of the nuclei \cite{Kuhlmann2013NPHYS}. The charge noise arises from the presence of a fluctuating distribution of charges in the vicinity of the quantum dot and leads to slow (longer than milliseconds) spectral diffusion of the quantum-dot resonance. To a large extent, this can be overcome by applying resonant excitation on electrically contacted quantum dots whereby near-transform-limited optical transitions have been observed \cite{Kuhlmann2013Arxiv}. Residual spectral diffusion can be overcome by locking the quantum dot to a stable reference  \cite{Prechtel2013PRX}. It should be emphasized that an adequate description of these dephasing processes would include how the functional form of the emission spectrum is modified by the dephasing, which is not included in the pure-dephasing model where the emission spectrum remains Lorentzian. Furthermore, phonon broadening is generally much faster (picosecond time scales) than the emitter lifetime and therefore also not captured by the pure-dephasing model. The phonon sidebands of the emission spectra are discussed below.

\begin{figure}
\includegraphics[width= \columnwidth]{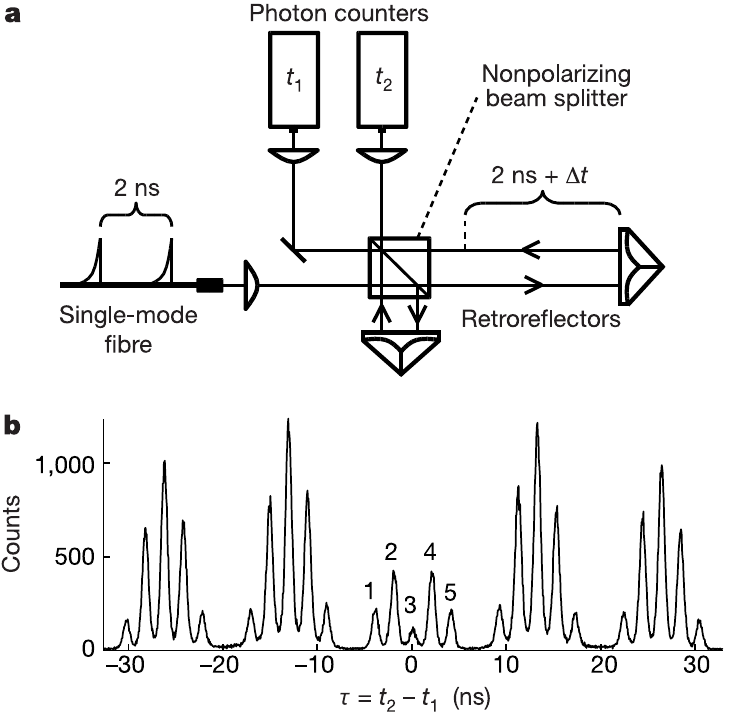}
\caption{a) Sketch of a Hong-Ou-Mandel (HOM) interferometer capable of measuring the degree of indistinguishability of two consecutively emitted photons from a quantum dot. The setup employs an asymmetric Michelson-Morley interferometer for compensating the path-length difference between the two photons enabling them to meet in one quarter of all incidences at the second passage through the beam splitter and thus to interfere. The quantum interference is gauged by recording the correlation function $g^{(2)}(\tau)$. (b) Example of HOM measurements for a quantum dot in a micropillar cavity. For distinguishable pairs of photons a five-peak spectrum with peak area ratios of 1:2:2:2:1 from the two encounters of the beam splitter is expected. A strong suppression of the central peak (labelled 3) relative to the neighboring peaks 1-2 and 4-5 is the experimental evidence for quantum interference of the two photons that is possible only for (partly) indistinguishable photons. Reprinted with permission from \citet{Santori2002Nature}.}
 \label{HOM-measurements}
\end{figure}

Another experimental test of single-photon coherence exploits a Hong-Ou-Mandel interferometer. In such an experiment, two single photons are interfered on a 50/50 beam splitter and the two outputs are recorded with single-photon detectors as shown in Fig.\ \ref{HOM-measurements}(a). This constitutes a generalization of the Hanbury-Brown-Twiss setup to a case where two input beams are incident on the beam splitter as opposed to just one. The degree of indistinguishability for a photon source that can be extracted with a Hong-Ou-Mandel interferometer is defined as \cite{Bylander2003EPJD,Kiraz2004PRA,Kaer2013PRB}
\bea
I &=&\frac{\int_0^\infty\mathrm{d}t\int_0^\infty\mathrm{d}\tau \left| \langle \hat{a}^\dagger(t+\tau)\hat{a}(t) \rangle \right|^2}
{\int_0^\infty\mathrm{d}t\int_0^\infty\mathrm{d}\tau \langle \hat{a}^\dagger(t+\tau)\hat{a}(t+\tau) \rangle \langle \hat{a}^\dagger(t)\hat{a}(t) \rangle}\label{eq:sst:indistinguishability1} \nn \\
&=& \frac{\gamma_{\rm tot}}{\gamma_{\rm tot} + 2 \gamma_{\rm dp}} = \frac{T_2}{2 T_1},\label{eq:sst:indistinguishability2}
\eea
where the expressions in the second line hold for the case of pure (i.e., Markovian) dephasing. The lack of any coincidence detection events in a Hong-Ou-Mandel measurement signifies that the two photons are fully indistinguishable, which is only the case if they are fully coherent. Such single-photon quantum interference is a direct ingredient in many quantum-information protocols and therefore an important test of a single-photon source. With quantum dots, Hong-Ou-Mandel interferometry has been performed by interfering two consecutively emitted photons from the same quantum dots, see Fig.\ \ref{HOM-measurements}(b). The photons were found to be partly indistinguishable, which is a consequence of dephasing of the quantum-dot levels. The degree of indistinguishability is determined from the peak area of the coincidence events detected when two consecutively emitted photons meet on the beam splitter. As opposed to the emission spectrum, which is sensitive to the slow spectral-diffusion processes, indistinguishability measurements on two consecutively emitted photons from the same source are only sensitive to fast dephasing occurring on a time scale shorter than or similar to the time delay between the two photons, such as phonon processes. As a consequence, a quantum-dot transition suffering from slow spectral diffusion, which is often the case for non-resonant excitation schemes, may still be capable of emitting highly indistinguishable photons.

\begin{figure}
\includegraphics[width=\columnwidth]{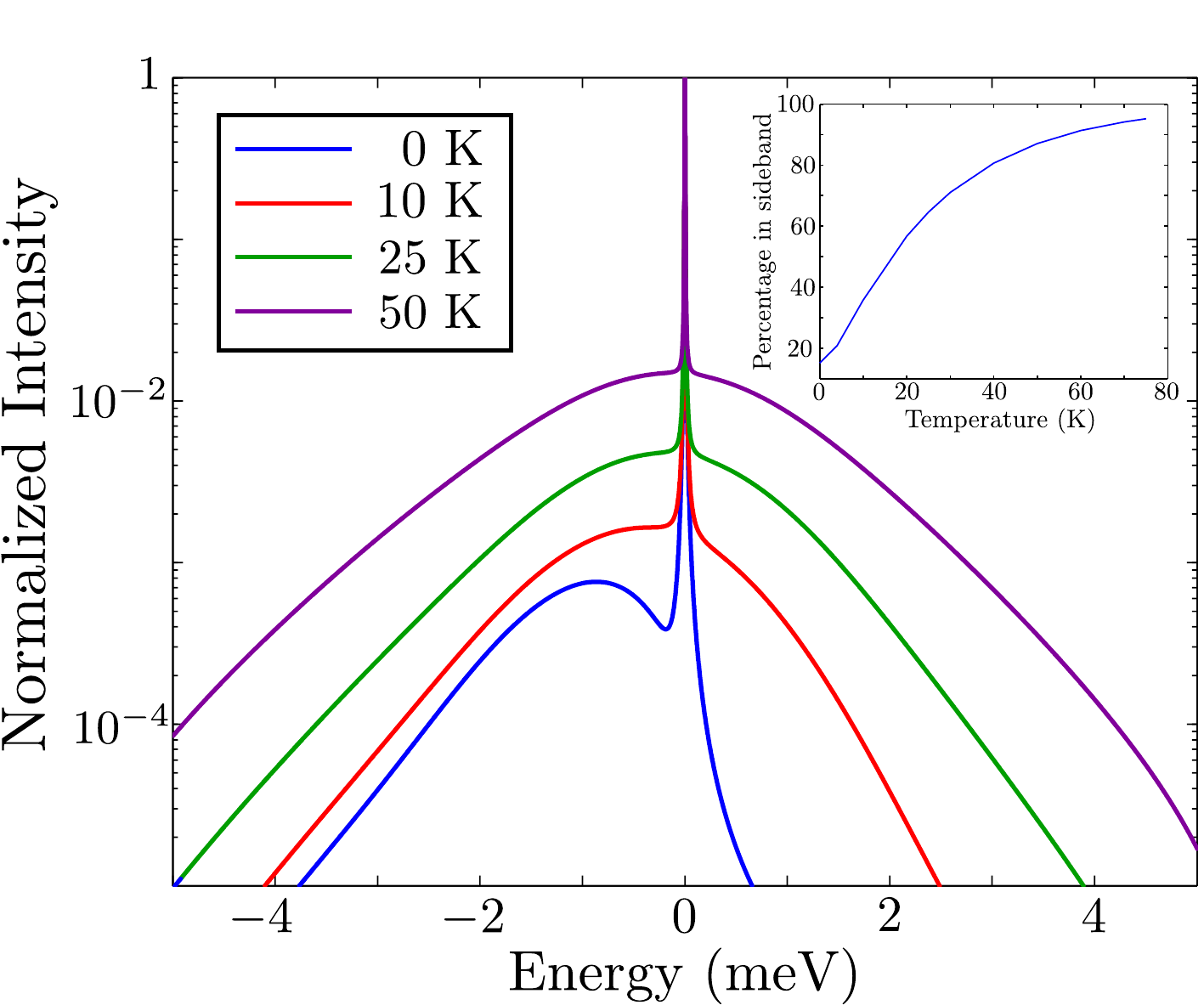}
\caption{Calculated normalized emission spectrum (logarithmic scale) for a quantum dot coupled to an LA-phonon reservoir at different temperatures. The phonon sidebands are highly asymmetric at low temperatures. The parameters used are experimental values reported in \citet{Madsen2013PRB}. Note that the peak of each spectrum is normalized to unity. The inset shows the fraction of the intensity in the sidebands as a function of temperature.}
\label{phonon-emission-spectrum}
\end{figure}

In the following we will discuss further dephasing induced by phonons that is unavoidable for quantum dots since they are embedded in a solid-state material. It should be emphasized that at elevated temperatures additional effects play a role, such as Coulomb-induced scattering to wetting layer states \cite{Steinhoff2012PRB}, but such effects are outside the scope of the present account and for further details, see \citet{Kira2011book}. By cooling down the quantum dots to close to zero absolute temperature, the influence of phonons can be strongly suppressed, although spontaneous emission of phonons prevails. Longitudinal acoustic (LA) phonons are a major broadband source of dephasing for quantum dots. A quantum dot can absorb or emit phonons through inelastic processes that generally require a description of coupling to a non-Markovian reservoir. LA phonons induce an asymmetric broadening of the emission spectrum reflecting that the probability of absorbing (emitting) a phonon of frequency $\Omega_{\bf k}$ is proportional to $n_{\bf k}$ ($n_{\bf k}+1$), where $n_{\bf k}$ is the phonon occupation at the relevant wave vector ${\bf k}$. A typical LA-phonon energy that leads to phonon sidebands of a quantum dot is about \SI{1}{\milli\electronvolt} \cite{Madsen2013PRB} where the relevant phonon energy decreases with the size of the exciton. At a temperature of \SI{10}{\kelvin}, we estimate $n_{\Omega_{\bf k}} = 1/\left(\exp\left[\hbar \Omega_{\bf k}/k_\text{B} T\right]-1 \right) \approx 0.46$. The coupling between an exciton in a quantum dot and a phonon reservoir can be described by the interaction Hamiltonian \cite{Krummheuer2002PRB, MahanBook, Kaer2012PRB}
\be
\hat{H}_{\mathrm{ph}} = \sum_{{\bf k}} \hbar \left( M_{g}^{\bf k}  \hat{\sigma}_- \hat{\sigma}_+ + M_{e}^{\bf k} \hat{\sigma}_+ \hat{\sigma}_-\right) \left(\hat{d}_{-{\bf k}}^{\dagger} + \hat{d}_{\bf k} \right),\label{eq:sst:phonon_interaction}
\ee
where $\hat{d}_{\bf k}$ and $\hat{d}_{\bf k}^{\dagger}$ are bosonic annihilation and creation operators for the phononic mode ${\bf k}$ and the raising and lowering operators $\hat{\sigma}_{\pm}$ for the quantum dot are introduced in Sec.\ \ref{LDOS-EOMs}. The quantum dot-phonon interaction strength is defined as
\be
M_{i}^{\bf k} = \sqrt{\frac{\hbar k}{2 d_\text{m} c_\text{s} V}} D_i \int \text{d}{\bf r} \left| \psi_i({\bf r}) \right|^2 e^{-i {\bf k} \cdot {\bf r}},
\label{phonon-interaction-strength}
\ee
with $i = \{ e, g \}$ and where a linear LA-phonon dispersion relation, $\Omega_{\bf k} = |{\bf k}| c_\text{s}$, is assumed for the relevant frequency range with $c_\text{s}$ the angular-averaged speed of sound. Furthermore, $d_\text{m}$ is the mass density, $V$ the quantization volume of the phonon modes, and $D_i$ and $\psi_i$ the deformation potential and electronic wave function for the state $i$, respectively. The quantum-dot-phonon interaction strength depends sensitively on the quantum-dot size since it is proportional to the spatial Fourier transform of the wavefunction for the electron in either the ground or excited state. As a consequence, large excitons are more robust to dephasing compared to smaller excitons. Figure \ref{phonon-emission-spectrum} shows examples of calculated spontaneous-emission spectra for a single quantum dot at various temperatures clearly demonstrating the asymmetric broadening due to LA phonons.

\section{Photonic nanostructures}

The essential challenge for quantum-optics experiments at the single-photon level is to strongly enhance and control the interaction between light and matter such that an emitted single photon  preferentially couples to one well-defined optical mode. To this end, photonic nanostructures are very well suited due to their ability to tailor the electromagnetic field on a length scale that can be made a fraction of the optical wavelength. Photonic nanostructures with excellent optical properties can be fabricated with modern nanofabrication methods such as molecular-beam epitaxy, electron-beam lithography, and etching. In the present section we introduce the various types of photonic nanostructures that have been employed in QED experiments and compare their relevant figures of merit.

\subsection{Photonic crystals \label{sec:PhotCrys}}

Photonic crystals are inhomogeneous dielectric materials where the refractive index is modulated periodically on a length scale determined by the optical wavelength. In such a structure, light propagation is controlled by  optical Bragg scattering of light, which is the optical analogue of electron Bragg diffraction employed in crystallographic experiments on solids. The strength of Bragg scattering increases with the refractive-index contrast of the materials composing the photonic crystal. Furthermore, any optical loss such as absorption is detrimental to the functionality of a photonic crystal since it suppresses light interference and therefore Bragg scattering.

A photonic crystal can be described by a spatially periodic dielectric permittivity $\epsilon({\bf r})$ that in general has both a real and an imaginary part where the latter is linked to the absorption. The most effective photonic crystals are typically made from semiconductor materials and constructed for a frequency range where absorption is sufficiently small such that it can be ignored to a good approximation, i.e., ${\rm Im}\left[\epsilon({\bf r})\right] \approx 0$. Furthermore, for the narrow band frequency applications that are usually considered in the context of quantum optics, any frequency dependence of the dielectric permittivity can be ignored. Here we will mainly consider semiconductor materials such as gallium arsenide or silicon with the large refractive index of $n = \sqrt{\epsilon} \sim 3.5$ where large-contrast photonic crystals can be obtained, e.g., by etching air voids.

\begin{figure}[t!]
\includegraphics[width=\columnwidth]{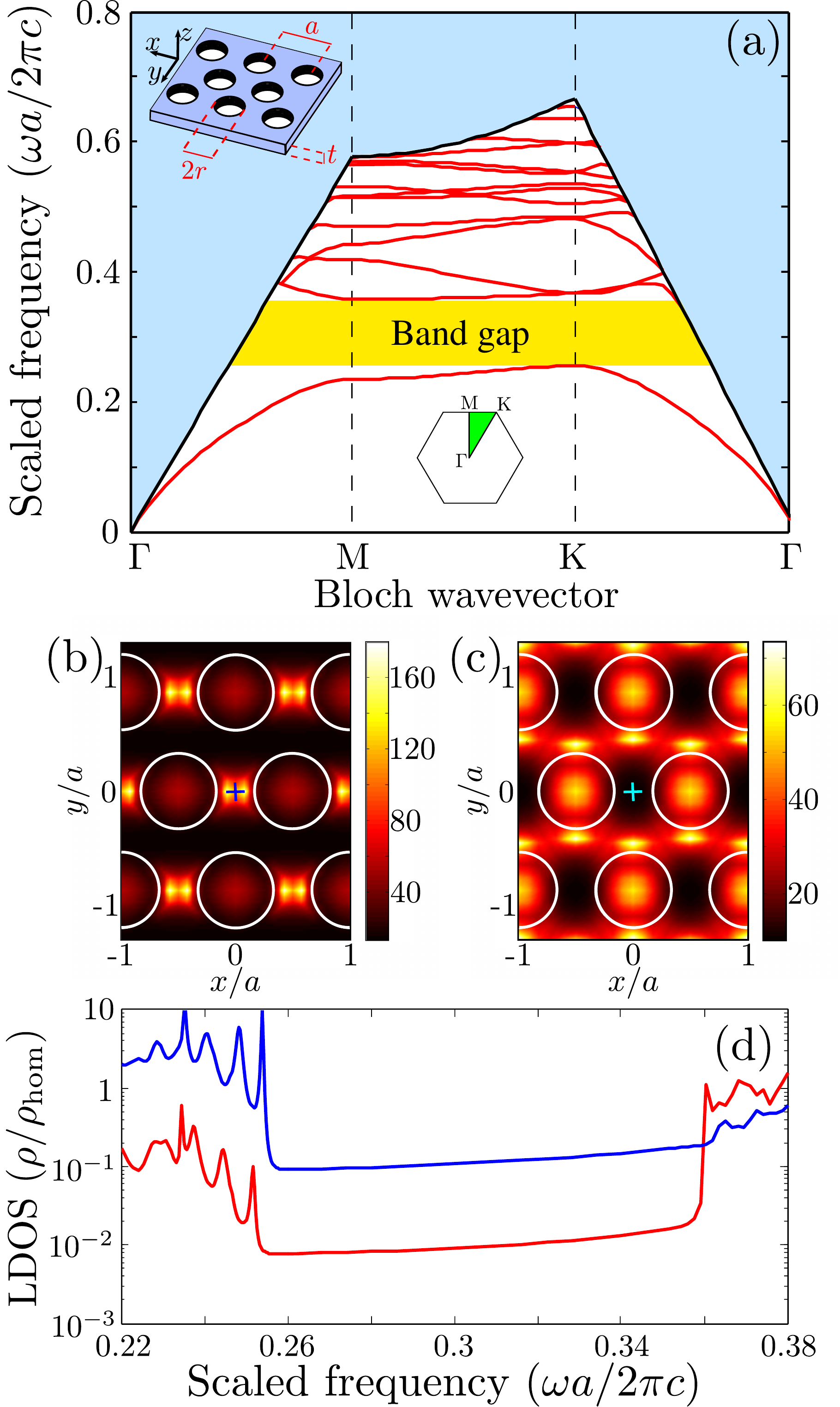}
 \caption{(a) Band diagram of a triangular-lattice photonic-crystal membrane (shown in the left inset) for TE-like modes with membrane refractive index $n=3.5$ (corresponding to GaAs), hole radius to lattice constant ratio of $r/a=1/3$, and membrane thickness $t=2a/3$. The high-symmetry points of the Brillouin zone are shown in the bottom inset. The light blue shaded area indicates the continuum of unbounded radiation modes inside the light cone, and the yellow shading shows the 2D photonic band-gap area. (b)-(c) Spatial map in the $x$-$y$ ($z=0$) plane of the inhibition factor $\rho_{\rm hom}/\rho$ at a scaled frequency of $\omega a/2\pi c= 0.2838$ for (b) an $x$-dipole and (c) a $y$-dipole  where $\rho_{\rm hom} = n \omega^2/3 \pi^2 c^3$ is the density of state for a homogeneous medium of GaAs. (d) Frequency dependence of the  LDOS $\rho/\rho_{\rm hom}$  plotted on a logarithmic scale for an $x$-dipole (red curve) and a $y$-dipole (blue curve) positioned in a photonic-crystal membrane at the crosses shown in (b) and (c). }
\label{2DPCall}
\end{figure}

A multitude of photonic crystals with different crystal symmetries or material compositions and based on different fabrication strategies have been studied over the years since photonic crystals were first proposed \cite{Bykov1975, Yablonovitch1987PRL, John1987PRL}. Here we are mainly concerned with photonic-crystal membranes made of GaAs as illustrated schematically in Fig.\ \ref{2DPCall}(a). These structures have proven very well suited for quantum-optics experiments since they can be fabricated with high precision and InGaAs quantum dots with good optical properties can be naturally  incorporated. For a thorough review of photonic crystals including details on fabrication, optical experiments, and numerical modeling, see \citet{Busch2007PhysRep} and \citet{JoannopoulosBook}.

The optical modes of a photonic crystal obey Bloch's theorem, i.e., the electric field satisfies \cite{JoannopoulosBook}
\begin{equation}
\mathbf E_\mathbf k (\mathbf r + \mathbf R) = \mathbf E_\mathbf k (\mathbf r) e^{i \mathbf k \cdot \mathbf R},
\end{equation}
where $\mathbf k$ is a Bloch wave vector and $\mathbf R$ is any vector in the Bravais lattice spanning the periodic photonic crystal. The Bloch modes are generally strongly dispersive, as can be seen in the dispersion diagram in Fig.\ \ref{2DPCall}(a) that plots the frequency versus $\mathbf k$ for various high-symmetry directions in the reciprocal lattice. This strong structural dispersion can be employed for tailoring light propagation, and, e.g., slow light can be obtained, as considered in further detail in Sec.\ \ref{Section-waveguide-QED}. A photonic band gap opens if Bragg scattering is so pronounced that no modes exist for a range of frequencies. A complete photonic band gap inhibiting all modes for any propagation direction and polarization can only be obtained in photonic crystals with periodicity in all three dimensions, and even then only for very high refractive-index contrast and certain crystal lattices. Importantly, even in lower-dimensional photonic-crystal structures, pronounced pseudo-gaps exist that modify major parts of the optical modes. Consequently, the density of optical states can be strongly modified, which is the basis for all quantum-optics applications of photonic crystals. Photonic-crystal membranes, cf.\ Fig.\ \ref{2DPCall}, have proven to be particularly well suited. In these structures the propagation of light in the plane of the membrane is suppressed due to the 2D photonic band gap, while leakage out of the structure is restricted to a narrow cone of wave vectors in the case of a  high-refractive-index-contrast semiconductor-air interface. In quantum-optics experiments the relevant leakage rate is that experienced by a quantum emitter in the nanostructure that is generally strongly dependent on emitter position and frequency. The photonic-crystal membranes have the obvious experimental advantage compared to full 3D photonic crystals that embedded quantum emitters can conveniently be probed by laser excitation from the top of the membrane, which enables spectroscopy on a single emitter. Furthermore, very mature planar fabrication methods can be readily employed to obtain high-quality photonic crystals.

\begin{table*}[t]
  \centering
  \begin{tabular}{ | l | l | l | l |}
    \hline
    (a) Micropillar cavity
    &
    (b) Photonic-crystal cavity
    &
    (c) Nanobeam cavity
    &
    (d) Microdisk cavity
    \\
    &&&
    \\
	\multicolumn{1}{ |c| }{	
    \begin{minipage}{0.16\textwidth}
      \includegraphics[height=2.6cm]{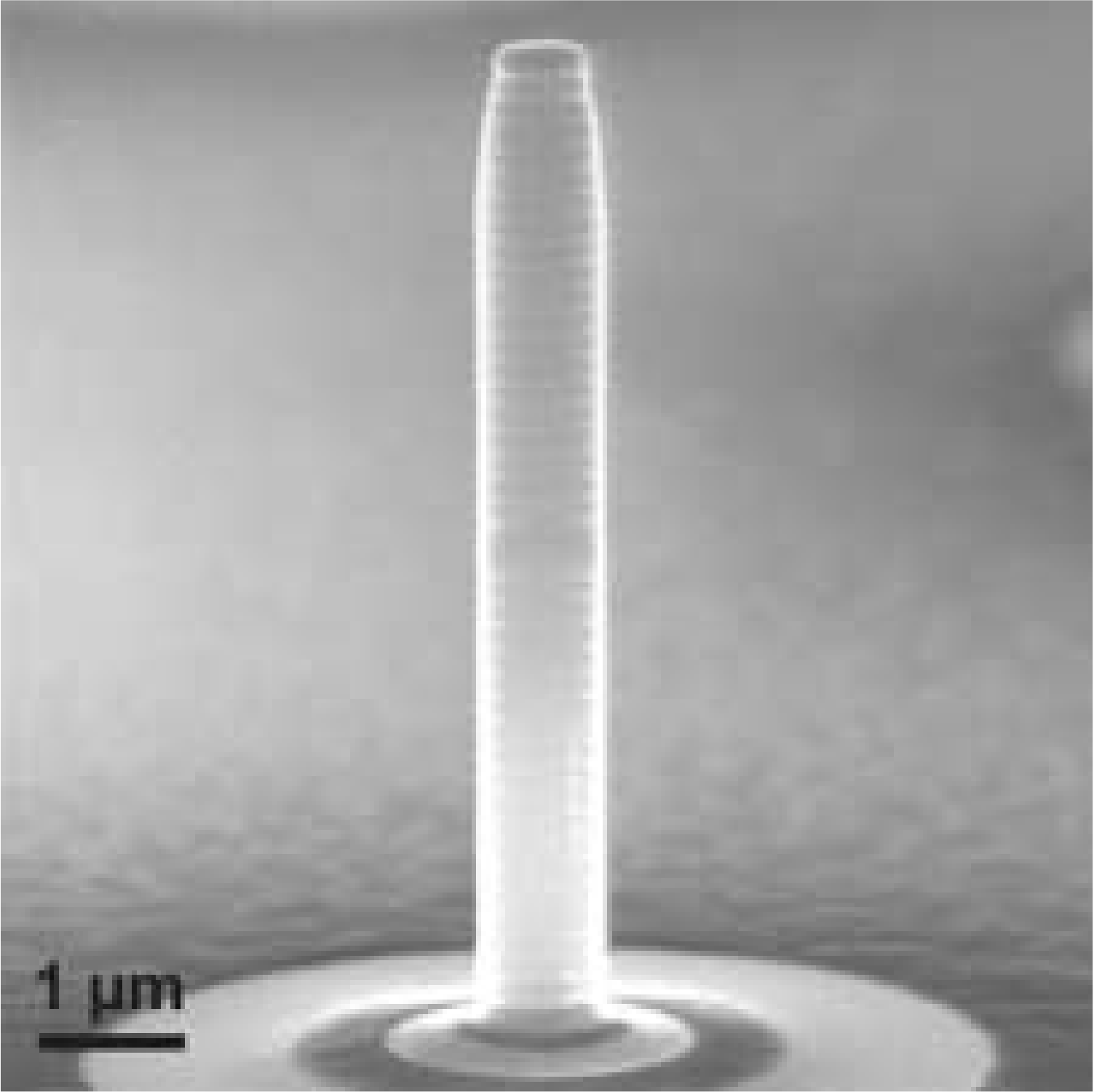}
    \end{minipage}
    }
    &
	\multicolumn{1}{ c| }{	
    \begin{minipage}{0.24\textwidth}
    \includegraphics[height=2.6cm]{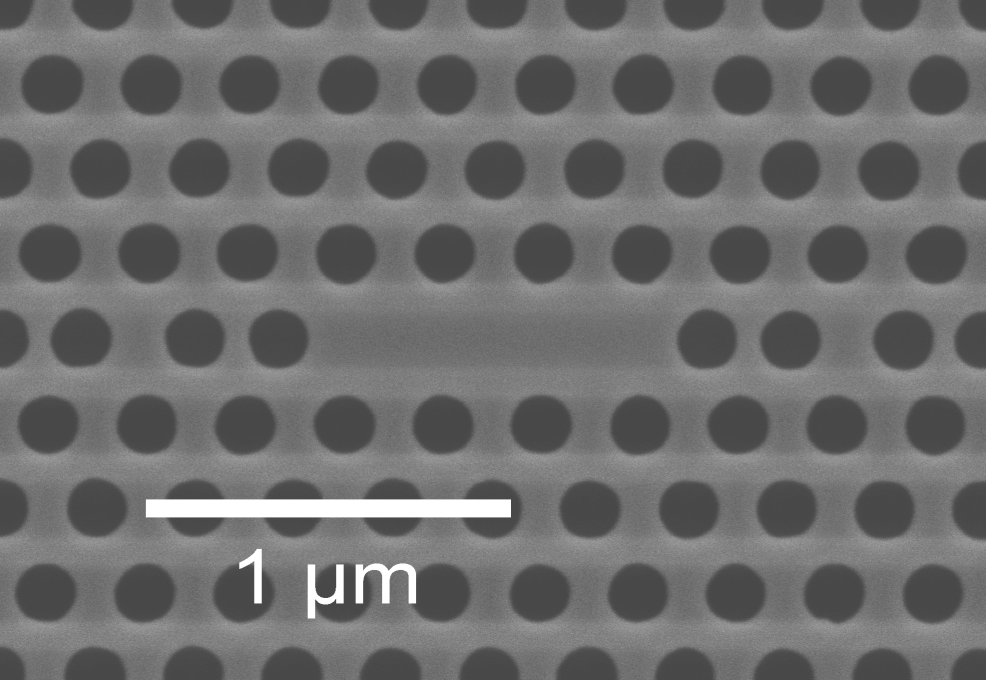}
    \end{minipage}
    }
    &
	\multicolumn{1}{ c| }{	
    \begin{minipage}{0.28\textwidth}
    \includegraphics[height=2.6cm]{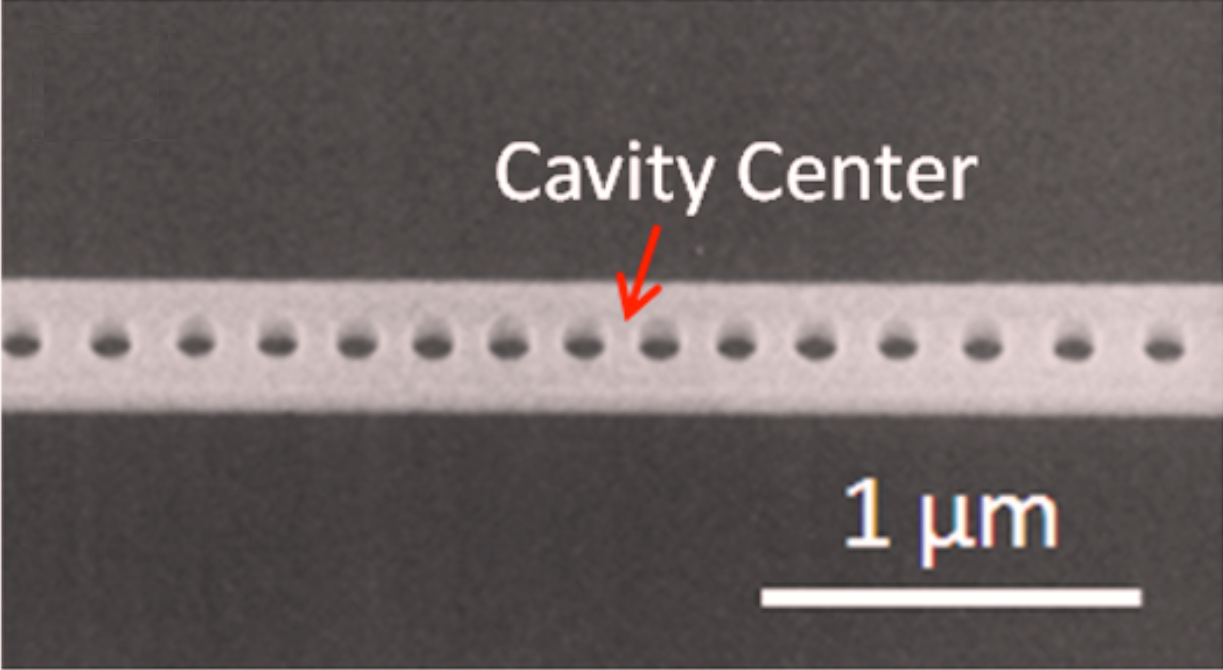}
    \end{minipage}
    }
    &
	\multicolumn{1}{ c| }{	
    \begin{minipage}{0.22\textwidth}
    \includegraphics[height=2.6cm]{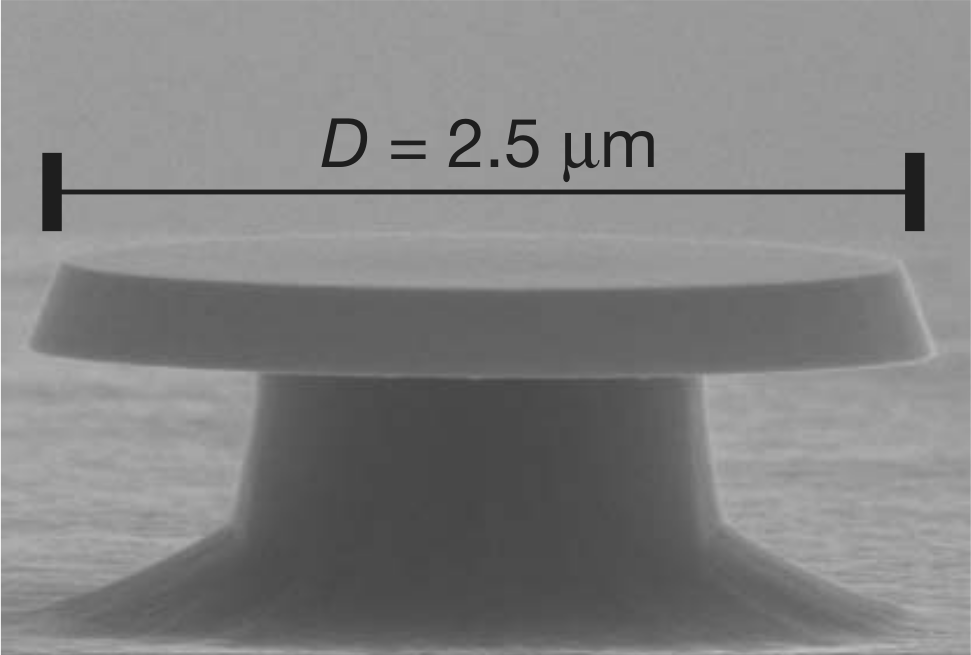}
    \end{minipage}
    }
    \\
    &&&
    \\
  	\multicolumn{1}{ |c| }{	
    \begin{minipage}{0.16\textwidth}
      \cite{Reithmaier2004Nature}
    \end{minipage}
    }
    &
    &
	\multicolumn{1}{ c| }{	
    \begin{minipage}{0.28\textwidth}
    \cite{Ohta2011APL}
    \end{minipage}
    }
    &
	\multicolumn{1}{ c| }{	
    \begin{minipage}{0.22\textwidth}
    \cite{Srinavasan2007Nature}
    \end{minipage}
    }
    \\
    &&&
    \\
	\multicolumn{1}{ |c| }{	
    \begin{minipage}{0.16\textwidth}
      \includegraphics[height=3cm]{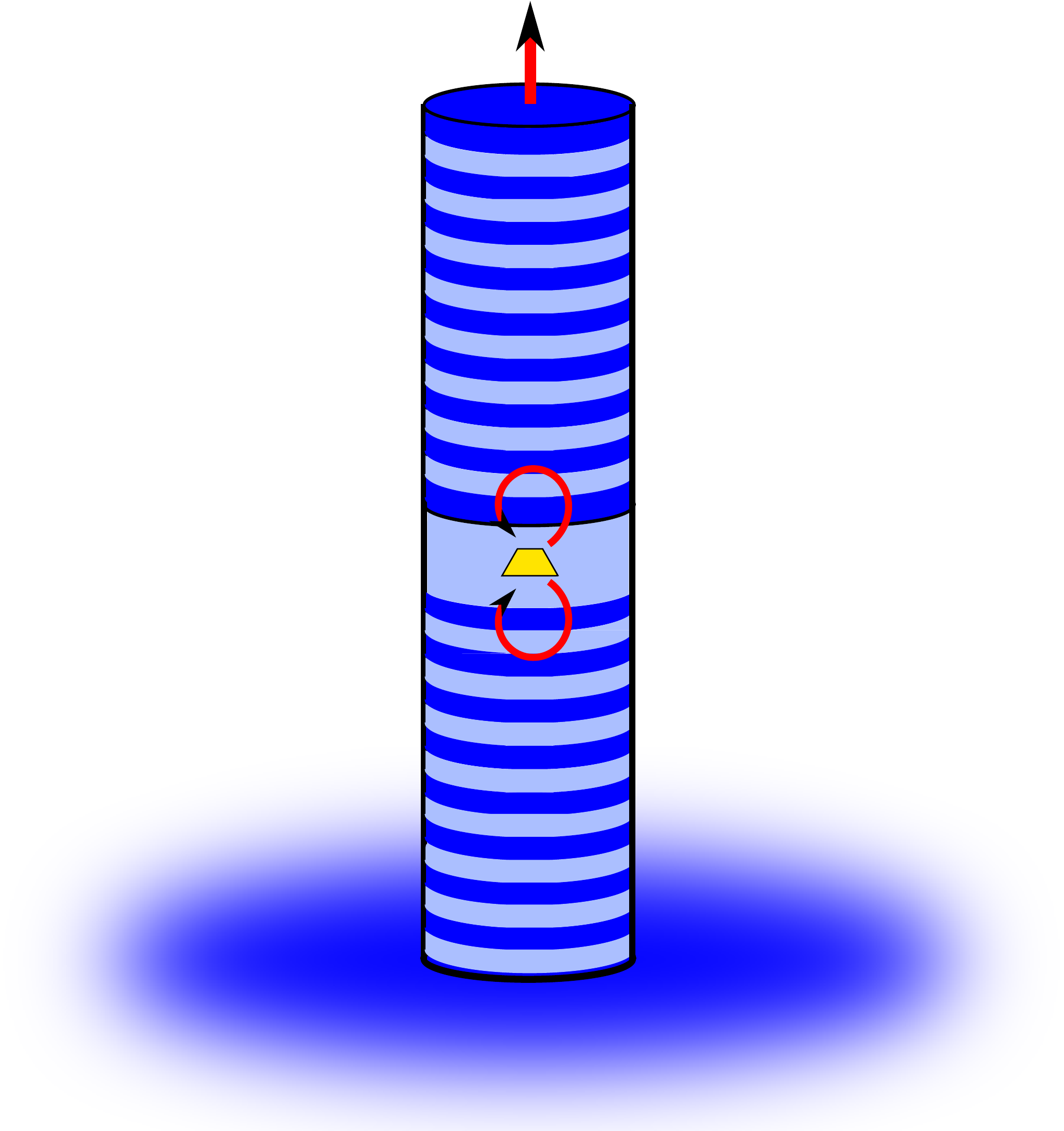}
    \end{minipage}
    }
    &
	\multicolumn{1}{ c| }{	
    \begin{minipage}{0.24\textwidth}
    \includegraphics[height=2.4cm]{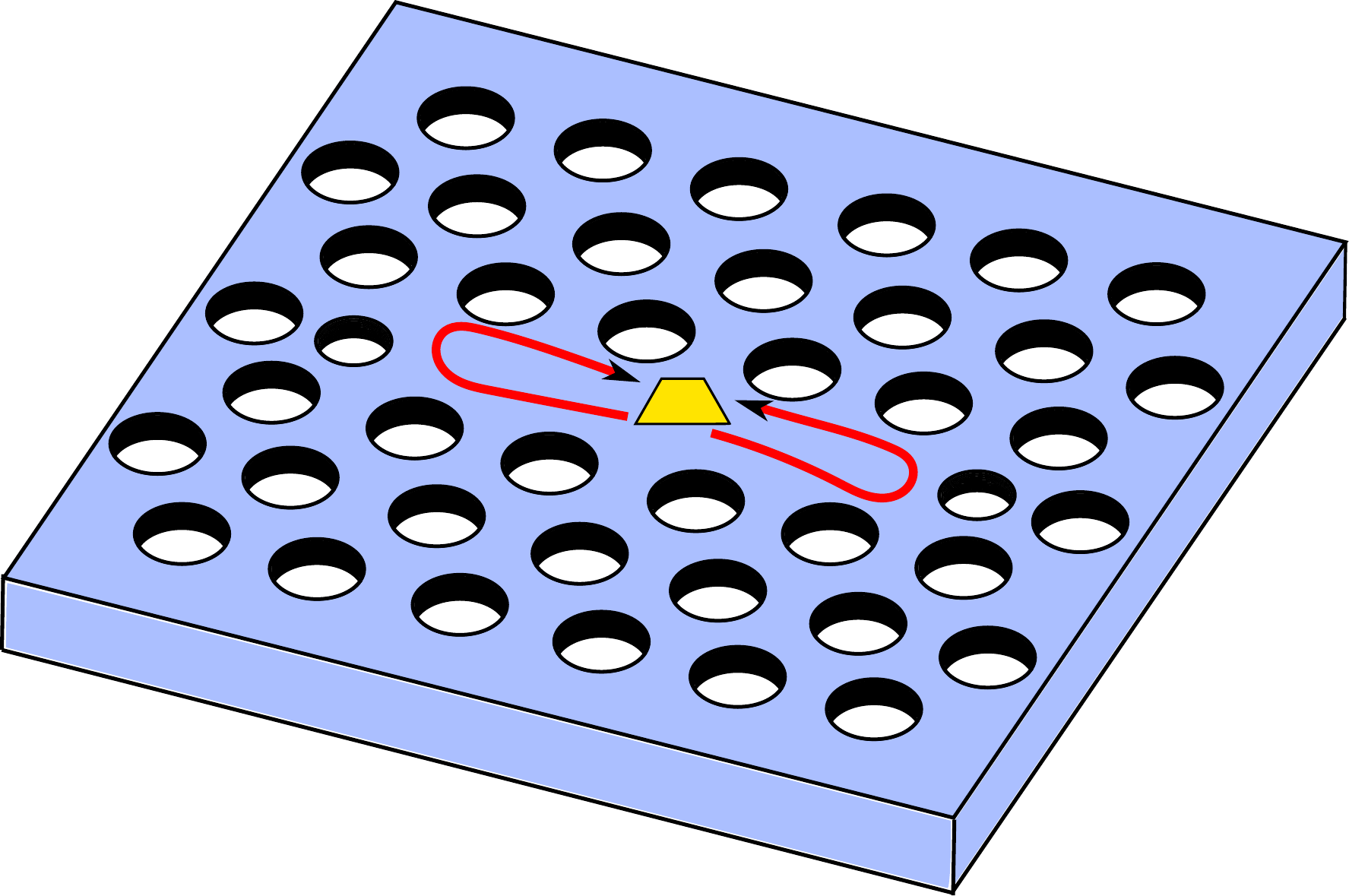}
    \end{minipage}
    }
    &
	\multicolumn{1}{ c| }{	
    \begin{minipage}{0.28\textwidth}
    \includegraphics[height=2.4cm]{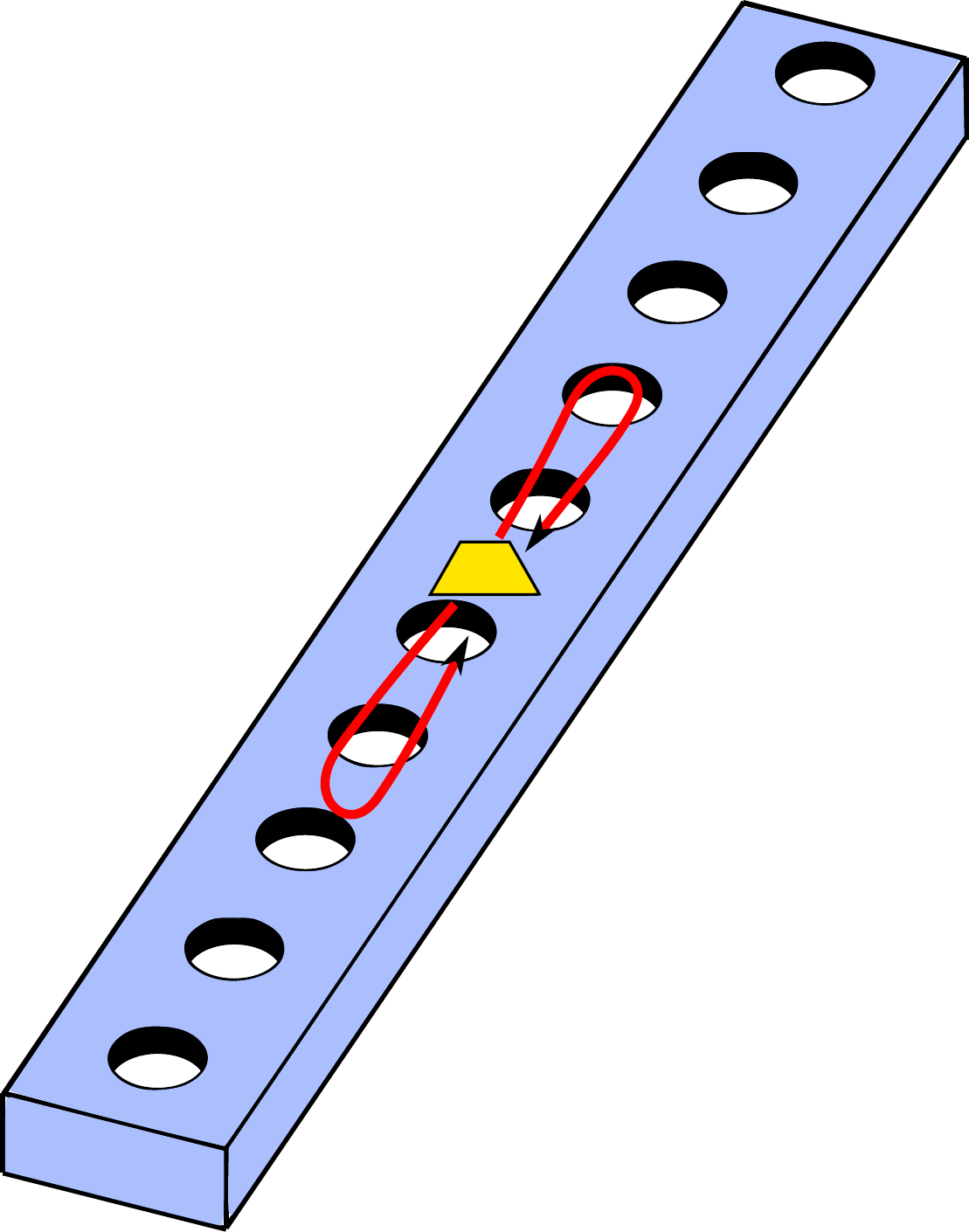}
    \end{minipage}
    }
    &
	\multicolumn{1}{ c| }{	
    \begin{minipage}{0.22\textwidth}
    \includegraphics[height=2.0cm]{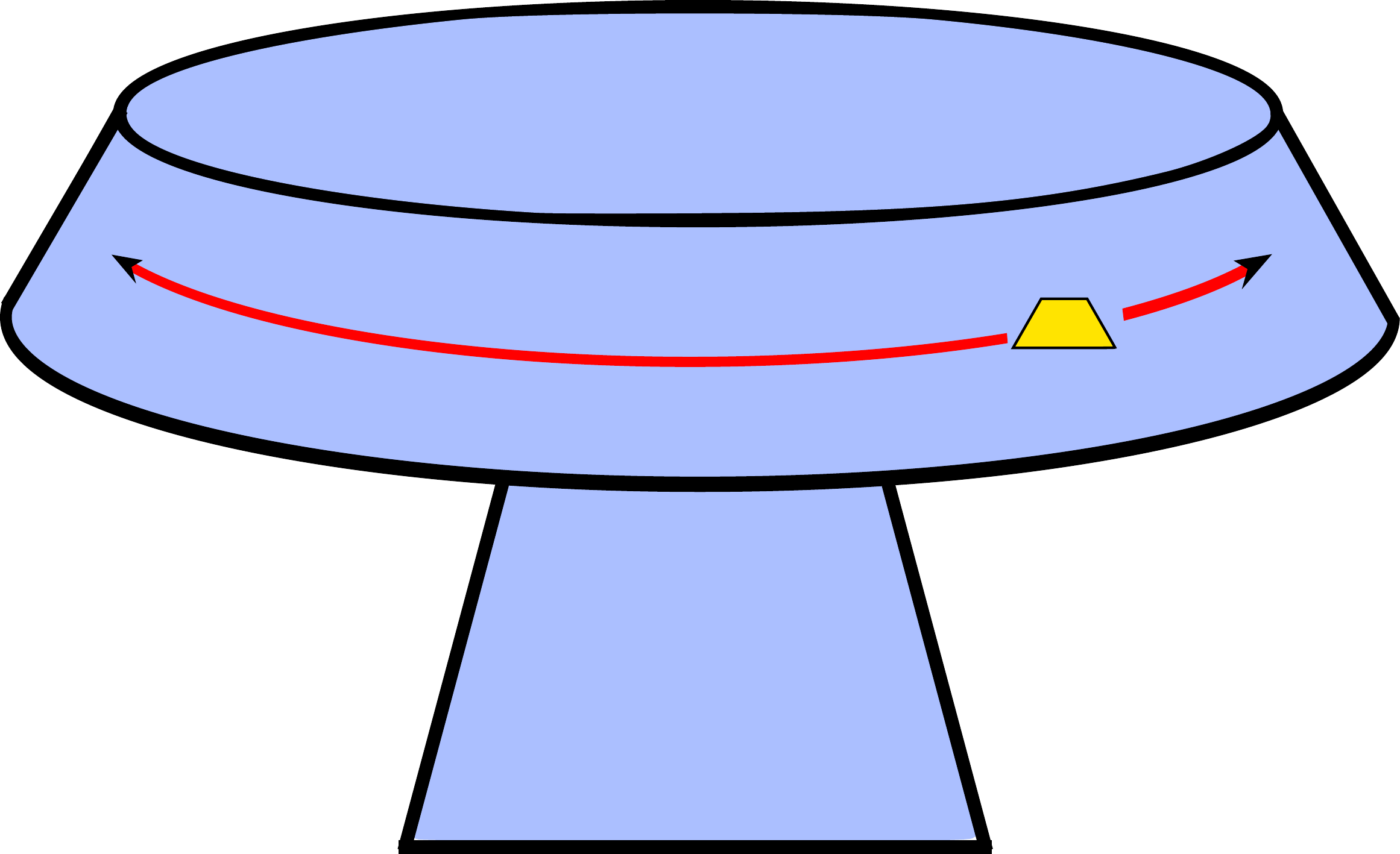}
    \end{minipage}
    }
    \\
	\multicolumn{1}{ |c| }{	
    \begin{minipage}{0.16\textwidth}
        \bea
            g/2\pi = 4~{\rm GHz}\nonumber\\
            \kappa/2\pi \sim 5~{\rm GHz}\nonumber\\
            \gamma/2\pi \sim 4~{\rm GHz}^*\nonumber\\
            Q \sim 6 \times 10^4\nonumber
        \eea
    \end{minipage}
    }
    &
	\multicolumn{1}{ c| }{	
    \begin{minipage}{0.24\textwidth}
        \bea
            g/2\pi = 22~{\rm GHz}\nonumber\\
            \kappa/2\pi \sim 11~{\rm GHz}\nonumber\\
            \gamma/2\pi < 0.1~{\rm GHz}\nonumber\\
            Q \sim 3 \times 10^4\nonumber
        \eea
    \end{minipage}
    }
    &
	\multicolumn{1}{ c| }{	
    \begin{minipage}{0.28\textwidth}
        \bea
            g/2\pi = 27~{\rm GHz}\nonumber\\
            \kappa/2\pi = 13~{\rm GHz}\nonumber\\
            \gamma/2\pi \sim 3~{\rm GHz}^*\nonumber\\
            Q = 3 \times 10^4\nonumber
        \eea
    \end{minipage}
    }
    &
	\multicolumn{1}{ c| }{	
    \begin{minipage}{0.22\textwidth}
        \bea
            g/2\pi = 3~{\rm GHz}\nonumber\\
            \kappa/2\pi = 1~{\rm GHz}\nonumber\\
            \gamma/2\pi \sim 0.6~{\rm GHz}\nonumber\\
            Q = 4 \times 10^5\nonumber
        \eea
    \end{minipage}
    }
    \\
    &&&
    \\
  	\multicolumn{1}{ |c| }{	
    \begin{minipage}{0.16\textwidth}
      \cite{Loo2010APL}
    \end{minipage}
    }
    &
	\multicolumn{1}{ c| }{	
    \begin{minipage}{0.24\textwidth}
    \cite{Hennessy2007Nature}
    \end{minipage}
    }
    &
	\multicolumn{1}{ c| }{	
    \begin{minipage}{0.28\textwidth}
    \cite{Ohta2011APL}
    \end{minipage}
    }
    &
	\multicolumn{1}{ c| }{	
    \begin{minipage}{0.22\textwidth}
    \cite{Srinavasan2007Nature}
    \end{minipage}
    }
    \\
    \hline
\end{tabular}
  \caption{Overview of nanophotonic cavities. Each panel displays a scanning electron micrograph of a real device along with a sketch illustrating the operational principle for a quantum emitter coupling to the structure. Furthermore, state-of-the-art experimental results are listed, to be discussed in Sec.\ \ref{Section-cavity-QED}. (a) Micropillar cavity (micrograph reprinted with permission from \citet{Reithmaier2004Nature}). The Bragg stack above and below the center of the pillar confines light to the central region as shown in the inset. (b) Modified photonic-crystal L3 cavity implemented in a membrane. The photonic band gap localizes light in the defect region and the schematic shows how a quantum dot preferentially emits into the cavity mode. (c) A nanobeam cavity (micrograph reprinted with permission from \citet{Ohta2011APL}). The cavity mode is confined by 1D Bragg diffraction in the high-refractive-index material of the nanorod. (d) Microdisk cavity (micrograph reprinted with permission from \citet{Srinavasan2007Nature}). The emitter couples to optical modes that travel circularly around the microdisk. The asterisk ($^*$) indicates that in these cases $\gamma$ was extracted from spectral rather than time-resolved data, i.e., it will be largened by dephasing processes.} \label{nanoCavities}
\end{table*}

\begin{table*}[t]
  \centering
  \begin{tabular}{ | l | l | l | }
    \hline
    (a) Photonic-crystal waveguide
    &
    (b) Photonic nanowire
    &
    (c) Plasmonic nanowire
    \\  & & \\
	\multicolumn{1}{ |l| }{	
    \begin{minipage}{0.4\textwidth}
      \includegraphics[height=0.6\textwidth]{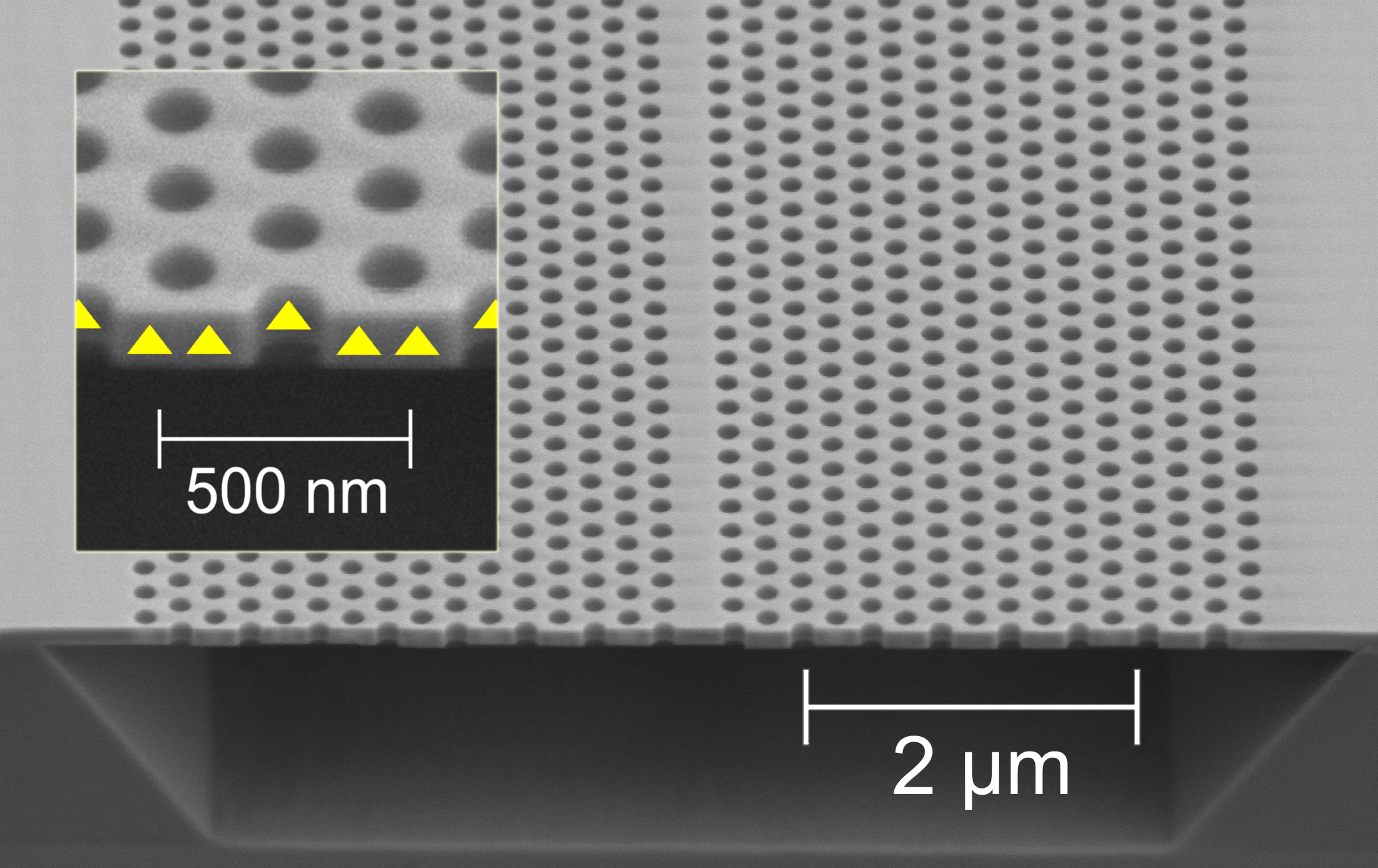}
    \end{minipage}
    }
    &
    \multicolumn{1}{ c| }{	
    \begin{minipage}{0.25\textwidth}
      \includegraphics[height=0.96\textwidth]{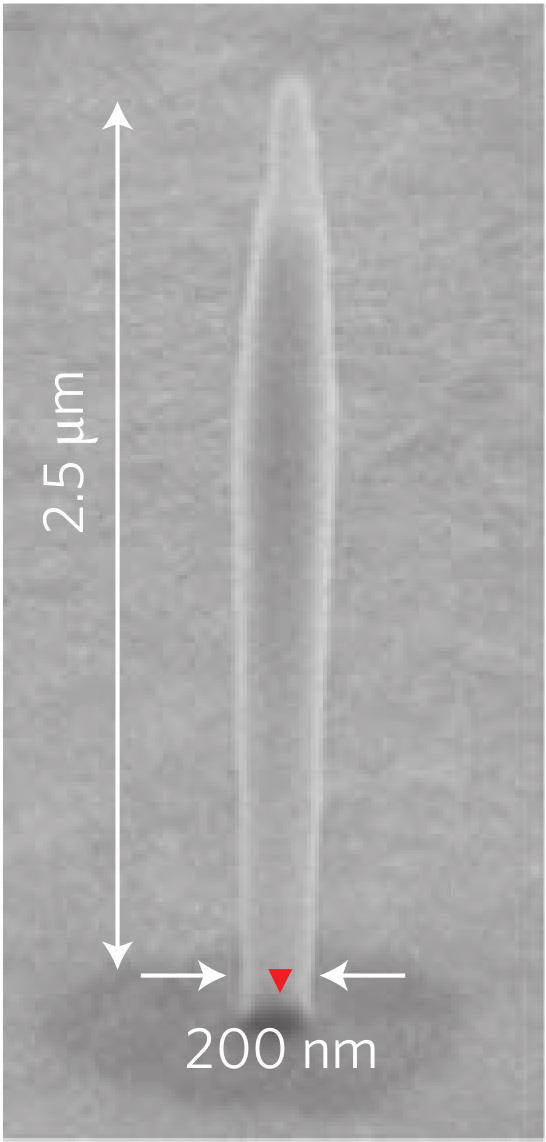}
    \end{minipage}
    }
    &
    \multicolumn{1}{ c| }{	
    \begin{minipage}{0.25\textwidth}
      \includegraphics[height=0.96\textwidth]{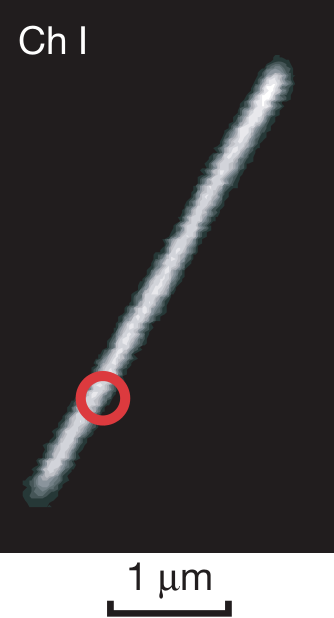}
    \end{minipage}
    }
    \\ & & \\
  	\multicolumn{1}{ |c| }{	
    \begin{minipage}{0.4\textwidth}
      \cite{Sapienza2010Science}
    \end{minipage}
    }
    &
	\multicolumn{1}{ c| }{	
    \begin{minipage}{0.25\textwidth}
    \cite{Claudon2010NPHOT}
    \end{minipage}
    }
    &
	\multicolumn{1}{ c| }{	
    \begin{minipage}{0.25\textwidth}
    \cite{Akimov2007Nature}
    \end{minipage}
    }
    \\ & & \\
	\multicolumn{1}{ |c| }{	
    \begin{minipage}{0.4\textwidth}
      \includegraphics[height=3cm]{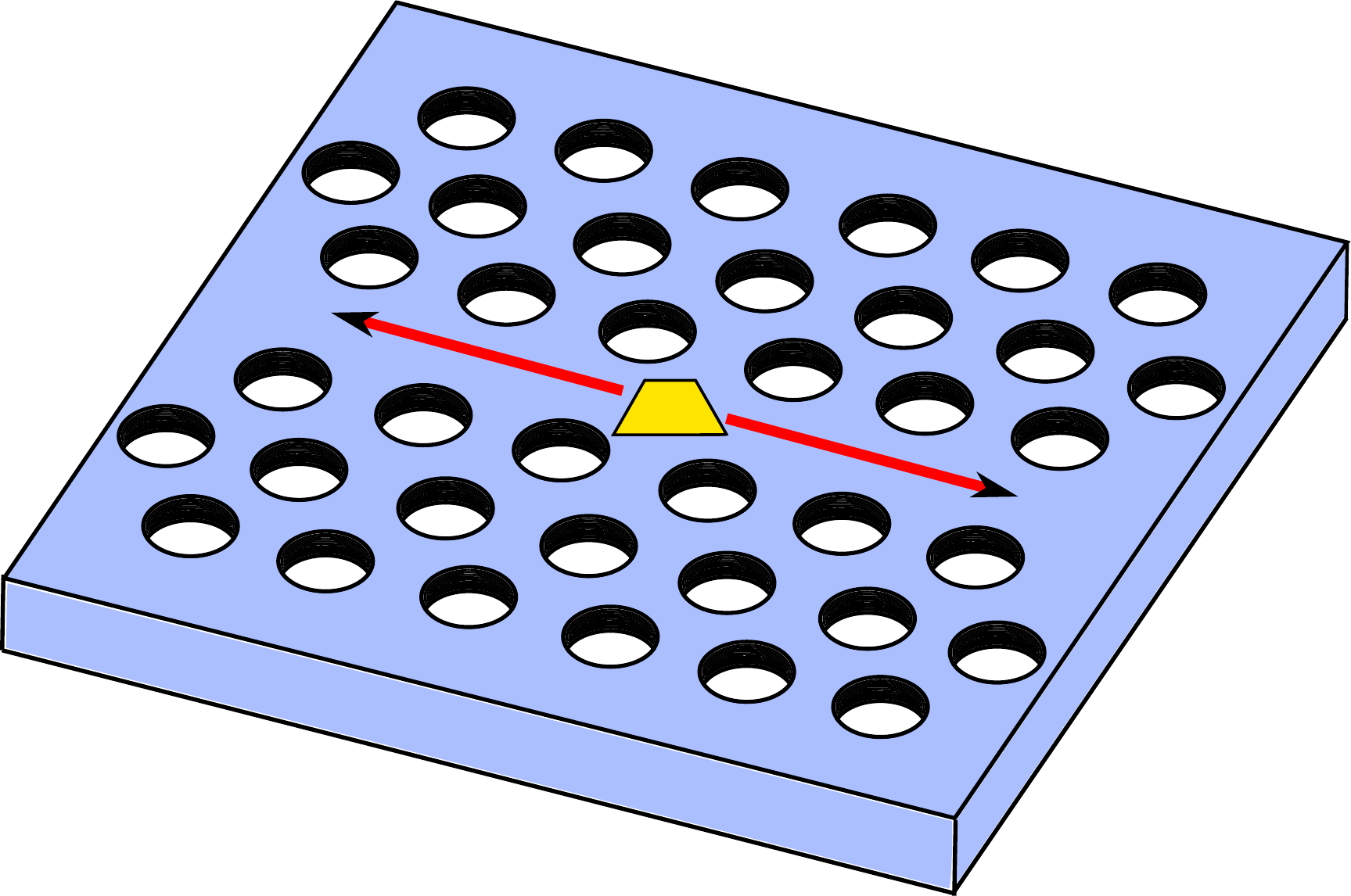}
    \end{minipage}
    }
    &
	\multicolumn{1}{ c| }{	
    \begin{minipage}{0.25\textwidth}
      \includegraphics[height=3cm]{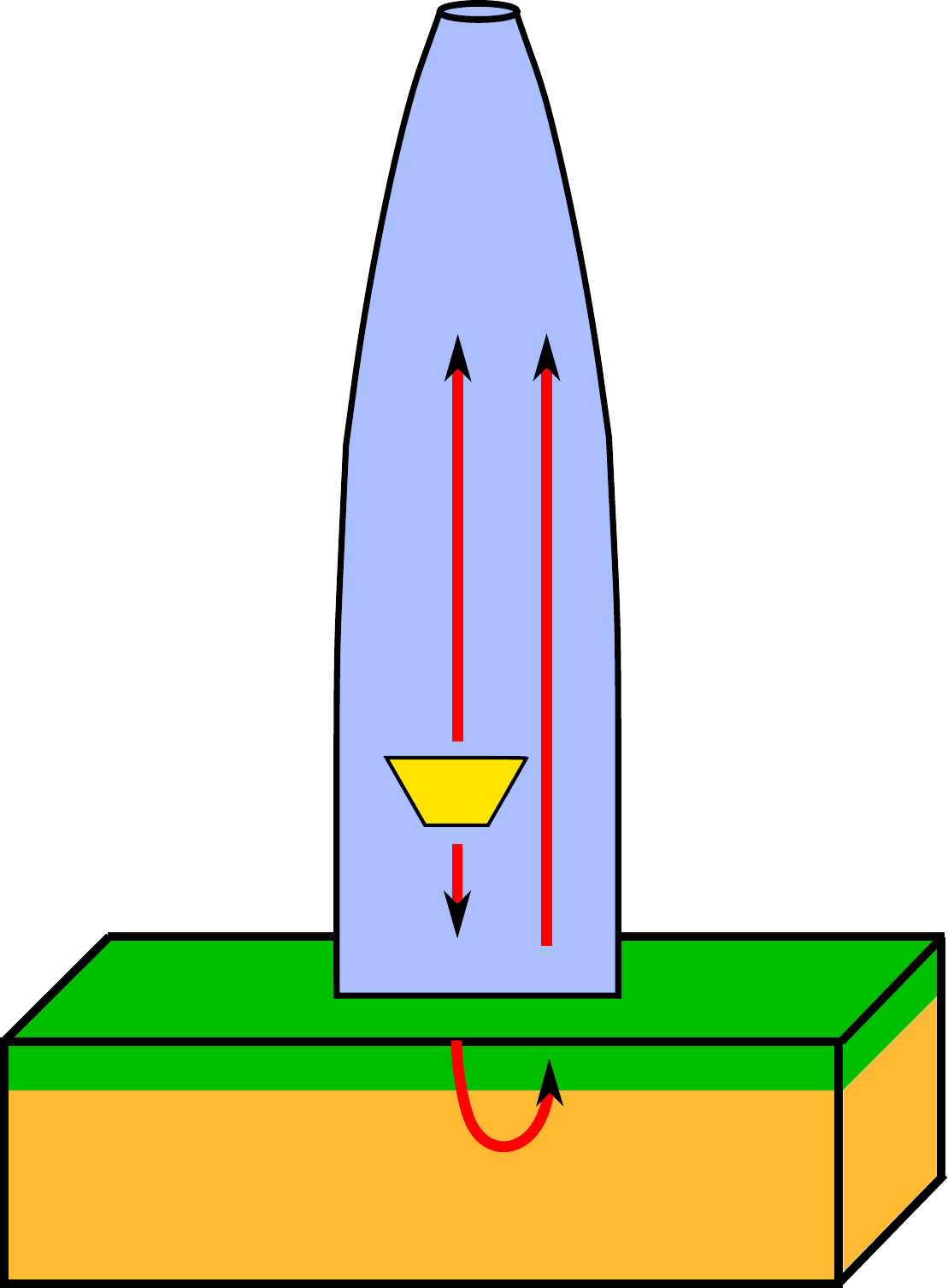}
    \end{minipage}
    }
    &
	\multicolumn{1}{ c| }{	
    \begin{minipage}{0.25\textwidth}
      \includegraphics[height=2.6cm]{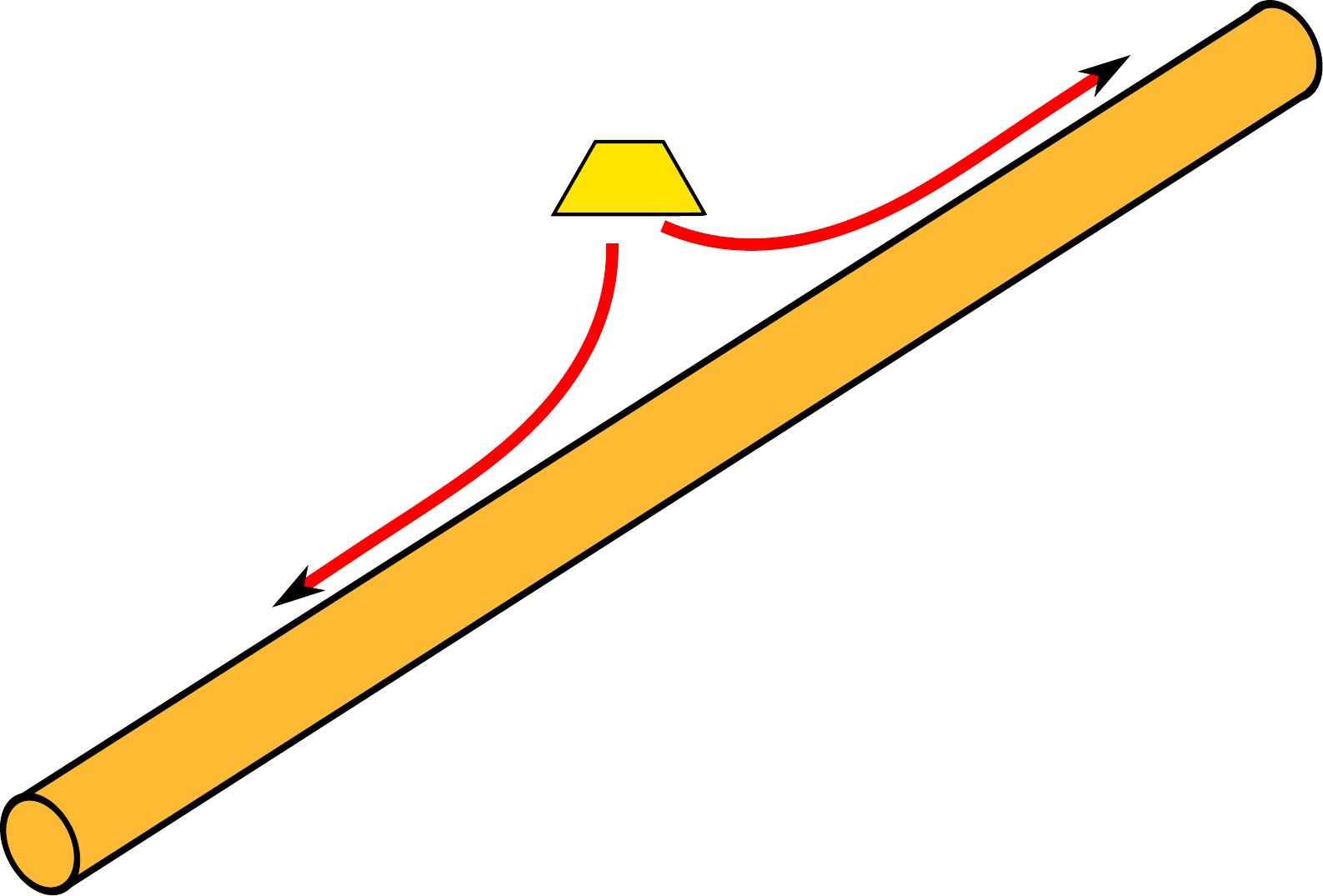}
    \end{minipage}
    }
    \\ & & \\
	\multicolumn{1}{ |c| }{	
    \begin{minipage}{0.4\textwidth}
    \bea
    \beta^e &= 0.98\,\dag\nonumber\\
    F_\text{P}^e &= 5 \,\dag\nonumber\\
    \beta^t &\rightarrow 1 \ddag\nonumber\\
    F_\text{P}^t &\rightarrow \infty \,\ddag\nonumber
    \eea
    \end{minipage}
    }
    &
	\multicolumn{1}{ c| }{	
    \begin{minipage}{0.25\textwidth}
    \bea
    \beta^e &\gtrsim 0.72 \,\dag\nonumber\\
    F_\text{P}^e &= 1.5 \,\dag\nonumber\\
    \beta^t &\sim 0.95 \,\ddag\nonumber\\
    F_\text{P}^t &\sim 1.7\,\ddag\nonumber
    \eea
    \end{minipage}
    }
    &
	\multicolumn{1}{ c| }{	
    \begin{minipage}{0.25\textwidth}
    \bea
    \beta^e &\sim 0.7 \,\dag\nonumber\\
    F_\text{P}^e &= 2.5 \,\dag\nonumber\\
    \beta^t &\sim 1\,\ddag\nonumber\\
    F_\text{P}^t   &\sim 500\,\ddag\nonumber
    \eea
    \end{minipage}
    }
    \\ & & \\
  	\multicolumn{1}{ |c| }{	
    \begin{minipage}{0.4\textwidth}
      \dag \cite{Arcari2013InPrep}\\
      \ddag \cite{MangaRao2007PRB}
    \end{minipage}
    }
    &
  	\multicolumn{1}{ c| }{	
    \begin{minipage}{0.25\textwidth}
      \dag \cite{Claudon2010NPHOT}\\
      \ddag \cite{Bleuse2011PRL}
    \end{minipage}
    }
    &
  	\multicolumn{1}{ c| }{	
    \begin{minipage}{0.25\textwidth}
      \dag \cite{Akimov2007Nature}\\
      \ddag \cite{Chang2006PRL}
    \end{minipage}
    }\\
    \hline
  \end{tabular}
  \caption{Overview of nanophotonic waveguides. Each panel displays a scanning electron micrograph of a real device with a sketch illustrating the operational principle for a quantum emitter coupling to the structure, as well as experimentally measured (theoretically calculated) $\beta$-factors and Purcell factors, $F_\text{P}^e$ and $\beta^e$ ($F_\text{P}^t$ and $\beta^t$), as discussed in Sec.\ \ref{Section-waveguide-QED}. The bottom panel shows the references for the Purcell and $\beta$-factors. (a) Photonic-crystal waveguide membrane containing a single layer of quantum dots in the center of the membrane (indicated as yellow triangles) (micrograph reprinted with permission from \citet{Sapienza2010Science}). (b) Photonic nanowire made of GaAs containing an InAs quantum dot (illustrated as a triangle). The upper part of the nanowire is tapered in order to maximize the outcoupling efficiency and a gold mirror is embedded in the structure below the nanowire (micrograph reprinted with permission from \citet{Claudon2010NPHOT}). The stated value of the experimental $\beta$-factor represents a lower bound determined from estimates of the total out-coupling efficiency. (c) Plasmonic nanowire made of silver that is coupled to a single quantum dot (placed within the red circle) (micrograph reprinted with permission from \cite{Akimov2007Nature}).}\label{nanoWaveguides}
\end{table*}

In quantum-optics experiments with dipole emitters, the LDOS is the relevant quantity that determines spontaneous emission and more generally the local light-matter interaction strength. It specifies the number of optical states at the frequency $\omega$ per frequency bandwidth and volume as experienced by the emitter, and is defined as
\be
 \rho(\mathbf{r}_0,\omega,{\bf \hat{e}_{d}}) =  \sum_{{\bf k}} \left| {\bf \hat{e}_{d}} \cdot {\bf u}_{\bf k}(\mathbf{r}_0) \right|^2 \delta \left(\omega - \omega_\mathbf k \right),\label{eq:LDOS-def}
\ee
where $\mathbf{r}_0$ is the position of the emitter and ${\bf \hat{e}_{d}}$ a unit vector specifying the orientation of the transition dipole moment. The normalized mode functions, ${\bf u}_{\bf k}({\bf r})$, in which the electric field is expanded, are described in Sec.\ \ref{LDOS-EOMs}. The LDOS is obtained by summing over all the mode functions with eigenfrequencies $ \omega_\mathbf k$ that enter through a Dirac delta function $\delta \left(\omega - \omega_\mathbf k \right)$. The LDOS was first introduced by \citet{Sprik1996EL} and subsequently evaluated for the case of photonic crystals by \citet{Busch1998PRB}.

The LDOS in a photonic crystal is strongly dependent on position and orientation. Figure \ref{2DPCall}(d) shows an example of a calculation of the frequency dependence of the LDOS for a GaAs photonic-crystal membrane. A frequency region with a strongly suppressed LDOS is revealed, which corresponds to the 2D photonic band gap for the transverse-electric-like (TE-like) fields in which case $(E_x, E_y, H_z)$ are the dominant field that are even functions with respect to $z=0$. In the band gap, a dipole oriented in the plane of the membrane will therefore only radiate weakly via the small residual coupling to the non-guided radiation modes that are found above the light line in the dispersion diagram in Fig.\ \ref{2DPCall}(a). The high refractive index of GaAs ensures that the coupling to radiation modes is strongly suppressed. Indeed Fig.\ \ref{2DPCall}(b) shows that the LDOS can be inhibited by up to a factor of $160$  relative to that of a homogenous medium of GaAs. Furthermore, the periodic spatial dependence of the inhibition factor in the photonic-crystal lattice is displayed for the two orthogonal in-plane dipole orientations that are relevant for quantum dots. While both dipoles are found to be suppressed at all positions compared to the case of a homogeneous medium, a strong anti-correlation is generally observed, i.e., two orthogonally polarized dipoles would observe a very different LDOS. The ability to suppress the coupling to unwanted leaky modes by more than two orders of magnitude is the main asset of photonic-crystal membranes in quantum-optics experiments, which was first realized by \citet{Koenderink2006JOSAB}. As discussed in Secs.\ \ref{Section-waveguide-QED} and \ref{Section-cavity-QED}, by introducing waveguides and cavities in the photonic crystals, the coupling to a preferred optical mode can be enhanced significantly. As a consequence, photonic crystals offer the possibility to tailor the local light-matter interaction strength by 3-4 orders of magnitude by combining suppression of unwanted leaky modes with enhancement of a single mode. With this approach a nearly perfect photon-emitter interface may be obtained; the physics of which is the main objective of the present review.

\subsection{Photonic cavities}
Resonating light in a cavity provides a way of enhancing light-matter interaction since the coupling to one localized mode can be strongly enhanced compared to all other modes. In photonic nanostructures the cavities can have very small mode volumes, which enhances the interaction strength. The two decisive cavity-QED figures of merit are the quality factor $Q$ and the effective mode volume $V$ of the localized quasi-mode. In the following, various nanophotonic approaches to high-$Q$ and small-$V$ cavities are reviewed.

Micropillar cavities have been widely exploited as high-$Q$ cavities. They can be fabricated by epitaxial growth of alternating layers of refractive indices $n_1$ and $n_2$ (e.g., GaAs and AlGaAs) each of thickness $\lambda/4 n_i$, $i=1,2,$ and by subsequently etching a micropillar with a typical diameter of a few microns and a height of around \SI{10}{\micro\meter}. The alternating layers form a Bragg mirror with a reflectivity that is controlled by the number of layers. An extended spacer layer of length, e.g., $\lambda/n_1$ in between two such Bragg mirrors forms a highly localized cavity mode where quantum dots can be positioned, see Table \ref{nanoCavities}(a). The two Bragg mirrors are often grown with different numbers of layers in order to make a one-sided cavity consisting of a highly-reflecting mirror and an out-coupling mirror with an optimized transmission. The diameter of the micropillar is chosen to restrict the lateral extension of the guided mode confined in the high-index material, which leads to a small mode volume. Choosing the diameter below 1-\SI{2}{\micro\meter}, however, is found to significantly reduce $Q$ due to the sensitivity to sidewall roughness \cite{Gazzano2013NCOM}. Typical mode volumes accessible in micropillar cavities are at the level of $\sim 10 (\lambda/n)^3$ and tapered cavities have recently been proposed as a way of localizing light even better \cite{Lermer2012PRL}.

Highly localized cavity modes can be obtained by introducing defects in photonic-crystal membranes and a multitude of different design possibilities have been explored in the literature. Indeed this flexibility of being able to tailor a cavity mode by controlling the geometry of the photonic crystal and the defect area provides a very important asset of the photonic-crystal platform. One of the most successful and important designs so far is the L3 cavity that is obtained from a 2D triangular lattice in a photonic-crystal membrane by leaving out three holes, cf.\ Table\ \ref{nanoCavities}(b), leading to a mode volume of less than $\sim (\lambda/n)^3$. It was realized that the $Q$-factor can be very significantly boosted by more than an order of magnitude by displacing the holes at each end of the cavity by just a fraction of a lattice constant, whereby leakage to radiation modes can be strongly suppressed \cite{Akahane2003Nature}.  This remarkable sensitivity to the detailed design reflects the large potential of photonic-crystal cavities. A cavity $Q$-factor of $2\times 10^6$ has been observed experimentally in modified L3 cavities in silicon at a wavelength of $\SI{1.55}{\micro\meter}$ \cite{Lai2014APL}. Lower $Q$-factors are generally observed for GaAs cavities since they are matched to the shorter wavelength of InGaAs quantum dots of around \SI{950}{\nano\meter} and therefore more sensitive to fabrication imperfections, residual scattering and absorption due to the embedded quantum dots, as well as bulk and surface absorption, which is more pronounced at shorter wavelengths \cite{Michael2007APL}. A cavity $Q$-factor of $3\times 10^4$ has been reported in a GaAs L3 cavity containing a single quantum dot \cite{Hennessy2007Nature}.

A number of different photonic-crystal cavity structures have been considered that potentially have higher $Q$-factors including waveguide heterostructures \cite{Song2005NatMat} for which a $Q$-factor of $5.5\times 10^4$ in a GaAs cavity with quantum dots has been reported \cite{Ota2011PRL}, nanobeam cavities \cite{Ohta2011APL}, or even random Anderson-localized modes in photonic-crystal waveguides due to naturally occurring fabrication imperfections \cite{Topolancik2007PRL,Smolka2011NJP}. The nanobeam cavity displayed in Table \ref{nanoCavities}(c) is a 1D photonic crystal consisting of etched holes in a narrow and thin membrane with a central defect area defining the cavity region. Finally microdisk cavities, see Table \ref{nanoCavities}(d), constitute another class of resonators where the trapping of light is due to total internal reflection rather than Bragg scattering. In a microdisk cavity the light is confined to the rim of the disk in a whispering-gallery mode that can have a very high $Q$, generally, however, at the expense of larger mode volume compared to cavities employing Bragg scattering. Table \ref{nanoCavities} summarizes the various cavity configurations including figures of merit in relation to their usage for cavity-QED experiments. Such cavity-QED experiments employing single quantum dots are reviewed in Sec.\ \ref{Section-cavity-QED}.

\subsection{Nanophotonic waveguides}

Nanophotonic waveguides enable routing photons between different locations on an optical chip, and could therefore be applied in integrated quantum networks for connecting stationary qubits (e.g., encoded in quantum dots) with flying qubits (photons). Nanophotonic waveguides can be highly dispersive, which may be employed for enhancing light-matter interaction as an alternative to the cavity case. In a waveguide, the quantum emitter can efficiently and over a wide bandwidth couple single photons directly to a propagating optical mode for immediate applications without the necessity of coupling out of a localized mode, which would be the case for high-$Q$ cavities. In reality low-$Q$ extended cavities may constitute the best compromise in order to obtain both a large Purcell enhancement and highly efficient and broadband coupling.

Waveguides can readily be implemented in photonic-crystal membranes.  The most simple design consists of leaving out a row of holes along the $\Gamma$-$K$ direction of a triangular lattice, which is referred to as a W1 waveguide, see Table \ref{nanoWaveguides}(a).  Figure \ref{Photonic-crystal-waveguide-dispersion-relation}(a) in Sec.\ \ref{Sec:Purcel-waveguide} displays the projected dispersion diagram for a photonic-crystal waveguide membrane. It is found that bands arise in the 2D band gap region corresponding to propagating modes that are spatially confined to the waveguide. These guided modes are highly dispersive; i.e., the group velocity of light $v_\text{g}(\omega) = \left| {\bf \nabla}_{\bf k} \omega \right| $ varies with frequency. The  group velocity can be strongly reduced in a photonic-crystal waveguide and in general be tailored by controlling the structural parameters of the photonic lattice. Furthermore, since the waveguide modes appear below the light line in the dispersion diagram (cf.\ Fig.\ \ref{Photonic-crystal-waveguide-dispersion-relation}), an ideal photonic-crystal waveguide features lossless propagation. In reality, unavoidable fabrication imperfections (cf.\ Sec.\ \ref{Section-fab-imperfections}) induce a finite leak rate of coupling vertically out of the waveguide. Two main features of the photonic-crystal waveguides are important for their quantum-optics applications: the ability to suppress the coupling to radiation modes due to the band gap and the simultaneous enhancement of coupling to the photonic-crystal waveguide mode that is enhanced by slow light. This is considered in detail in Sec.\ \ref{Section-waveguide-QED}. A number of other potential applications of photonic-crystal waveguides for integrated photonics have been envisioned that exploit photonic band-gap guiding of light. For a review of this topic see \citet{JoannopoulosBook}.

Photonic nanowires constitute another class of waveguide-type photonic nanostructure. They are extended cylindrical rods of a high-refractive-index dielectric material, see Table \ref{nanoWaveguides}(b). Photonic nanowires can either be fabricated by etching or be epitaxially grown, e.g., in the form of GaAs nanowires with embedded quantum dots. For diameters in the range of 150-\SI{300}{\nano\meter} the nanowire supports a single well-confined guided mode at $950 \: \mathrm{nm}$ \cite{Friedler2009OE}, which is a typical wavelength for InGaAs quantum dots. The bandwidth of this mode is up to \SI{70}{\nano\meter}, which makes nanowires insensitive to the wavelength of the employed emitter. As opposed to photonic-crystal waveguides, the mode of the nanowire is weakly dispersive, which means that slow-light enhancement of light-matter interaction cannot be obtained. Therefore applications of nanowires in quantum optics are primarily based on the ability to suppress leaky modes rather than enhancing a single mode, which makes them less sensitive to fabrication imperfections at the expense of a less efficient  light-matter coupling strength compared to photonic-crystal waveguides.

Finally, metallic waveguide structures can guide light even for a wire diameter that is just a tiny fraction of the optical wavelength in free space. Such sub-wavelength guidance can potentially lead to a very large electromagnetic field strength per photon, as required in quantum-optics experiments. In metallic waveguides, the light is guided in the form of surface-plasmon polaritons that are surface waves confined to a metal-dielectric interface that propagate along the interface while being exponentially damped in the direction perpendicular to the interface. An example of a plasmonic nanowire is shown in Table \ref{nanoWaveguides}(c). Just like dielectric nanowires, the response of plasmonic nanowires is broadband, with the added benefit that the strong field confinement and dispersion of plasmons imply that the light-matter interaction strength can be strongly enhanced. Unfortunately, plasmonic nanowaveguides are inherently lossy due to absorption in the metal, so while they may be useful for locally enhancing light-matter interaction, the subsequent guiding of photons is preferably carried out in dielectric structures. To accommodate this point, plasmon-based quantum circuits have been proposed where the plasmon mode is adiabatically coupled to a dielectric waveguide \cite{Chang2006PRL}. Another possibility is to use a slot waveguide created by placing two plasmon waveguides next to each other where a plasmon mode can form in the gap region between the waveguides with enhanced propagation length. Plasmon nanowires can be fabricated by, e.g., chemical synthesis of crystalline structures, metal evaporation in trenches written by electron-beam lithography, or etching of thin metal films. While the theoretical potential of plasmonic nanostructures seems very promising \cite{Chang2006PRL}, the fabrication of high-quality plasmonic waveguides still remains a challenge given the design, material, and fabrication restrictions presently valid \cite{Chen2010OE}. Consequently, at present dielectric waveguides are advantageous for quantum-optics applications.

\subsection{The role of fabrication imperfections}
\label{Section-fab-imperfections}

Photonic nanostructures can be fabricated by a range of different techniques using either top-down approaches such as electron-beam lithography and etching or bottom-up approaches based on self-assembly of nanoscale objects.  Almost all the examples of nanostructures discussed here are fabricated by patterning semiconductor substrates by electron-beam lithography followed by reactive-ion and/or wet-chemical etching. These methods are compatible with self-assembled quantum dots and are also most commonly employed for fabricating integrated photonic circuits. A detailed account of the various fabrication methods is outside the scope of the present review, but for further details we refer to \citet{Busch2007PhysRep} and references therein. Common for all methods is that statistical imperfections are unavoidably introduced during the fabrication process, which potentially can reduce the functionality of the device by, e.g., inducing unwanted optical loss. The fabrication imperfections can be of various sorts including roughness, polydispersity, displacements, vacancies, etc. Such random imperfections are treated by means of statistical physics and eventually can lead to pronounced random multiple scattering of light even for sub-wavelength features because high-refractive-index-contrast composites induce large scattering cross sections. In state-of-the-art dielectric photonic nanostructures the amount of fabrication imperfections can be reduced to a standard deviation at the level of a few nanometers \cite{Garcia2013APL}.  The relevance of such an amount of imperfections is fully dependent on the actual device under consideration, and the role of fabrication imperfections should be addressed individually for each application.

Extensive theoretical and experimental work has been carried out on the role of fabrication imperfections in photonic-crystal waveguides \cite{Hughes2005PRL, Mazoyer2009PRL, Garcia2013APL}. Imperfections lead to two  effects: the backscattering of the propagating mode into the counter-propagating direction in the waveguide and out-of-plane scattering due to coupling to leaky modes that are above the light line. The former (latter) is characterized by an average length $\ell_{\rm back}$  ($\ell_{\rm leak}$) leading to a total extinction length $1/{\ell_{\rm ext}} = 1/\ell_{\rm back} + 1/\ell_{\rm leak}$ that determines how the average intensity decays along the waveguide. It should be emphasized that while a single realization of imperfections leads to a complex intensity speckle pattern, the extinction length predicts how light decays on average and is obtained after an ensemble average over configurations of disorder. The back-scattering length is strongly dispersive in the photonic-crystal waveguide and in a perturbative description was predicted to scale as $\ell_{\rm back} \propto v_\text{g}^2$ \cite{Hughes2005PRL}. In the context of 1D Anderson localization the back-scattering length is referred to as the localization length and has been measured to be below \SI{10}{\micro\meter} in the slow-light regime of state-of-the-art photonic-crystal waveguides \cite{Smolka2011NJP}, which has been the basis for experiments on cavity QED with Anderson-localized modes \cite{Sapienza2010Science}. The resulting average loss length associated with out-of-plane leakage was found to be around \SI{700}{\micro\meter}. By operating outside the slow-light regime, the propagation distance can be significantly increased. Alternatively, waveguides shorter than the back-scattering length can be applied, such that the strong light-matter interaction achievable in photonic waveguides can be exploited without suffering from Anderson-localization effects. This will be the setting of the description of waveguide quantum optics of Sec.\ \ref{Section-waveguide-QED}.

\section{Spontaneous emission of single photons from solid-state emitters in photonic nanostructures}
\label{Section-spontaneous-emission}

\begin{figure}
\includegraphics[width= \columnwidth]{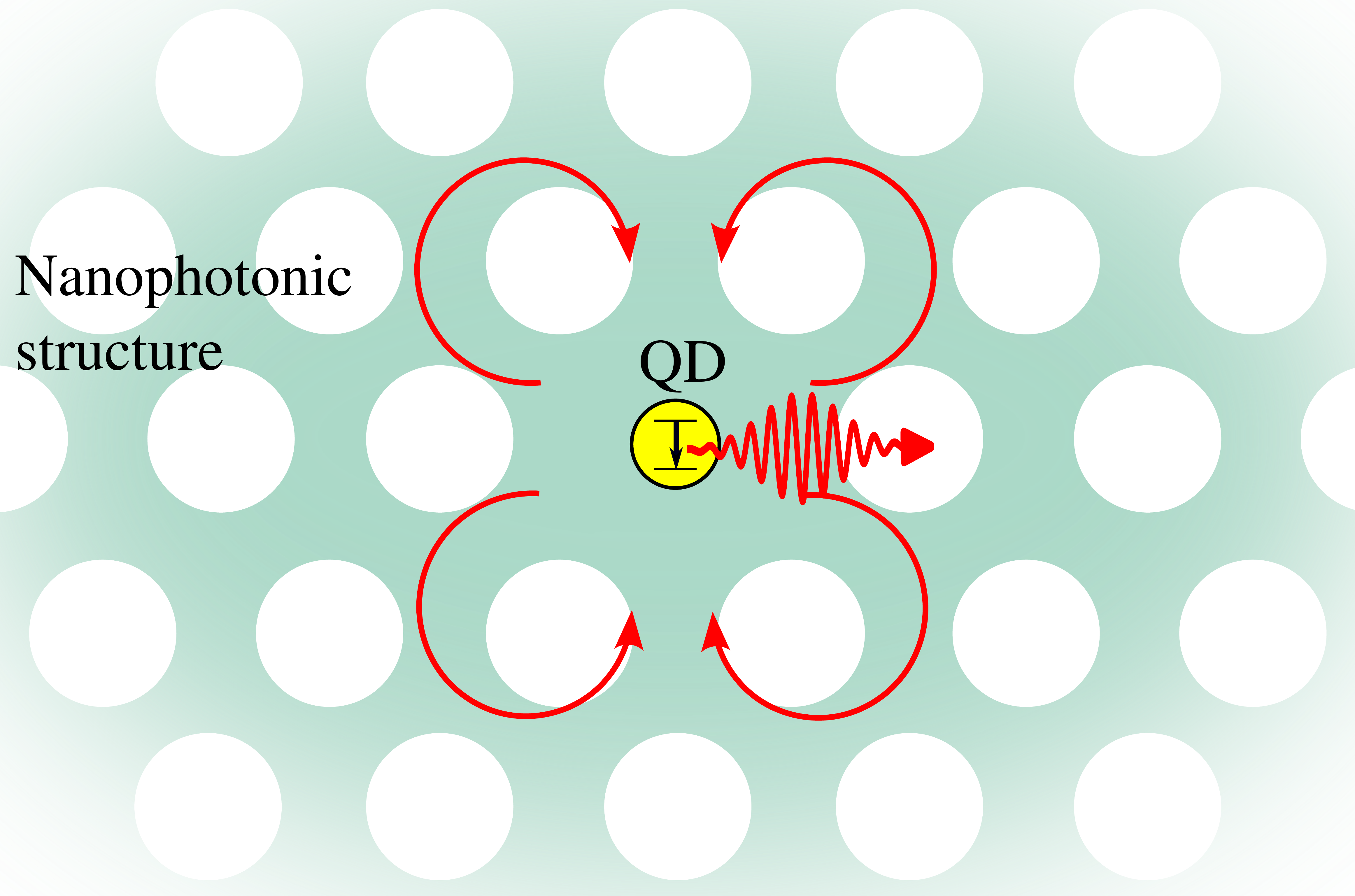}
 \caption{Illustration of spontaneous emission from a two-level emitter in a photonic medium. Single photons (red wave packets) are emitted one at a time from a quantum dot (QD) placed in a nanophotonic structure. For a full description of spontaneous emission in nanophotonics, the potential influence of the back action (indicated by red arrows) of the vacuum electric field on the quantum emitter must be accounted for, which can lead to intricate non-exponential dynamics and strongly modified emission spectra.}
\label{SE_backActionSchem}
\end{figure}

Much of the physics discussed in the present review concerns a single quantum emitter that is emitting (or absorbing) a single quantum of light, a photon, into (or from) an engineered photonic environment.
In this section we describe a general theoretical framework for this setting, which is illustrated in Fig.\ \ref{SE_backActionSchem}, and encompasses photon emission in any inhomogeneous photonic nanostructure. The quantum emitter is described as a two-level dipole emitter but the formalism can be readily generalized to describe more complex level schemes, cf.\ the discussion of quantum-dot level schemes in Sec.\ \ref{sec:sst:decaydynamics}. The coupling to the photonic nanostructure is described by the projected LDOS, which quantifies the magnitude of vacuum fluctuations responsible for spontaneous emission of photons. The LDOS is linked to the electric-field Green's tensor that contains all details about the spatio-temporal properties of the electromagnetic field.

In the following, the basics of the LDOS formalism is discussed in some detail, since it is a powerful and generally applicable framework for QED in any nanophotonic structure and appears not to be textbook material. Key quantities addressed include the spontaneous-emission spectrum and dynamics beyond the Markov approximation. The Markov approximation holds when radiation back action from the reservoir is negligible, but may break down in, e.g., photonic crystals and nanocavities due to the large local electromagnetic field strength and the strongly dispersive behavior. Photonic nanostructures can be employed for tailoring the quantum vacuum and thereby to control fundamental QED-processes and interactions like spontaneous emission, the Lamb shift, Casimir forces, or dipole-dipole interactions \cite{Milonni}.  In particular we explain in the following how the Lamb shift of electronic transitions can be strongly modified in photonic nanostructures. Another benefit of the LDOS formalism is that the spontaneous-emission rates are rigorously introduced. This contrasts more specific QED models, such as the commonly used dissipative Jaynes-Cummings model of cavity-QED discussed in Sec.\ \ref{Section-cavity-QED}, where spontaneous-emission rates are included as phenomenological parameters.

\subsection{Equations of motion for spontaneous emission in the local-density-of-states description }
\label{LDOS-EOMs}

We consider a single two-level dipole emitter with the excited state $\left| e \right>$ and ground state $\left| g \right>$ separated by the transition frequency $\omega_0$ and coupled to a continuum of optical modes at frequencies $\omega_\mathbf k$. The total Hamiltonian in the rotating-wave approximation is
%
\bea \hat{H}_{\mathrm{tot}}= &&\sum_{{\bf k}} \hbar \omega_\mathbf k  \hat{a}_{{\bf
k}}^{\dagger} \hat{a}_{{\bf k}}  + \frac{1}{2}
\hbar \omega_{0} \hat{\sigma}_z  \nn \\
&& +  \sum_{{\bf k}}  \left[\hbar g_{{\bf k}} \hat{\sigma}_+ \hat{a}_{{\bf k}}e^{i (\omega_0 -
\omega_\mathbf k) t}  + \text{h.c.} \right], \label{H-tot} \eea
where $\hat{a}_{{\bf k}}$, $\hat{a}_{{\bf k}}^{\dagger}$ are annihilation and creation operators, respectively, for photons in the mode ${\bf k}$ and $\hat{\sigma}_z = \left| e \right> \left< e \right| - \left| g \right> \left< g \right|$, $\hat{\sigma}_+ = \left| e \right> \left< g \right| $ are the emitter-population inversion and raising operators, respectively. The coupling rate to each optical mode is $g_{{\bf k}}(\mathbf r_0) = i {\bf d} \cdot {\bf E}^*_{{\bf k}}(\mathbf r_0)/\hbar$, which contains the transition dipole moment of the emitter and the local electric field at the position of the emitter, $\mathbf{E_k}(\mathbf r_0)$, for the mode specified by $\mathbf k$ and there are two polarization components for each wave vector. $\text{h.c.}$ denotes the Hermitian conjugate of the preceding term. As mentioned in Sec.\ \ref{sec:sst:TransitionMatrixElement}, the $\mathbf{p}\cdot\mathbf{A}$ interaction Hamiltonian is often used in solid-state quantum optics and leads also to Eq.\ (\ref{H-tot}). Dephasing processes are omitted in the following but Markovian dephasing can be readily implemented through a non-Hermitian term in the Hamiltonian, which is valid in a Monte-Carlo wave-function description of spontaneous emission \cite{Meystre2007elements}. Furthermore, a thorough discussion of field quantization in a dielectric medium is outside the scope of the present account, but further information can be found in \cite{Wubs2003PRA}.

The equation of motion for the excited state of the emitter can be formulated as
\bea
\frac{\partial c_\text{e}}{\partial t} = &-& \sum_{{\bf k}}  \left| g_{{\bf k}}(\mathbf r_0) \right|^2  \int_0^t \text{d}t' c_\text{e}(t') e^{i \Delta_{{\mathbf k}}
(t-t')} \nn \\
&-& i \sum_{{\bf k}} g_{{\bf k}}^*(\mathbf r_0)
c_{\text{g},{\bf k}}(0) e^{i \Delta_{{\mathbf k}} t}, \label{ce-equation}
\eea
with $\Delta_{{\mathbf k}} = \omega_{0} - \omega_{{\mathbf k}}$ and where the combined quantum state of light and matter has been expanded as $\left| \Psi(t) \right> = c_\text{e}(t) \left|e,\{0 \} \right> + \sum_{{\bf k}} c_{\text{g},{\bf k}}(t) \left|g,\{{1_{\bf k}} \} \right>$. Here $\left|\{0 \} \right> $ is the collective vacuum state of all modes in the radiation reservoir and $\left|\{1_{{\bf k}_j} \} \right> = \left|0_{{\bf k}_1},0_{{\bf k}_2},\ldots,1_{{\bf k}_j},\ldots \right>$ corresponds to one photon in the mode with wave vector $\mathbf{k}_j$. By the restriction of having only a single excitation in the system, the formalism is suitable for describing spontaneous emission or single-photon absorption. In photonic nanostructures the interaction strength and thus the dynamics depends on the position of the emitter, $\mathbf r_0$, but this argument has been omitted in $c_\text{e}(t)$ and $c_{\text{g},{\bf k}}(t)$ for simplicity of notation. The last term in Eq.\ (\ref{ce-equation}) vanishes in the case of spontaneous emission $(c_{\text{g},{\bf k}}(0)=0)$, but it can readily be included for describing single-photon absorption, as was detailed in \citet{Chen2011NJP} for the case of a 1D waveguide geometry.

The quantized electromagnetic field is expanded according to
\be
\hat{{\bf E}}({\bf r}, t) = \sum_{\bf k} \left[{\bf E}_{\bf k}({\bf r}) { \hat{a}_{\bf k}}e^{-i \omega_\mathbf k t } + {\rm h.c.} \right],
\ee
where ${\bf E}_{\bf k}({\bf r}) = \sqrt{\frac{\hbar \omega_k}{2 \epsilon_0}} {\bf u}_{\bf k}({\bf r})$ defines the field amplitude entering in the local coupling rate $g_{{\bf k}}(\mathbf r_0)$. The mode functions ${\bf u}_{\bf k}({\bf r})$ constitute a normalized set of basis functions  used to expand the field and obey the wave equation \cite{Yao2010LPR, NanoOpticsBook}
\be
{\bf \nabla} \times {\bf \nabla} \times {\bf u}_{\bf k}({\bf r}) - \frac{\omega_\mathbf k^2}{c^2} \epsilon({\bf r}) {\bf u}_{\bf k}({\bf r}) = 0,
\ee
with a generalized normalization condition given by $ \int \text{d}^3 {\bf r} \, \epsilon({\bf r}) {\bf u}_{\bf k}({\bf r}) \cdot {\bf u}_{\bf k'}^*({\bf r}) = \delta_{\bf k k'}.$ In photonic crystals a convenient choice of basis functions is Bloch modes. It is useful to express the electric field in terms of Green's tensor $\overleftrightarrow{G}$, which solve Maxwell's wave equation in the photonic medium for a delta-function source term \cite{Yao2010LPR, NanoOpticsBook}, i.e.,
\be
{\bf \nabla} \times {\bf \nabla} \times \overleftrightarrow{G}({\bf r},{\bf r'},\omega) - \frac{\omega^2}{c^2} \epsilon({\bf r}) \overleftrightarrow{G}({\bf r},{\bf r'},\omega) = \frac{\omega^2}{c^2}  \overleftrightarrow{I} \delta \left({\bf r} - {\bf r'} \right),
\ee
where $\overleftrightarrow{I}$ is the unit tensor. Here the case of no free charges is considered where the generalized transversal condition $\nabla \cdot (\epsilon({\bf r})\overleftrightarrow{G}) = {\bf 0}$ applies. Furthermore, longitudinal components in Green's tensor only contribute in absorptive media $(\mathrm{Im}(\epsilon) > 0)$ leading to non-radiative decay, which generally should be minimized in quantum-photonics applications. Green's tensor can be expanded in the mode functions according to
\be
 \overleftrightarrow{G}({\bf r},{\bf r}',\omega) = \sum_{\bf k} \omega^2 \frac{{\bf u}_{\bf k}({\bf r}) \otimes {\bf u}_{\bf k}^*({\bf r'})}{\omega_\mathbf k^2-\omega^2}, \label{Greens-def}
\ee
where $\otimes$ denotes the outer product. The electric field radiated by a dipole source at ${\mathbf r}_0$ can be obtained in the Green-tensor formalism from
\be
\hat{\mathbf E}(\mathbf{r},\omega) = \frac{1}{\epsilon_0}\overleftrightarrow{G}(\mathbf{r},\mathbf{r}_0,\omega) \cdot \hat{\mathbf d}(\omega),
\ee
where $\hat{\mathbf d}(\omega) = \mathbf d \, \left[\hat{\sigma}_+(\omega) + {\hat \sigma}_-(\omega)\right]$.

The equation of motion for the emitter, Eq.\ (\ref{ce-equation}), can now be expressed in terms of Green's tensor. It follows from Eq.~(\ref{Greens-def}) that \cite{Barnett1996}
\bea
\mathrm{Im} \left[ \hat{\bf e}_{\mathbf{d}}^* \cdot \overleftrightarrow{G}({\bf r},{\bf r},\omega) \cdot \hat{\mathbf{e}}_\mathbf{d} \right]  &=& \frac{\pi \omega}{2} \sum_{\bf k}  \left| \hat{\mathbf{e}}_\mathbf{d} \cdot {\bf u}_{\bf k}({\bf r}) \right|^2  \delta(\omega - \omega_\mathbf k)\nn\\
&=& \frac{\pi \omega}{2}\rho({\bf r},\omega,\hat{\mathbf{e}}_\mathbf{d}) \label{Im-G}
\eea
which links the summation over the radiation reservoir in Eq.\ (\ref{ce-equation}) to the LDOS, which was defined in Eq.\ (\ref{eq:LDOS-def}). It follows consequently that
\be
\frac{\partial c_\text{e}}{\partial t} = - \frac{d^2}{2 \epsilon_0 \hbar} \int_0^{\infty} \text{d} \omega \, \omega \rho(\mathbf{r_0},\omega,\hat{\mathbf{e}}_\mathbf{d}) \int_0^t \text{d}t' c_\text{e}(t') e^{i \Delta_{{\mathbf k}}
(t-t')}. \label{c_e-LDOS}
\ee
%
%
%
Equation (\ref{c_e-LDOS}) provides a complete description of spontaneous emission in any inhomogeneous photonic environment and accounts fully for the back-action of the vacuum electric field of the environment, which enters through the LDOS. The LDOS is the mode density that is ``seen'' by a dipole emitter. It is a classical quantity obtained by solving Maxwell's equations but it also determines the mode density of the vacuum electromagnetic field that is required to describe spontaneous emission. Thus, the variance of the projected vacuum electric field (the `vacuum fluctuations') is given by
\bea
\Delta (\hat{\mathbf E} (\mathbf r, \omega) \cdot \hat{\mathbf{e}}_\mathbf{d}) &=& \sum_\mathbf k |\hat{\mathbf{e}}_\mathbf{d} \cdot \mathbf E_\mathbf k (\mathbf r)|^2 \delta (\omega - \omega_\mathbf k)\\
&=& \frac{\hbar \omega}{2 \epsilon_0} \rho(\mathbf r, \omega, \hat{\mathbf{e}}_\mathbf{d}),
\eea
which is indeed proportional to the LDOS. It should be emphasized that while the applied wavefunction approach fully accounts for non-Markovian coupling to the photonic reservoir, any non-Markovian dephasing processes cannot be captured by such a formalism. Non-Markovian dephasing is most appropriately treated by density operator theory, cf.\ Sec.\ \ref{Section-Phonon-dephasing}.

It is instructive  to express Eq.\ (\ref{c_e-LDOS}) as
\be
\frac{\partial c_\text{e}(t)}{\partial t} =  -\int_0^t \text{d}t'
c_\text{e}(t') K({\mathbf r_0},t-t',\hat{\mathbf{e}}_\mathbf{d}), \label{eq-of-motion}\ee
with the introduction of the memory kernel \cite{Vats2002PRA}
\be
 K({{\mathbf r},t-t',\mathbf{e}}_\mathbf{d}) = \frac{d^2}{2 \epsilon_0 \hbar }  \int_0^{\infty} \text{d}\omega e^{i( \omega_0-\omega)(t-t')}
\omega \rho({\mathbf r},\omega,\hat{\mathbf{e}}_\mathbf{d}). \label{memory-kernel}\ee
The kernel expresses the memory of the radiation reservoir, i.e., to what extent the state of the reservoir at previous times $t'$ influences $c_\text{e}(t).$ A special and simple case is that of Wigner-Weisskopf theory that holds when $\omega \rho(\omega)$ varies insignificantly over the linewidth of the emitter such that the memory kernel can be approximated as
\be
K_{\mathrm{WW}}({\mathbf r},t-t',\hat{\mathbf{e}}_\mathbf{d}) \approx \frac{\pi d^2 \omega_0 \rho({\mathbf r},\omega_0,\hat{\mathbf{e}}_\mathbf{d})}{ \epsilon_0 \hbar }  \delta(t-t').
\ee
In this case the memory kernel is singular, i.e., the radiation reservoir is memoryless. This is also referred to as the Markov approximation where back-action from the radiation reservoir is negligible and the population of the excited state of the emitter decays exponentially in time with a radiative rate of
\be \gamma_{\mathrm{rad}}({\mathbf r_0},\omega_0, {\bf d}) =  \frac{\pi d^2}{\epsilon_0 \hbar} \omega_0 \rho({\mathbf r_0},\omega_0,\hat{\mathbf{e}}_\mathbf{d}).
\label{rate}\ee
The corresponding emission spectrum is a Lorentzian of width $\gamma_{\mathrm{rad}}$, as explained in Sec.\ \ref{sec:Lamb}. The power emitted by the dipole is also proportional to the LDOS because for continuous-wave excitation the radiated power is $P=\hbar\omega_0\gamma_\text{rad}$. Equation (\ref{rate}) shows that spontaneous emission can be controlled by modifying the LDOS, which is essential in quantum nanophotonics for engineering the light-matter interaction strength. This will be elaborated in further detail in Sec.\ \ref{Section-spontaneous-emission-control}.

\subsection{The spontaneous-emission spectrum and the Lamb shift\label{sec:Lamb}}

The spontaneous-emission spectrum is another important measurable quantity modified by interaction with the radiation reservoir. Following the approach by \citet{Vats2002PRA}, the Fourier transform of Eq.\ (\ref{eq-of-motion}) is
\be \tilde{c}_\text{e}(\Omega-\omega_0) =
\frac{1}{\tilde{K}(\Omega-\omega_0)-i \left(\Omega - \omega_0
\right)}, \label{ca-laplace}\ee
where $c_\text{e}(t \rightarrow \infty)=0$ is assumed, i.e., the population decays to the radiation reservoir at very long times. The Fourier transform of the memory kernel is
\bea \tilde{K}(\Omega-\omega_0) & = & \int_0^{\infty} \text{d}\tau
K(\tau) e^{i(\Omega - \omega_0) \tau} \\
&=& \frac{d^2}{2 \epsilon_0 \hbar} \int_0^\infty \text{d}\omega \omega \rho(\omega) \int_0^{\infty} \text{d}\tau e^{i (\Omega - \omega) \tau}, \nn
\eea
where the explicit dependence of the LDOS on position and dipole orientation (cf.\ Sec.\ \ref{sec:PhotCrys}) is omitted for brevity. The emission spectrum is obtained after using the Wiener-Khintchine theorem that applies since the radiation reservoir is described by a stationary and ergodic statistical process \cite{Mandel1995optical, Meystre2007elements,Cui2006PRA} \footnote{This equation is derived using the mathematical identity $\int_0^{\infty} \text{d}\tau e^{i (\Omega - \omega) \tau} = \pi \delta(\Omega - \omega) + i \mathcal{P} \left( \frac{1}{\Omega - \omega} \right)$, where $\mathcal{P} \left(\cdots \right)$ denotes the principal-value part.}
\bea S_\text{e}(\Omega) &\propto& \left| \tilde{c}_\text{e}(\Omega - \omega_0) \right|^2 \nn \\
&=& \frac{1}{\left[\Omega - \omega_0 - \Delta_L(\Omega) \right]^2 + \chi^2 \Omega^2 \rho^2(\Omega)},
\label{spectrum}\eea
where $\chi = \pi d^2/2 \epsilon_0 \hbar.$  This expression for the emission spectrum is valid to all orders in the light-matter coupling strength and thus extends beyond the Markovian Wigner-Weisskopf theory that is usually considered.

The emission spectrum contains a Lamb shift
\be
  \Delta_L(\Omega) = \frac{d^2}{2 \epsilon_0 \hbar} \mathcal{P} \left[ \int_0^{\infty} \text{d}\omega \frac{\omega \rho(\omega)}{\Omega - \omega} \right],
\label{Lamb}
\ee
that is obtained as a principal-value integration over all frequencies of a function that contains the LDOS. By using the Kramers-Kronig relations, the Lamb shift can alternatively be shown to be proportional to $\mathrm{Re} \left[ \hat{\bf e}_{\mathbf{d}}^* \cdot \overleftrightarrow{G}({\bf r},{\bf r},\omega) \cdot \hat{\mathbf{e}}_\mathbf{d} \right]$. The theory illustrates that the Lamb shift can be modified in photonic nanostructures, and an anomalous Lamb shift has first been predicted in highly dispersive photonic crystals by John and coworkers \cite{John1991PRB}. It should be mentioned that the total Lamb shift diverges because for sufficiently high frequencies all photonic crystals may be approximated as homogeneous media where $\rho(\omega) \propto \omega^2$ and the integration extends to infinity in Eq.\ (\ref{Lamb}). One approach for obtaining the absolute value of the Lamb shift applies a cutoff on the divergent integrals at the relativistic Compton frequency; a discussion of this issue can be found in \citet{Vats2002PRA} and will not be considered further here where we will focus on the relative modifications of the Lamb shift induced by LDOS variations. We note also that measurements of the absolute Lamb shift with quantum dots seem outside immediate reach because it would require a very accurate reference calculation without the vacuum effects, which likely would require very precise knowledge of the exciton energy levels that would be sensitive to the detailed configuration of the many atoms constituting the quantum dot. The relative Lamb shift, however, is usually finite since the LDOS of an inhomogeneous photonic structure would only differ from that of a homogeneous medium over the finite frequency range that it is tailored for. Interestingly, in the full non-Markovian theory considered here, the Lamb shift is seen not just to be a single-valued detuning of the optical transition, but rather a function of the observation frequency $\Omega$. Consequently, different frequency components of the emission spectrum are shifted by different amounts and the ``Lamb-shift function'' would generally be a more adequate term. Examples of the modification of the Lamb shift in photonic-crystal waveguides and cavities are considered in Secs.\ \ref{Section-waveguide-QED} and \ref{Section-cavity-QED}.

The frequency-dependent Lamb shift and LDOS entering in the emission spectrum of Eq.~(\ref{spectrum}) lead to non-Lorentzian lineshapes, which is the spectral signature of coupling to a non-Markovian radiation reservoir. The corresponding dynamic signature is a non-exponential decay of the emitter.  As a limiting case, the Wigner-Weisskopf result is found when the Lamb shift and $\Omega \rho(\Omega)$ both can be assumed to vary insignificantly over the linewidth of the emitter,
\be S_\text{e}^{\mathrm{WW}}(\Omega) \propto  \frac{1}{\left[ {\Omega - \omega}_0 -  \Delta_L(\omega_0) \right]^2 + \gamma_{\mathrm{rad}}^2/4 }.
\label{spectrum-WW}
\ee
In this case the Lamb shift is a single-valued number that merely redefines the quantum-dot transition frequency, while the emission spectrum is of the well-known Lorentzian functional form.

\subsection{Control of spontaneous emission}
\label{Section-spontaneous-emission-control}
The established link between the spontaneous-emission dynamics and the LDOS opens a way of controlling spontaneous emission by altering the surrounding medium of the emitter. To this end, photonic crystals are particularly well suited since the LDOS may be strongly modulated. In many cases, the Wigner-Weisskopf approximation is excellent such that a two-level emitter decays exponentially in time with the rate of Eq.\ (\ref{rate}). Figure \ref{2DPCall} shows examples of simulations of the LDOS for a photonic-crystal membrane made of GaAs where very pronounced spatial and spectral variations are found. The spatial variations reflect the strong modulation of the local electric field in the photonic crystal and a very sensitive dependence on dipole orientation is also found. The frequency-dependent LDOS reveals the presence of a 2D band gap in the scaled-frequency interval of $\omega a/2 \pi c = a/\lambda = [0.254, 0.361]$ where the LDOS is strongly suppressed for all positions. In the band gap the LDOS can be inhibited by up to a factor of 160 relative to the level of a homogeneous medium while at the edge of the band gap the LDOS rises drastically and can be enhanced.  Photonic crystals with periodicity in all three dimensions could potentially induce even stronger LDOS modifications than in the membrane structures since ideally a band gap with a vanishing LDOS could open in the case of a sufficiently high refractive-index contrast. So far, quantum-optics experiments using single quantum emitters have been lacking in 3D photonic crystals since it is challenging to optically address a single emitter inside these structures. One exception has been the work of \citet{Barth2006PRL} on weakly-scattering 3D photonic crystals where single-emitter spectroscopy was achieved.

The experimental progress on spontaneous-emission control has been significant during the last decade. Early work demonstrated both inhibition and enhancement of the decay dynamics for varying lattice constants in 3D inverse-opal photonic crystals infiltrated by an ensemble of colloidal quantum dots operated at room temperature \cite{Lodahl2004Nature}. This work has subsequently been extended to the case of 3D inverse-woodpile silicon photonic crystals where a ten-fold inhibition was reported \cite{Leistikow2011PRL}. Subsequently, inhibition of spontaneous emission in GaAs photonic-crystal membranes was demonstrated with a single quantum dot \cite{Kaniber2008PRB} and recently an inhibition factor of $70$ was achieved \cite{Wang2011PRL}.

The relation between the LDOS and the radiative decay rate that is valid for a dipole emitter (Eq.\ (\ref{rate})) provides a way of experimentally recording the LDOS. To this end, single InGaAs quantum dots are very suitable LDOS probes since it has been explicitly demonstrated how radiative and non-radiative processes can be separated by taking advantage of the exciton fine structure, as discussed in detail in Sec.\ \ref{sec:sst:decaydynamics}. This is especially appealing since unavoidable imperfections induced during fabrication of the photonic nanostructure (e.g., surface roughness) may alter the LDOS. Using quantum dots for probing the LDOS applies in situations where Wigner-Weisskopf theory and the dipole approximation are valid. The former is often a good approximation apart from the case of high-$Q$ cavities and potentially near photonic-band edges. For the latter, the extended size of quantum dots and lack of rotational symmetry imply that the dipole approximation may fail in photonic nanostructures. It turns out that for photonic-crystal membranes with quantum dots positioned in the center of the membrane, the dipole approximation is very accurate even for large quantum dots. We discuss effects beyond the dipole approximation in further detail in Sec.\ \ref{Section-beyond-dipole}.

\begin{figure}
\includegraphics[width= 0.8\columnwidth]{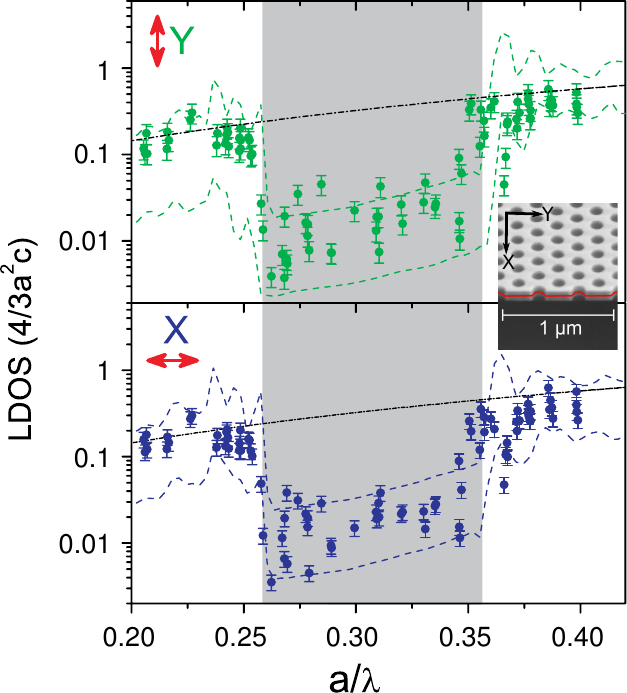}
 \caption{Experimental mapping of the frequency variation of the LDOS of a photonic-crystal membrane. The LDOS is measured for two different dipole projections X and Y and for various scaled frequencies, $a/\lambda$. The dashed curves are calculated for two positions with relatively high and low LDOS and the experimental points are therefore expected to fall within the two dashed curves, which is consistent with observations. The dashed-dotted curve is the corresponding LDOS for a homogeneous medium (GaAs). The inset shows the dipole orientations with respect to the photonic-crystal lattice. The data are reproduced from \citet{Wang2011PRL}. }
 \label{LDOS-mapping}
\end{figure}

Based on Eq.\ (\ref{rate}), the LDOS can be expressed as
\be
\rho({\bf r},\omega,\hat{\mathbf{e}}_\mathbf{d}) = \frac{n \omega^2}{3 \pi^2 c^3} F_{\mathrm{P}}({\bf r},\omega,\hat{\mathbf{e}}_\mathbf{d}),
\label{LDOSpurcell}
\ee
where the Purcell factor is defined as
\be
F_{\mathrm{P}}({\bf r},\omega,\hat{\mathbf{e}}_\mathbf{d}) = \frac{ \gamma_{\mathrm{rad}}({\bf r},\omega,\hat{\mathbf{e}}_\mathbf{d})}{\gamma_{\mathrm{rad}}^{\mathrm{hom}}(\omega)},
\label{ratePurcell}
 \ee
i.e., the ratio between the radiative decay rate of the dipole emitter at position ${\bf r}$ to the radiative rate of an identical emitter placed in a homogeneous medium of refractive index $n$. This constitutes a generalization of the Purcell factor originally formulated for cavities \cite{Purcell1946PR}, cf.\ Sec.\ \ref{Section-cavity-QED}, to the case of arbitrary nanophotonic structures. In photonic crystals the Purcell factor can be either below unity (suppression of spontaneous emission) or above unity (enhancement of spontaneous emission) while in a cavity the Purcell factor is usually above unity due to coupling to the cavity mode. The LDOS can be obtained from measurements of the Purcell factor and Fig.\ \ref{LDOS-mapping} displays the experimental mapping of the frequency dependence of the LDOS of a photonic-crystal membrane. In this experiment individual quantum dots with similar emission wavelengths were spectrally selected while varying the lattice constant $a$ of the photonic crystal in order to record the LDOS as a function of the scaled frequency $a/\lambda$. A very pronounced suppression of the LDOS for two different dipole projections was observed in the region of the 2D photonic band gap. The largest recorded spontaneous-emission inhibition factor (the inverse of the Purcell factor) was $70$.  The point-to-point fluctuations reveal the spatial variation of the LDOS within the unit cell of the photonic crystal as probed by quantum dots at different positions, and are in accordance with numerical simulations of the LDOS extrema \cite{Koenderink2006JOSAB}. The good agreement between experiment and theory illustrates that inhibition of spontaneous emission in photonic-crystal membranes is robust to fabrication imperfections, which has been confirmed in systematic experiments where imperfections were deliberately introduced \cite{Garcia2012PRL}.

The experimental mapping of the spatial variation of the LDOS is challenging since techniques for positioning or locating a single quantum dot with nanometer precision relative to a photonic nanostructure are tedious, cf.\ the discussion in Sec.\ \ref{sec:sst:growth}. Another approach would be to vary the location of the emitter, which is only possible for emitters positioned on the surface of the nanostructures and therefore not suitable for epitaxially grown quantum dots. To this end, experimental progress has been obtained by attaching a dielectric bead containing fluorescent molecules to a scanning probe \cite{Frimmer2011PRL}, although the single-emitter sensitivity generally required for full LDOS mapping was not yet achieved. In the context of quantum plasmonics, nanocrystals containing a single NV-defect center were successfully manipulated with an atomic-force microscope \cite{Huck2011PRL, Schell2011RevSciInstrum}.

\subsection{Non-Markovian spontaneous-emission dynamics}
In many cases the Wigner-Weisskopf approximation suffices for describing the spontaneous-emission dynamics and predicts an exponential decay with time. However, if the spectral variation of the LDOS is very pronounced over a frequency interval comparable to the emitter linewidth, the Wigner-Weisskopf approximation breaks down. This can potentially occur near a photonic-crystal band edge or in a high-$Q$ cavity. In this case the memory kernel of Eq.\ (\ref{memory-kernel}) is non-singular implying that non-Markovian memory effects change the dynamics. Accordingly, the  signature of non-Markovian dynamics is a non-exponential decay induced by the nanophotonic environment, which corresponds to a non-Lorentzian emission spectrum. Non-Markovian photon-emitter interactions in photonic crystals have been investigated extensively theoretically, predicting exotic quantum optics effects such as the fractional decay \cite{John1994PRA}. Fractional decay may occur for emitters tuned to the sharp edge of a band gap. The emitter decays by spontaneous emission to a nonzero fractional population that under idealized conditions would prevail even in the steady-state limit of $t \rightarrow \infty$, which represents an entangled photon-emitter bound state. In any experimental implementations, however, the unavoidable fabrication imperfections and the finite size of the photonic crystal would imply that the emitter eventually decays fully to the ground state. A detailed study of realistic photonic-crystal structures containing quantum dots has shown that significant non-Markovian dynamics may be expected \cite{Kristensen2008OL}.

So far the observations of non-Markovian dynamics have been sparse and limited only to the cases of micropillar cavities \cite{Madsen2011PRL} and pulsed transmission experiments in photonic-crystal cavities \cite{Majumdar2012PRA}  since the measurements are challenging due to the finite time resolution of available single-photon detectors.  Measurements in the spectral domain may overcome the resolution limitation but any dephasing processes would also modify the spectra and would need to be clearly distinguished from non-Markovian photon-emitter effects. The observations of the vacuum Rabi splitting of a single quantum dot embedded in a micropillar cavity \cite{Reithmaier2004Nature} or a photonic-crystal cavity \cite{Yoshie2004Nature} constitute examples of non-Markovian coupling to a radiation reservoir that includes the cavity quasi-mode. At microwave frequencies, non-Markovian coupling was observed in a 3D photonic crystal by using a magnetic dipole source \cite{Hoeppe2012PRL}.

\subsection{Light-matter interaction beyond the dipole approximation}
\label{Section-beyond-dipole}

\begin{figure}
\begin{center}
\includegraphics[width=\columnwidth]{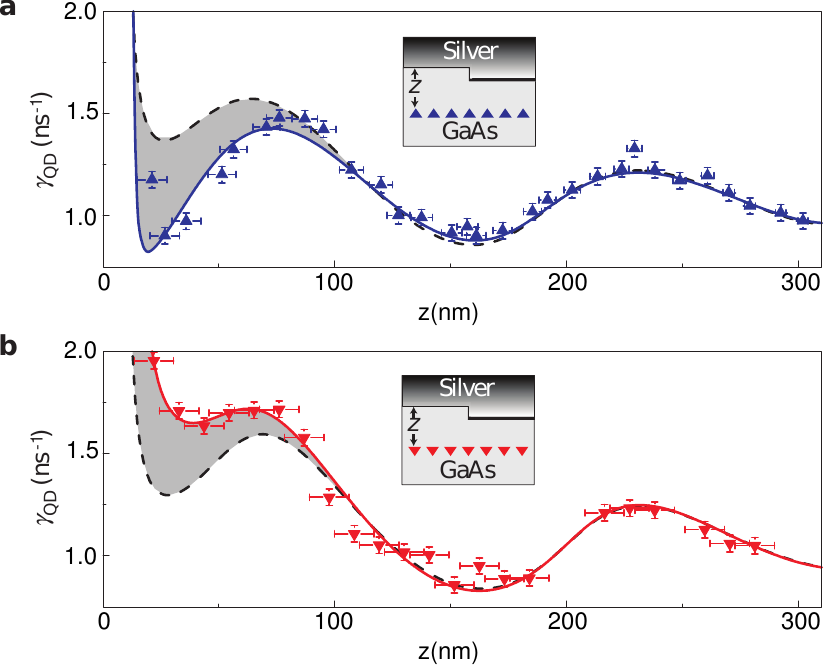}
\caption{Experimental demonstration of the breakdown of the dipole approximation. The points show measured decay rates from ensembles of quantum dots placed near a semiconductor-silver interface for (a) the as-grown structure and (b) with the quantum dots placed upside down. The orientations of the quantum dots are indicated in the insets. The prediction of dipole theory is shown as the dashed curve that describes both data sets very well far from the interface. Close to the interface the data deviate from dipole theory showing either (a) suppression or (b) enhancement of the decay rate meaning that surface plasmons are excited less or more efficiently than expected for dipoles, respectively. These effects agree with predictions of the theory beyond the dipole approximation (solid curves) and allow extracting an experimental value of the mesoscopic moment of the quantum dots. Reprinted with permission from \citet{Andersen2011NPHYS}.
\label{fig:sst:Mads}}
\end{center}
\end{figure}

A crucial assumption underlying the LDOS formalism is that the spatial extent of the exciton wave function in the quantum dot, $L$, is much smaller than the optical wavelength, which is the electric dipole approximation that is valid if $|\mathbf{k}|L\ll 1$. The dipole approximation is very accurate in atomic physics and leads to great simplifications but turns out not to be generally valid for quantum dots. For standard-sized InGaAs quantum dot in GaAs, $|\mathbf{k}|L\approx 0.5$, which is evaluated for a homogeneous dielectric medium. It turns out that the symmetries of both the nanophotonic environment and the quantum dot play a crucial role for the validity of the dipole approximation and large deviations from dipole theory may be found in particular in structures with large electromagnetic-field gradients since $\mathbf{k}$ is modified.

A quantum theory of light-matter interaction beyond the dipole approximation may be derived using the $\mathbf{p}\cdot\mathbf{A}$ Hamiltonian discussed in Sec.\ \ref{sec:sst:TransitionMatrixElement} and the Markov approximation. This theory can be readily extended to include non-Markovian effects because the description beyond the dipole approximation only modifies the spatial parts of the electromagnetic field and wavefunctions while non-Markovian effects concern only the time dependence. The generalized spontaneous-emission rate for an exciton with envelope wave function $\phi (\mathbf{r}_0,\mathbf{r}_\text{e},\mathbf{r}_\text{h})$, which is centered at $\mathbf{r}_0$ and where $\mathbf{r}_\text{e}$ ($\mathbf{r}_\text{h}$) is the electron (hole) coordinate, can be written in a Green-tensor formalism as \cite{Stobbe2012PRB}
\begin{align}
\gamma_\text{rad}(\mathbf{r}_0,\omega, \hat{\mathbf{e}}_\mathbf{d}) = \frac{\pi q^2}{\hbar m_0 \epsilon_0} \frac{\rho_{\mathrm{NL}}(\mathbf{r}_0,\omega, \hat{\mathbf{e}}_\mathbf{d})}{\omega},\label{eq:GammaBDA}
\end{align}
where the non-local interaction function is defined as
\begin{align}
\begin{split}
\rho_{\mathrm{NL}}(\mathbf{r}_0,\omega,\hat{\mathbf{e}}_\mathbf{d}) = & \frac{\omega E_\text{p}}{\pi c^2} \int \mathrm{d}^3\mathbf{r} \int \mathrm{d}^3\mathbf{r}'
\phi (\mathbf{r}_0,\mathbf{r},\mathbf{r})  \, \phi^\ast(\mathbf{r}_0,\mathbf{r}',\mathbf{r}')
\\ &\times
 \mathrm{Im}\left\{ \hat{\mathbf{e}}_\mathbf{d}^* \cdot \overleftrightarrow{G}( \mathbf{r},\mathbf{r}',\omega ) \cdot \hat{\mathbf{e}}_\mathbf{d} \right\}.\label{eq:NLIF_in_terms_of_G}
 \end{split}
\end{align}
A similar result has been derived in a semi-classical model \cite{Ahn2003PRB}. It is clear from this expression that the convenient separation between light-and-matter variables found in dipole theory does not hold beyond the dipole approximation, i.e., light and matter are intertwined. This leads to counterintuitive effects. For example, for a spherically symmetric exciton in a homogeneous medium, the spontaneous-emission rate diverges for large $L$ due to the giant-oscillator strength effect discussed in Sec.\ \ref{sec:sst:OS} but in a model beyond the dipole approximation it vanishes for large $L$ since the contributions to the total decay rate from various spatial parts of the exciton wavefunction average to zero in Eq.\ (\ref{eq:NLIF_in_terms_of_G}). Nonetheless, when computing the Purcell factor for spherically symmetric excitons it turns out that the result obtained in the dipole approximation is exact for any nanophotonic structure and any $L$ \cite{Kristensen2013PRB}. The intertwining of light and matter beyond the dipole approximation gives new opportunities for tailoring the light-matter interaction strength in quantum-optics applications where the quantum-dot wavefunction and the photonic nanostructure are engineered in mutual accordance. The optimum design strategies remain largely unexplored, but could be performed with the above formalism.

Equation (\ref{eq:GammaBDA}) is exact to any order in the light-matter coupling, but generally it must be solved numerically and therefore cannot be used to gain physical insight. To this end it is instead useful to Taylor expand the vector potential, which gives rise to additional coupling terms of which many are vanishing due to symmetry arguments. However, due to the shape of the confinement potential discussed in Sec.\ \ref{sec:sst:growth}, Stranski-Krastanov quantum dots have no parity along the growth direction and it turns out that a single parameter, the mesoscopic moment, is responsible for all significant effects beyond the dipole approximation.
The mesoscopic moment is an intrinsic property of the emitter and vanishes for atoms due to parity. In a first-order Taylor expansion of the vector potential, the correction to the total radiative rate of the dipole transition due to the mesoscopic moment is given by \cite{Tighineanu2014PRL}.
\begin{align}
\begin{split}
&\gamma_{\text{rad},\Lambda}(\mathbf{r}_0,\omega) = \\
&\frac{4 q^2}{\epsilon_0 \hbar m_0^2 c_0^2} \text{Re}\left[ \Lambda P^\ast \right]
\frac{\partial}{\partial x} \text{Im}
\left[
\hat{\mathbf{e}}^\ast_\mathbf{x}\cdot\overleftrightarrow{G}_{zx}(\mathbf{r},\mathbf{r}_0,\omega)\cdot\hat{\mathbf{e}}_\mathbf{x}
\right]
\Big|_{\mathbf{r}=\mathbf{r}_0},
\end{split} \label{eq:gamma-BDA}
\end{align}
where the mesoscopic moment is defined as
\begin{align}
\Lambda = \frac{q}{m_0}\langle \Psi_\text{v} | \hat{x} \hat{p}_z |  \Psi_\text{c} \rangle,
\end{align}
and the bright exciton with dipole moment along $x$ is considered, i.e., $\mathbf{P}_x = P \hat{\mathbf{e}}_x$, cf.\ the discussion in Sec.\ \ref{sec:sst:TransitionMatrixElement}. By symmetry the other bright exciton is governed by the same dipole and mesoscopic moments. The mesoscopic moment depends sensitively on exciton symmetry and the in-plane exciton size \cite{Tighineanu2014Arxiv}. Eq.\ (\ref{eq:gamma-BDA}) reveals how the coupling to the electromagnetic field enters through the gradient of certain Green's tensor components.

Mesoscopic effects beyond the dipole approximation can be revealed in well-controlled experiments on decay dynamics in simple nanostructures where the spatial dependence of Green's tensor can be calculated exactly enabling a direct comparison between experiment and theory. Figure\ \ref{fig:sst:Mads} shows the results of such an experiment where the decay rate of an ensemble of quantum dots was recorded as a function of distance to a silver-GaAs interface~\cite{Andersen2011NPHYS}. Close to the interface the quantum dots excite surface-plasmon polaritons and the strong field gradients governing this interaction leads to the breakdown of the dipole approximation. The experimental data are well explained by the theory and is tested for two different orientations of the quantum dots relative to the mirror in order to validate the predictions from theory. In this experiment, the mesoscopic moment either add to or subtract from the rate of plasmon excitation for a dipole meaning that plasmon excitation can either be promoted or suppressed. Even larger effects are expected in metallic nanowires, and the mesocopic effects therefore provide a way of enhancing plasmon-emitter interactions for applications in quantum plasmonics. In photonic-crystal membranes, there may also be strong field gradients but the TE-like symmetry of the electromagnetic field implies that the in-plane derivatives of the electric field component perpendicular to the membrane vanishes in the center of the photonic-crystal membrane, which implies that the mesoscopic term vanishes according to Eq.\ (\ref{eq:gamma-BDA}). For this reason the dipole approximation was found to be valid in the experimental work on probing the LDOS with quantum dots as discussed in Sec.\ \ref{Section-spontaneous-emission-control}.

\section{Resonance fluorescence from a quantum dot}
\label{Section-Resonance-Fluorescence}

Resonant excitation of a quantum dot leads to novel opportunities compared to the traditional non-resonant excitation schemes. It enables coherent manipulation of the excitonic states in a quantum dot without creating additional carriers or phonons that inevitable are generated with non-resonant excitation and may cause undesirable dephasing processes. The fluorescence from a resonantly driven quantum emitter produces highly non-classical light and depending on the excitation condition, detuning, and spectral filtering, e.g., cascaded-photon emission can be generated.  Recently, also highly coherent single-photon emission from quantum dots was experimentally demonstrated.
In this section we briefly review the basic theory of resonance fluorescence of quantum emitters including the role of dephasing treated within the Markov approximation. Resonant excitation of quantum dots in photonic nanostructures (e.g., cavities or waveguides) provides a very promising pathway to highly coherent interaction between photons and excitons, which is a key resource for optical quantum-information processing. Resonant excitation of quantum dots in photonic waveguides is considered in Sec.\ \ref{Section-waveguide-QED}.

\subsection{Coherent and incoherent scattering}
\label{Section-coherent-and-incoherent-scattering}

\begin{figure}
\includegraphics[width= \columnwidth]{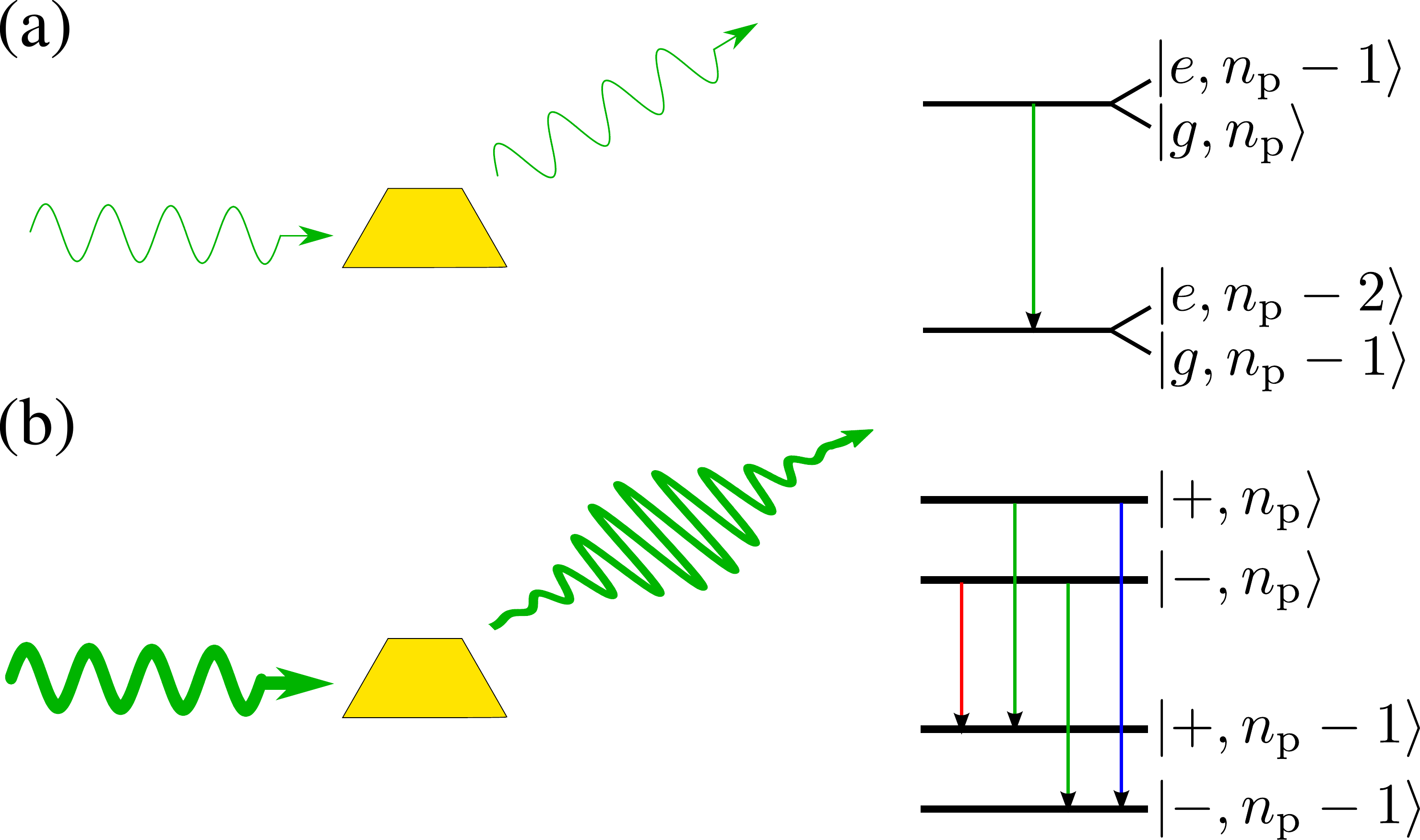}
 \caption{(a) Resonance fluorescence of a quantum dot in the weak-excitation limit showing the coherent scattering of a single frequency $\omega_\text{p}$. The emitted pulse inherits the temporal shape of the excitation field. The resulting energy levels are the bare states that are doubly degenerate. (b) The Mollow regime where a high-intensity driving field with frequency $\omega_\text{p}$ is scattered by a quantum dot. The fluorescence is emitted in pulses that are determined by the lifetime of the excited state and the amplitude of the driving field. The quantum-dot levels are dressed by the driving field leading to a ladder of eigenstates of the form $| \pm, n_\text{p}  \rangle =  (|g, n_\text{p}\rangle \pm |e, n_\text{p}-1\rangle)/\sqrt 2$. The dressed states have transitions with three different energies $\omega_\text{p}$ (green), and $\omega_\text{p} \pm \mu$ (blue and red).}
\label{Resonant-scat}
\end{figure}

The basic setting of resonance fluorescence is a two-level dipole emitter initially prepared in the ground state that is driven by an electromagnetic field. Furthermore the emitter is coupled to a radiation reservoir into which single photons from the driving field can be scattered, as is illustrated in Fig.\ \ref{Resonant-scat}. A two-level emitter can only scatter a single photon at a time, which is a consequence of the Pauli exclusion principle for fermions. Consequently resonant scattering can induce non-classical states of light. In the following a continuous drive term is considered but the results can readily be generalized to the case of pulsed excitation. The coupling between a dipole emitter and a radiation reservoir was discussed in Sec.\ \ref{Section-spontaneous-emission} and the Hamiltonian for resonance fluorescence is given by Eq.\ (\ref{H-tot}) with an additional driving term,
\be
\hat{V}(t) =  \hbar \Omega_\text{p} \hat{\sigma}_+ e^{i(\omega_0-\omega_\text{p})t} + \mathrm{h.c.}
\ee
In the following, dephasing is included by adding Markovian-dephasing terms to the Hamiltonian as considered in detail in Sec.\ \ref{The dissipative Jaynes-Cummings model}. The excitation field at frequency $\omega_\text{p}$ is assumed to be in a coherent state, which implies that the annihilation operator for this mode can be replaced by a complex number \cite{Mandel1995optical}. The amplitude of the driving term, $\Omega_{\mathrm{p}} = \sqrt{n_\text{p}} {\bf d} \cdot {\bf E}_\text{p}/\hbar,$ contains the electric field strength per photon, ${\bf E}_\text{p}$, the transition dipole moment ${\bf d}$, and the average number of photons $n_\text{p}$. The radiation reservoir is treated within Wigner-Weisskopf theory.

The general behavior of resonance fluorescence depends on the amplitude of the driving term relative to the dissipation and decoherence rates. For simplicity, non-radiative recombination is neglected in the following and when this approximation is used we replace $\gamma_\text{rad}$ by $\gamma$. At a low excitation amplitude the emitter will operate as a passive scatterer. By increasing the excitation amplitude $\Omega_\text{p}$ the population of the quantum dot will Rabi oscillate between the ground and excited states, which is an example of a nonlinear light-matter interaction induced by the saturation of the emitter. The Rabi oscillations are damped by dephasing and spontaneous emission. The scattered intensity reaches a steady-state value that can be expressed as the sum of two parts \cite{Meystre2007elements}
\bea
I_{\mathrm{coh}} &=& I_0 \frac{4 \gamma^2 \left| \Omega_\text{p} \right|^2}{\left( \gamma^2 + 2 \gamma \gamma_\text{dp} + 8 \left| \Omega_\text{p} \right|^2 \right)^2}, \\
I_{\mathrm{inc}} &=&
I_0 \frac{4 \left| \Omega_\text{p} \right|^2 \left(2 \gamma \gamma_\text{dp} + 8\left| \Omega_\text{p} \right|^2 \right)}{\left( \gamma^2 + 2 \gamma \gamma_\text{dp} + 8 \left| \Omega_\text{p} \right|^2 \right)^2},
\eea
\label{Resonant-scattering-intensity}%
where $I_0$ is an overall scaling amplitude, $\gamma_\text{dp}$ denotes the dephasing rate, and it is assumed that the excitation field is resonant with the quantum dot. $I_{\mathrm{coh}}$ dominates at weak excitation and is referred to as the coherent part of the intensity. It can be calculated in a semi-classical model where the quantum fluctuations of the dipole emitter are neglected. The incoherent part $I_{\mathrm{inc}}$ dominates at strong excitation and originates from the quantum fluctuations of the driven dipole. The prevalent terminology of referring to this as coherent and incoherent intensity can be somewhat misleading since both terms depend on both the coherent driving field $\Omega_\text{p}$ and the incoherent dephasing rate $\gamma_\text{dp}$. For a moderate amount of dephasing $(\gamma_\text{dp} \lesssim \gamma)$ the coherent and incoherent intensities dominate at low and high excitation intensities, respectively. In the former case, the quantum dot remains weakly excited so that the excited-state population is small while in the latter case, an increasing excitation intensity eventually saturates the emitter leading to a decrease of the coherent term.

At all excitation levels the emitter can only scatter a single photon at a time meaning that $g^{(2)}(0)=0$ for the scattered light. The full expression for the photon auto-correlation function is obtained by extending the textbook calculation of \citet{Scully_and_Zubairy} to include dephasing \cite{Flagg2009NPHYS},
\bea
& &g^{(2)}(\tau) = \label{g2-res-excitation} \\
& & 1 - e^{-(3\gamma/4 + \gamma_\text{dp}/2) \tau} \left( \cos(\mu \tau) + \frac{3 \gamma + 2 \gamma_\text{dp}}{4 \mu} \sin(\mu \tau) \right), \nn
\eea
where $\mu = \sqrt{4 \left| \Omega_\text{p} \right|^2 - \left( \gamma/4 - \gamma_\text{dp}/2 \right)^2}$ is the effective Rabi frequency.  In the the limit of weak excitation, i.e., $\left| \Omega_\text{p} \right| \ll \gamma , \gamma_\text{dp} $, the auto-correlation function increases monotonously from zero to unity with time delay $\tau$, while for an excitation rate exceeding the dissipation rates, coherent Rabi oscillations appear.

Further insight into resonance fluorescence is obtained from the emission spectrum, which similarly can be divided into a coherent and an incoherent part: $S(\omega ) = S_{\mathrm{coh}}(\omega)+ S_{\mathrm{inc}}(\omega ).$ In the resonant case, $\omega_0 = \omega_\text{p}$, we have
\bs
\bea
&&S_{\mathrm{coh}}(\omega ) = \frac{n_\text{s}^2 \gamma^2}{4 \left| \Omega_\text{p} \right|^2} \delta(\Delta_\text{p}),  \\
&&S_{\mathrm{inc}}(\omega ) = \frac{n_\text{s}}{2} \frac{\gamma_\text{dp}+\gamma/2}{\Delta_\text{p}^2 + (\gamma_\text{dp} + \gamma/2)^2}  \label{inc-spectrum} \\
&& + \frac{n_\text{s}^2}{4 \left| \Omega_\text{p} \right|^2}\mathrm{Re} \left[ \left( \frac{A}{2} + \frac{B}{8 i \mu} \right) \frac{1}{i (\Delta_\text{p} - \mu) + (\gamma_\text{dp}/2 + 3 \gamma/4)} \right] \nn \\
&& + \frac{n_\text{s}^2}{4 \left| \Omega_\text{p} \right|^2}\mathrm{Re} \left[ \left( \frac{A}{2} - \frac{B}{8 i \mu} \right) \frac{1}{i (\Delta_\text{p} + \mu) + (\gamma_\text{dp}/2 + 3 \gamma/4)} \right], \nn
\eea
\es
where $\Delta_\text{p} = \omega - \omega_\text{p}$, $A = 4 \left| \Omega_\text{p} \right|^2 - \gamma (\gamma/2 - \gamma_\text{dp})$, and $B = -\left(4 \left| \Omega_\text{p} \right|^2 \left[2 \gamma_\text{dp}-5 \gamma \right] + 2 \gamma \gamma_\text{dp}^2 - 2 \gamma^2 \gamma_\text{dp} + \gamma^3/2 \right)$, and the steady-state population of the emitter is
\be
n_\text{s} = \frac{4 \left| \Omega_\text{p} \right|^2}{\gamma^2 + 2 \gamma \gamma_\text{dp} + 8 \left| \Omega_\text{p} \right|^2}.
\ee
The coherent part of the spectrum is dominating at low excitation intensities and is proportional to a Dirac delta function in frequency. In an experiment, the width of this coherent peak is determined by the linewidth of the laser driving the emitter. This process can therefore be exploited for generating highly coherent single photons since they inherit the narrow linewidth of the excitation laser, as illustrated in Fig.\ \ref{Resonant-scat}(a). The incoherent three-peak spectrum dominates at high excitation power where the quantum emitter is saturated. The central peak is resonant with the excitation field while the two satellite peaks are positioned symmetrically around the central peak at the frequencies $\omega = \omega_\text{p} \pm \mu \approx \omega_\text{p} \pm 2 \left| \Omega_\text{p} \right|$, where the approximation holds for a very strong excitation field. This structure of the saturated spectrum is referred to as the Mollow triplet \cite{Mollow1969PR} and is a result of the dressing of the emitter by the driving field forming pairs of dressed states, $\left| \pm , n_\text{p} \right> = \left( \left| g,n_\text{p} \right> \pm  \left|e,n_\text{p}-1 \right> \right) / \sqrt{2}$, cf.\ Fig.\ \ref{Resonant-scat}(b), and gives rise to three different transitions between manifolds of states.

Resonant scattering of a coherent state on a quantum emitter can be employed as an alternative to spontaneous emission for generating single photons. The spatial mode structure of the scattered photon is thus of importance in order to enhance the single-photon efficiency. In the weak-excitation regime where stimulated-emission processes are negligible, the radiation pattern of the scattered photon is identical to that of spontaneous emission since the emitter couples to the same radiation reservoir in the two cases \cite{AllenEberlyOpticalResonance}.   Consequently, important concepts from the theory of spontaneous emission, e.g., the LDOS introduced in Sec.\ \ref{Section-spontaneous-emission} may equally well be applied for describing weak resonance fluorescence. In Sec.\ \ref{Section-single-photon-nonlinearity} it is explained how resonant scattering in a photonic waveguide can be used to achieve a large nonlinearity capable of operating at the single-photon level.

\subsection{Coherent optical manipulation of a single quantum dot}

\begin{figure}
\includegraphics[width= 0.84\columnwidth]{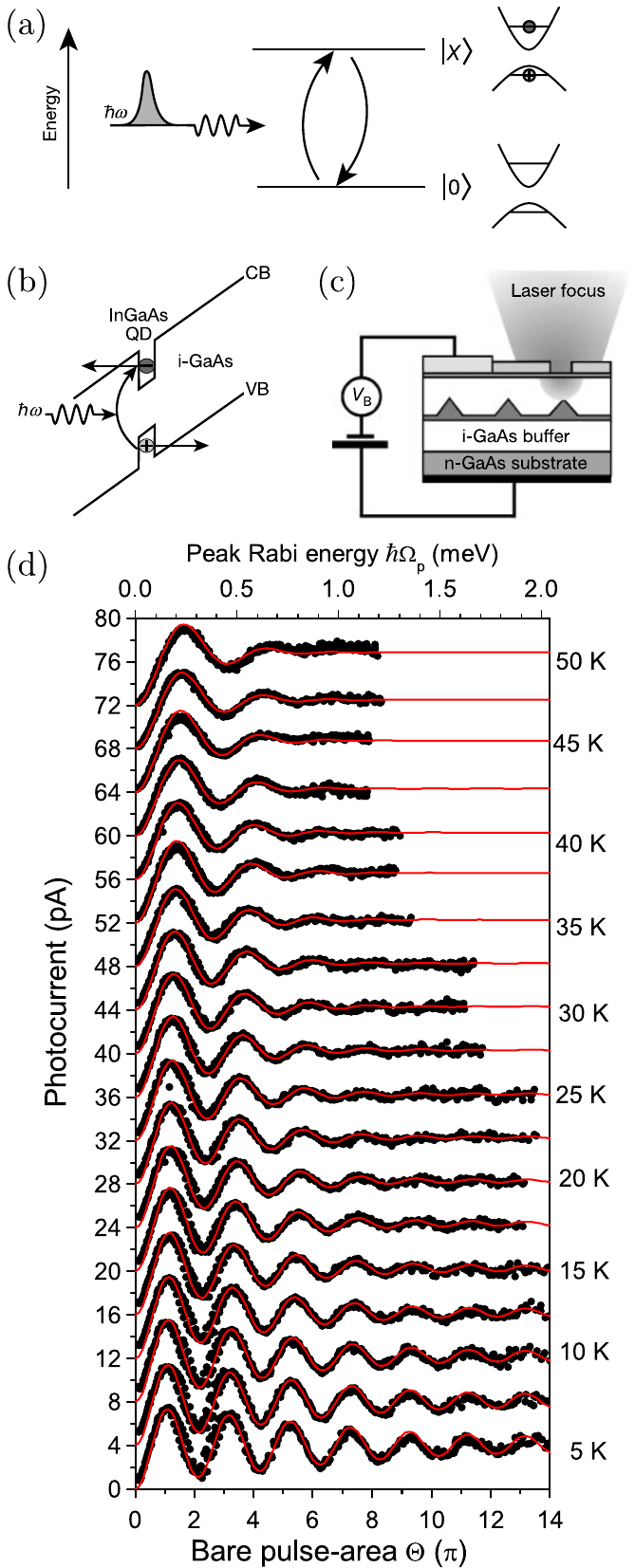}
\caption{Experimental demonstration of Rabi oscillations of a quantum dot. (a) Rabi oscillations between an exciton $\left| e \right> = \left|X \right>$ and no exciton $\left| g \right> = \left|0 \right>$ in the quantum dot when driven by a short optical pulse. (b) The quantum-dot population can be read out as a photocurrent in the Schottky diode by (c) applying an electrical bias across the quantum dot. (d) Examples of Rabi oscillations recorded in the photocurent for various excitation amplitudes as quantified by the overall pulse area $\Theta.$ The excitation pulse duration was kept constant at \SI{4}{\pico\second} and the pulse energy was varied. An increased damping is observed with temperature in accordance with the theory of LA-phonon dephasing (red curves). (a)-(c) Reprinted with permission from \citet{Zrenner2002Nature}. (d) Reprinted with permission from \citet{Ramsay2010PRL}.}
\label{Rabi-oscillations}
\end{figure}

Resonant excitation can be applied as a  versatile way of controlling the population of a quantum dot whereby superposition states between the ground and excited state can be encoded. Recently resonant excitation was employed for mapping the polarization state of a light pulse onto the exciton spin, followed by the subsequent read out of the spin state through a biexciton transition \cite{Benny2011PRL}. The observation of resonance fluorescence has been challenging since the weak optical signal appears at the same frequency as the strong excitation field, thus all-optical excitation and read-out pose strict demands on the ability to filter the fluorescence from the residual excitation light. This can be particularly challenging in photonic nanostructures since their inhomogeneous structure implies that residual light scattering can be difficult to suppress. Rabi oscillations have been observed experimentally for a single quantum dot through a number of different approaches. Early work applied time-resolved pump-probe spectroscopy to record the dynamics of the optically induced polarization of the two-level system, and the first signatures of Rabi oscillations were reported in \citet{Stievater2001PRL}. Subsequently, electrically gated quantum dots were implemented in resonant-excitation experiments where the population of the quantum dot was read out from a photocurrent  \cite{Zrenner2002Nature}. Figures \ref{Rabi-oscillations}(a)-(c) illustrate the operating principle of such an experiment: a picosecond optical pulse induces Rabi oscillations in a quantum dot by an amount that is controlled by the overall pulse area $\Theta$ that for a fixed pulse duration can be varied by the pulse energy. The population of the quantum dot was read out as a tunnel photocurrent from the quantum dot by embedding it in a Schottky diode structure and applying an electric field.  Rabi oscillations of up to seven full cycles have been observed in such an experiment, see Fig.\ \ref{Rabi-oscillations}(d) \cite{Ramsay2010PRL}. The damping of the Rabi oscillations was well explained by dephasing from LA phonons, and a quantitative agreement between theory and experiment for the temperature-dependent data was found. Dephasing due to LA phonons is described in Sec.\ \ref{Section-Phonon-dephasing} in the context of cavity QED. Rabi oscillations have also been recorded directly in the fluorescence by applying picosecond-pulsed excitation \cite{Muller2007PRL, Melet2008PRB}. In these experiments, highly efficient spatial filtering could be implemented since the excitation light was guided in a planar waveguide structure and the fluorescence collected perpendicularly to the excitation beam. Another successful filtering approach has applied a cross-polarized excitation-detection configuration for the case of quantum dots in photonic-crystal cavities \cite{Englund2007Nature}, where the residual excitation light could be strongly suppressed by high-extinction polarizers enabling the observation of resonance fluorescence.

\subsection{Generation of coherent single photons by resonant scattering on a quantum dot}
\label{Section-coherent-single-photons}

\begin{figure}[t!]
\includegraphics[width= \columnwidth]{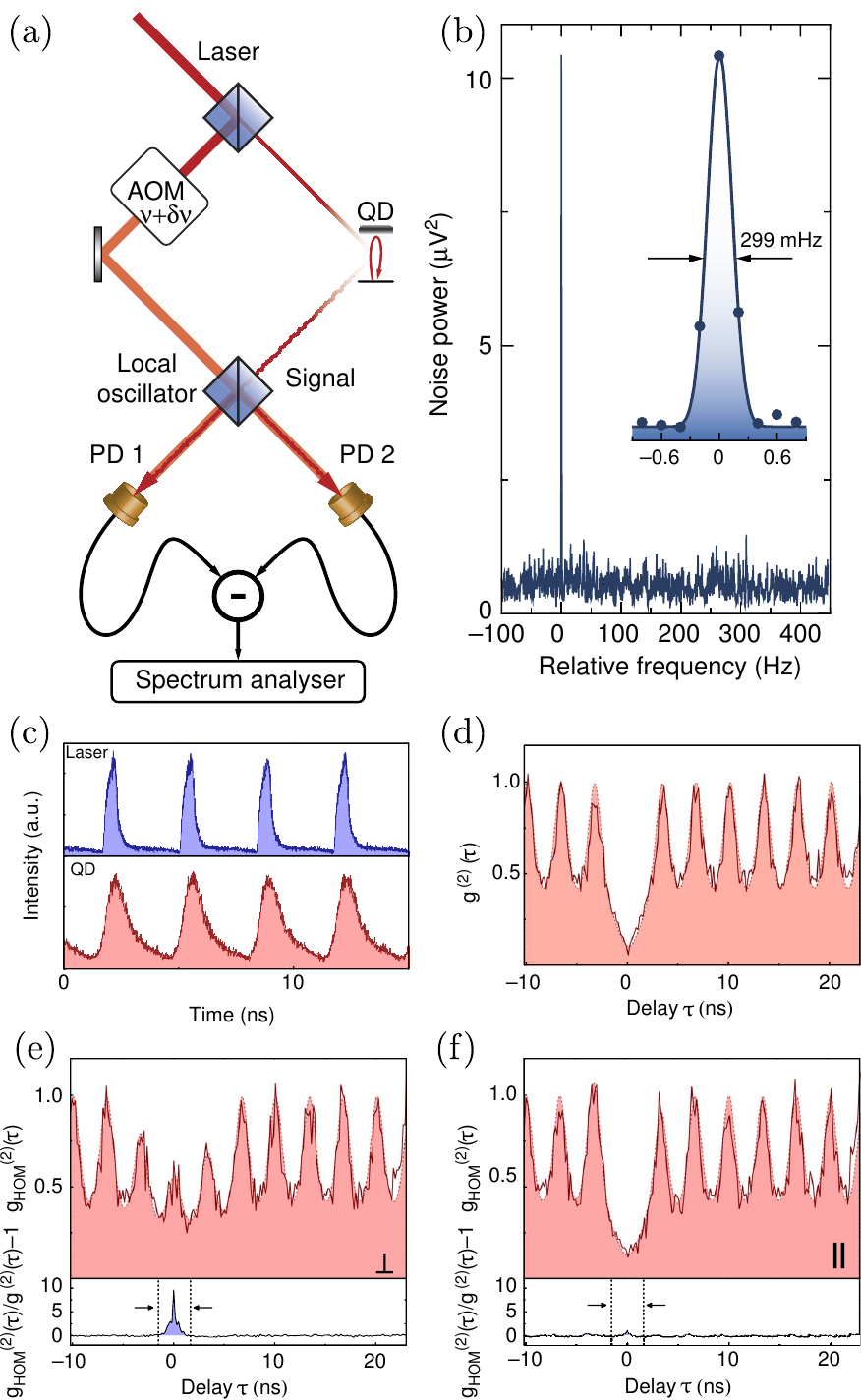}
 \caption{ Mutual coherence measurements between the driving field and a photon generated by weak coherent scattering on a quantum dot. (a) Illustration of the optical heterodyne setup used for detection of a beatnote at $\delta \nu = 210 \: \mathrm{kHz}$. (b) Observed beat-note spectrum displaying a peak with a full width half maximum of \SI{299}{\milli\hertz}. (c)-(d) Synthesized train of single photons. (c) The pulse train of the ex\-ci\-ta\-tion laser (upper plot) and of the generated single photons (lower plot). (d) Measurement of the auto-correlation function for the scattered photons. (e)-(f) Hong-Ou-Mandel measurements of the degree of indistinguishability (e) in the absence of two-photon interference (orthogonal polarization) and (f) with two-photon interference.  The upper plots show the raw data of the normalized auto-correlation function measured with the Hong-Ou-Mandel interferometer and the lower plots quantify the degree of indistinguishability by analyzing the amount of coincidence counts in the peak at zero time delay. Reprinted with permission and adapted from \citet{Matthiesen2013NCOM}.}
\label{Synthesized-photons}
\end{figure}

Coherent scattering in the weak-excitation regime has recently been exploited as a  promising way of generating highly coherent single photons from quantum dots \cite{Nguyen2011APL, Matthiesen2012PRL}, thereby overcoming most of the inherent dephasing encountered in solid-state systems when using different excitation schemes. In the experiment of \citet{Matthiesen2012PRL} a weak and narrow-linewidth laser was resonantly driving a single quantum dot electrostatically tuned to resonance with the laser. From the electric-field correlation function, a coherence time as long as $T_2 = \SI{22}{\nano\second}$ was observed, which corresponds to a linewidth of $\Delta \omega / 2 \pi = \SI{7}{\mega\hertz}$ (equal to $\SI{0.03}{\micro\eV}$), which is $15$ times narrower than the natural linewidth of the transition set by spontaneous emission. An almost ideal anti-bunching of $g^{(2)}(0) = (1 \pm 1) \%$ showed that high-purity single-photon scattering could be obtained. A limitation of this approach, however, is that since the quantum dot is driven by a continuous-wave laser, the photons are not emitted deterministically.

Weak coherent scattering can also be exploited in a pulsed regime for generation of triggered single photons. The synthesis of various shapes of single-photon wavepackets  was recently demonstrated \cite{Matthiesen2013NCOM}. In these experiments it was utilized that in the weak coherent-scattering regime the scattered single photon is phase-locked to the driving field and therefore inherits its coherence. The phase locking was demonstrated by measuring the mutual coherence between the scattered single photon and the drive field, see Fig.\ \ref{Synthesized-photons}(a) and (b). A beat-note frequency with a linewidth of $\Delta \omega / 2 \pi = \SI{299}{\milli\hertz}$ was observed, which proves that dephasing of the quantum-dot spectrum can be overcome using weak resonant excitation. This ability to generate single photons at a frequency that is slaved to the excitation laser may be applicable for interfering photons from different quantum dots since they can be driven by the same laser source. The physical origin of the long mutual coherence time stems from the fact that the quantum-dot population remains on average small for weak excitation meaning that the transition experiences little disturbance. In the experiment the quantum dot was incorporated in a Schottky diode enabling suppression of charge fluctuations in the environment. An example of the generation of a train of synthesized single photons is reproduced in Fig.\ \ref{Synthesized-photons}(c). A tailored train of excitation pulses of \SI{500}{\pico\second} width and \SI{300}{\mega\hertz} repetition rate was generated by electro-optical modulation of a continuous-wave laser. The temporal shape of the single photons was found to resemble that of the driving pulse with only a weak exponential tail attributed to spontaneous emission. The employed method would enable generating arbitrarily-shaped single-photon pulses with a bandwidth limited by the spontaneous-emission decay time. The generated single photons were found to be of excellent quality with $g^{(2)}(0) < 5 \%$ (Fig.\ \ref{Synthesized-photons}(d), and photon indistinguishability of $(96 \pm 4) \%$ (Fig.\ \ref{Synthesized-photons}(e) and (f)), as was obtained after spectrally filtering the LA-phonon sidebands.

Single-photon generation by weak coherent scattering has the drawback that the photons are generated with a relatively modest efficiency since the quantum dot cannot be driven into saturation where it emits a single photon every time it is triggered. The generation efficiency is controlled by increasing the excitation intensity, although this also increases the amount of incoherent scattering and thus gradually destroys the mutual coherence of the photon with the drive field. Recently, highly coherent single photons have been observed with pulsed resonant excitation in the strong-excitation regime \cite{He2013NNANO}. By applying resonant $\pi$-pulses, on-demand single-photon generation with $99.7\%$ purity and $97\%$ indistinguishability was obtained, which makes this source very well suited for proof-of-concept linear-optics quantum-computing applications \cite{Kiraz2004PRA}. These achievements demonstrate that resonant excitation may be employed for eliminating the abundant dephasing processes often encountered with non-resonant excitation schemes. So far experiments have been performed mainly in homogeneous dielectric media, but extending this work to the realm of photonic nanostructures seems highly appealing since much higher single-photon generation efficiencies could be obtained. This would lead to new standards for highly efficient and coherent interfacing of light and matter, as discussed in further detail in Sec.\ \ref{Section-single-photon-nonlinearity}.

Finally it should be mentioned that the above discussion focused on the generation of coherent single photons for two-level solid-state emitters. The availability of a three-level $\Lambda$-system with two stable ground states, which can be obtained with trion states in quantum dots, opens new opportunities for the deterministic generation of coherent single photons. In this case, a cavity-enhanced Raman transition is employed to stimulate the deterministic emission of a photon  \cite{Bergmann1998RMP, Imamoglu1999PRL}. Interestingly, the frequency of the emitted photon is tunable over a range determined by the emitter-cavity coupling, which could be employed as a way of overcoming effects of inhomogeneous broadening that are unavoidable in solid-state implementations. Single-photon cavity-stimulated Raman emission has been reported with single atoms in a cavity \cite{Kuhn2002PRL} and very recently also with single quantum dots \cite{Sweeney2014NPHOT}.

\subsection{Observation of the Mollow triplet with a quantum dot}

\begin{figure}
\includegraphics[width= 0.9\columnwidth]{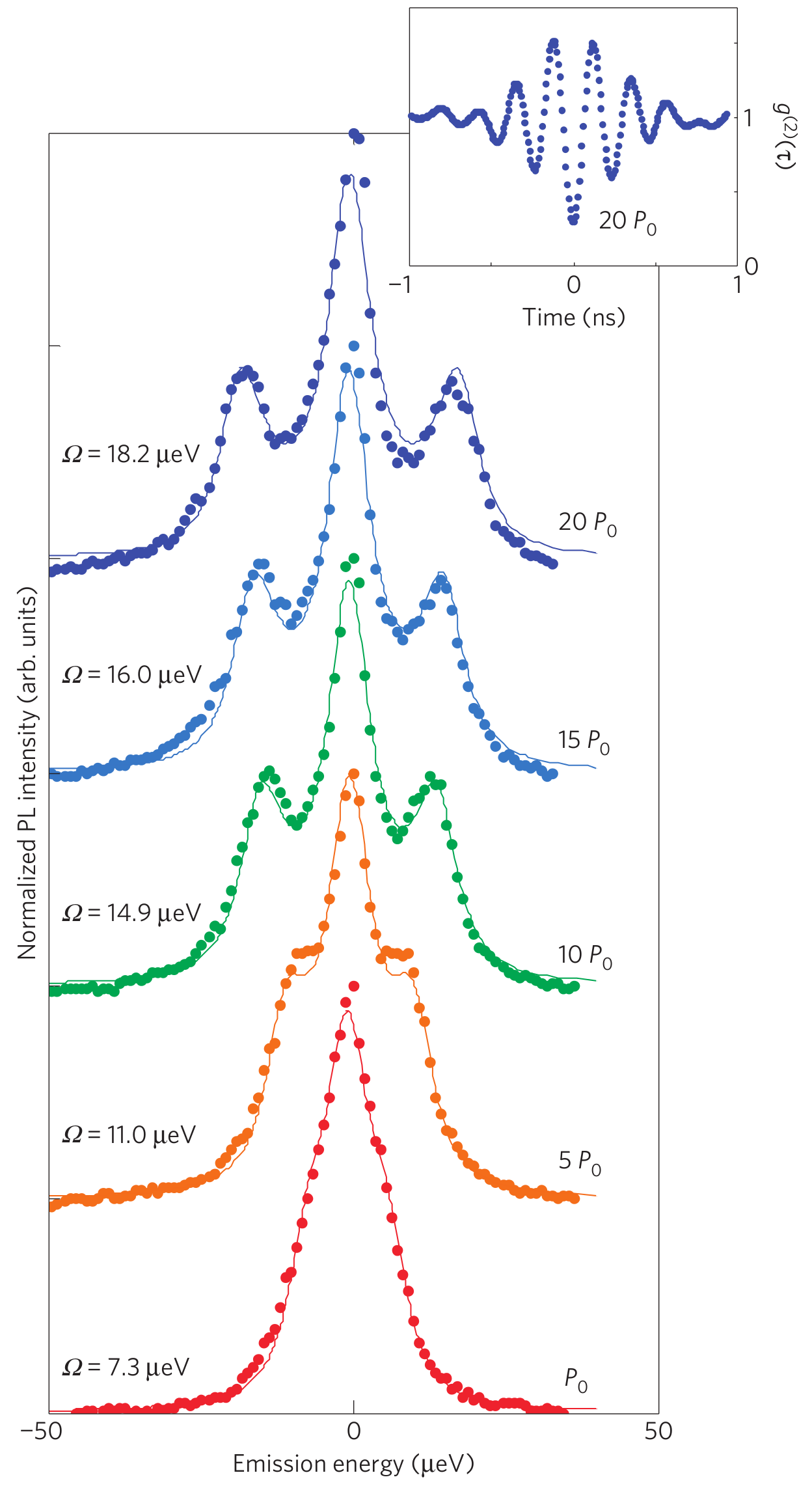}
 \caption{Observation of the Mollow triplet for a resonantly driven quantum dot for different excitation powers where $P_0 = 0.2 \: \mathrm{mW}$ and $\Omega \equiv 2 \left| \Omega_\text{p}\right|$. The experimental data can be modeled well with the theoretical expression for the incoherent spectrum of Eq.\ (\ref{inc-spectrum}). The inset shows the recorded auto-correlation function that displays Rabi oscillations in accordance with the prediction from Eq.\ (\ref{g2-res-excitation}). Reprinted with permission from \citet{Flagg2009NPHYS}.}
 \label{Mollow-triplet}
\end{figure}

In the regime of strong resonant coherent excitation, the two-level emitter is dressed and saturated by the excitation field leading to the Mollow-triplet emission spectrum, cf.\ Fig.\ \ref{Resonant-scat}(b). The Mollow spectrum has been observed for quantum dots in absorption measurements  \cite{Xu2007Science}. Here also an Autler-Townes splitting was found for a bright exciton transition when driving simultaneously the orthogonally polarized bright exciton state with a strong excitation field. The experimental observation of the Mollow-triplet emission spectrum has been reported in \citet{Flagg2009NPHYS}, \citet{Vamivakas2009NPHYS}, and \citet{Ates2009PRL}. Figure \ref{Mollow-triplet} shows an example of the Mollow spectrum obtained by spectrally resolving the resonant fluorescence with a Fabry-Perot interferometer. The sidebands emerge when the excitation power is increased and the Rabi splitting between the peaks was found to increase proportionally to the square root of the excitation power in accordance with Eq.\ (\ref{inc-spectrum}). In addition, Rabi oscillations were observed in the second-order correlation function, as also displayed in Fig.\ \ref{Mollow-triplet}. The Mollow sideband peaks are broadened by phonons and a detailed analysis of the dependence on excitation power has pinpointed the importance of excitation-induced dephasing  \cite{Ulrich2011PRL,Roy2011PRL}. Another recent prediction has been that a multitude of Mollow-like peaks can appear in the regime where the quantum dot is driven by relatively long and strong optical pulses \cite{Moelbjerg2012PRL}.

Interesting photon correlations are predicted in the Mollow spectrum when the excitation field is detuned away from the quantum-dot transition and certain parts of the spectrum are spectrally selected. Under these conditions, cascaded photon emission can occur since two or more photons can be emitted in succession from the relaxation of the coupled system through the dressed-state ladder. The time delay between the emitted photons is determined by the filter bandwidth and the excitation power \cite{Nienhuis1993PRA}. Such photon bunching has been observed experimentally by second-order correlation measurements on the two Mollow sidebands or by cross-correlation measurements between two different sidebands  \cite{Ulhaq2012NPHOT}. Furthermore, anti-bunching signifying single-photon emission was observed when just a single sideband was probed. In addition, long-timescale bunching (longer than \SI{10}{\nano\second}) was observed and attributed to the effect of quantum-dot blinking, as a consequence of slow charge-tunneling processes in and out of the quantum dot. In contrast, in the weak-excitation regime considered above, the influence of blinking was eliminated as evident from the extremely narrow bandwidth obtained from mutual-coherence measurements.
\section{Quantum electrodynamics in nanophotonic waveguides}
\label{Section-waveguide-QED}

\begin{figure}
\includegraphics[width= 0.9\columnwidth]{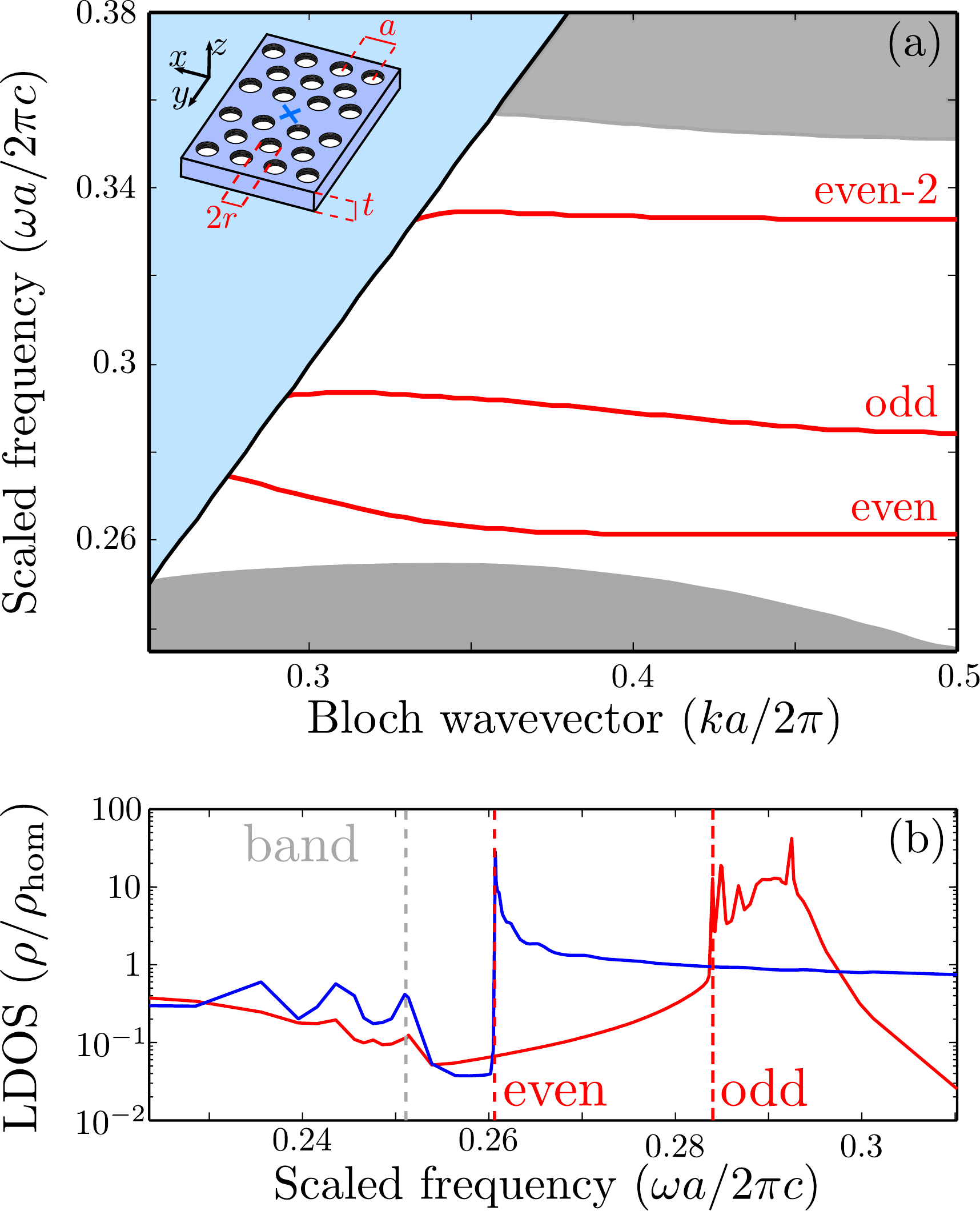}
\caption{Dispersion diagram and frequency dependence of the LDOS in a photonic-crystal waveguide.
(a) Projected band diagram of the TE-like modes of a W1 photonic-crystal waveguide of GaAs (with refractive index $n=3.5$, hole radius $r=a/3$ and membrane thickness $t= 2a/3$ where $a$ is the lattice period) displaying the scaled frequency as a function of the Bloch wave vector. Three waveguide modes (red lines) appear in the band gap and are labelled according to the symmetry of $E_y$. The lowest-frequency mode is usually the mode of interest. The grey and blue areas correspond to the regions outside the TE band gap where extended Bloch modes and a continuum of leaky radiation modes are found, respectively. The inset shows a schematic of the modeled waveguide. (b) Frequency dependence of the LDOS (normalized to a homogeneous medium with $n=3.5$) for an $x$-dipole (red curve) and a $y$-dipole (blue curve) positioned at the blue cross shown in the inset of (a). The vertical dashed lines mark the frequency regions corresponding to the extended Bloch modes in the photonic-crystal band, the even mode, and the odd mode.}
\label{Photonic-crystal-waveguide-dispersion-relation}
\end{figure}

Quantum emitters in nanophotonic waveguides provide a very promising way of enhancing light-matter interaction at the single-photon level as was originally proposed by \citet{Kleppner1981PRL}. Since waveguides are open systems, a single photon emitted by a quantum emitter in a waveguide can be channeled directly into a propagating mode and employed as a flying qubit in quantum-information processing. This contrasts waveguides with cavities, which are discussed in Sec.\ \ref{Section-cavity-QED}, where single photons are coupled to localized modes and subsequently need to be coupled out of the resonator for applications. In nanophotonic waveguides, the photon-matter coupling can be enhanced if the waveguide mode is strongly confined and has a low group velocity. Furthermore, in photonic crystals the coupling to lossy non-guided modes can be efficiently suppressed. This section reviews the underlying theory as well as experimental progress on QED with single-photon emitters in photonic waveguides, including plasmonic and dielectric nanowires and photonic-crystal waveguides.

\subsection{Purcell effect in a nanophotonic waveguide}
\label{Sec:Purcel-waveguide}

In the following, the theory of spontaneous emission in a photonic waveguide is presented. As a specific example, a photonic-crystal waveguide is considered in detail \cite{Hughes2004OL, LeCamp2007PRL, MangaRao2007PRB} but the theoretical framework is of general validity and may be applied to, e.g., dielectric or metallic nanowires as well. We consider a W1 waveguide, which is a photonic-crystal membrane with one missing row of holes, cf.\ Table \ref{nanoWaveguides}(a). The dispersion relation for the TE-polarized modes of a W1 waveguide is displayed in Fig.\ \ref{Photonic-crystal-waveguide-dispersion-relation}(a). The basis functions for a single band of the photonic waveguide are Bloch modes of the form \cite{Yao2010LPR}
\be
{\mathbf u}_k({\mathbf r}) = \sqrt{\frac{a}{L}} {\mathbf b}_k(\mathbf r) e^{i k x},
\ee
where ${\bf k} = k \,{\bf \hat{e}_x}$ is the Bloch wave vector, $L$ is the length of the waveguide, $a$ is the lattice constant, and ${\bf{b}}_{k}(\bf r)$ is a function  that is 1D periodic along the axis of the waveguide. Here we restrict the calculation of the LDOS to only considering the contribution of a single transverse guided mode, which is a good approximation since the three guided bands are separated in frequency and the coupling rate to the guided modes largely dominates the rate of coupling to non-guided radiation modes, which is quantified by the $\beta$-factor discussed in Sec.\ \ref{Sec:Photon-det-eff}. The waveguide constitutes a truly one-dimensional optical system in which light propagation is characterized by a single Bloch wave vector ${\bf k}$, and from Eq.\ (\ref{Greens-def}), the waveguide Green's tensor can be expressed as
\be
\overleftrightarrow{G}({\bf r},{\bf r'},\omega) \approx \frac{L \, \omega}{4 \pi} \int_{-\infty}^{\infty} \frac{d \omega_k}{v_\text{g}} \frac{  {\bf u}_{{k}}({\bf r}) \otimes {\bf u}_{{k}}^*({\bf r'})}{\omega_k - \omega - i \delta}, \nn
\ee
where a sum has been converted into integration according to $\sum_k \rightarrow (L/2 \pi)\int \text{d}k$, and $\delta$ is an in\-fi\-ni\-te\-si\-mal number added to the denominator allowing for the evaluation of the integral giving
\bea
&& \overleftrightarrow{G}({\bf r},{\bf r'},\omega) =
\frac{i a \omega}{2 v_\text{g}}  \Theta(x-x'){\bf b}_{k}({\bf r}) {\bf b}_{ k}^*({\bf r'}) e^{ik(x-x')} \nn \\
   + && \frac{i a \omega}{2 v_\text{g}} \Theta(x'-x){\bf b}_{k}^*({\bf r}) {\bf b}_{k}({\bf r'}) e^{-ik(x-x')} ,
\label{Greens-fct-waveguide}
\eea
where the Heaviside function $\Theta$ determines the two branches of forward- and backward-propagating modes. The associated LDOS can be cast in the form
\be
\rho\left(\mathbf r,\omega, \hat{\mathbf{e}}_\mathbf{d} \right) = \frac{a}{\pi v_\text{g}} \frac{f (\mathbf r)}{\epsilon(\mathbf r) V_{\mathrm{eff}}} \left| \hat{\mathbf e}_k(\mathbf r) \cdot \hat{\mathbf{e}}_\mathbf{d} \right|^2,
\ee
where $\hat{\mathbf e}_k(\mathbf r)$ is the unit vector of the electric field of the waveguide mode, $V_{\mathrm{eff}}^{-1} = \mathrm{max} \left[ \epsilon({\bf r}) \left| {\bf b}_k({\bf r}) \right|^2 \right]$ is the inverse of the effective mode volume per unit cell and $\mathrm{max} \left[\ldots \right]$ evaluates the maximum value within one unit cell of the photonic crystal.  The dimensionless function $f(\mathbf r) = \epsilon(\mathbf r) \left| \mathbf b_\mathbf k(\mathbf r) \right|^2 V_{\mathrm{eff}}$ varies between zero and unity and expresses the spatial mismatch between the emitter and the waveguide-mode field maximum. $\left| \hat{\mathbf e}_k(\mathbf r) \cdot \hat{\mathbf{e}}_\mathbf{d} \right|^2$ quantifies the alignment of the dipole with respect to the waveguide Bloch mode.

\begin{figure}
\includegraphics[width= 0.8\columnwidth]{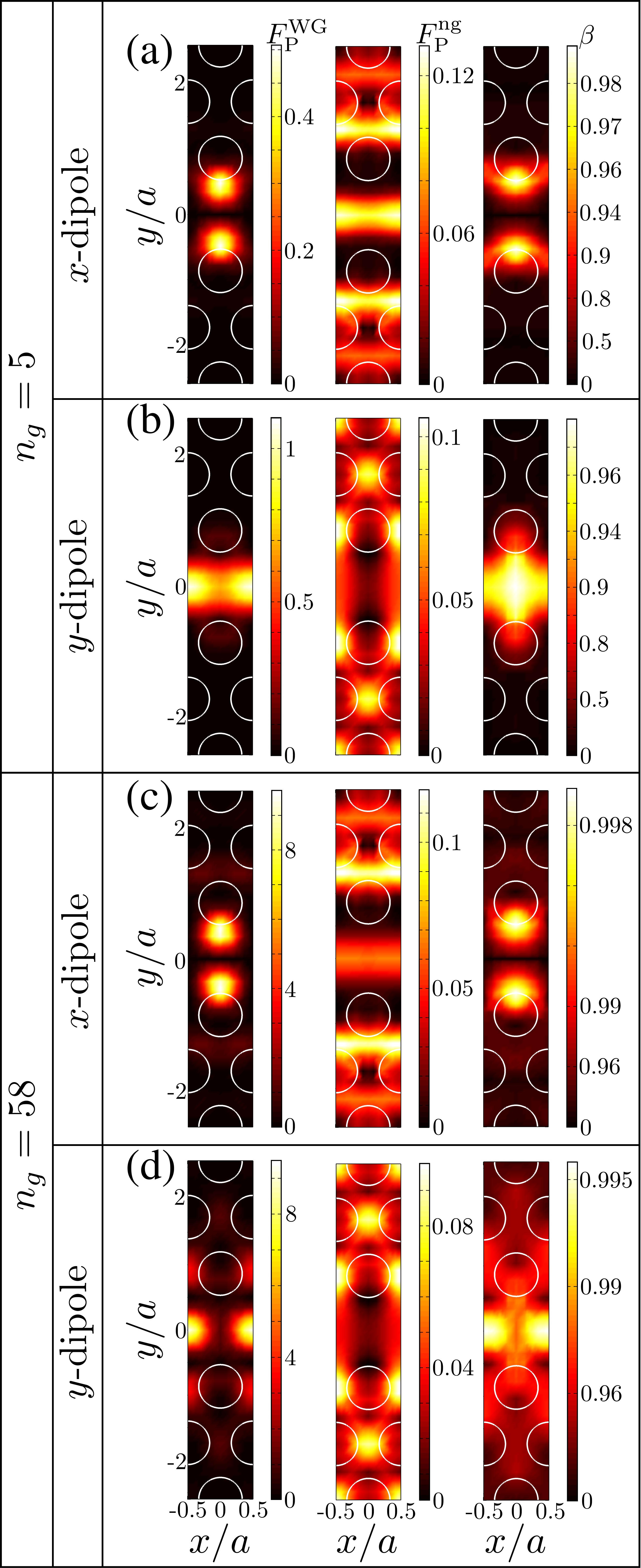}
 \caption{ Spatial dependence of the Purcell factor and the $\beta$-factor in a photonic-crystal waveguide. The photonic-crystal waveguide is made in a membrane with refractive index $n=3.5$, hole radius $r= a/3$, where $a$ is the period, and membrane thickness $t=2a/3$. The left column shows the overall Purcell factor, the middle column the Purcell factor for emission into non-guided radiation modes, and the right column the radiative $\beta$-factor ($\gamma_{\rm nrad}=0$) for coupling into the waveguide mode.  Note the nonlinear color scale. (a) Spatial maps for a dipole oriented along the $x$-direction at a frequency where the even photonic waveguide has a group index of $n_\text{g}=5$. (b) Same as (a), but for a $y$-dipole. (c) Spatial maps for an $x$-dipole with a frequency corresponding to the waveguide mode having a group index of $n_\text{g} = 58$. (d) Same as (c), but for a $y$-dipole.}
 \label{LDOS-WG-maps}
\end{figure}

In a photonic-crystal waveguide the mode is confined to a small spatial area implying that light propagation does not satisfy the paraxial approximation. As a consequence the guided mode generally has an electric-field component also along the propagation direction. Figure \ref{Photonic-crystal-waveguide-dispersion-relation}(b) shows the frequency-dependent LDOS for a W1 photonic-crystal waveguide for $x$ and $y$-dipoles, respectively, positioned at the cross in the inset of Fig. \ref{Photonic-crystal-waveguide-dispersion-relation}(a). Since the dipole is placed at a high symmetry point, the $x$-dipole only couples to the odd mode, while the $y$-dipole only couples to even modes. For the $y$-dipole a broad range of frequencies with an enhanced LDOS is observed (for scaled frequencies above 0.26) as a result of the coupling to the highly dispersive even mode. This mode has been employed in experiments for the observation of broadband Purcell enhancement, as is discussed in Sec.\ \ref{Sec:waveguide-experiments}. Ideally the LDOS is predicted to diverge at the band edge of the waveguide mode where the group velocity vanishes. This divergence is unresolved in the numerical simulations presented in Fig.\ \ref{Photonic-crystal-waveguide-dispersion-relation}(b) due to the finite spectral resolution of the calculations that focus on broadband features. In experiments the predicted divergence is smoothed due to fabrication imperfections, cf.\ Sec.\ \ref{Section-fab-imperfections}. A dipole oriented along the $x$-axis is predicted to experience several sharp peaks from the LDOS, which are due to the group velocity of the odd mode approaching zero at several frequencies. Finally, the frequency interval $\omega a/ 2 \pi c = (0.255, 0.26)$ in Fig.\ \ref{Photonic-crystal-waveguide-dispersion-relation}(b) is below the onset of the waveguide modes and  therefore in the band-gap region. Here the LDOS for both dipoles is strongly suppressed, which quantifies the suppression of radiation modes that can be achieved in photonic-crystal waveguides.

A dipole optimally positioned at an antinode of the photonic-crystal waveguide mode ($f(\mathbf r)=1$) and oriented along the electric field has a maximum Purcell factor of
\be
F_{\mathrm{P}}^{\mathrm{max}}(\omega) = \left(\frac{3}{4\pi n} \frac{\lambda^2/n^2}{V_{\rm eff}/a} \right)n_\text{g}(\omega),
\label{PCWpurcell}
\ee
where $n_\text{g}(\omega) = c/v_\text{g}(\omega)$ is the group index that specifies the slow-down factor of the photonic-crystal waveguide. This expression illustrates how the light-matter enhancement is accommodated in a photonic waveguide by two effects: a slow group velocity as can be obtained in dispersive waveguides and a tight confinement of the mode as expressed by the effective mode volume. In a photonic-crystal waveguide both effects are employed, i.e., the structural dispersion of the Bloch modes gives rise to slow light and the mode is tightly confined to the diffraction-limited defect area. In plasmonic nanowires, subwavelength confinement combined with the slow propagation of the lowest-order guided mode lead to potentially large Purcell factors \cite{Chang2007PRB}. In contrast, the confinement and slow-down is less pronounced in dielectric photonic nanowires than in plasmonic structures and in the relevant regime of single-mode operation, the spontaneous-emission rate is typically suppressed relative to the value in a homogeneous medium corresponding to a Purcell factor of less than unity \cite{Bleuse2011PRL}.

In a W1 photonic-crystal waveguide, the effective mode volume is $V_{\rm eff} \sim a (\lambda/n)^2/3$ and is found to vary weakly over the waveguide band. The maximum achievable Purcell factor is thus determined by the group index $n_\text{g}$.  Experimentally,  $n_\text{g} \sim 300$ has been reported for W1 photonic-crystal waveguides in silicon \cite{Vlasov2005Nature}, which means that a Purcell factor approaching $60$ should be achievable.  In active GaAs photonic-crystal membranes containing quantum dots, typical slow-down factors of $n_\text{g} \sim 50$ were observed \cite{Arcari2013InPrep}. Even larger slow-down factors can potentially be obtained by improving fabrication quality or by designing photonic-crystal-waveguide bands that are more robust to imperfections. Figure \ref{LDOS-WG-maps} illustrates the spatial dependence of the Purcell factor in a photonic-crystal waveguide both in the fast-light $(n_\text{g}=5)$ and slow-light $(n_\text{g}=58)$ regimes corresponding to experimentally relevant parameters. For  $n_\text{g}=58$, cf.\ Figs.\ \ref{LDOS-WG-maps}(c) and (d), $F_\text{P}^{\mathrm WG} \sim 9$ is expected for both $x$- and $y$-dipole orientations. The spatial profile of the Purcell factor is determined by the Bloch function of the waveguide mode.

\begin{figure}
\includegraphics[width= \columnwidth]{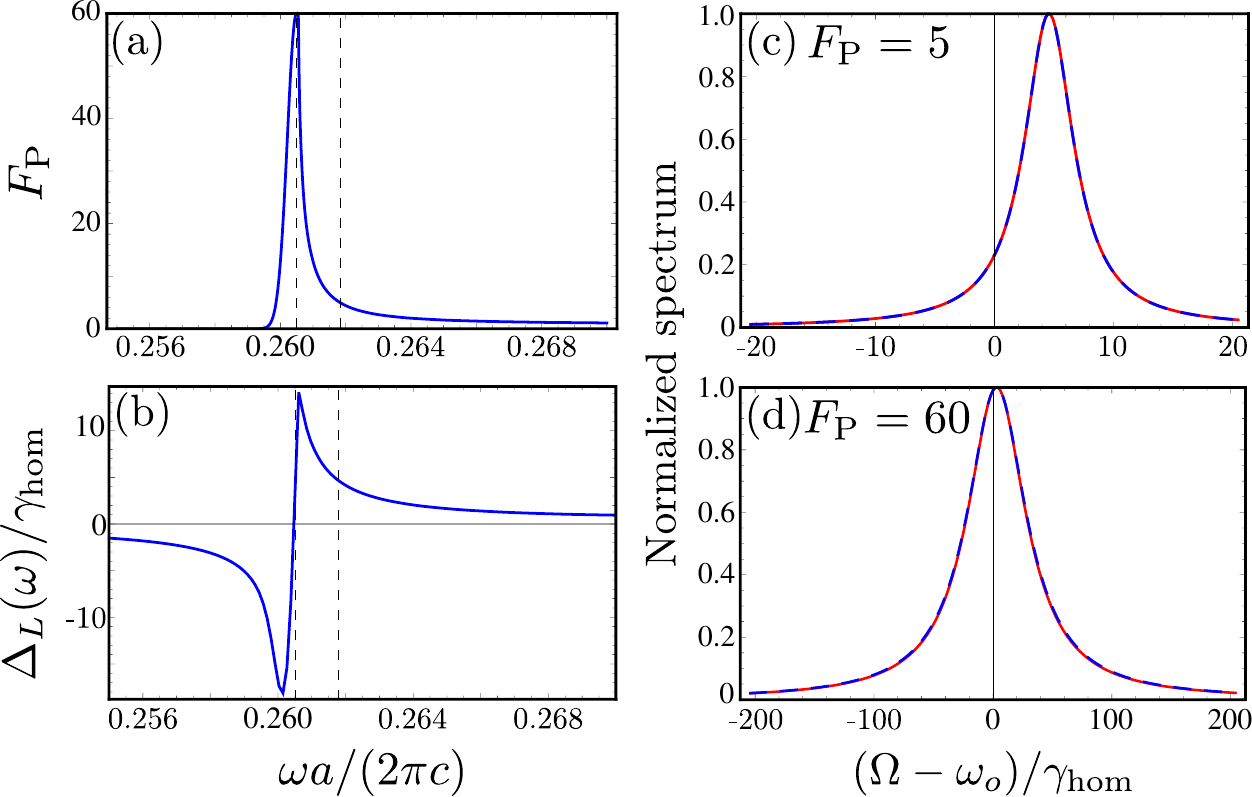}
 \caption{Spectral properties of emission in a photonic-crystal waveguide  from a quantum dot with homogeneous decay rate $\gamma_\text{hom} = \SI{1}{\nano\second^{-1}}$  and wavelength $\lambda = \SI{950}{\nano\meter}$. (a) A model Purcell factor for a photonic-crystal waveguide with a peak value of $F_\text{P} \sim 60$ and assuming a Gaussian roll-off below the peak in order to resemble broadening of the band edge due to fabrication imperfections. (b) Plot of the frequency dependence of the Lamb shift for the parameters used in (a). (c) The emission spectrum of an emitter centered at the frequency where $F_\text{P} \sim 5$ (indicated by dashed vertical line in (a) and (b)).  (d) same as (c) but for $F_\text{P} \sim 60$. In (c) and (d) the red (blue) curves are computed by applying (not applying) the Wigner-Weisskopf approximation and the two curves are found to be identical on the plotted scale. Note the different scales on the abscissae in (c) and (d). }
\label{Purcell-factor-waveguide}
\end{figure}

For large Purcell factors in the photonic-crystal waveguide non-Markovian dynamics from the coupling to the radiation reservoir could play a role. The theoretical framework for this is presented in Sec.\ \ref{Section-spontaneous-emission}. For applications as a single-photon source such quantum back action may be a nuisance since it  effectively can extend the lifetime of the excited state of the emitter, and therefore reduce the achievable rate of single-photon generation. Figure \ref{Purcell-factor-waveguide} exploits the Lamb shift and the corresponding emission spectra for photonic-crystal waveguides with experimentally relevant values of the Purcell factors of up to $60$. The Lamb shift is found to be strongly modified by the photonic-crystal waveguide, which should be experimentally observable. Nonetheless the emission spectra are  extremely well described by Wigner-Weisskopf theory testifying that non-Markovian dynamical processes are absent.

\subsection{Efficiency of a single-photon source}
\label{Sec:Photon-det-eff}

While the Purcell factor quantifies the enhancement of the single-photon emission rate, another important figure of merit for a single-photon source is the overall efficiency, which depends on generation, collection, and subsequent detection efficiencies. It can be expressed as
\be
\eta_{\rm tot} = \eta_{\rm det} \times \beta \times \eta_{\rm gen},
\label{overall-eff}
\ee
where $\eta_{\rm gen}$ is the probability that the excitation of the quantum dot leads to the preparation of a bright exciton state, $\beta$ is the probability that an exciton recombination leads to a photon in the desired waveguide mode, and $\eta_\text{det}$ is the probability of collecting and detecting the photon once is has been launched in the waveguide. In the following these terms are discussed in detail.

The generation efficiency $\eta_{\rm gen}$ depends on the way the quantum dot is excited and is sensitive to charge fluctuations in the nearby environment associated with defect sites. By pumping the ground-state transition into saturation, by, e.g., non-resonant excitation schemes, a single exciton can be prepared with near-unity probability. Depending on the pumping conditions, however, quantum dots may suffer from blinking processes, e.g., by spin flips that turn bright excitons into dark excitons ~\cite{Johansen2010PRB} or by tunneling of carriers in or out of the quantum dot. As a consequence the emission will turn on and off, which can occur at various time scales of, e.g., 100 nanoseconds or longer \cite{Santori2004PRB}. The spin-flip processes as well as the coupling to other charged exciton complexes can be modified by applying a DC electric field across the quantum dot \cite{Smith2005PRL}, which may be employed for optimizing $\eta_{\rm gen}$.

Nanophotonic waveguides are very well suited for obtaining a large $\beta$-factor. The $\beta$-factor is defined as the rate of spontaneous-photon emission into the waveguide mode, $\gamma_{\rm wg}$, relative to the total recombination rate of the emitter by all possible decay processes, i.e.,
\be
\beta(\omega) = \frac{\gamma_{\rm wg}}{\gamma_{\rm wg} + \gamma_{\rm ng} + \gamma_{\rm nrad}},
\label{eq:beta-factor-definition}
\ee
where $\gamma_\mathrm{ng}$ is the loss rate of coupling to all non-guided modes and $\gamma_\mathrm{nrad}$ is the rate of intrinsic non-radiative recombination in the quantum dot as discussed in Sec.\ \ref{sec:sst:decaydynamics}. The radiative rates in the $\beta$-factor can be computed from the LDOS associated with the guided and nonguided modes by applying Eq.\ (\ref{rate}). In dielectric nanostructures (photonic-crystal waveguides and nanowires) the non-guided modes are extended radiation modes. For plasmonic nanowires, however, the imaginary part of the dielectric permittivity is non-negligible, which leads to an additional non-radiative decay channel due to resistive heating in the nanowire. Although the resistive heating is fully described by the projected LDOS and as such is an effect of the photonic environment, it does not lead to emission of photons. The calculated spatial dependence of the $\beta$-factor for a W1 photonic-crystal waveguide including the fraction of the emission that couples respectively to the guided mode and the non-guided modes are shown in Figure \ref{LDOS-WG-maps} for the case of $\gamma_{\rm nrad}=0$. The $\beta$-factor is predicted to be remarkably close to unity even in the fast-light regime where the coupling to the waveguide is not Purcell enhanced. This robustness stems from the fact that the 2D photonic band gap suppresses $\gamma_{\rm ng}$, i.e., the leakage to unwanted modes is strongly inhibited. Importantly, near-unity $\beta$-factors are expected in essentially any spatial position within the photonic-crystal waveguide since either an $x$-oriented dipole or a $y$-oriented dipole couples well to the waveguide mode. These characteristics make photonic-crystal waveguides very appealing for realizing efficient single-photon sources and giant photon nonlinearities. It is an interesting observation that $\beta$ in practice will be limited rather by intrinsic non-radiative processes $(\gamma_{\rm nrad})$ in the quantum dot rather than the actual waveguide. The large bandwidth of a photonic-crystal waveguide could in certain situations be a drawback since it could imply that phonon sidebands are not suppressed, which would be the case in a narrow bandwidth cavity. However, the ability to tailor the dispersion may enable engineering photonic-crystal waveguides with optimized bandwidths.

Finally, the detection efficiency $\eta_{\rm det}$ depends both on the collection of  photons from the waveguide and the subsequent propagation loss and detector efficiency. This requires shaping the optical mode from the waveguide such that it can be efficiently collected by a microscope objective with a given numerical aperture. Quantitative measurements of the collection efficiency are challenging since the outcoupling from a photonic nanostructure is generally sensitive to fabrication imperfections and precise alignment of the optical setup. One approach has been to infer the collection efficiency by comparing the target quantum dot to another quantum dot that is positioned in a non-structured medium and where the collection efficiency is readily calculated, although such an approach does not include the potential influence of blinking effects.  In many present experiments the detection efficiency is rather low (typically about a few percent) but can potentially be optimized significantly by elaborate outcoupling designs from the nanostructure.

So far, the highest reported rate of detecting single photons was \SI{4}{\mega\hertz} for a quantum dot embedded in a microcavity and driven by pulsed laser at the repetition rate of \SI{82}{\mega\hertz} \cite{Strauf2007NPHOT}, i.e., an overall efficiency of $\eta_{\rm tot} \sim 5 \%$. We anticipate that the simultaneous optimization of all efficiencies in Eq.\ (\ref{overall-eff}) in one optical device will enable observing higher efficiencies in the near future. The following section addresses the experimental progress on nanophotonic waveguides for increasing the efficiency of quantum-dot single-photon sources.

\subsection{Experimental progress on waveguide single-photon sources}
\label{Sec:waveguide-experiments}

Considerable efforts have been devoted to the development of single-photon sources based on nanophotonic waveguides. In this section the experimental progress on photonic-crystal waveguides as well as dielectric and metallic nanowires is reviewed.

\begin{figure}
\includegraphics[width= 0.8\columnwidth]{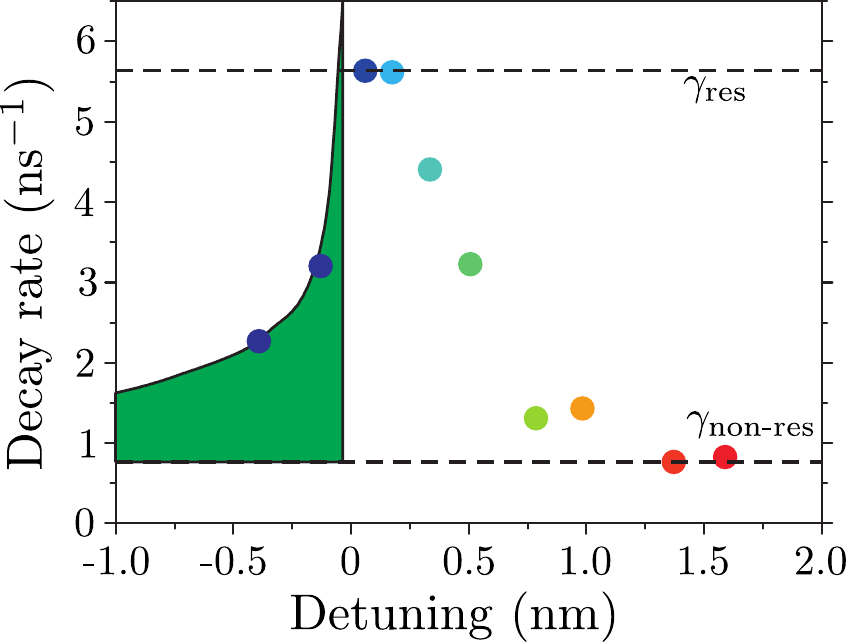}
 \caption{
Examples of measurements of the spontaneous-emission rate of a quantum dot while temperature tuning it into resonance with a photonic-crystal waveguide. The temperature was varied between \SI{10}{\kelvin} (dark blue color) and \SI{60}{\kelvin} (red color) in steps of \SI{5}{\kelvin} whereby the quantum-dot transition red shifted. The green curve traces the frequency variation of the LDOS for an ideal photonic-crystal waveguide without fabrication imperfections. Reprinted with permission from \citet{Thyrrestrup2010APL}.}
 \label{beta-factor-measurements}
\end{figure}

\subsubsection{Photonic-crystal waveguides}

\begin{figure}
\includegraphics[width= 0.8\columnwidth]{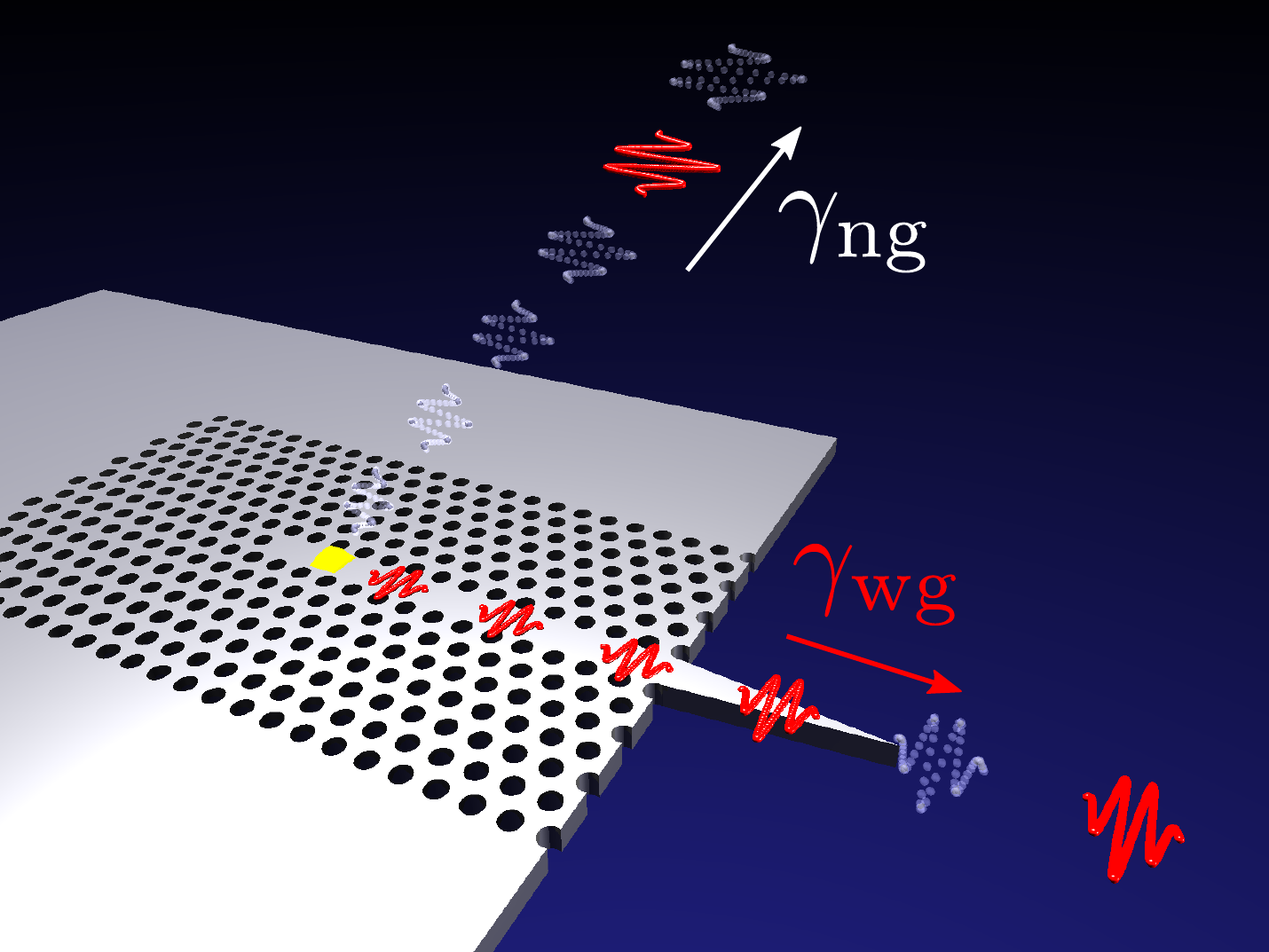}
 \caption{Illustration of a highly efficient single-photon source, where photon wave packets are being emitted from a quantum dot in a photonic-crystal waveguide.  The high $\beta$-factor means that almost all photons couple to the waveguide mode and may subsequently be coupled efficiently off chip via a tapered tip protruding from the waveguide. Reprinted with permission from \citet{Arcari2013InPrep}.
}
 \label{PCW_povray}
\end{figure}

Photonic-crystal waveguides fabricated in GaAs and containing InGaAs quantum dots have proven to be well suited for Purcell enhancement and record-high $\beta$-factors.  The first experimental demonstration of Purcell enhancement in a photonic-crystal waveguide used quantum dots embedded in a W3 (three rows of missing holes) photonic-crystal waveguide \cite{Viasnoff-Schwoob2005PRL} where a $16 \%$ Purcell enhancement   was found. Better enhancement and suppression of leaky radiation modes can be obtained in membrane structures and by using the more confined mode of a W1 waveguide, cf.\ the electron micrograph displayed in Table \ \ref{nanoWaveguides}(a). The first experiments on single quantum dots in such samples showed that the $\beta$-factor could approach $90 \%$ and reach above $50 \%$ in a wide bandwidth of $20 \: \mathrm{nm}$ \cite{Lund-Hansen2008PRL}. Subsequent experiments used temperature tuning of a single quantum dot relative to the cut-off frequency of the waveguide mode. Figure \ref{beta-factor-measurements} shows an example of such measurements where a quantum dot was tuned into resonance with the slow-light region of the waveguide mode from which a Purcell factor of $F_\text{P}=5.2$ was observed \cite{Thyrrestrup2010APL}. This corresponds to a decay time of \SI{175}{\pico\second}, which is sufficiently fast to aid in overcoming dephasing processes, as is required for the generation of indistinguishable photons. With increasing temperature the quantum dot red shifted into the band-gap region of the photonic crystal giving rise to a gradual decrease of the decay rate reflecting the frequency tail of the waveguide LDOS that is broadened by fabrication imperfections.
The $\beta$-factor could be estimated from $\beta = \left( \gamma_\textrm{res}-\gamma_\textrm{non-res} \right) / \gamma_\textrm{res}$ where $\gamma_\textrm{res}$ and $\gamma_\textrm{non-res}$ are the respective resonant and non-resonant decay rates, cf.\ Fig.\ \ref{beta-factor-measurements}. From these measurements, $\beta=85 \%$ was extracted, which importantly constitutes a very conservative estimate since the non-resonant rate was obtained at an elevated temperature of \SI{60}{\kelvin} where the population of excited states and increased non-radiative recombination \cite{Tighineanu2013PRB}, and phonon-assisted coupling to the waveguide \cite{Madsen2013PRB} may also increase the rate significantly.

The full potential of photonic-crystal waveguides can be realized by noting that the suppression of coupling to non-guided radiation modes is typically suppressed to less than $10 \%$ of the decay rate in a homogeneous medium \cite{LeCamp2007PRL}, cf.\ Fig.\ \ref{LDOS-WG-maps}. Therefore a quantum dot with a Purcell factor of $F_\text{P}=5.2$ is expected to have a $\beta$-factor very close to unity, which is also evident from the calculations shown in Fig.\ \ref{LDOS-WG-maps}. Indeed, in a recent experiment $\beta = 98.4  \%$ has been reported \cite{Arcari2013InPrep}, which demonstrates that this promising potential of photonic-crystal waveguides can be realized experimentally.

The near-unity $\beta$-factors make photonic-crystal waveguides very promising for highly-efficient single-photon sources or single-photon nonlinearities (see Sec.\ \ref{Section-single-photon-nonlinearity}). With $\beta = 98.4 \%$ a device would emit a deterministic train of single photons with a failure probability of only $1.6 \%$ that an excited bright exciton in the quantum dot does not lead to a photon in the waveguide. Such a device is illustrated in Fig.\ \ref{PCW_povray}. In this case it is essential to implement short waveguides so that multiple scattering due to fabrication imperfections (discussed in Sec.\ \ref{Section-fab-imperfections}) does not hinder light propagation. This turns out to be feasible since a length of only 10-20 unit cells (approximately $\SI{5}{\micro\meter}$) is enough to achieve very large Purcell effects in a photonic-crystal waveguide \cite{MangaRao2007PRL}. Experimental progress on short photonic-crystal waveguides was reported in \citet{Dewhurst2010APL} and \citet{Hoang2012APL}. The single-photon purity of a quantum-dot photonic-crystal waveguide source has been studied in \citet{Schwagmann2011APL,Laucht2012PRX,Arcari2013InPrep}, where the best reported value so far of $\sim 5 \%$ is expected to be further improved by implementing quasi-resonant or resonant excitation schemes. Another important issue for immediate applications is to efficiently couple the single photons generated in the waveguide off the chip for detection. To this end, experimental efforts on implementing an adiabatic taper tip on the waveguide have been reported demonstrating $\sim 80\%$ out coupling from a photonic-crystal waveguide to free space \cite{Tran2009APL}.  Gratings that couple photonic-crystal-waveguide modes vertically out of the structure with high efficiency have also been demonstrated \cite{Faraon2008OE, Wasley2012APL}, although due to the symmetry of the structures half of the light is emitted downwards and not directly collected. This could potentially be improved by incorporating a distributed Bragg reflector below the air gap under the photonic-crystal membrane. Another possibility is to couple the photons in the photonic-crystal waveguides to ridge waveguides. This is efficient only for low-group-index photonic-crystal waveguide modes, meaning that a high-$n_\text{g}$ mode must be converted to a low-$n_\text{g}$ mode, which can be done efficiently by using a photonic-crystal waveguide-transition region \cite{Hugonin2007OL} or even in a direct coupling from a slow to a fast waveguide due to the existence of strong evanescent modes \cite{deSterke2009OE}. Once light is coupled from a photonic-crystal waveguide to a ridge waveguide it can propagate longer distances with minimal scattering losses. Joining multiple photonic-crystal waveguides via ridge waveguides could enable the construction of planar photonic circuits directly suitable for quantum-information processing.

\subsubsection{Dielectric and plasmonic nanowires}

\begin{figure}
\includegraphics[width= \columnwidth]{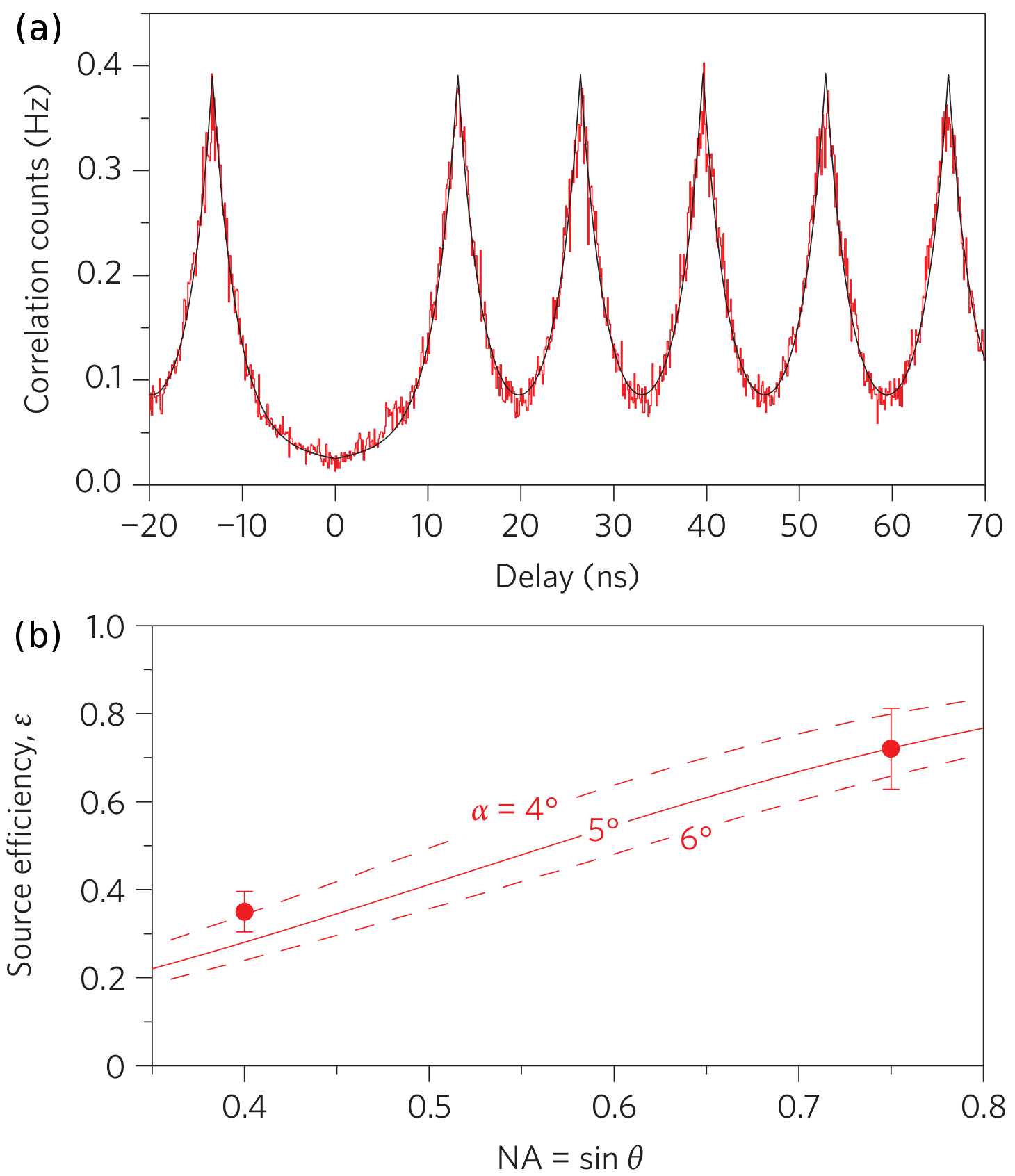}
 \caption{Optical characterization of a quantum dot in a dielectric photonic nanowire for the structure shown in Table \ref{nanoWaveguides}(b). (a) Measured second-order correlation function demonstrating excellent single-photon purity when driving the quantum dot with a pulsed laser. From the data, $g^{(2)}(0) < 0.8 \%$ was extracted. (b) Estimated collection efficiency from the source by the first lens as a function of the numerical aperture (NA) of the collection lens and for different values of the tapering angle $\alpha$ of the nanowire. Reprinted with permission from \citet{Claudon2010NPHOT}.}
 \label{Nanowire}
\end{figure}

Dielectric nanowires are promising alternatives for achieving large $\beta$-factors over a wide bandwidth. In a nanowire with a diameter at the scale of the wavelength, see Table\ \ref{nanoWaveguides}(b), the spontaneous emission to radiation modes can be strongly suppressed while the coupling to a single guided mode in the nanowire can be sizable. For instance, $\beta$-factors approaching $95 \%$ were predicted for a GaAs nanowire with a diameter of about a quarter of the vacuum wavelength, $\sim \lambda/4$, with an operation bandwidth as large as \SI{70}{\nano\meter} since the guided mode is only weakly dispersive. The experimental demonstration of spontaneous-emission inhibition in GaAs nanowires was reported in \citet{Bleuse2011PRL} where an inhibition factor of $16$ was measured on narrow nanowires with a diameter of $0.13 \lambda$. For the wider nanowires (diameter $\sim \lambda/4$) where large $\beta$-factors are predicted, a Purcell factor of $1.5$ was observed. A potential asset of dielectric nanowires compared to photonic-crystal waveguides is that their structural simplicity may make them robust towards fabrication imperfections. Important progress has been made on out coupling single photons from a dielectric nanowire with very high efficiency. This can be achieved by fabricating a tapering of the nanowire tip and integrating a gold mirror beneath the quantum dot \cite{Claudon2010NPHOT}. For a lens with $\mathrm{NA} = 0.75$, an overall single-photon collection efficiency of up to $72 \%$ was inferred, see Fig.\ \ref{Nanowire}(b). In this work the $\beta$-factor was not recorded explicitly, but the inferred efficiency constitutes a conservative lower bound. The excellent purity of the single-photon generation was demonstrated by pulsed auto-correlation measurements even when driving the quantum dot into saturation, cf.\ Fig.\ \ref{Nanowire}(a). These experiments illustrate the very promising potential of dielectric nanowires for highly-efficient single-photon generation. One potential limitation of this method is that only modest Purcell enhancement can be achieved, which limits the rate at which single photons is generated and potentially also the indistinguishability of the photons.

Metallic nanowires offer an interesting alternative to dielectric nanowires since the spontaneous-emission rate can potentially be strongly Purcell enhanced. A quantum emitter placed in the vicinity of a metallic nanowire can decay by exciting surface-plasmon polaritons that propagate along the metal wire. One challenge for plasmonic photonic circuits is that they suffer from inherent absorptive losses and an enhanced sensitivity to fabrication imperfections due to the very strong confinement of the optical modes. The experimental progress on coupling quantum dots to plasmonic nanowires has so far been limited; a detailed theoretical investigation of experimentally realistic structures is presented in \citet{Chen2010OE}. Here it was found that since self-assembled quantum dots must be overgrown with a semiconductor capping layer of typically 20-\SI{30}{\nano\meter} in order to be optically active, design strategies are constrained to rather leaky structures. As a consequence, the achievable $\beta$-factor for dipole transitions is typically below $50 \%$ even without accounting for effects due to fabrication imperfections. Indeed the experimental demonstration of efficient coupling of self-assembled quantum dots to plasmon nanowires has been lacking so far.
Important progress has been obtained with other types of solid-state emitters in room-temperature experiments: the first experimental demonstration of emitter-plasmon coupling in a metallic nanowire was performed with a colloidal CdSe quantum dot coupled to a chemically grown crystalline silver nanowire, where Purcell enhancement of $2.5$ was observed \cite{Akimov2007Nature}. These results were later improved by coupling a single nitrogen-vacancy center to a plasmonic nanowire \cite{Schietinger2009NanoLett} and a Purcell factor of up to $8.3$ was obtained with a propagating plasmonic gap mode residing in between two parallel silver nanowires \cite{Kumar2013NL}. Furthermore, the wave-particle duality of a single surface-plasmon-polariton excitation has been experimentally demonstrated \cite{Kolsesov2009NPHYS}. Recently, plasmonic nanocavities have been constructed \cite{Russell2012NPHOT, deLeon2012PRL} in attempting to enhance light-matter interaction beyond the level possible in nanowire geometries. Recent interest has also been on antenna structures where resonant plasmonic nanostructures can help directing photons with high efficiency \cite{Curto2010Science}. Also alternatives based on planar dielectric layers have been considered where an impressive single-photon collection efficiency of $96 \%$ was reported \cite{Lee2011NPHOT}.

\subsection{Single-photon nonlinearity}
\label{Section-single-photon-nonlinearity}

\begin{figure}
\includegraphics[width= \columnwidth]{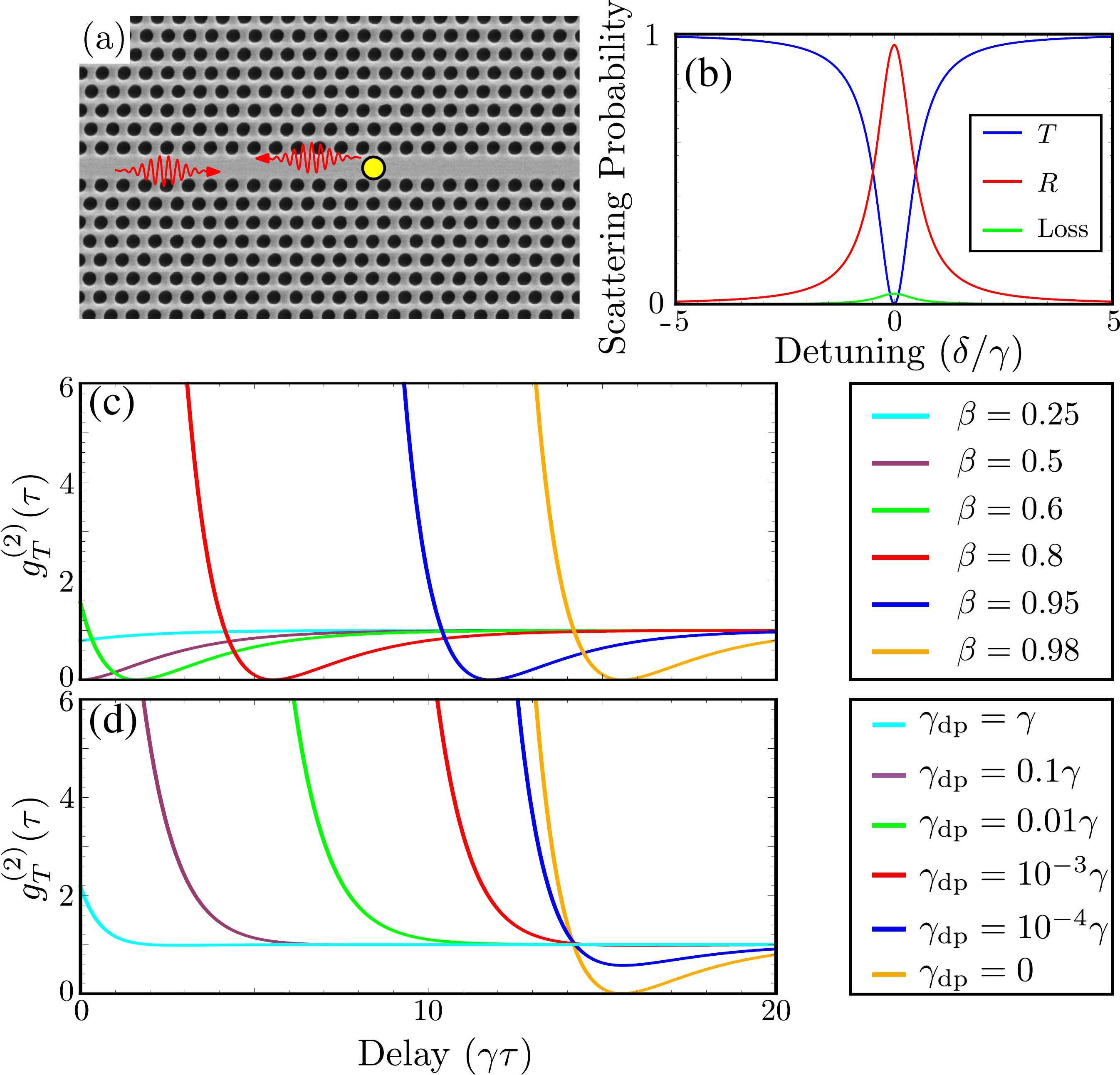}
 \caption{Single-photon nonlinearity with a single quantum dot in a photonic-crystal waveguide by scattering weak coherent pulses. (a) Schematic illustration that a single quantum dot (yellow circle) in a photonic-crystal waveguide may reflect a single-photon wavepacket with near-unity probability. (b) The reflection and transmission probabilities as a function of detuning for $\beta = 98 \%$, which is the value obtained experimentally in a photonic-crystal waveguide \cite{Arcari2013InPrep}. In this calculation dephasing was neglected, i.e., $\gamma_{\rm dp}=0.$ (c)-(d) Second-order correlation function $g_T^{(2)}(\tau)$ for the transmitted field scattered from a single quantum emitter in a nanophotonic waveguide with (c) $\gamma_{\rm dp}=0$ and different values of $\beta$ and for (d) $\beta = 98 \%$ and different values for the pure-dephasing rate $\gamma_{\rm dp}$.}
\label{spnl_g2}
\end{figure}

A single quantum emitter efficiently coupled to a propagating mode in a nanophotonic waveguide can operate as a giant photon nonlinearity that for $\beta \rightarrow 1$ is sensitive at the single-photon level. Figure \ref{spnl_g2} illustrates the basics of a photonic-crystal-waveguide single-photon nonlinearity. As discussed previously, a unity $\beta$-factor implies that an excited quantum dot channels every single photon into the waveguide. Conversely, a narrow bandwidth photon pulse (relative to the linewidth of the emitter) that is launched into the waveguide will scatter on the emitter with unity probability. Since the emitter can only scatter a single photon at a time, a single photon suffices in saturating the emitter, which provides a nonlinear response where the single-photon component of a pulse is fully reflected  while higher-order photon components have increased probability of being transmitted. The theory of resonance fluorescence (cf.\ Sec.\ \ref{Section-Resonance-Fluorescence}) restricted to a 1D geometry describes this situation, where the photons can only scatter back or forth or leak out of the waveguide at a suppressed rate that is related to  $1-\beta$.  Such a setting can be referred to as the ``one-dimensional artificial atom'' by paraphrasing the terminology introduced by Kimble and co-workers in the context of atomic quantum optics  \cite{Kimble1998PhysScripta}. The transmission and reflection properties of an emitter in a 1D waveguide have been considered in \citet{Rice1988IEEE}, \citet{Chang2007NPHYS}, \citet{Shen2005OL}, and \citet{Shen2007PRA}. For a narrow-bandwidth single-photon pulse on resonance with the emitter,  the transmission and reflection coefficients are given by
\bea
T &=& \frac{1}{1 + \gamma/2 \gamma_{\rm dp}}+ \frac{1}{1 + 2 \gamma_{\rm dp}/ \gamma} \left( 1-\beta\right)^2,\label{Trans+ref-coeffs1}\\
R &=& \frac{1}{1 + 2 \gamma_{\rm dp}/ \gamma} \beta^2,\label{Trans+ref-coeffs2}
\eea
where Markovian dephasing with a rate $\gamma_{\rm dp}$ has been included. Figure \ref{spnl_g2}(b) shows the frequency dependence of the reflection and transmission coefficients. In the case of highly-efficient coupling, $\beta \rightarrow 1$, and a coherent emitter, $\gamma_{\rm dp}/ \gamma \rightarrow 0$, a single photon is perfectly reflected by the emitter on resonance.

In the case of launching low-power pulses $(\Omega_{\rm p} \ll \gamma )$ in a coherent state, Eqs. (\ref{Trans+ref-coeffs1}) and (\ref{Trans+ref-coeffs2}) are approximately valid. In this case a quantum emitter can be employed for reflecting the single-photon component of the coherent state while higher order components have increased probability to be transmitted. This constitutes the basis for a nonlinear switch: for weak excitation the emitter in the waveguide reflects the coherent state with large probability while for strong excitation it is saturated and the light is transmitted. The quantum character of this nonlinear response is revealed from the photon statistics of the transmitted and reflected light. The reflected field is described by the theory of resonance fluorescence, reviewed in Sec.\ \ref{Section-Resonance-Fluorescence}, where antibunching is predicted, cf.\ Eq.\ (\ref{g2-res-excitation}). The transmission in the waveguide contains unique quantum correlations that arise from the quantum interference between the scattered and incident fields. In the weak-excitation limit and for vanishing dephasing, $(\gamma_{\rm dp}=0)$, the second-order correlation function is given by \cite{Chang2007NPHYS}
\be
g_{\rm T}^{(2)}(\tau) = e^{-\gamma \tau} \left( \frac{\beta^2}{(1 - \beta)^2} - e^{\gamma \tau/2}\right)^2.
\ee
Very pronounced modifications of the photon statistics are predicted and Figs.\ \ref{spnl_g2}(c)-(d) plot the correlation function for various realistic values of $\beta$ and $\gamma_{\rm dp}$. For $\beta$ approaching unity, pronounced bunching $(g_{\rm T}^{(2)}>1)$ is predicted at $\tau = 0$ expressing  that two and higher-order photon components of the coherent state are transmitted with increased probability while the single-photon component is reflected. The pronounced bunching is followed by antibunching at subsequent times, which is a quantum interference phenomenon that is very sensitive to dephasing. Using the experimental value of $\beta = 98\%$ reported for photonic-crystal waveguides \cite{Arcari2013InPrep} together with a realistic dephasing rate of $\gamma_{\rm dp} = 0.1 \gamma$ \cite{Matthiesen2012PRL} leads to $g^{(2)}(0) = 33$ for the transmission. This illustrates the dramatic potential of quantum dots in nanophotonic waveguides for generation of non-classical photonic quantum states.

The giant photon nonlinearity may find a number of important applications in quantum-information processing, some of which are discussed in further detail in Sec.\ \ref{Section-QIP}. Interestingly, a more elaborate dynamical theory than the one discussed above predicts that by scattering short few-photon pulses on an emitter with a high $\beta$-factor in a waveguide, polaritonic photon-emitter bound states may be excited whereby the photons become trapped in entangled light-matter states \cite{Longo2010PRL}. So far experimental realizations of photon nonlinearities in a waveguide have been reported in the microwave regime where superconducting transmon qubits in a waveguide were used for generating non-classical photon states \cite{Hoi2012PRL}. Finally, it should be noted that a cavity operating in the weak-coupling regime with a high $\beta$-factor could also be used to implement the type of single-emitter nonlinearity considered here \cite{Rice1988IEEE}, although the operation bandwidth would be limited by the narrow cavity linewidth. Another and more widely explored option for cavities is to generate a photon nonlinearity by scattering photons from strongly-coupled quantum dot-cavity system. Cavity nonlinearities are discussed in Sec.\ \ref{Section-cavity-nonlinearity}.

\subsection{Dipole-dipole interaction in a photonic waveguide}
\label{Section-dipole-dipole-interaction}

Another exciting application of nanophotonic waveguides is to mediate the interaction between separate quantum emitters over extended distances. Such a waveguide-mediated dipole-dipole interaction may allow constructing deterministic two-emitter quantum-phase gates \cite{Dzsotjan2010PRB} or entanglement between distant emitters \cite{Gonzalez-Tudela2011PRL} with the performance ultimately determined by the potentially very large emitter-waveguide coupling efficiency.

The theory of the dispersive dipole-dipole interaction between two quantum emitters in a photonic waveguide provides an extension of the one-emitter case considered in Sec.\ \ref{Sec:Purcel-waveguide}. Consider the two dipole emitters A and B positioned at ${\bf r}_\text{A}$  and ${\bf r}_\text{B}$ and decaying with rates $\gamma_\text{A}$ and $\gamma_\text{B}$ that are assumed to be dominated by the coupling to the waveguide, i.e., the $\beta$-factors are close to unity. The dipole-dipole interaction mediated by the waveguide field gives rise to an additional decay rate due to the presence of the other emitter that by the use of Eq.\ (\ref{Greens-fct-waveguide}) can be expressed as \cite{Dzsotjan2010PRB}
\bea
\gamma_\text{AB} &=& \frac{2 d^2}{\hbar \epsilon_0} \mathrm{Im} \left[ \hat{\mathbf{e}}_\text{A}^* \cdot \overleftrightarrow{G}({\bf r}_\text{A},{\bf r}_\text{B},\omega_0)  \cdot \hat{\mathbf{e}}_\text{B} \right]  \label{dipole-dipole-rate}\\
 &=& \frac{a d^2 \omega_0}{\hbar \epsilon_0 v_\text{g}}  \left|\hat{\bf e}^*_\text{A} \cdot {\bf b}_{k}({\bf r}_\text{A}) \right| \left|\hat{\bf e}_\text{B} \cdot {\bf b}_{k}^*({\bf r}_\text{B}) \right|  \cos \left[ k r_\text{AB}+\phi \right], \nn
\eea
 where $r_\text{AB} = \left| {\bf r}_\text{A} - {\bf r}_\text{B} \right|$ and $\phi$ is a phase depending on the projections of the two dipoles on the local electric fields. Here, the two emitters are assumed to have the same resonance frequency $\omega_0$ and transition dipole moment $d$, while their orientations are denoted $\hat{\mathbf{e}}_\text{A}$ and $\hat{\mathbf{e}}_\text{B}$, respectively. It is observed that the dipole-dipole interaction rate between two emitters can be considered a natural extension of the spontaneous-emission rate of a single emitter, cf.\ Eqs.\ (\ref{Im-G}) and (\ref{rate}), where the latter can be interpreted as the self-interaction of the radiated field on the emitter. The second equation of (\ref{dipole-dipole-rate}) holds for the specific example of a photonic-crystal waveguide. It follows from symmetry that $\gamma_{BA} = \gamma_\text{AB}$.

The benefit of high-$\beta$-factor nanophotonic waveguides for mediating the dipole-dipole interaction is immediately clear from Eq.\ (\ref{dipole-dipole-rate}). Thus, the dipole-dipole interaction in an ideal lossless photonic-crystal waveguide is infinitely ranging since it varies sinusoidally with the distance between the emitters, which is a consequence of the one-dimensional nature of the interaction. For comparison, the dipole-dipole interaction between two emitters in a homogeneous medium decays rapidly with a scaling of $1/\left|{\mathbf{r}_1} - {\mathbf{r}_2} \right|^3$ \cite{NanoOpticsBook}. In reality any loss in the waveguide due to fabrication imperfections will result in residual light leakage and the exponential damping of the range of the dipole-dipole coupling strength \cite{Minkov2013PRB}. In photonic-crystal waveguides the extinction length is found to be in the range of 20-\SI{200}{\micro\meter} \cite{Garcia2010PRB} depending on whether the waveguide is operated in the slow- or fast-light regime meaning that the dipole-dipole interaction can extend as far as several hundreds of optical wavelengths.

The dipole-dipole interaction can be used to entangle two quantum emitters. The three eigenstates for the two coupled emitters in the absence of dissipation are $\left|e_\text{A}, e_\text{B} \right>$, $\left| \pm \right> = \left( \left|g_\text{A}, e_\text{B} \right> \pm \left|e_\text{A}, g_\text{B} \right> \right)/\sqrt{2}$, and $\left|g_\text{A}, g_\text{B} \right>$. The states $\left| \pm \right>$ are Bell states with maximal entanglement between the two spatially separated emitters that share an excitation that is either located on emitter A or B. The two entangled states decay with rates $\gamma_{\pm} = \gamma \pm \gamma_\text{AB}$ ($\gamma_\text{A} = \gamma_\text{B} = \gamma$) and depending on the mutual distance between the emitters, $\gamma_\text{AB}$ alternates between $- \gamma$ and $+ \gamma$. Consequently the two entangled states are either sub- or super-radiant depending on whether their decay is slower or faster than the spontaneous-emission rate of a single emitter $\gamma$. A detailed account of the amount of emitter-emitter entanglement predicted in the case of a plasmon nanowire can be found in \citet{Gonzalez-Tudela2011PRL}. So far, experimental demonstrations have been lacking. In photonic cavities, however, two quantum dots have been coupled through their interaction with the same cavity mode \cite{Reitzenstein2006OL, Laucht2010PRB}.

\section{Cavity quantum electrodynamics with single quantum dots}
\label{Section-cavity-QED}

Cavities have traditionally been widely used in quantum optics as a means to enhance the interaction between light and matter by resonating the electromagnetic field. In solid-state systems this approach has also been very successful and in the present section we review the progress on cavity QED with quantum emitters in photonic nanocavities.

\subsection{Local-density-of-states theory}
\label{Section-LDOS-theory}

We first review the theory of cavity QED using the LDOS formalism presented in Sec.\ \ref{Section-spontaneous-emission}. This approach is unconventional in the context of cavity QED, where the Jaynes-Cummings model discussed in Sec.\ \ref{The dissipative Jaynes-Cummings model} is more commonly used. In the LDOS description, the field of cavity QED is found to merge naturally with the broader class of QED systems studied in quantum nanophotonics such as photonic crystals and photonic waveguides.

Nanophotonic cavities enable the pronounced enhancement of a single optical mode due to a strong spatial and spectral confinement of light so that the radiative decay rate into the cavity mode, $\gamma_\text{cav}$, is much faster than the decay rate into all other (non-guided) modes, $\gamma_\text{ng}$. It is convenient to express the Purcell factor introduced in Eq.\ (\ref{ratePurcell}) as $F_\text{P}=F_\text{P}^\text{cav}+F_\text{P}^\text{ng}$, where $F_\text{P}^\text{cav}=\gamma_\text{cav}/\gamma_\text{hom}$ and $F_\text{P}^\text{ng}=\gamma_\text{ng}/\gamma_\text{hom}$ are the Purcell factors for the cavity and for all other modes, respectively. The emission efficiency into the cavity mode follows from the $\beta$-factor of Eq.\ (\ref{eq:beta-factor-definition}).  In a typical nanocavity QED experiment, $\gamma_\text{cav}\gg \gamma_\text{ng}$, and in the following only $F_\text{P}^\text{cav}$ is considered.

In the case of a cavity with a single transverse non-degenerate mode, it follows from Eq.\ (\ref{Greens-def}) that the cavity Green's tensor can be written in the form \cite{Yao2010LPR}
\be
 \overleftrightarrow{G}({\bf r},{\bf r'},\omega) = \omega^2 \frac{{\bf u}_{c}({\bf r'}) \otimes {\bf u}_{c}^*({\bf r})}{\omega_\text{c}^2-\omega^2- i \omega \Gamma_0},
 \ee
where $\omega_\text{c}$ is the cavity resonance frequency, $\Gamma_0 = \omega_\text{c}/Q$, and $Q$ is the cavity quality factor. The LDOS follows from Eq.~(\ref{Im-G}):
\be
\rho\left({\bf r},\omega, \hat{\mathbf{e}}_\text{d} \right) = \frac{2}{\pi} \frac{\omega^2 \Gamma_0}{(\omega_\text{c}^2 - \omega^2)^2 + \omega^2 \Gamma_0^2} \frac{f({\bf r})}{\epsilon({\bf r}) V_{\text{eff}}}  \left|\hat{\mathbf{e}}_\text{d} \cdot \hat{\mathbf{e}}_\text{c} \right|^2,
\label{LDOS-cavity}
\ee
where $\hat{\mathbf{e}}_\text{c}$ is the unit polarization vector of the cavity mode, $V_\text{eff} = 1/\operatorname{max}\left[ \epsilon({\mathbf r}) \left| {\mathbf u}_\text{c}({\mathbf r}) \right|^2 \right]$ is the effective mode volume,
and $f({\mathbf r}) =  \epsilon({\mathbf r}) \left| {\mathbf u}_\text{c}({\mathbf r}) \right|^2 / \operatorname{max} \left[ \epsilon({\mathbf r}) \left| {\mathbf u}_\text{c}({\mathbf r}) \right|^2 \right]$ defines the spatial mismatch between the emitter and the cavity field.
This expression for the effective mode volume is not strictly valid for the case of open leaky cavities supporting quasi modes \cite{Kristensen2012OL} but is approximately valid if spatial cut offs are applied.

\begin{figure*}
\includegraphics[width= \textwidth]{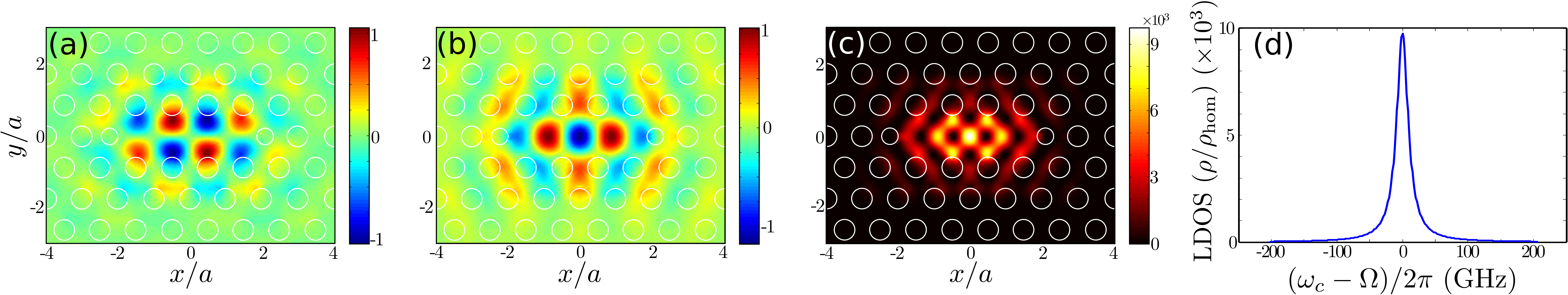}
\caption{ Simulations of the LDOS of a photonic-crystal L3 cavity ($n=3.46$, $a=\SI{240}{\nano\meter}$, $r=0.3a$, slab thickness $t=0.6 a$, and with holes adjacent to the cavity shifted by $\Delta x = 0.16 a$ and their radius changed to $r_1 = 0.24a$) following the design in \citet{Hennessy2007Nature}. The cavity mode has a quality factor of $Q=9.6\times 10^4$ with an effective mode volume of $V_\text{eff}=0.75(\lambda/n)^3$. (a) $E_x$ and (b) $E_y$ field components of the fundamental cavity mode. The white circles indicate the positions of the air holes making up the photonic crystal. (c) The projected LDOS at a frequency resonant with the cavity mode for a dipole at each position oriented along the local electric field of the cavity mode. The LDOS is scaled to the value in a homogeneous medium, i.e., $\rho_{\rm hom} = n \omega^2/3 \pi^2 c^3.$  (d) The detuning-dependent projected LDOS for a dipole positioned at the field maximum and aligned with the field, i.e., $f(\mathbf r)=1$ and $\left|\hat{\mathbf{e}}_\text{d} \cdot \hat{\mathbf{e}}_\text{c} \right|^2=1$.}
\label{FpMaps}
\end{figure*}

The calculated spatial and spectral dependence of the LDOS for a photonic-crystal nanocavity is plotted in Fig.\ \ref{FpMaps}. Figures \ref{FpMaps}(a)-(b) show the $E_x$ and $E_y$ electric-field components of the fundamental mode of an L3 cavity. The components $E_x$ and $E_y$ determine the cavity polarization vector $\hat{\mathbf{e}}_\text{c}$ on which the LDOS depends. Figure \ref{FpMaps}(c) illustrates the spatial dependence of the LDOS for an emitter aligned with the field of the cavity and on resonance, i.e., $\left|\hat{\mathbf{e}}_\text{d} \cdot \hat{\mathbf{e}}_\text{c} \right|^2=1$ and $\omega=\omega_\text{c}$. This plot therefore illustrates the role of the function $f(\mathbf r)$ in determining the LDOS. Finally, Fig.\ \ref{FpMaps}(d) shows the variation of the LDOS with detuning for an optimally placed and oriented emitter. The width of this curve is determined by the Q-factor of the cavity.

\begin{figure}
\includegraphics[width= \columnwidth]{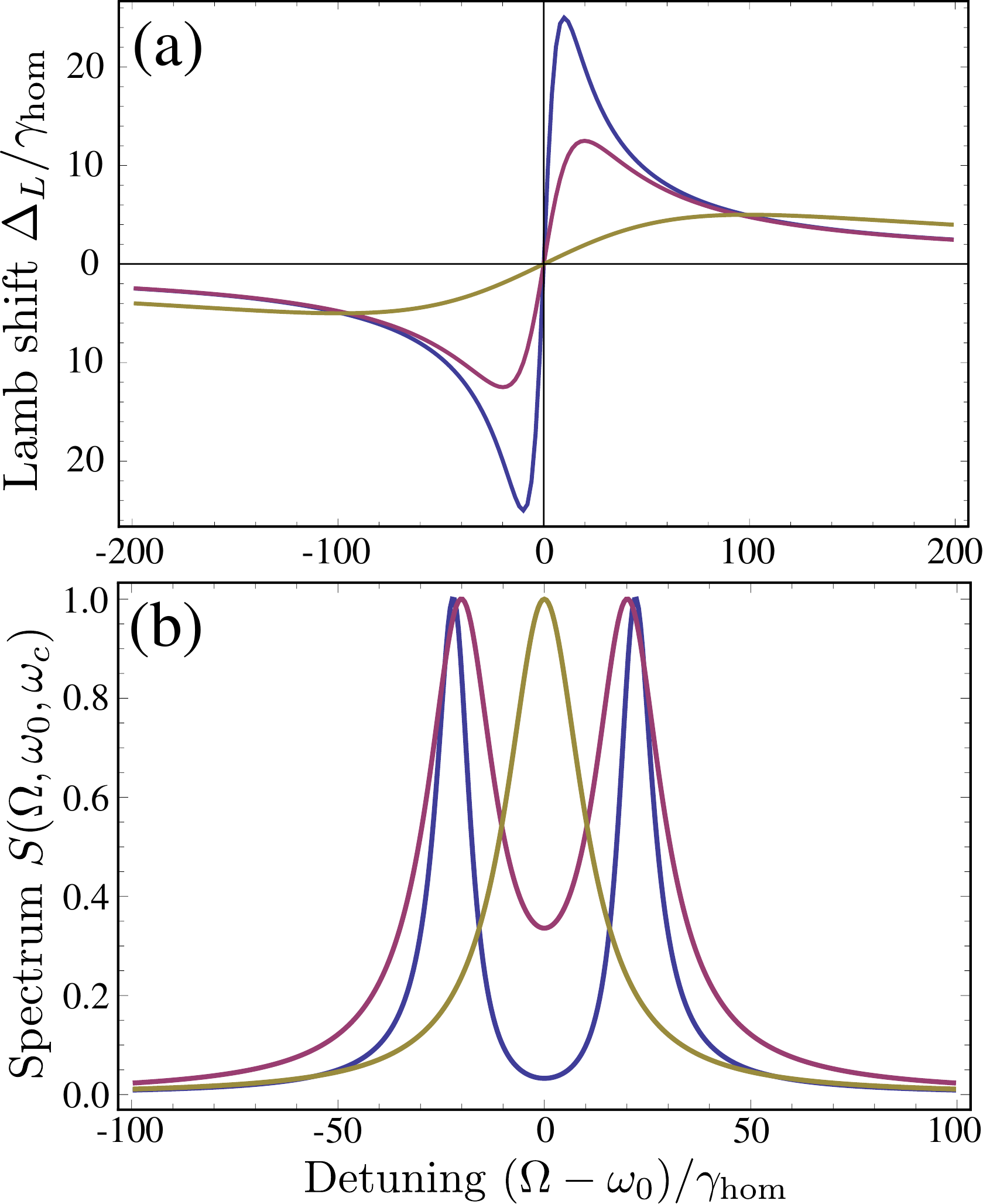}
\caption{Emission spectra for an emitter in a cavity for parameters that are realistic with quantum dots. (a) Lamb shift and (b) emission spectrum versus frequency detuning from the bare emitter frequency $\omega_0=\omega_c$ and scaled to the decay rate of the emitter in a homogeneous medium $\gamma_\text{hom}$ and for $\{Q, F_\text{P}^\text{res} \} = \{10^4, 20 \}$ (brown curve), $\{Q, F_\text{P}^\text{res} \} = \{5 \times 10^4, 50 \}$ (purple curve), and $\{Q, F_\text{P}^\text{res} \} = \{10^5, 100 \}$ (blue curve) . It is assumed that $\gamma_\text{hom} = \SI{1}{\nano\second^{-1}}$  at the wavelength $\lambda = \SI{950}{\nano\meter}$, which are typical parameters for a quantum dot.  }
\label{emission-spectra-cavity}
\end{figure}


The $Q$-factor and the mode volume are two figures of merit for the cavity. However, since the mode volume is not a readily measurable quantity for nanophotonic cavities, we choose instead to specify the maximum achievable Purcell factor that a quantum emitter in the cavity experiences when being resonant with the cavity, $F_{\text{P}}^{\text{res}}$. Note that while $Q$ and $F_{\text{P}}^{\text{res}}$ are not independent parameters they are readily recorded experimentally, where the latter can be probed directly from the exponential emission rate in time-resolved emission measurements in the weak-coupling regime. The maximum Purcell factor loses its direct physical meaning in the strong-coupling regime, butit can still be used as the governing parameter and since most experiments so far are either in the weak-coupling regime or slightly into the strong-coupling regime this is a viable choice. From Eqs.\ (\ref{LDOSpurcell}) and (\ref{LDOS-cavity}) it follows that
\be
F_\text{P}^\text{cav}\left({\mathbf r},\Delta, \hat{\mathbf{e}}_\text{d} \right)= F_\text{P}^{\text{max}} f(\mathbf r)  \left|\hat{\mathbf{e}}_\text{d} \cdot \hat{\mathbf{e}}_\text{c} \right|^2 \frac{\omega_\text{c}^2/4 Q^2}{\Delta^2 + \omega_\text{c}^2/4Q^2},
\label{Purcell-factor-general}
\ee
where $\Delta = \omega_\text{c} - \omega$, $F_{\text{P}}^{\text{max}} = \frac{3 (\lambda/n)^3}{4 \pi^2} \frac{Q}{V_{\text{eff}}}$ is the optimum Purcell factor for an ideally positioned emitter \cite{Purcell1946PR}, and $F_{\text{P}}^{\text{res}}=F_{\text{P}}^{\text{max}} f(\mathbf r)  \left|\hat{\mathbf{e}}_\text{d} \cdot \hat{\mathbf{e}}_\text{c} \right|^2.$

In the photonic-crystal cavity of Fig.\ \ref{FpMaps}, the mode volume is $V_{\text{eff}} = 0.75 (\lambda/n)^3$ implying that $F_{\text{P}}^{\text{max}} \sim 0.1 Q$. Quality factors of $Q \sim 10^4$ can be routinely observed experimentally in such cavities, which corresponds to an optimum Purcell factor of $F_{\text{P}}^{\text{max}} \sim 1000$ that would typically be deep in the strong-coupling regime. Curiously, this expected Purcell factor is much larger than the values typically observed in experiments, which is mainly attributed to spatial mismatch of the emitter relative to the cavity mode, but could also be partly related to other effects such as fabrication imperfections. Consequently, unlike the spontaneous-emission inhibition in the photonic band gap discussed in Sec.\ \ref{Section-spontaneous-emission-control}, the full potential of photonic nanocavities has not yet been obtained experimentally. A detailed account of the experimental status of cavity QED in photonic-crystal nanocavities is given in Secs.\ \ref{Section-Purcell-effect} and \ref{Section-observation-strong-coupling}.

The expressions for the emission spectrum and Lamb shift follow from Eqs.\ (\ref{spectrum}) and (\ref{Lamb})
  \bea S(\Omega,\Delta_0,\Delta_\text{c}) &\propto& \frac{1}{\left[ \Delta_0 + \Delta_L^{\text{cav}}(\tilde{\Omega}) \right]^2 +  \frac{(\gamma_{\text{hom}} F_\text{P}^{\text{res}})^2/4}{\left[1+4 Q^2 \Delta_\text{c}^2/\omega_\text{c}^2 \right]^2}}, \label{spectrum-cavity-LDOS} \\
 \frac{\Delta_L^{\text{cav}}(\tilde{\Omega})}{\gamma_{\text{hom}}} &=& \frac{F_\text{P}^\text{res}}{2 \pi}  I(\tilde{\Omega}),
\label{spectrum+Lamb-cavity-LDOS}\eea
where $\Delta_0 = \omega_0 - \Omega$, $\Delta_\text{c} = \omega_\text{c} - \Omega$, and $\gamma_{\text{hom}}$ is the radiative decay rate of the emitter in the homogeneous medium. Furthermore, we have defined the integral $I(\tilde{\Omega}) = \mathcal{P} \int_0^{\infty} \text{d} \tilde{\omega} \frac{\tilde{\omega}}{(\tilde{\Omega}-\tilde{\omega})\left[ 1+4Q^2(1-\tilde{\omega})^2 \right]}$ with $\tilde{\Omega} = \Omega/\omega_\text{c}$ that can be evaluated numerically and is convergent.
These expressions hold for any light-matter interaction strength and can thus be used to describe both the weak- and strong-coupling regimes of cavity QED.

Figure \ref{emission-spectra-cavity}(b) plots the emission spectra in the case of realistic experimental parameters. As the $Q$-factor and Purcell factor are increased, the transition from weak to strong coupling is observed through the spectrum changing from a single- to a double-peaked spectrum (vacuum Rabi splitting). Strong coupling is experimentally reachable for instance in photonic-crystal cavities, as discussed in Sec.\ \ref{Section-observation-strong-coupling}. Figure \ref{emission-spectra-cavity}(a) shows the Lamb shift versus observation frequency. We note that the Lamb shift has a broad spectral response relative to the quantum-dot linewidth and can attain a size that is an order of magnitude larger than $\gamma_{\text{hom}}$. The modified Lamb shift could be extracted in experimental measurements of the emission spectra by applying Eq.\ (\ref{spectrum-cavity-LDOS}), and the Lamb shift is partly responsible for the vacuum Rabi splitting displayed in Fig.\ \ref{emission-spectra-cavity}(b).

\subsection{The dissipative Jaynes-Cummings model }
\label{The dissipative Jaynes-Cummings model}

\begin{figure}
\includegraphics[width=0.6\columnwidth]{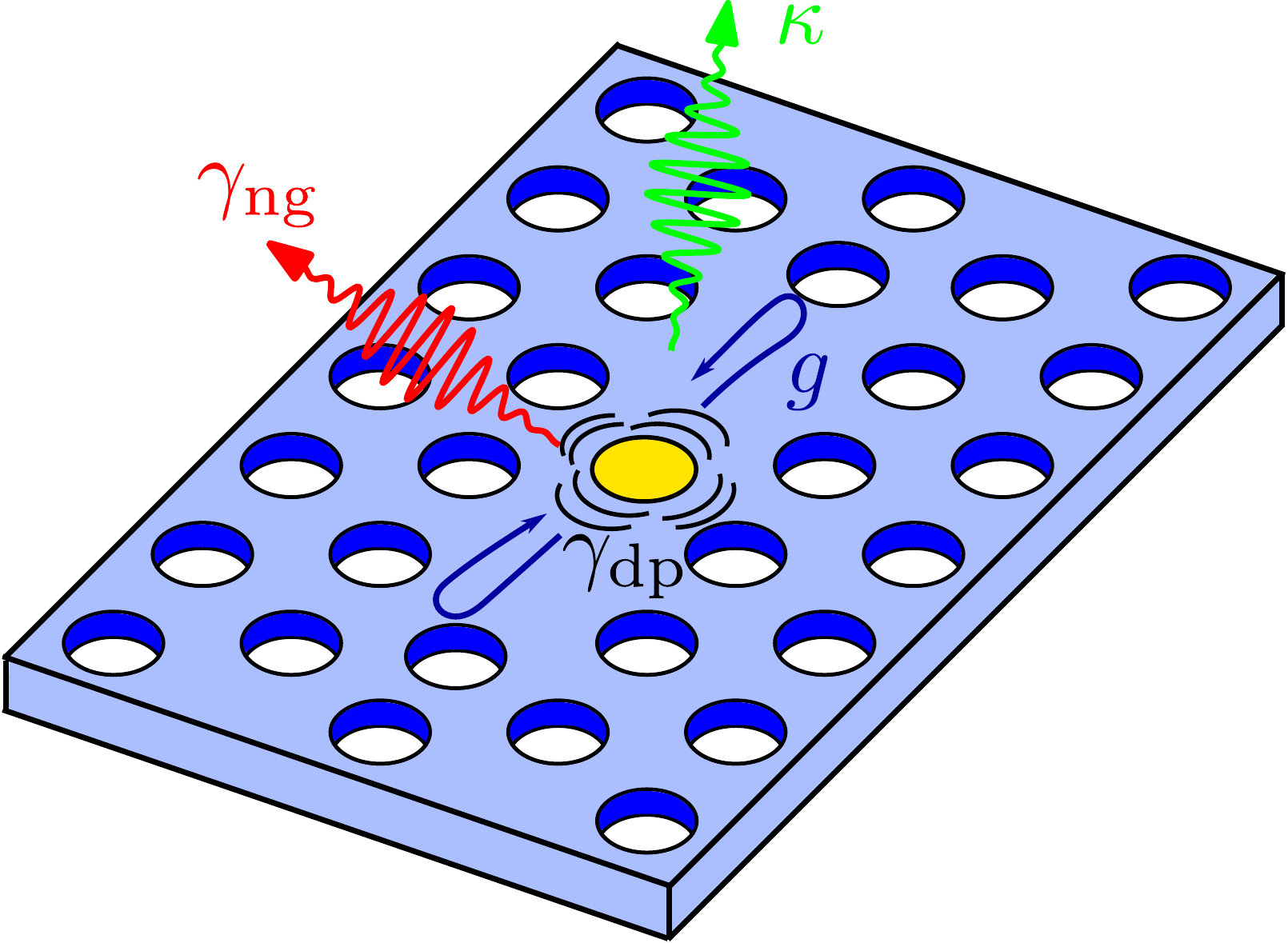}
\caption{Illustration of the various rates in the Jaynes-Cummings model. The quantum dot couples to the cavity mode with a rate $g$, while it emits photons into other non-guided modes at a rate $\gamma_\text{ng}$. The cavity loss rate is $\kappa$ and the pure-dephasing rate of the quantum dot is $\gamma_\text{dp}$. }
\label{Cavity-QED-generic-sketch}
\end{figure}

The preceding section concerned cavity QED in the LDOS description, which is a very adequate description for spontaneous-emission dynamics. In extending the description to account for dephasing mechanisms, a density-operator formalism is useful. Figure \ref{Cavity-QED-generic-sketch} sketches the basic physical processes under consideration: a single quantum emitter is positioned in a leaky cavity and coupled to a reservoir of radiation modes. Furthermore, the emitter may dephase. The Jaynes-Cummings model describes the system in the Markov approximation where the two reservoirs are spectrally broad, which is the equivalent of assuming that they have ``no memory'', i.e., the temporal correlation function between two different times vanishes \cite{Meystre2007elements}. In contrast the full dynamics of the single cavity mode is accounted for. The master equation for the reduced density operator is of the Lindblad form \cite{Carmichael1993PRL, Carmichael1989PRA},
\bea  \frac{\text{d} \hat{\rho}}{\text{d}t} = &-&\frac{i}{\hbar} \left[\hat{H}_{\mathrm JC},\hat{\rho} \right] + \frac{\gamma_{\text{ng}}}{2} \left( 2 \hat{\sigma}_- \hat{\rho} \hat{\sigma}_+ - \{\hat{\sigma}_+ \hat{\sigma}_-,\hat{\rho} \}  \right) \nn \\
&+&  \frac{\kappa}{2} \left( 2 \hat{a} \hat{\rho} \hat{a}^{\dagger} - \{\hat{a}^{\dagger} \hat{a},\hat{\rho} \}  \right) + \frac{\gamma_{\text{dp}}}{2} \left(\hat{\sigma}_z \hat{\rho} \hat{\sigma}_z - \hat{\rho} \right), \label{JC-master-eq}\eea
where $\hat{H}_{\text{JC}} =  \hbar \left[ g \hat{\sigma}_+ \hat{a} e^{i \Delta t}  + \text{h.c.}\right]$ is the Jaynes-Cummings Hamiltonian, $\hat{a}$ is the annihilation operator for the mode in the cavity that is fed by the emitter with a rate $g$, $\Delta = \omega_0 - \omega$ is the detuning, and $\gamma_{\text{ng}}$, $\kappa$, and $\gamma_{\text{dp}}$ are the rates for spontaneous emission out of the cavity, decay of the cavity, and dephasing, respectively as illustrated in Fig.\ \ref{Cavity-QED-generic-sketch}. The Jaynes-Cummings Hamiltonian is written in the rotating-wave approximation, i.e., only energy-conserving processes are included, which is often a good approximation. It breaks down in the ultrastrong-coupling regime \cite{Ciuti2006PRA} but this is usually beyond reach for single emitters at optical frequencies. For reference, the relation to the quantities introduced in the LDOS formalism is
\bea
g &=& \sqrt{\frac{F_\text{P}^\text{cav} \gamma_{\text{hom}} \omega}{4 Q}},\label{eq:relation_between_LDOS_and_JC}\\
\kappa &=& \frac{\omega}{Q},\\
\gamma_{\text{ng}} &=& \gamma_{\text{hom}} F_\text{P}^\text{cav} \left(\frac{1}{\beta} - 1 \right),
\eea
where $F_\text{P}^\text{cav}$ of Eq.\ (\ref{Purcell-factor-general}) has been evaluated at $\Delta = 0$ and non-radiative processes have been neglected.
The dephasing term in Eq.\ (\ref{JC-master-eq}) accounts for fast elastic scattering processes and could describe the broadening of the zero-phonon line of the quantum dot. The formation of LA phonon sidebands is not captured by such elastic Markovian scattering processes, and a more elaborate (non-Markovian) model is discussed in Sec.\ \ref{Section-Phonon-dephasing}. It should also be noted that a Lamb shift of the quantum-dot transition is induced by both phonon and photon reservoirs that is assumed to be implicitly incorporated in the emitter frequency.

Equation\ (\ref{JC-master-eq}) can be solved by restricting to the case of one excitation in the system, which suffices for describing spontaneous emission and vacuum Rabi oscillations.  The emitter can be either in the ground state $|g\rangle$ or the excited state $|e\rangle$ and the cavity in the vacuum state $|0\rangle$ or a single-photon state $|1_\text{c} \rangle$, i.e., $\left| 1 \right > = \left| e,0 \right >$, $\left| 2 \right > = \left|
g,1_\text{c} \right >$, and $\left| 3 \right > = \left| g,0 \right >$. The equations of motion for the decisive entries of the density matrix are
\begin{equation}
\begin{split}
\frac{\text{d} \rho_{11}}{\text{d}t} &= i g^* \rho_{12} - i g\rho_{12}^* - \gamma_{\text{ng}} \rho_{11}, \\
\frac{\text{d} \rho_{22}}{\text{d}t} &=  i g \rho_{12}^* - i g^* \rho_{12} - \kappa \rho_{22},  \\
\frac{\text{d} \rho_{12}}{\text{d}t} &=  i g \left(\rho_{11} -
\rho_{22} \right) - \frac 1 2 \left( 2 i\Delta + \gamma_{\text{ng}} + \kappa + 2 \gamma_{\text{dp}}   \right) \rho_{12}, \\
\frac{\text{d} \rho_{13}}{\text{d}t} &= - i g \rho_{23} - \left( \frac{ i \Delta+\gamma_{\text{ng}} + 2 \gamma_{\text{dp}} }{2} \right) \rho_{13}, \\
\frac{\text{d} \rho_{23}}{\text{d}t} &=  - i g^* \rho_{13} -  \frac{\kappa - i \Delta}{2} \rho_{23},
\label{density-matrix-elements}
\end{split}
\end{equation}
where the cross terms have been transformed according to: $\rho_{12} \rightarrow
\rho_{12} e^{i \Delta t}$, $\rho_{23} \rightarrow
\rho_{23} e^{-i \Delta t/2}$ and $\rho_{13} \rightarrow
\rho_{13} e^{-i \Delta t/2}$. The additional terms in the matrix are obtained from $\rho_{ji} = \rho_{ij}^*.$ The physical significance of the various elements is as follows: $\rho_{11} = \left<\hat{\sigma}_+ \hat{\sigma}_- \right> $ is the population of the emitter, $\rho_{22} = \left<\hat{a}^{\dagger} \hat{a} \right>$ is the number of photons in the cavity, $\rho_{12} = \left<\hat{a}^{\dagger} \hat{\sigma}_- \right>$ is the cavity-assisted polarization, while $\rho_{13} = \left<\hat{\sigma}_- \right>$ and $\rho_{23} = \left<\hat{a}\right>$. From this set of equations the dynamics and emission spectra of the Jaynes-Cummings system are readily obtained.

\subsubsection{Dynamics and emission spectra}

In this section we  present expressions for the dynamics and the spectra of the dissipative Jaynes-Cummings model. In the most general cases the equations of motion are solved numerically but in special cases simple analytical expressions are obtained.
In the absence of dephasing ($\gamma_\text{dp}=0$), on resonance ($\Delta=0$), and for a real coupling rate ($g=g^*$, which is valid by fixing the absolute phase of the local electric field), the first three equations in (\ref{density-matrix-elements}) admit analytic solutions of the form
\begin{equation}
{\boldsymbol \rho}(t) = c_1 e^{\lambda_1 t} \mathbf u_1 + c_+ e^{\lambda_+ t} \mathbf u_+ + c_- e^{\lambda_- t} \mathbf u_-,
\label{JC-full-solution}
\end{equation}
%
%
where ${\boldsymbol \rho}(t)= [\rho_{11} \,\,\, \rho_{12} \,\,\, \rho_{22}]^T$, $c_i$ are constants, and $\lambda_1=(-\gamma_{\text{ng}}-\kappa)/2$ and $\lambda_{\pm} = (-\gamma_{\text{ng}}-\kappa)/2 \pm \Omega_\text{R}$ with $\Omega_\text{R} = \sqrt{(\gamma_{\text{ng}} - \kappa)^2/4 -4g^2}$ are the eigenvalues for the three coupled equations with corresponding eigenvectors $\mathbf u_i$. By assuming the emitter to be initially in the excited state, the time evolution of the emitter population follows
%
%
\begin{equation}
\begin{split}
\rho_{11}(t) =  &\frac {e^{- (\gamma_{\text{ng}} + \kappa)t/2}}{4 \Omega_\text{R}^2} \big[\cosh(\Omega_\text{R} t) (8 g^2-(\gamma_{\text{ng}}-\kappa)^2)\\
& + 2 \Omega_\text{R} (\gamma_{\text{ng}}-\kappa) \sinh(\Omega_\text{R} t) + 8 g^2\big].
\label{JCemitterSol}
\end{split}
\end{equation}
The nature of the solutions depends on the relative values of $\kappa$, $\gamma_{\text{ng}},$ and $g$. For $g \ll |\gamma_{\text{ng}} - \kappa|/4$,  $\Omega_\text{R}$ is real and the emitter decays exponentially, which is the weak-coupling regime. For quantum dots in photonic cavities the cavity decay rate is often dominating, i.e., $\kappa \gg \gamma_{\text{ng}}$ in which case the decay follows the simple expression
\begin{equation}
\rho_{11}(t) \sim e^{- \left(\gamma_{\text{ng}} + \frac{4 g^2}{\kappa} \right) t}.
\label{rho11}
\end{equation}
This leads to the expression for the Purcell factor in Eq.\ (\ref{Purcell-factor-general}) and is valid in the Wigner-Weisskopf approximation.
For  $g > |\gamma_{\text{ng}} - \kappa|/4$,  $\Omega_\text{R}$ is purely imaginary and all eigenvalues  have the same real part. Consequently, the excitation oscillates in time corresponding to Rabi oscillations with a frequency of $\left|\Omega_\text{R} \right|$ between the emitter and the cavity that are eventually damped at a rate $(\gamma_{\text{ng}} + \kappa)/2$. This is the strong-coupling regime that occurs when the emitter-cavity excitation oscillates at a rate that exceeds the dissipative rates. Interestingly, also an intermediate-coupling regime can be defined before the onset of strong coupling ($g \lesssim |\gamma_{\text{ng}} - \kappa|/4$) where a non-exponential decay of spontaneous emission is found, which can also be expected from the general multi-exponential form of the solution, cf.\ Eq.\ (\ref{JC-full-solution}).

The emission spectra are given by \cite{Meystre2007elements} $S(\omega) \propto  \int_{-\infty}^{\infty} \text{d} \tau\, e^{i \omega \tau} \int_{0}^{\infty} \text{d}t
\left< \hat{x}^{\dagger}(t) \hat{x}(t+\tau) \right>$
and by applying the quantum-regression theorem \cite{Carmichael1993PRL}. Using the last two of Eqs.\ (\ref{density-matrix-elements}), the following closed expressions for the spectra of the emitter, $S_\text{em}$, ($\hat{x} = \hat{\sigma}_-$) and cavity, $S_\text{cav}$ ($\hat{x} = \hat{a}$), are derived
%
%
%
%
%
\begin{equation}
\begin{split}
S_\text{em}(\omega) &\propto {\rm Re}\left[ \frac{-2 i g\overline{\rho_{11}} + \left[\gamma_{\text{ng}} + 2 \gamma_{\text{dp}} + i\Delta_- \right] \overline{\rho_{21}}}{4g^2 - i(\gamma_{\text{ng}} + 2 \gamma_{\text{dp}} + i \Delta_-)(\Delta_+ +  i \kappa)}\right],\\
S_\text{cav}(\omega) &\propto {\rm Re}\left[ \frac{i \, \left[ 2 g \overline{\rho_{22}} + (\Delta_+ + i \kappa)  \overline{\rho_{12}} \right]}{4 g^2 - i(\gamma_{\text{ng}} + 2\gamma_{\text{dp}} + i \Delta_-)(\Delta_+ + i \kappa)} \right],
\end{split}
\label{spectra}
\end{equation}
where $\Delta_\pm = \Delta \pm 2 \omega$, and $\overline{\rho_{ij}} = \int_0^\infty \, \text{d}t \, \rho_{ij}(t)$. In an experiment, the emitter and cavity spectra would be measured by recording the light leaking from the cavity mode or from the emitter, respectively. Notably, in the strong-coupling regime emitter and cavity are entangled and both quantities would therefore have light and matter character. These spectra provide a generalization of the LDOS theory of the emitter spectrum and dynamics, presented in Sec.\ \ref{Section-LDOS-theory}, to include also dephasing. The two methods are found to agree in the limit of vanishing dephasing despite the fact that the radiation reservoir is treated rather differently in the dissipative Jaynes-Cummings model compared to the LDOS theory.  The vacuum Rabi splitting, cf.\ Fig.\ \ref{emission-spectra-cavity}(b), constitutes an important figure of merit for a strongly-coupled cavity-emitter system that is often extracted in experiments. Deep in the strong coupling regime ($g \gg |\gamma_{\text{ng}} - \kappa|/4$) and on resonance, the Rabi splitting is approximately given by $|\Omega_\text{R}| = \sqrt{4 g^2 - (\gamma_{\text{ng}} - \kappa)^2/4 }$ while the full widths of each spectral peak is $(\gamma_\text{ng} +\gamma_\text{dp} +\kappa)/2$. Note that the assumption of being deep into the strong-coupling regime is usually not met in experiments on photonic nanocavities.

\subsubsection{Influence of LA phonon dephasing}
\label{Section-Phonon-dephasing}

\begin{figure}
\includegraphics[width=1\columnwidth]{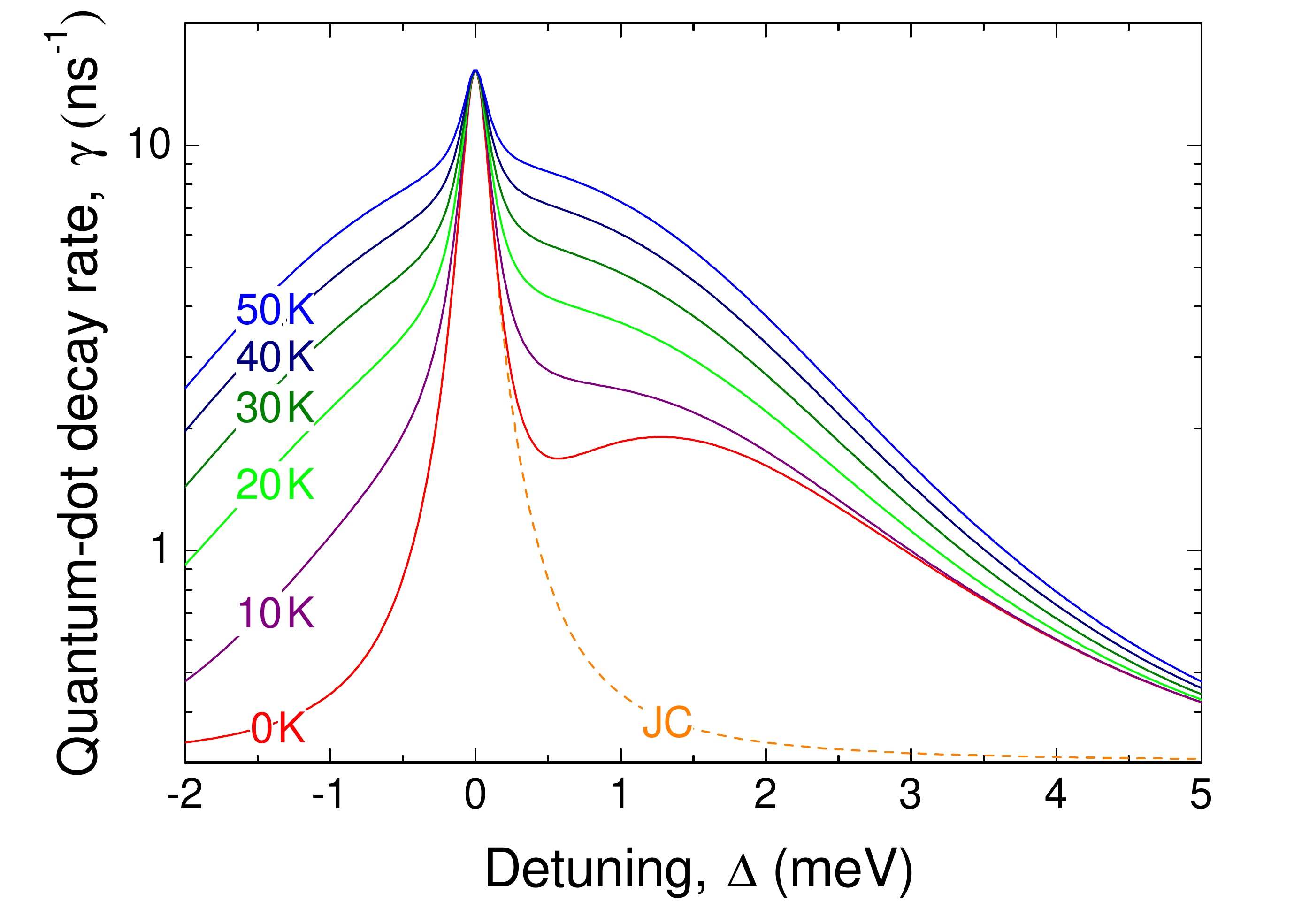}
\caption{Calculated decay rate of a quantum dot weakly coupled to a photonic-crystal nanocavity. The phonon sidebands due to LA phonons greatly enhance the coupling to the cavity for large detunings as compared to the prediction from the conventional Jaynes-Cummings model (JC). At zero temperature the decay rate is highly asymmetric since the phonon population vanishes and only spontaneous phonon emission is possible. At finite temperatures (indicated in the plot) the phonon population plays an increasingly dominant role and partly wash out the asymmetry of the broadband phonon-assisted Purcell effect. The parameters used in this plot correspond to the experimental values of \citet{Madsen2013PRB}.}
\label{KristianTheory}
\end{figure}

Dephasing from LA phonons is an important source of decoherence for quantum dots that is not well described by a Markovian pure-dephasing rate. As discussed in Sec.\ \ref{sec:sst:singlephotonsources}, LA phonons give rise to broad sidebands that in the context of cavity QED may be enhanced or filtered by the cavity. LA phonons can readily be modeled microscopically with the dissipative Jaynes-Cummings model by incorporating the interaction Hamiltonian in Eq.\ (\ref{eq:sst:phonon_interaction}) \cite{Hohenester2010PRB, Kaer2010PRL, Wilson-Rae2002PRB}. This leads to a modification of  Eqs.\ (\ref{density-matrix-elements}) such that the equation of motion for the photon-assisted polarization term generalizes to \cite{Kaer2012PRB}
\bea
\frac{\text{d} \rho_{12}}{\text{d}t} &=&  - \left(i \Delta + \frac{\gamma_{\text{ng}} + \kappa }{2} + \gamma_{12}(t) \right) \rho_{12} \nn \\
&-&  i \left[ g + \mathcal{G}^{<}(t) \right] \rho_{22} + i \left[g + \mathcal{G}^{>}(t)  \right]
\rho_{11}, \nn \\
&&
\eea
which contains three new effects:  a time-dependent dephasing rate $\gamma_{12}(t)$ emerges, the coupling strength $g$ is renormalized by the real part of the phonon-reservoir memory Kernel functions $\mathcal{G}^{<}(t)$, and $\mathcal{G}^{>}(t)$ (see \citet{Kaer2010PRL} for their explicit definitions), and an additional decay rate is introduced by the imaginary part of the same functions. Furthermore, an overall polaron frequency shift has been absorbed into the detuning. The magnitude and thus importance of the phonon-dephasing terms are fully characterized by the temperature $T$ and the quantum-dot-phonon interaction strength of Eq.\ (\ref{phonon-interaction-strength}). The typical memory depth of LA phonons is about \SI{5}{\pico\second} and when describing dynamics it is usually an excellent approximation to take the long-time limit $t \rightarrow \infty$ for the time-dependent terms, while in contrast the full time dependence is generally required when describing coherent properties such as the degree of indistinguishability of single photons \cite{Kaer2013PRB}. We emphasize that even when the long-time limit is valid, the interaction with the phonon reservoir is memory dependent since $\mathcal{G}^{<}(t)$, $\mathcal{G}^{>}(t)$, and $\gamma_{12}(t)$ are obtained as integrations over all previous times \cite{Kaer2012PRB}. This constitutes the phononic analogue of the photonic non-Markovian effects considered in Sec.\ \ref{Section-spontaneous-emission}. Interestingly this non-Markovian phonon reservoir was predicted to be able to stabilize coherent quantum dynamics in a cavity QED setup \cite{Carmele2013NJP}. This is in opposition to the common perception that dephasing tends to suppresses quantum behavior, which only is generally valid for a Markovian reservoir.

While the effective photon-emitter coupling strength is reduced by the presence of phonons, a salient feature is that the bandwidth of the coupling can be significantly increased beyond the linewidth of the cavity. This can be understood as being due to a phonon-assisted Purcell effect: a blue- (red-) detuned quantum dot can emit a photon to the cavity after a single LA phonon is  emitted (absorbed). The broadband nature of LA phonons (meV range) implies that this effect can very significantly increase the coupling range of the cavity. Figure \ref{KristianTheory} plots the decay rate of a quantum dot in the weak-coupling regime when varying the detuning relative to the cavity and for different temperatures. The broadband phonon-assisted Purcell effect is very pronounced and found to display an asymmetry with detuning reflecting the imbalance between phonon emission and absorption due to the existence of spontaneous phonon emission. By increasing the temperature this asymmetry is gradually leveled out.

Approximate analytical results can be derived in the limit of $\Delta \gg g$, i.e., the large-detuning tails of the Purcell enhanced coupling range. The detuning-dependent decay rate of a quantum dot in the cavity can be expressed as
\be
\gamma(\Delta) = \gamma_{\text{ng}} + 2 g^2 \frac{\gamma_{\text{dis}}}{\gamma_{\text{dis}}^2 + \Delta^2} \left[1 + \frac{\Phi(\Delta)}{\hbar^2 \gamma_{\text{dis}}} \right], \label{Purcell-rate-with-phonons}
\ee
where $\gamma_{\text{dis}} = (\gamma_{\text{ng}} + \kappa)/2$ is the total dissipation rate, and
\be
\Phi(\Omega) = \pi \sum_{\bf k} \left| M^{\bf k} \right|^2 \left( n_{\bf k} \delta(\Omega + \Omega_{\bf k}) + \left[ n_{\bf k} + 1 \right] \delta(\Omega - \Omega_{\bf k}) \right), \label{phonon-eff-density}
\ee
with $M^{\bf k} = M_e^{\bf k} - M_g^{\bf k}$ and $n_{\bf k}$ is the phonon occupation as introduced in Sec.\ \ref{sec:sst:singlephotonsources}. This constitutes a generalization of the Purcell-enhanced decay rate of Eq.\ (\ref{rho11}) to include the phonon-assisted coupling to the cavity mode. The phonon interaction is described by $\Phi(\Omega)$, which is the effective phonon density that couples to the quantum dot; it is displayed in the inset of Fig.\ \ref{phononDosFigure}. An effective phonon density of states may be defined from Eq.\ (\ref{phonon-eff-density}) by dividing out the phonon occupation factors giving
\be
\rho_\text{ph}(\Omega) = \pi \sum_{\bf k} \left| M^{\bf k} \right|^2 \left[ \delta(\Omega + \Omega_{\bf k}) +  \delta(\Omega - \Omega_{\bf k}) \right], \label{phonon-eff-DOS}
\ee
which can be considered the phonon analogue to the photonic LDOS defined in Eq.\ (\ref{eq:LDOS-def}), since the sum counts the density of phonon modes weighted by the phonon-emitter coupling strength. Since the coupling strength depends on the exciton wave function, an effective DOS will exist for each individual quantum dot. The applied formalism holds in the case of bulk phonon modes, which turns out to be a good approximation for most of the energies relevant for quantum-dot dephasing by phonons, as considered in further detail below.

\subsection{The Purcell effect in photonic cavities}
\label{Section-Purcell-effect}

Photonic nanocavities are very well suited for large Purcell factors since they confine light in a small mode volume while achieving high $Q$-factors that approximately match the internal linewidth of quantum dots. The first experimental demonstration of the Purcell effect for quantum dots was presented in \citet{Gerard1998PRL} where a Purcell factor of 5 was observed for an ensemble of emitters in a micropillar cavity. The beneficial role of the Purcell effect for overcoming dephasing was explicitly demonstrated in \citet{Santori2002Nature} by recording the degree of indistinguishability of the emitted single photons in two-photon interference measurements, cf.\ the experimental data in Fig.\  \ref{HOM-measurements} of Sec.\ \ref{sec:sst:singlephotonsources}. Subsequently, photonic-crystal cavities have been investigated intensely due to the small mode volumes that can be obtained and typical Purcell factors of about 10 \cite{Kress2005PRB,Englund2005PRL} have been reported in addition to the generation of indistinguishable photons \cite{Laurent2005APL}. We note that the claim of significantly larger Purcell factors can be found in the literature, but those are often based on spectral measurements that do not directly probe the dynamics and therefore may be influenced by, e.g., multi-exciton effects.

\begin{figure}
\includegraphics[width=\columnwidth]{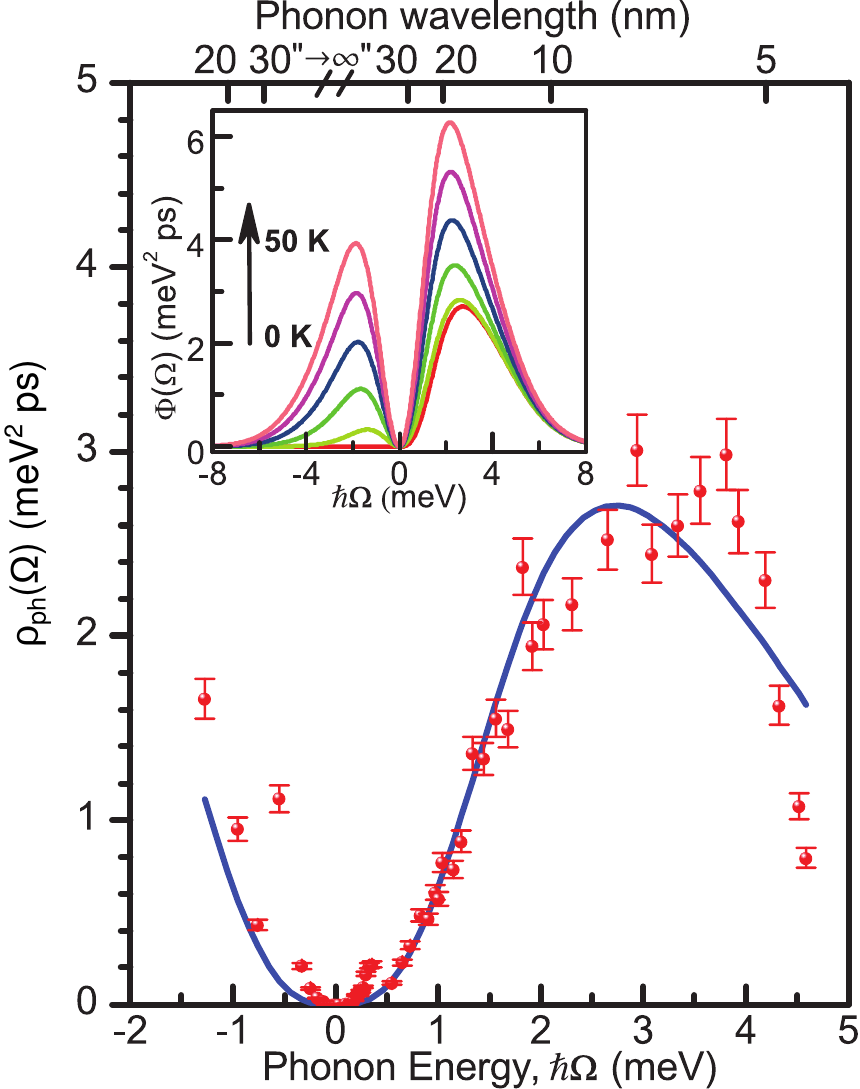}
\caption{Measured effective phonon DOS for a quantum dot versus phonon energy obtained by embedding the quantum dot in a photonic-crystal cavity and measuring the detuning-dependent dynamics (data not shown). The blue curve shows the theoretical effective phonon DOS when assuming LA phonons in bulk GaAs. The inset shows the phonon density for different temperatures. Reprinted with permission from \citet{Madsen2013PRB}.}
\label{phononDosFigure}
\end{figure}

The experimental observation that a single quantum dot can be coupled to a photonic-crystal cavity even when detuned many linewidths away from resonance was reported in \citet{Hennessy2007Nature}, and \citet{Hohenester2009PRB} considered the specific example of phonon-assisted Purcell enhancement as discussed Sec.\ \ref{Section-Phonon-dephasing}. Purcell-enhanced phonon-assisted processes have also been studied by applying resonant excitation \cite{Ates2009NPHOT}, which has enabled the demonstration of phonon-mediated coupling between two different quantum dots that were radiatively coupled to the same cavity mode \cite{Majumdar2012PRB}. \citet{Madsen2013PRB} presented a detailed comparison between experiment and theory and the theory of phonon-assisted Purcell enhancement, expressed by Eq.\ (\ref{Purcell-rate-with-phonons}), was found to model the experimental data quantitatively. In this experiment, the effective phonon DOS as experienced by the quantum dot could be recorded over a broad energy range, which was controlled experimentally by varying the quantum dot detuning relative to the cavity. The data are reproduced in Fig.\ \ \ref{phononDosFigure}. A maximum in the effective phonon DOS is observed for a phonon wavelength of around 7 nm followed by a roll off at shorter wavelengths reflecting that the wavelength becomes comparable to the size of the exciton wavefunction. Interestingly, the experimental data are explained well by the theory of bulk phonons since at most energies the phonon wavelength is significantly smaller than the lattice period of the photonic crystal ($a = \SI{240}{\nano\meter}$) apart from at very small energies where the method is not conclusive since the single exciton line could not be distinguished from other emitters that may feed the cavity. This experiment constitutes an example of the powerful potential of cavity QED to enhance light-matter interaction to such an extent that weak phonon processes can be studied in detail. In a broader context these studies connect to the research field of quantum optomechanics where the quantum properties of phononic degrees of freedom are exploited \cite{Kippenberg2008Science} and where the ability to combine phononic and photonic band gap structures has been demonstrated \cite{Eichenfield2009Nature}. An interesting proposal is to use the stress induced by the excitation of an exciton in a single quantum dot as a mechanism for cooling the mechanical motion of a nanomembrane \cite{Wilson-Rae2004PRL}.


\subsection{Observation of strong coupling}
\label{Section-observation-strong-coupling}

\begin{figure}
\includegraphics[width=\columnwidth]{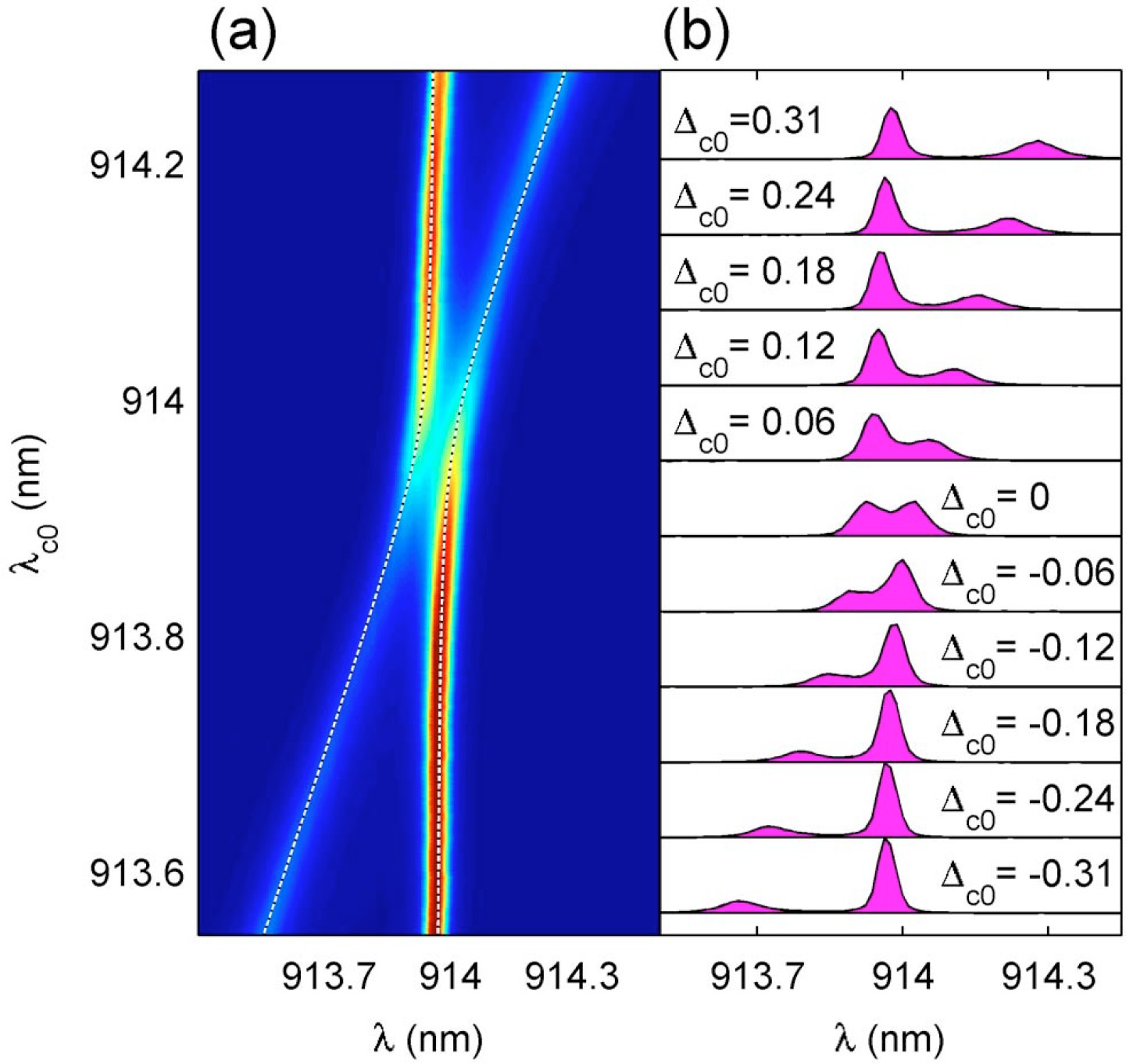}
\caption{Experiments on strong coupling between an exciton in a quantum dot and a photonic-crystal cavity. (a) Density plot of spontaneous-emission spectra measured when tuning the cavity by gas deposition across a quantum-dot line. (b) Selected spectra with detunings indicated in the plot. A clear anti-crossing is observed and on resonance the area and linewidth of the two peaks are very similar as expected from theory. $\lambda_{\rm c0}$ is the wavelength of the cavity for the case of no coupling,  $\lambda$ is the observation wavelength, and $\Delta_{\rm c0} = \lambda - \lambda_{\rm c0} $ in nanometer. Reprinted with permission from \citet{Thon2009APL}.}
\label{anti-crossing}
\end{figure}

By increasing the coherent light-matter interaction strength between a single quantum emitter and a photon in a cavity, the transition to the strong-coupling regime eventually occurs. Here the light and matter degrees of freedom become quantum entangled leading to the formation of a cavity polariton. The new dressed eigenstates and eigenenergies of the coupled emitter-cavity system in the absence of any dissipation and on resonance with the cavity are
\bs
\bea
\left|\pm, n\right> &=& \frac{1}{\sqrt{2}} \left| e,n-1\right> \pm \frac{1}{\sqrt{2}} \left| g,n\right>, \\
 E_{\pm, n} &=& (n+1/2) \hbar \omega \pm \hbar \Omega_n/2,
\eea
\label{eq:eigenstates+energies-JC-ladder}
\es
where the Rabi frequency is $\Omega_n = 2g \sqrt{n}$.
The experimental signature of strong coupling is the observation of an anti-crossing when detuning a single quantum emitter through resonance of a cavity mode. This phenomenon was first observed experimentally with quantum dots in 2004 both in a micropillar cavity \cite{Reithmaier2004Nature} and a photonic-crystal cavity, \cite{Yoshie2004Nature} and subsequently also in a microdisk cavity \cite{Peter2005PRL}. Figure \ref{anti-crossing} shows an example of the observed anti-crossing in a photonic-crystal cavity that was deterministically aligned to a single quantum dot. The quantum-dot tuning relative to the cavity was controlled by gas deposition, which adsorbs on the photonic-crystal surfaces and thus changes the resonance frequency. An avoided crossing was observed on resonance where the two peaks had approximately the same width. This is a consequence of the fact that the quantum dot and cavity modes become inseparable, i.e., the light and matter degrees of freedom entangled.

Since the first observations of strong coupling with quantum dots it has been realized that these solid-state systems contain distinct properties that differ from their atomic counterparts. For instance, it was observed that even when a single quantum dot is detuned \SI{4}{\nano\meter} away from the cavity resonance, corresponding to more than 50 cavity linewidths, it can feed the cavity mode, which was proven by a pronounced anti correlation between the quantum-dot peak and the cavity peak \cite{Hennessy2007Nature}. While the phonon-assisted processes discussed in the previous section can be responsible for coupling ranging up to  approximately \SI{4}{\nano\meter}, the coupling at even larger detunings has been attributed to the formation of a continuum of exciton states induced by the hybridization of the quantum-dot states and wetting-layer states \cite{Winger2009PRL}. These states are populated in non-resonant excitation experiments in particular at strong continuous-wave excitation conditions where more carriers are populating the quantum dot and its surroundings. Such states were also found to be important in photonic-crystal lasers with quantum-dot gain media, see \citet{Strauf2011LasPhotRev} for a detailed review. Furthermore, it was found that an anti-crossing could be observed in spectral measurements for a photonic-crystal cavity even though it was proven from the time-resolved measurements that the most efficiently coupled quantum dot was not strongly coupled to the cavity \cite{Madsen2013NJP}. This observation was attributed to collective effects due to the presence of other quantum dots or exciton transitions feeding the cavity, which effectively can increase the cooperativity and therefore the Rabi splitting \cite{Diniz2011PRA}.

\subsection{Photon nonlinearities in cavity QED}
\label{Section-cavity-nonlinearity}

\begin{figure}
\includegraphics[width=\columnwidth]{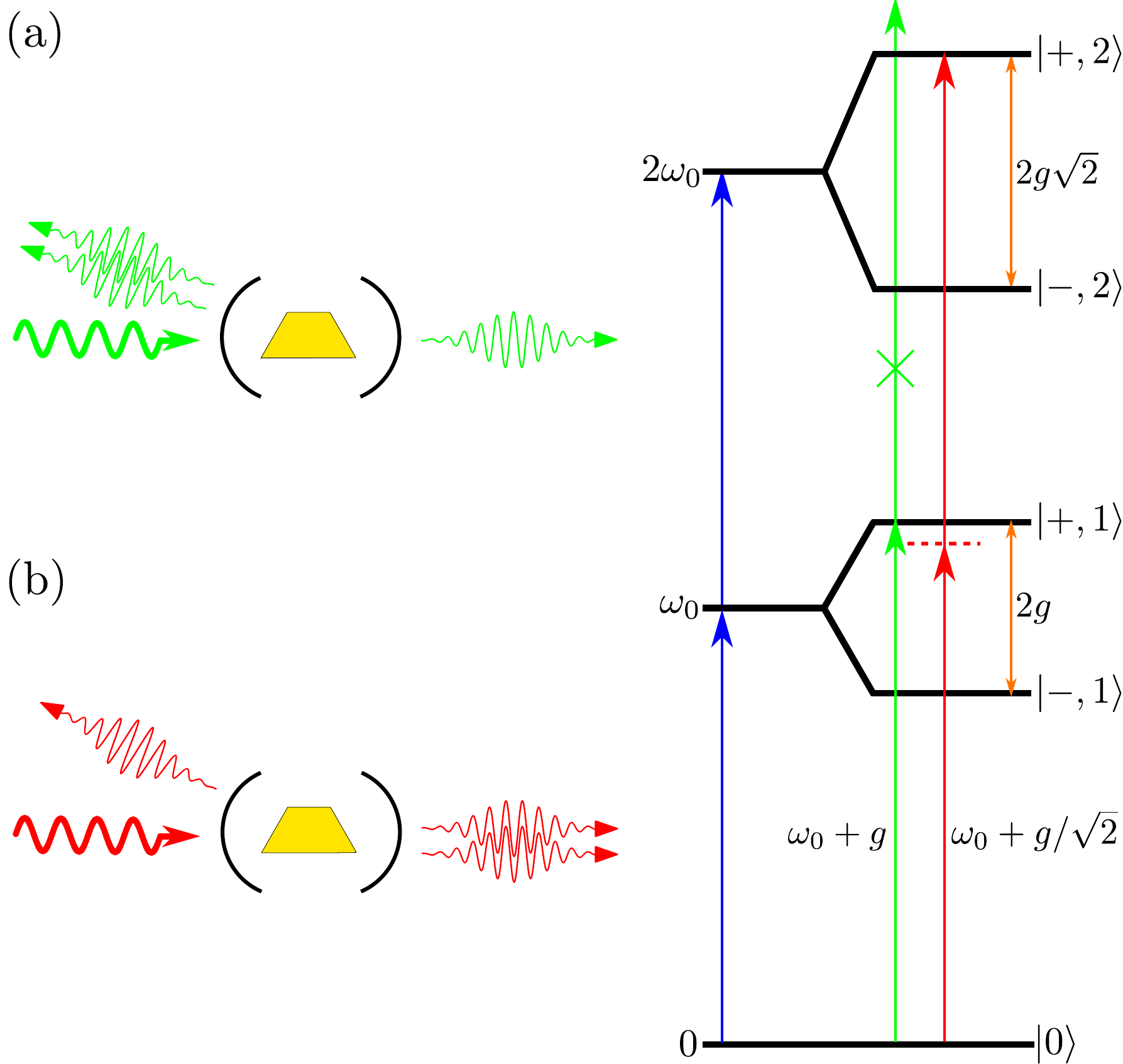}
\caption{ Illustration of the mechanisms behind photon blockade (a) and photon tunneling (b) in a strongly coupled emitter-cavity system. The bare energy of the transition is $\omega_0$ and when the cavity and emitter are strongly coupled, the energy levels are split forming the dressed eigenstates of Eqs.\ (\ref{eq:eigenstates+energies-JC-ladder}). (a) When the excitation laser is tuned to the energy of the $|+, 1\rangle$ state (green line), the emission from the cavity is antibunched since only $n=1$ photon Fock states are populated as higher order states are not allowed in the cavity and therefore reflected. (b) When the laser is tuned to half the energy of  $|+, 2\rangle$ (red line) strongly bunched emission is observed as only $n=2$ Fock states are populated.}
\label{JC-ladder}
\end{figure}

A quantum emitter strongly coupled to a cavity can be exploited as a highly nonlinear medium potentially operating down to the single-photon level. A nonlinearity can be induced by the anharmonicity of the Jaynes-Cummings energy levels, enabling the generation of light with either sub-Poissonian or super-Poissonian photon statistics when varying the detuning of a weak probe coherent state relative to the coupled emitter-cavity system. Figure\ \ref{JC-ladder} shows the Jaynes-Cummings energy ladder and illustrates the nonlinearity. Photon blockade occurs when the probe beam is tuned to resonance with one of the two dressed polariton states. If the emitter-cavity system is initially not excited, the anharmonicity of the Jaynes-Cummings ladder implies that only a single photon can be stored  since transitions from the first to the second manifold will be off-resonant and therefore blocked. As a consequence, the field transmitted through the cavity will be antibunched. This effect can be considered the photonic analogue of the Coulomb blockade of electron transport where the photon-photon repulsion is mediated by the strongly-coupled cavity.

\begin{figure}
\includegraphics[width=\columnwidth]{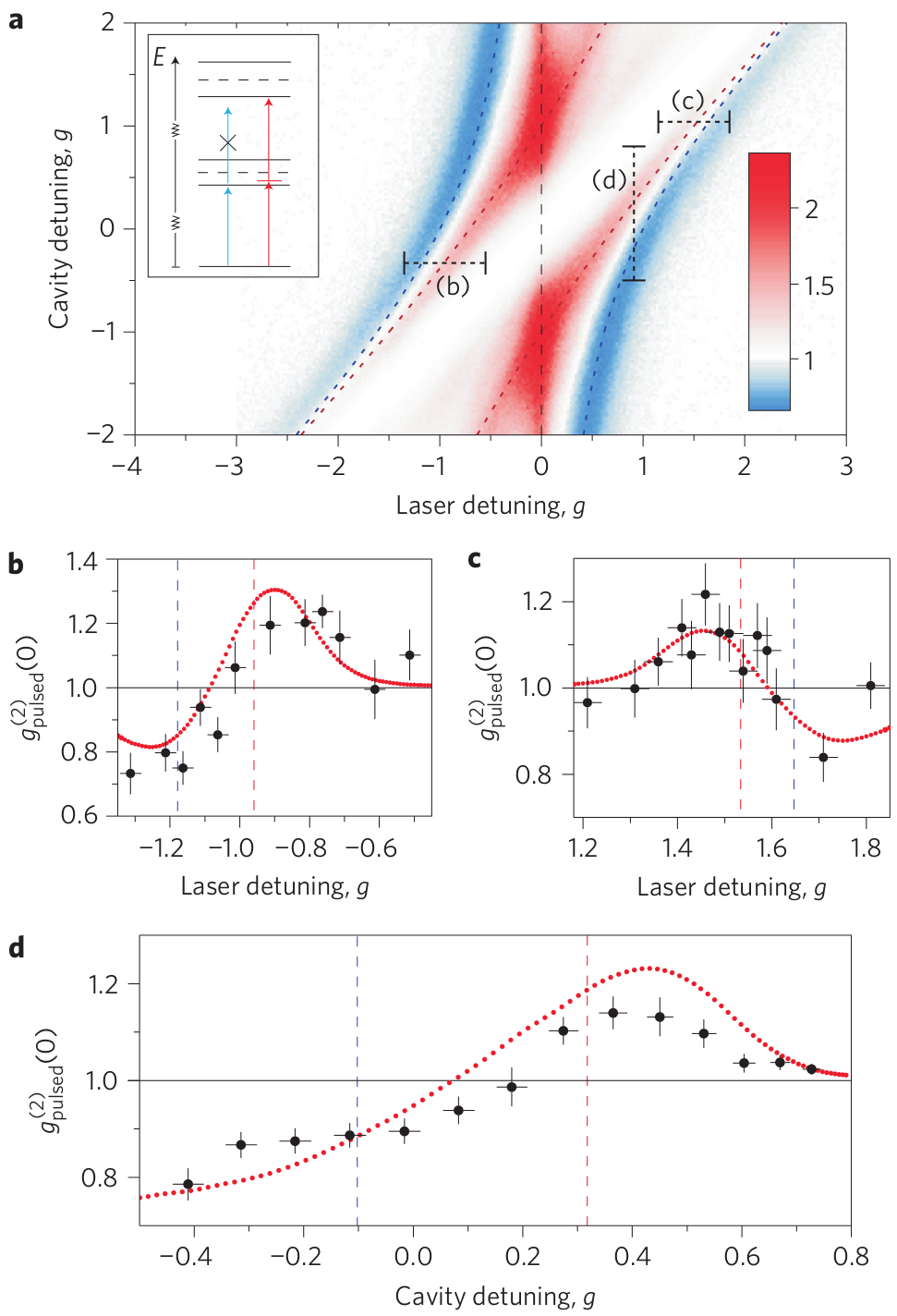}
\caption{Experimental observation of a photon blockade for a quantum dot in a photonic-crystal nanocavity. (a) Calculated second-order correlation function $(g^{(2)}(0))$ versus cavity and laser detuning from the quantum dot in units of the coupling strength $g$. Photon bunching (antibunching) is predicted in the red (blue) areas corresponding to the processes sketched in the inset and in Fig.\ \ref{JC-ladder}. (b)-(d) Detailed comparison between the measured and predicted values. Reprinted with permission from \citet{Reinhard2012NPHOT}.}
\label{Photon-blockade}
\end{figure}

Photon blockade with a single quantum dot in a photonic-crystal nanocavity was observed first in \citet{Faraon2008NPHYS}. A detailed experimental study of the detuning-dependent photon statistics for a strongly-coupled quantum dot-cavity system is presented in Fig.\ \ref{Photon-blockade} \cite{Reinhard2012NPHOT}. A clear photon-blockade effect was observed when tuning a probe laser to the dressed state of the first Jaynes-Cummings manifold (blue areas in Fig.\ \ref{Photon-blockade}(a) while two-photon processes to the second Jaynes-Cummings manifold (red areas in Fig.\ \ref{Photon-blockade}(a) are found to lead to photon bunching, which is the photon tunneling effect illustrated in Fig.\ \ref{JC-ladder}(b). This nonlinear response can be ultrafast, and in a two-colour experiment a signal field could be switched on and off on a timescale of $\sim \SI{50}{\pico\second}$ \cite{Volz2012NPHOT}, which is ultimately determined by the anharmonicity of the Jaynes-Cummings ladder. So far, rather modest photon-blockade effects have been observed with a reduction of the second-order correlation function of approximately 25\%, which could potentially be improved by increasing the cooperativity $(g^2/\kappa \gamma_{\mathrm {ng}})$ further to take the cavity deeper into the strong-coupling regime. The photon bunching was also found to be sensitive to blinking of the quantum-dot emission line, which to a large extent could be counteracted by applying an above band repumping laser. The experimental data were found to be explained well by Monte Carlo wavefunction simulations that include the effects of quantum-dot blinking (cf.\ Fig.\ \ref{Photon-blockade}(b)-(d)).

The above mentioned progress on photon nonlinearities in photonic cavities concerned the situation where few photons inside the cavity were sufficient to induce strong nonlinarities, while the actual driving fields outside the cavity contained many photons due to a low in-coupling efficiency into the cavity mode. For many applications of photon nonlinearities, however, such a loss will limit the performance. To this end, a nonlinearity threshold of only eight incident photons was demonstrated in a micropillar cavity by constructing a cavity with large input-coupling efficiency \cite{Loo2012PRL}, which is an important step towards a true single-photon nonlinearity. The considerable experimental progress on few-photon nonlinearities in solid-state cavity-QED systems may lead to applications within quantum-information processing, e.g., for controlled-phase gates or single-photon transistors. These applications are considered in further detail in Sec.\ \ref{Section-QIP}.

\section{Photonic quantum-information processing}
\label{Section-QIP}
In the present manuscript we have reviewed the physics of light-matter interaction with single quantum dots in photonic nanostructures with special emphasis on photonic crystals. These systems have matured very significantly over the last decade and they now constitute a toolbox of very promising components for all-solid-state quantum-information processing. Consequently, researchers worldwide are gradually starting to apply nanophotonic systems for proof-of-concept quantum-information processing, and are considering how to construct more advanced photonic quantum architectures. Various protocols have been put forward of using the spin of single electrons in quantum dots for quantum-information processing where the coupling is mediated  either by electron exchange interaction \cite{Loss1998PRA} or by two spins common interaction with a high-$Q$ cavity \cite{Imamoglu1999PRL}. Another approach utilizes the large light-matter interaction strength achievable in photonic nanostructures to generate  efficient single photons on demand and photon nonlinearities for photonic quantum-information processing protocols. Hybrid systems that interface the spin of single electrons or holes in quantum dots with photons seem particularly promising since a single electron or hole spin can have coherence times much longer (i.e., microseconds) \cite{Brunner2009Science} than excitons while the photons can carry quantum information with low loss over long distances. In the present section we focus primarily on the current experimental status and progress on photonic quantum-information processing with quantum dots in photonic nanostructures and address the future potential of scaling these systems to larger photonic quantum networks.

\subsection{Photonic quantum-information processing with quantum dot sources}
The essential prerequisite for many photonic quantum-information protocols is the ability to generate highly efficient and coherent single photons. In the following a brief summary of the various quantum-dot sources reported in the literature is given. The current state-of-the-art single-photon source is a micropillar cavity where the deterministic coupling of a single quantum dot to an optimized cavity \cite{Dousse2008PRL} has led to  82\% indistinguishable photons collected with an efficiency of 65\% \cite{Gazzano2013NCOM}. Pulsed resonant excitation has been found to enable even larger degrees of indistinguishability and nearly perfectly pure single photons with multiphoton probability of only $1.2\%$ and indistinguishability of $97\%$ was reported in \citet{He2013NNANO} but this source presented a collection efficiency limited to $1.3 \%$.  Such a high degree of indistinguishability corresponds to a pure-dephasing time of $T_2^* = \SI{5.7}{\nano\second}$ that was extracted from resonance-fluorescence measurements. Implementing these methods on a photonic-crystal platform appears very appealing; $T_2^* = \SI{0.6}{\nano\second}$  was so far observed in Hong-Ou-Mandel interferometry in a photonic-crystal cavity. This was obtained by exciting the quantum dot via an LA-phonon sideband  and an out-coupling collection efficiency of $44 \%$ was demonstrated \cite{Madsen2013InPrep}. With such a long coherence time, indistinguishability exceeding 75\% is readily anticipated in photonic crystals for a moderate and experimentally achievable Purcell factor of $10$, and this coherence is expected to increase even further with resonant excitation. In combination with the near-unity $\beta$-factor observed in the photonic-crystal waveguide \cite{Arcari2013InPrep} and the ability to couple photons efficiently off chip \cite{Tran2009APL} such a source could be an almost ideal source of on-demand and coherent single photons with immediate applications in quantum-information processing.

An efficient source of entangled photons is another essential quantum resource required for photonic quantum-information processing. One approach uses a single quantum dot consecutively triggered to emit two photons. Both photons are send to a Mach-Zehnder interferometer with unequal path lengths for compensation of the time delay, and half-wave plates and non-polarizing beam splitters for polarization entangling of the two photons. Such a source was reported in \citet{Fattal2004PRL_ENTANGLE} where the violation of Bell's inequality was demonstrated. Another approach uses the cascaded recombination process of biexcitons that occurs through the two bright exciton states either by emitting two horizontally or two vertically polarized photons in succession \cite{Benson2000PRL}, cf.\ Fig.\ \ref{fig:sst:Bright-dark_level_scheme}(c) in Sec.\ \ref{sec:sst:decaydynamics}. Entanglement can be obtained in this way if the two different decay paths of the biexciton are made indistinguishable, which has been obtained by spectral filtering \cite{Akopian2006PRL} or by growing quantum dots with small fine-structure splitting and applying a small in-plane magnetic field for tuning \cite{Young2006NJP}. Such a source can even be driven electrically enabling a light-emitting diode of entangled photons \cite{Salter2010Nature}. The brightness of the biexciton entanglement source was increased to $12 \%$ collected photons per excitation pulse by incorporating the quantum dot in a micropillar cavity \cite{Dousse2010Nature}. To achieve this result, a specially designed coupled cavity was constructed that consisted of two optically coupled micropillar cavities in order to remove any which-path information for the two emitted photons. The entanglement sources based on biexcitons are maturing as an exciting alternative to heralded entanglement sources based on spontaneous parametric downconversion that suffer inherently from multi-photon processes \cite{Lounis2005RPP}, although no obvious way of scaling this approach to larger number of photons exists. For that purpose, a single quantum dot that emits a train of single-photon pulses into a single optical mode, which, e.g., can be achieved in nanophotonic waveguides, seems very attractive. By implementing electro-optical pulse picking and compensation of the fixed delay between the pulses, a large number of single photons could potentially be obtained, which would constitute and interesting quantum resource for photonics quantum-information processing.

A number of proof-of-concept quantum-information protocols have already been implemented with  quantum-dot single-photon sources. Early work has demonstrated quantum cryptography following the BB84 protocol of encoding quantum information in the polarization of single photons \cite{Waks2002Nature}. An alternative approach has been to implement two-qubit gates in exciton and biexciton states in the quantum dot \cite{Li2003Science} but this method is not scalable to more qubits. Quantum information can also be encoded in the propagation path of a single photon, and single photons from quantum dots were used to implement a Deutsch-Jozsa two-qubit algorithm \cite{Scholz2006PRL} and a probabilistic linear-optics CNOT gate \cite{Pooley2012APL,He2013NNANO} while the entangling capability was fully realized in \citet{Gazzano2013PRL}. Recently also a CNOT gate operation between a single-photon qubit and a qubit consisting of a quantum dot strongly coupled to a photonic-crystal cavity was demonstrated \cite{Kim2013NPHOT}. Quantum teleportation of a single-photon qubit was reported in \citet{Fattal2004PRL_TELEPORT} and \citet{Nilsson2013NPHOT}.

The effective detection of photonic quantum states is another essential requirement for optical quantum-information processing. A detailed account is outside the scope of the present account; for a recent review see \citet{Hadfield2009NaturePhot}. Essential features of a good photon detector include high speed, near-unity quantum efficiency, low dark-count rates and after-pulsing probabilities, broadband operation, and ideally the ability to resolve the number of photons in a pulse. Significant progress on all these directions has been reported. The most common detector applied is the single-photon-counting avalanche photodiode that is simple and robust and even can be applied for resolving the photon number \cite{Kardynal2008NPHOT}. Superconducting nanowire single-photon detectors have recently been developed and they are potentially fast and broadband and can be integrated on chip \cite{Reithmaier2013SciRep}. Another useful functionality is the frequency transduction of a single photon from one wavelength to another, which, e.g., would allow transducing a single-photon source between the visible part of the spectrum and the telecommunication band, since present detectors have the highest efficiency in the visible range while superior low-loss optical circuits have been developed for telecom applications. Efficient transduction of single photons from a quantum dot were reported in \citet{Rakher2010NPHOT} by nonlinear wavelength conversion.

An alternative approach to quantum-information processing applies a spin-photon interface by connecting the spin of an electron or hole confined in a quantum dot to photons emitted by spontaneous emission. This topic has recently been reviewed thoroughly \cite{Urbaszek2013RMP,DeGreve2013RPP} and here we will just highlight a few recent breakthroughs in the context of quantum-information processing. An essential resource is the deterministic preparation of trion states by adding a single electron or a hole by controlling the gate voltage across the quantum dot. Spontaneous emission from the trion state prepares a single electron spin in the quantum dot that is directed either up or down, and spin-photon entanglement has recently been reported from such a decay \cite{DeGreve2012Nature,Gao2012Nature}. This spin-photon interface was subsequently exploited for quantum teleportation of a qubit encoded in a photon onto the spin of the electron in the quantum dot  \cite{Gao2013NCOM}. Furthermore, the possibility of using connected photonic waveguides for coupling a quantum-dot spin to a path-encoded photon has been demonstrated \cite{Luxmoore2013PRL}. In photonic-crystal waveguides the spin-photon interface can be made deterministic, and a deterministic and scalable on-chip CNOT gate with an operation fidelity exceeding 90 \% has recently been proposed \cite{Sollner2014arxiv}. This could form the basis for advanced quantum architectures that interface quantum-dot spins with photonic quantum circuits.

\subsection{Towards scalable photonic quantum networks}

\begin{figure}
\includegraphics[width=1\columnwidth]{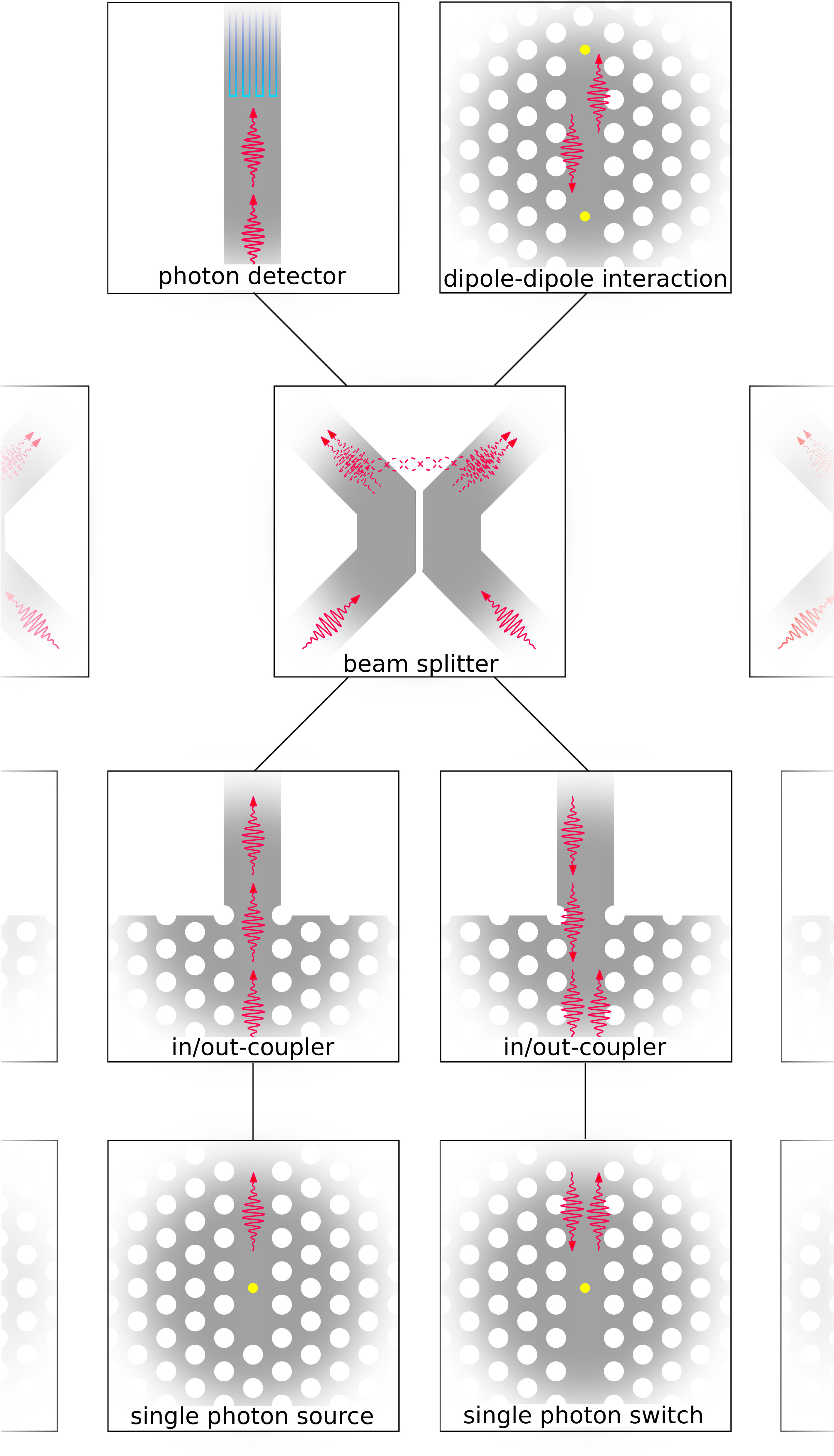}
\caption{Schematics of basic building blocks of an integrated quantum network based on quantum dots and photonic crystals. From the bottom and up: A single quantum dot in a photonic-crystal waveguide is used  as the quantum resource for a highly efficient single-photon source (left image) or a nonlinear single-photon switch (right image). The generated quantum light can be coupled in and out of the photonic-crystal waveguide sections with high efficiency by engineering the interface to dielectric waveguides. The routing and processing of photons can be carried out with low loss photonic circuits constructed from  dielectric waveguides. As an example a beam splitter is illustrated where the interfering of two single photons creates path-entangled photons. The efficient detection can be done on chip by integrating superconducting single-photon detectors (left upper image). Multiple quantum dots can be controllably coupled by the extended dipole-dipole interaction present in photonic-crystal waveguides (right upper image). Potentially many of these building blocks could be merged together into a complex quantum network.}
\label{fig:QuantumNetwork}
\end{figure}

Most of the examples discussed so far consider just a single photon at a time emitted from or interacting with a single quantum dot. Obviously, in order to implement advanced quantum-information processing and computing these simple systems need to be scaled to larger quantum architectures. In this context, a single quantum dot can be considered a stationary node in a quantum network that may be connected to other quantum dots by photons representing flying qubits. So far, experimental progress using single photons to connect separate emitters has been reported with single molecules \cite{Rezus2012PRL}, quantum dots \cite{Gao2013NCOM},  and atoms in separate cavities \cite{Ritter2012Nature}    An interesting alternative to scaling up the number of quantum dots is to exploit the  near-unity on-demand single-photon generation efficiency of quantum dots in nanophotonic structures to generate a deterministic train of coherent single-photon sources. In reality, the combination of these two methods is likely to be most promising, i.e., the quantum network would contain several spatially distinct quantum-dot nodes that can be individually addressed and coherently connected by trains of single photons. Figure\ \ref{fig:QuantumNetwork} illustrates the basics of this vision and displays the various building blocks and functionalities that a photonic quantum network could consist of. As thoroughly discussed in the present manuscript, photonic-crystal waveguides are very well suited for generating single photons on chip and giant photon nonlinearities. These quantum-optics resources may form the backbone of the photonic quantum network, where individual, (e.g., electrical) tuning and deterministic positioning of each quantum dot relative to the photonic-crystal waveguide would be essential requirements. The generated photons may subsequently be transferred to dielectric waveguides by engineered out couplers since dielectric waveguides can be fabricated with very low propagation loss due to tolerance to fabrication imperfections, and therefore photonic circuitry is preferably implemented in these rather than in photonic-crystal waveguides. Finally, detection could be performed on-chip as well by nanowire superconducting detectors. The benefit of the nanophotonics approach is that all functionalities could be highly integrated on a photonic chip. The number of obtainable nodes in the quantum network and the fidelity of photonic gates would depend sensitively on the ability to engineer and subsequently control the local coupling of quantum dots to the photonic crystals. The present manuscript has reviewed the recent significant progress and current status on these issues; continuous spectacular  progress is expected as the research field develops even further. It remains an exciting research topic to exploit how large and complex quantum networks that can be achieved experimentally, including the dependence on qubit control, fabrication imperfections, and decoherence processes.

Important experimental progress has already been obtained towards scaling quantum-dot systems. The quantum interference of single photons emitted from two separate quantum dots has been reported both in pulsed \cite{Flagg2010PRL} and continuous-wave \cite{Patel2010NPHOT} operation. In these experiments the ability to identify two different quantum dots with similar decay and dephasing rates and subsequently to locally tune them into mutual resonance is essential in order to remove any which-path information of the photons. In \citet{Flagg2010PRL} the two quantum dots were located in two different samples and one of the quantum dots was tuned by applying strain. In \citet{Patel2010NPHOT} electrical tuning was applied to a quantum dot that was tuned into resonance with an electrically driven target quantum dot mounted in another cryostat \SI{1.1}{\meter} away whereby post-selected two-photon interference was observed.  The method of electrical tuning is scalable to connecting more quantum dots on the same sample, where local gates could be defined on the optical chip, e.g., by etching trenches for in-plane electrical isolation or by local implantation of dopant ions \cite{Ellis2011NPHOT} while the quantum dots could be optically connected by waveguides.

Another important point to consider is the level-scheme of the emitter. A quantum dot populated by a single neutral exciton forms a three-level V-scheme consisting of the two optically active bright states and the stable ground state, cf.\ Fig.\ \ref{fig:sst:Bright-dark_level_scheme}(b) in Sec.\ \ref{sec:sst:decaydynamics}. For many applications it is favorable to have one of these states efficiently coupled, e.g., to a waveguide mode, while the other transition should be metastable. This can be realized in photonic crystals that typically act as a ``highly anisotropic vacuum'', i.e., if one dipole orientation is efficiently coupled, the perpendicularly oriented dipole would be weakly coupled, as can be seen from the calculations presented in Fig.\ \ref{LDOS-WG-maps} of Sec.\ \ref{Sec:waveguide-experiments} and this was experimentally demonstrated in \citet{Wang2010OL}.  Further flexibility is offered by forming quantum-dot molecules consisting of two or more coupled quantum dots that, e.g., can be stacked vertically. In this way, very long-lived indirect exciton states can be formed by an electron residing in one quantum dot and the hole in the other, and the lifetime can be tuned by an applied electric field. Controlled coupling of two stacked quantum dots has been reported in \citet{Krenner2005PRL}.

A variety of different quantum-information protocols may be considered with quantum dots in photonic nanostructures. The following will just give a brief outline of some of the exciting directions that appear to be best suited given the current experimental progress on quantum-dot photon sources.

In the context of linear-optics quantum-information processing \cite{Kok2007RMP}, highly efficient quantum dot single-photon sources are very appealing. Such a source would constitute an important alternative to heralded spontaneous-parametric-downconversion sources that despite impressive recent advancements \cite{Brida2012APL} remain probabilistic in nature. Linear-optics protocols are interesting for quantum simulators that harness quantum parallelism to efficiently compute properties of complex quantum systems, which were originally envisioned by Richard Feynman \cite{Feynman1982IJTP}. Photonic quantum simulators hold exciting promises for tackling problems in, e.g., quantum chemistry and would strongly benefit from deterministic photon sources \cite{AspuruGuzik2012NPHYS}. Another example is the so-called ``boson-sampling problem'' \cite{Broome2013Science} that potentially could demonstrate quantum-enhanced speed-up in a foreseeable future if the system can be scaled to more photons. A scheme for universal linear optical quantum computing based on time-bin encoded qubits propagating in a single spatial mode was recently put forward \cite{Humphreys2013PRL} and this way of encoding quantum information seems very well suited for quantum-dot light sources since they can emit a train of photons. Another very promising proposal for quantum dots is to apply a deterministic single-photon source for preparing a string of entangled photons (a cluster state) \cite{Lindner2009PRL} that is a resource for one-way quantum computing \cite{Raussendorf2001PRL, Walther2005Nature}. This proposal would require near-unity coupling of single photons to a single propagating mode and the ability to manipulate the electron spin after photon emission, which would be feasible by addressing a trion state in a quantum dot embedded in a waveguide.

Adding efficient photon nonlinearities to the toolbox leads to a number of additional opportunities. For instance, a single-photon transistor has been proposed where an emitter with a near-unity $\beta$-factor in a photonic waveguide can either reflect or transmit a control pulse with the operation gated by the presence or absence of a single photon \cite{Chang2007NPHYS}. A similar scheme was subsequently proposed as a photon sorter capable of distinguishing one- and two-photon pulses enabling efficient Bell-state detection \cite{Witthaut2012EL}. Furthermore, scalable quantum-computing architectures have been put forward based on controlled phase shifts induced by a single emitter in a cavity \cite{Duan2004PRL} or photon-photon interactions mediated by a four-level emitter in a photonic waveguide \cite{Zheng2013PRL}. Embedding many identical quantum emitters in a photonic waveguide each subjecting a nonlinear response on the propagating photons with single-photon sensitivity could potentially induce a strongly correlated ``crystal of photons'' that would constitute a new quantum state for photons \cite{Chang2008NPHYS}. Related ideas of quantum-phase transitions of light have been explored for the case of arrays of strongly coupled cavities \cite{Hartmann2006NPHYS, Greentree2006NPHYS}.

The present manuscript has reviewed the progress of all-solid-state photonic approaches towards quantum-information processing at optical frequencies with main emphasis on self-assembled quantum dots. Significant experimental progress has also been obtained in other solid state quantum systems, notably using defect vacancy centers \cite{Aharonovich2011NPHOT}. Recently also hybrid approaches have started to emerge. One important frontier aims at coupling a single trapped atom to a photonic-crystal cavity or waveguide. A recent experimental breakthrough has been the trapping of an atom by an optical tweezer in the evanescent tail of a photonic-crystal waveguide mode \cite{Thompson2013Science}, and waveguides have been engineered for the purpose of trapping atoms inside the waveguide to achieve larger coupling efficiencies \cite{Hung2013NJP}.

\section{Concluding remarks and outlook}
In the last decade, remarkable progress has been achieved in the use of single quantum dots in photonic nanostructures for quantum-optics experiments. The general approach has been to confine and engineer light at the nanoscale whereby the photon-emitter interaction strength can be tremendously enhanced and very well controlled. This has led to, e.g., highly efficient and deterministic single-photon sources and large photon nonlinearities. Interestingly, the achievable photon-emitter coupling efficiencies in nanophotonic systems now start to approach the impressive level that can be achieved in superconducting microwave circuits\footnote{For a quantitative comparison of coupling coefficients in photonic nanostructures and superconducting circuits, compare the $\beta$-factors in \citet{Arcari2013InPrep} and \citet{Hoi2012PRL}.} with the additional benefit that optical circuits can be scaled to very small sizes. Furthermore, highly efficient single-photon detectors are rapidly being developed. It thus seems timely and very promising to start the quest of merging the simple quantum building blocks into larger and more complex quantum architectures. It is an important research challenge for the future to identify how large quantum systems can be assembled and controlled given the level of imperfections present. To this end, the photonic nanostructures have now matured to such a degree that quantum functionalities are limited by the emitter (decoherence, non-radiative processes) rather than the nanostructure. The overall vision of solid-state quantum photonics is to have all functionalities integrated on a single photonic chip, i.e., single-photon source, quantum circuit, and photon detection. Importantly, even a small-scale integration leads to a number of exciting new opportunities and applications on the road towards full integration. The basis for solid-state quantum-information processing with quantum dots in photonic nanostructures is established and the potential is very promising; we look forward to witnessing the exciting progress in the research field in the years to come.

In the present manuscript, we have reviewed the research field of quantum nanophotonics and in particular focused on the exciting prospects for applications in quantum-information processing. It is important to emphasize, however, that the described concepts and methods are rather general and of much wider applicability. The ability to engineer the light-matter interface with nanophotonics has, e.g., been proposed for improving photovoltaic devices  \cite{Atwater2010NMAT}. Furthermore, access to ultimate photon nonlinearities may lead to novel opportunities for optical logic circuits that encode and process classical information \cite{Miller2010NPHOT}. The emerging research disciplines of cavity optomechanics \cite{Aspelmeyer2014RMP} and metamaterials \cite{Soukoulis2011NPHOT} also heavily rely on the fabrication of novel nanostructures. From the prospect of fundamental physics, the control over photon emission and propagation will open new possibilities of studying photonic realizations of exotic phenomena orgiginally developed in condensed-matter physics, such as photonic topological insulators \cite{Lu2014Arxiv}. Photons are often considered elusive since they interact weakly and are consequently difficult to generate, trap, and route. The progress on photonic nanostructures can potentially change this conception leading to whole new avenues for fundamental research and applied photonics.

\begin{acknowledgments}
We are indebted to Alisa Javadi for carrying out the local-density of states calculations of Fig. \ref{2DPCall} and \ref{LDOS-WG-maps} as well as to Kristian H{\o}eg Madsen, Anders S. S{\o}rensen, Immo S\"{o}llner, and Petru Tighineanu for help during preparation of the manuscript and for illuminating discussions. The manuscript was completed during an extended stay in the group of Andrew G. White at University of Queensland and P.L. gratefully acknowledges the hospitality and inspiring scientific discussions. We gratefully acknowledge the financial support over the years from the following sources: The Danish Council for Independent Research (FNU and FTP), The European Research Council (through ERC consolidator grant "ALLQUANTUM"), The Villum Foundation, The Carlsberg Foundation, The Lundbeck Foundation, The Augustinus Foundation, and The A.P. M{\o}ller and Chastine Mc-Kinney M{\o}ller Foundation for General Purposes.
\end{acknowledgments}

\section*{List of symbols}

\settowidth{\tenchars}{0123456789}
\setlength{\restofcolumn}{\columnwidth-\tenchars}

\noindent
\begin{longtable}{@{}p{\tenchars} @{}p{\restofcolumn}}
$\mathbf{A}$ & vector potential in the generalized Coulomb gauge \\
$\mathbf{E}$ & electric field \\
$E_{\rm c/v}$ & conduction-/valence-band-edge energy \\
$E_\text{P}$ & Kane energy\\
$F_{\rm P}$ & total Purcell factor \\
$F^{\rm WG}_{\rm P}$ & waveguide-mode Purcell factor \\
$F^{\rm max}_{\rm P}$ & maximum Purcell factor \\
$F^{\rm res}_{\rm P}$ & Purcell factor on resonance \\
$F^{\rm ng}_{\rm P}$ & Purcell factor of non-guided modes\\
$F_{\rm c/v}$ & envelope function for the conduction/valence band \\
$\overleftrightarrow{G}$ & Green's tensor \\
$I$ & Indistinguishability\\
$K$ & memory kernel \\
$Q$ & Quality factor \\
$S_{\rm cav}$ & cavity emission spectrum \\
$S$ & emission spectrum \\
$S_{\rm em}$ & emitter emission spectrum \\
$T_1$ & total lifetime \\
$T_2$ & total coherence time\\
$T_2^*$ & pure-dephasing time\\
$V$ & confinement potential \\
$V_{\rm eff}$ & effective mode volume \\
$\mathbf P$ & momentum matrix element \\
$X^-$ & negatively charged trion \\
$X^+$ & positively charged trion \\
$X_\text{b}$ & $x$-polarized bright exciton \\
$X_\text{d}$ & dark exciton \\
$XX$ & biexciton \\
$Y_\text{b}$ & $y$-polarized bright exciton \\
$Y_\text{d}$ & dark exciton \\
$a$ & lattice period \\
$c_{\rm e}$ & excited-state amplitude \\
$c_{{\rm g}, \mathbf k}$ & amplitude of ground state with photon in mode with wave vector $\mathbf k$ \\
$\mathbf d$ & dipole moment\\
$\hat{\mathbf e}_\textrm{c}$ & cavity-mode polarization unit vector \\
$\hat{\mathbf e}_\textrm{d}$ & dipole-moment unit vector \\
$\hat{\mathbf e}_k$ & waveguide-mode polarization unit vector \\
$f$ & oscillator strength \\
$g$ & coupling rate \\
$g^{(2)}$ & Second-order correlation function\\
$g^{+/-}$ & hole in valence band/electron in conduction band\\
$\mathbf k$ & Bloch wave vector or plane-wave vector depending on context \\
$n_\text{g}$ & group index \\
$m^*$ & effective mass of electron \\
$m_0$ & rest mass of electron \\
$u_{\rm c/v}$ & electronic Bloch function at the $\Gamma$-point for the conduction/valence band \\
$\mathbf u_\mathbf k$ & normalized electromagnetic Bloch function at Bloch wave vector $\mathbf k$ \\
$t$ & membrane thickness or time, depending on context \\
$\Delta_L$ & Lamb shift \\
$\Psi_{\rm c/v}$ & electron/hole wave function \\
$\Omega_{\rm p}$ & driving field rate \\
$\alpha_{\rm c/v}$ & spin of conduction/valence band \\
$\beta$ & $\beta$-factor \\
$\gamma$ & total decay rate under the approximation $\gamma_{\rm tot} \sim \gamma_{\rm rad}$ \\
$\gamma_{\rm db/bd}$ & dark-to-bright/bright-to-dark spin-flip rate\\
$\gamma_{\rm dp}$ & dephasing rate \\
$\gamma_{\rm f/s}$ & fast/slow decay rates \\
$\gamma_{\rm ng}$ & rate of coupling to non-guided modes \\
$\gamma_{\rm nrad,b/d}$ & non-radiative decay rate of bright/dark exciton \\
$\gamma_{\rm rad}$ & total radiative decay rate \\
$\gamma^{\rm hom}_{\rm rad}$ & radiative decay rate in homogeneous medium\\
$\gamma_{\rm rad,b}$ & radiative decay rate of bright exciton\\
$\gamma_{\rm tot}$ & total decay rate\\
$\eta$ & quantum efficiency \\
$\kappa$ & cavity loss rate \\
$\lambda$ & electromagnetic wavelength in vacuum \\
$\mu$ & effective Rabi frequency \\
$\rho$ & projected local density of optical states \\
$\rho_{\rm NL}$ & non-local interaction function \\
$\rho_{\rm hom}$ & projected local density of optical states of a homogeneous medium\\
$\rho_{\rm b/d}$ & bright/dark exciton population \\
$\omega_0$ & exciton recombination frequency \\
$\left|\uparrow\right>$ or $\left|\downarrow\right>$ & electron pseudo-spin state state in $z$-basis \\
$\left|\Uparrow\right>$ or $\left|\Downarrow\right>$ & hole pseudo-spin state state in $z$-basis \\
\end{longtable}

\bibliography{RMP_Chap12_References}

\end{document}